\documentclass[12pt]{article}
\usepackage{amsmath}
\usepackage{graphicx}
\usepackage{enumerate}
\usepackage{natbib}
\usepackage{url} 
\usepackage{booktabs} 
\usepackage{outlines}
\usepackage{subcaption} 
\usepackage{bm}
\usepackage{bbm}
\usepackage{amssymb,amsfonts}
\usepackage{textcomp}
\usepackage{xcolor}
\usepackage[nocomma]{optidef}
\usepackage{tabularx}
\usepackage{algorithm, amsmath}
\usepackage{algpseudocode}
\usepackage{comment}
\usepackage{tikz}
\usepackage[justification=centering]{caption}
\usepackage{multicol}
\usepackage{textpos}
\usetikzlibrary{positioning}
\usepackage{amsthm}    
\newtheorem{newthm*}{\normalfont\scshape Theorem}
\newtheorem{lemma}{Lemma}
\newtheorem{coroll*}{Corollary}
\newtheorem{cond*}{Condition}
\newtheorem{definition*}{Definition}
\newtheorem{remark*}{Remark}
\newtheorem{result*}{Result}
\usepackage{xr-hyper}
\usepackage{hyperref}
\usepackage{dsfont} 
\hypersetup{
    colorlinks=true,
    linkcolor=blue,
    filecolor=blue,
    urlcolor=blue,
    citecolor=blue,
    }
\makeatletter

\newcommand{\blind}{1}
\newcommand{\bs}[1]{\boldsymbol{#1}}

\begin{document}

\def\spacingset#1{\renewcommand{\baselinestretch}%
{#1}\small\normalsize} \spacingset{1}


\if1\blind
{
 \title{\bf Dependence-Aware False Discovery Rate Control in Two-Sided Gaussian Mean Testing}
  \author{\\ Deepra Ghosh \\
    Department of Statistics, Operations and Data Science, Temple University\\
    and \\ Sanat K. Sarkar
    \thanks{
    The reserach is supported by \textit{NSF grant DMS 2210687}}\hspace{.2cm}\\
    Department of Statistics, Operations and Data Science, Temple University}
    \date{July 2024}
  \maketitle
} \fi

\if0\blind
  \medskip
 \fi

\begin{abstract} This paper develops a general framework for controlling the false discovery rate (FDR) in multiple testing of Gaussian means against two-sided alternatives. The widely used Benjamini–Hochberg (BH) procedure provides exact FDR control under independence or conservative control under specific one-sided dependence structures, but its validity for correlated two-sided tests has remained an open question. We introduce the notion of positive left-tail dependence under the null (PLTDN), extending classical dependence assumptions to two-sided settings, and show that it ensures valid FDR control for BH-type procedures. Building on this framework, we propose a family of generalized shifted BH (GSBH) methods that incorporate correlation information through simple $p$-value adjustments. Simulation results demonstrate reliable FDR control and improved power across a range of dependence structures, while an application to an HIV gene expression dataset illustrates the practical effectiveness of the proposed approach.
\end{abstract}

\noindent%
\vfill

\newpage
\spacingset{1.9} 
\section{Introduction}
\label{sec:intro}
Multiplicity continues to pose a fundamental challenge across modern scientific research, ranging from small-scale confirmatory experiments to large-scale high-dimensional studies. Even in moderate-sized investigations, simultaneous testing of several related hypotheses often arises, for instance, in clinical trials, neuroscience, or controlled laboratory experiments. At larger scales, the proliferation of data in genomics, neuroimaging, and other high-throughput domains has made the simultaneous testing of thousands of hypotheses routine. In both regimes, ensuring rigorous control of false discoveries remains a central statistical concern.

The false discovery rate (FDR), introduced by \cite{bh1995}, has become the dominant framework for Type I error control in multiple testing. Their linear step-up procedure, now universally known as the BH method, remains one of the most influential and widely applied FDR-controlling methods in scientific practice.

Despite its widespread use, the theoretical validity of the BH method crucially depends on the dependence structure among test statistics or, equivalently, the corresponding $p$-values. For Gaussian test statistics, the BH procedure is known to control the FDR under independence or specific positive dependence structures, notably the positive regression dependence on a subset (PRDS) condition (\cite{by2001, sks2002, sks2008, fdr2007, br2008}). The Gaussian model, in particular, provides a natural and widely used framework for studying multiple testing problems, both because many classical test statistics (e.g., $z$- and $t$-statistics) are asymptotically Gaussian, and because it offers analytical tractability for modeling dependence among tests.

However, this theoretical assurance holds primarily for one-sided tests, while in practice, such as in gene expression (\cite{Efron2001, StoreyTibshirani2003, Smyth2004}), neuroimaging (\cite{Friston1994, NicholsHayasaka2003, EklundNicholsKnutsson2016}), or variable selection (\cite{by2005b, bc2015, CandesFanJansonLv2018}), 
the hypotheses of interest are almost always two-sided. When the test statistics are correlated, typically following a Gaussian distribution with a known or unknown positive-definite covariance matrix, no general proof currently guarantees that the BH method controls the FDR in the two-sided setting.

Nevertheless, empirical findings have long suggested that the BH procedure performs well even for correlated two-sided Gaussian tests. As \cite{benjamini2010fdr} noted, “convincing simutheoretical evidence indicates that the same holds for two-sided $z$-tests with any correlation structure $\cdots$ but the theory awaits a complete proof.” Simulation studies (\cite{ReinerBenaim2007, Farcomeni2006, KimVandeWiel2008} support this conjecture, showing reasonable FDR control across a range of dependence structures. Yet, in the absence of rigorous theoretical justification, the continued use of BH in correlated two-sided contexts rests on heuristic and empirical confidence rather than formal proof.

Motivated by the challenges of controlling the false discovery rate (FDR) under dependence, recent research has advanced multiple-testing methodology along several complementary directions. Dependence-adjusted BH-type procedures explicitly incorporate covariance structure, including the conditional-calibration dBH, the Bonferroni–BH hybrid (BBH), and the Shifted BH (SBH) procedures (\cite{fl2022, st2022, sz2025}), with BBH’s selective-inference perspective conceptually analogous to fissioning a correlated Gaussian vector into conditionally independent components (\cite{sarkar2025comments, leiner2025data}).

A unified class of symmetry-based methods, encompassing both the knockoff and Gaussian mirror frameworks (\cite{bc2015, CandesFanJansonLv2018, xing2023gaussian}), constructs synthetic or mirrored variables that reproduce the dependence structure of the original data. By exploiting distributional symmetry or exchangeability under the null, these approaches enable rigorous, often finite-sample, FDR control in Gaussian and related models.

Complementing these, resampling-based methods—including permutation and bootstrap procedures (\cite{yekutieli1999resampling, romano2008bootstrap})—empirically estimate null distributions of correlated test statistics, offering flexible alternatives when explicit modeling of dependence is challenging. Finally, structure-adaptive approaches, often developed beyond Gaussian settings, leverage auxiliary information such as grouped or hierarchical organization (\cite{yekutieli2008hierarchical, benjamini2014selective, liu2016grouped, hu2010fdrgroups, lei2020interactive, GuoSarkar2020, nandi2021classified, guo2018graphicalFDR, miecznikowski2023treeFDR, li2022bottomup, loper2022smoothed, li2019sabha}).

Together, these developments provide a rich and increasingly unified toolkit for controlling false discoveries under complex dependence and structural constraints.

A more recent and conceptually distinct direction moves beyond the $p$-value paradigm entirely, using $e$-values, nonnegative random variables whose expectation under the null does not exceed one. The resulting $e$-BH procedure (\cite{wang2022}) enables FDR control under arbitrary dependence. Although $e$-values offer appealing robustness and generality, their practical use in multivariate Gaussian mean testing remains unclear, motivating simpler, theoretically grounded, $p$-value–based alternatives.

The present paper contributes to this ongoing discourse by developing a rigorous theoretical framework for FDR control in Gaussian mean testing against two-sided alternatives. Building on insights from dependence-adjusted and shifted-$p$-value methods, we introduce a class of structured dependencies under two-sided testing. Specifically, we define and study positive left-tail dependence under the null (PLTDN), a property capturing conditional left-tail monotonicity of null test statistics. This framework extends beyond the classical PRDS assumption, which is limited to one-sided alternatives, and enables exact FDR control for BH-type procedures in two-sided Gaussian mean testing.

Our framework naturally leads to a family of generalized shifted BH (GSBH) procedures, unifying and extending recent dependence-aware constructions. Empirical studies demonstrate that these procedures reliably control the FDR and achieve substantial power gains across diverse dependence structures. Median- and harmonic-mean–based GSBH variants perform particularly well, and their Bonferroni–BH extensions (SBBH) retain these advantages in regression and knockoff-assisted settings. In real-data analysis of HIV drug resistance, the proposed methods identify scientifically meaningful signals while maintaining nominal error control, underscoring their robustness and practical utility for high-dimensional inference under dependence.

The following section develops the mathematical foundation of this framework, establishing exact expressions for the FDR of general step-up procedures and a formal characterization of PLTDN that guarantees FDR control. These results lay the groundwork for constructing BH-type methods that are analytically tractable, computationally efficient, and theoretically valid for correlated two-sided Gaussian tests, with applications spanning confirmatory to large-scale exploratory studies.

\section{Methodological Foundation}
\label{sec:methods}
This section presents and establishes foundational results that underpin the development of novel, theoretically valid FDR controlling BH-type methods in testing Gaussian means against two-sided alternatives with known correlation structures. A key feature of these developments is that they rely on a form of positive dependence fundamentally different from the widely used PRDS condition, which is typically invoked only for one-sided alternatives. This distinction is crucial, as it enables the extension of FDR control to settings where the standard PRDS-based framework is not applicable.

Consider $d$ null hypotheses $H_1, \ldots, H_d$ with corresponding statistics (not necessarily $p$-values) ${P}_1, \ldots, {P}_d$. A general step-up procedure with critical constants ${\alpha}_1 \le \ldots \le {\alpha}_d$ operates as follows: Order the $\tilde{P}_i$’s as ${P}_{(1)} \le \cdots \le {P}_{(d)}$, set $R = \max_{1 \le i \le d} \{i: {P}_{(i)} \le {\alpha}_i \}$, and reject $H_i$ for all $i$ such that ${P}_i \le {P}_{(R)}$, provided $R$ exists; otherwise, reject none.

For this step-up test, the FDR, defined in \cite{bh1995} as $E(V/\max(R,1))$, where $V$ is the number of false rejections, can be expressed as follows.
\vskip 6pt
\begin {lemma}\label{Lemma1} Let $I_0 = \{j: H_j\; \mbox{is true} \}$. Then,
\begin {eqnarray*} \textrm{FDR} & = & \sum_{i \in I_0} \sum_{r=0}^{d-1} (r+1)^{-1} E \left \{ \textrm{Pr} \left ({P}_i \le {\alpha}_{r+1} \mid {\boldsymbol{P}}_{-i} \right ) \mathbbm{1} \left (R({\boldsymbol{P}}_{-i}) = r \right )  \right \} \\
& = & \sum_{i \in I_0} \sum_{r=0}^{d-1} E \left [ \left \{(r+1)^{-1} \textrm{Pr}\left ({P}_i \le \tilde{\alpha}_{r+1} \mid R({\boldsymbol{P}}_{-i}) \ge r \right ) - \right. \right.  \nonumber \\
& & \qquad \left. \left. r^{-1} \textrm{Pr}\left ({P}_i \le {\alpha}_{r} \mid R({\boldsymbol{P}}_{-i}) \ge r \right ) \right \} \textrm {Pr} \left \{ R({\boldsymbol{P}}_{-i}) \ge r \right \} \right ] , \end{eqnarray*}
where $ {0}/{0} = 0$, ${\boldsymbol{P}}_{-i} = ({P}_1, \ldots, {P}_d)\setminus\{{P}_i\}$, and $ R_{-i}\equiv R({\boldsymbol{P}}_{-i}) = \max_{1\le j \le d-1}\{j:{P}_{(j)\setminus\{i\}} \le \alpha_{j+1} \},$ with ${P}_{(1)\setminus\{i\}} \le \cdots \le {P}_{(d-1)\setminus\{i\}}$ being the ordered components of ${\boldsymbol{P}}_{-i}$. \end{lemma}
See \cite{sks2002, sks2008} for these formulas; similar results appear in other works \citep{by2001, br2008, fdr2007}.

Controlling the FDR at any level $\alpha$ requires algebraic manipulation of any of the expressions in Lemma \ref{Lemma1}. At the heart of it lies the dependency of $P_i$ on $\bs{P}_{-i}$, rather than of $\bs{P}_{-i}$ on $P_i$ as in PRDS. We formally define such dependency in the following: 

\begin{cond*} \label{Cond1} For each $i \in I_0$ and a fixed $c \in (0,1)$,

\noindent a. $\textrm{Pr} ( {P}_i \le u \mid {\boldsymbol{P}}_{-i})/u$ is decreasing in $u \in (0,c)$, or

\noindent b. $\textrm{Pr} ({P}_i \le u \mid g({\boldsymbol{P}}_{-i}) \le t)/u$ is decreasing in $u \in (0,c)$, for any increasing function $g$ of ${\boldsymbol{P}}_{-i}$ and fixed constant $t > 0$.

\end{cond*}

The above condition is a weaker form of the concavity of $\textrm{Pr} ( {P}_i \le u \mid {\boldsymbol{P}}_{-i})$ in $u \in (0,c)$, for each $i \in I_0$. It
characterizes a positive dependency between ${P}_i$ and ${\boldsymbol{P}}_{-i}$, with Condition \ref{Cond1}b representing a weaker form of Condition \ref{Cond1}a. Specifically, it implies that for each $i \in I_0$, ${P}_i$  stochastically dominates $U(0,1)$ toward smaller values in $(0,c)$ when conditioned on ${\boldsymbol{P}}_{-i}$ or on a set of smaller values of ${\boldsymbol{P}}_{-i}$. We refer to this property as positive left-tail dependence of ${P}_i$ under the null (PLTDN) on ${\boldsymbol{P}}_{-i}$ over $(0,c)$, and simply as PLTDN on ${\boldsymbol{P}}_{-i}$ if $c=1$.

These observations lead directly to the following result: {\it For a stepup test based on ${P}_i$'s and critical constants ${\alpha_i} = i \tilde{\alpha}/d$, $i=1,\ldots,d$, for some fixed $\tilde{\alpha} \in (0,1)$, FDR $\le \sum_{i \in I_0} \textrm{Pr} ({P}_i \le \tilde{\alpha}/d)$, if ${P}_i$ is PLTDN on ${\boldsymbol{P}}_{-i}$ over $(0,c)$, for any fixed $c \in (\tilde{\alpha}, 1]$.}

What if the PLTDN condition does not hold for some ${P}_i$? In this case, we note the following:
\begin {remark*} \label{Remark1} Let $\hat{P}_i$ be stochastically smaller, increasing function of ${P}_i$, conditionally given ${\boldsymbol{P}}_{-i}$, for some $i$. Then, defining $\mathcal{S} = \{i: \hat{P}_i \overset{\text{st}}{\preceq} {P}_i\}$, we have the following result: $\textrm{FDR} \le \textrm{FDR}_{\mathcal{S}} + \textrm{FDR}_{\mathcal{S}^c}$,
where   \begin {eqnarray*} \textrm{FDR}_{\mathcal{S}} & = & \sum_{i \in I_0 \bigcap{S}}\sum_{r=0}^{d-1} (r+1)^{-1} E \left \{ \textrm{Pr} \left (\hat{P}_i \le {\alpha}_{r+1} \mid {\boldsymbol{P}}_{-i} \right ) \mathbbm{1} \left (R({\boldsymbol{P}}_{-i}) = r \right )  \right \} \\
& = & \sum_{i \in I_0 \bigcap{S}} \sum_{r=0}^{d-1} E \left [ \left \{(r+1)^{-1} \textrm{Pr}\left (\hat{P}_i \le {\alpha}_{r+1} \mid R({\boldsymbol{P}}_{-i}) \ge r \right ) - \right. \right.  \nonumber \\
& & \qquad \left. \left. r^{-1} \textrm{Pr}\left (\hat{P}_i \le {\alpha}_{r} \mid R({\boldsymbol{P}}_{-i}) \ge r \right ) \right \} \textrm {Pr} \left \{ R({\boldsymbol{P}}_{-i}) \ge r \right \} \right ], \end{eqnarray*} and $\textrm{FDR}_{\mathcal{S}^c}$ refers to the original expressions for the FDR in the lemma with $\sum_{i \in I_0}$ being replaced by $\sum_{i \in I_0 \bigcap{S}^c}$. \end{remark*}

Based on this remark, we have the following theorem, stated in a more general form than the result stated before Remark \ref{Remark1} and providing a key theoretical foundation for the novel FDR-controlling procedures developed in the next section

\begin{newthm*} \label{Theorem1} Consider a stepup test based on ${P}_i$'s and critical constants ${\alpha_i} = i \tilde{\alpha}/d$, $i=1,\ldots,d$, for some fixed $\tilde{\alpha} \in (0,1)$. Let, for each $i \in I_0$, either ${P}_i$ is LTPDN on ${\boldsymbol{P}}_{-i}$ over $(0,c)$, or if not, there is a stochastically smaller and increasing function of ${P}_i$, conditionally given ${\boldsymbol{P}}_{-i}$, say $\hat{P}_i$, which is LTPDN on ${\boldsymbol{P}}_{-i}$ over $(0,c)$, for some $c \in (\tilde{\alpha}, 1]$.  Then, the FDR of this step-up test is bounded above by $\sum_{i \in I_0 \bigcap \mathcal{S}^{c}} \textrm{Pr} ({P}_i \le \tilde{\alpha}/d) + \sum_{i \in I_0 \bigcap \mathcal{S}} \textrm{Pr} (\hat{P}_i \le \tilde{\alpha}/d)$. \end{newthm*}

Proof. Let us first suppose that Condition \ref{Cond1}b holds in the definition of PLTDN for $\hat{P}_i$. Since $R({\boldsymbol{P}}_{-i})$ is decreasing in ${\boldsymbol{P}}_{-i}$ for any fixed $r =1, \ldots d-1$, we have the following  result under this condition: $\textrm{Pr} \left (\hat {P}_i \le u \mid R({\boldsymbol{P}}_{-i}) \ge r \right )/u$ is decreasing in $u \in (0,1)$, implying that $\textrm{Pr} \left (\hat{P}_i \le (r+1) \tilde{\alpha}/d \mid R ({\boldsymbol{P}}_{-i}) \ge r \right )/(r+1) \le \textrm{Pr} \left (\hat{P}_i \le r \tilde{\alpha}/d \mid R({\boldsymbol{P}}_{-i}) \ge r \right )/r,$ for each $i \in I_0 \bigcap \mathcal{S}$ and $r =1, \dots, d-1$. Using this in the second expression for FDR$_{\mathcal{S}}$ (in Remark \ref{Remark1}), we have the second term in the desired upper bound on FDR in the theorem. The first  term in this upper bound can be derived similarly using the second expression of the FDR formula in Lemma \ref{Lemma1}. The bound follows under Condition \ref{Cond1}a as well since it is stronger than Condition \ref{Cond1}b.  Thus, the theorem is proved.

\section{Testing Gaussian Means Against Two-Sided Alternatives}\label{sec:framework}

Given a $d$-dimensional random vector $\boldsymbol{X} = (X_1, \ldots, X_d)^{\prime} \sim N_d(\boldsymbol{\mu}, \eta^2\boldsymbol{\Sigma})$, with unknown mean vector $\boldsymbol{\mu} = (\mu_1, \ldots, \mu_d)^{\prime}$ and known positive definite covariance matrix $\boldsymbol{\Sigma} = ((\sigma_{ij}))$, we consider the problem of testing the coordinate-wise hypotheses $H_i: \mu_i=0$ versus $\bar{H}_i: \mu_i \neq 0$, $i=1, \ldots, d$, subject to a control of the FDR under the following two different scenarios:

\begin{enumerate}
\item[]\textbf{\hypertarget{Setting1}{Setting 1} (Known $\eta^2$).} In this case, each $X_i$ can be standardized as  $\tilde{X}_i=X_i/\eta\sqrt{\sigma_{ii}}$, allowing us, without loss of generality, to treat $X_i$ itself as the marginal test statistic for $H_i$, with $\boldsymbol{\Sigma}$ regarded as the correlation matrix. The standard two-sided $z$-tests then apply, yielding $p$-values $P_i = \bar{\Psi}_{1}(X_i^2)$, $i=1, \ldots, d$, where $\Psi_1 = 1 - \bar{\Psi}_1$ is the cumulative distribution function of the $\chi_1^2$ distribution.\

\item[]\textbf{\hypertarget{Setting2}{Setting 2} (Unknown $\eta^2$).} Suppose an independent estimator $V/\nu$ is available for $\eta^2$ with $V \sim \eta^2 \chi^2_\nu$, where $\nu > 0$ is known. Then, without loss of generality, we can take $\eta^2 =1$  and use $T_i = X_i / \sqrt{V}$ as the marginal test statistic for $H_i$, again with $\boldsymbol{\Sigma}$ treated as the correlation matrix. The standard two-sided $t$-tests are then  applied, with $p$-values $P_i = \bar{\Psi}_{1, \nu}(T_i^2)$, $i=1, \ldots, d$, where $\Psi_{1,\nu} = 1 - \bar{\Psi}_{1, \nu}$ denotes the cumulative distribution function of the $F_{1, \nu}/\nu$ distribution.\
\end{enumerate}

The problems described under Settings \hyperlink{Setting1}{1} and \hyperlink{Setting2}{2} were most recently investigated in \cite{sz2025}, where the authors proposed BH-type procedures with provably valid FDR control. Their approach shifts each $p$-value to the left and then applies a step-up test to the shifted values, using critical constants ${\alpha}_i = i\tilde{\alpha}/d$, $i = 1, \ldots, d$. The magnitude of the shift and the choice of $\tilde{{\alpha}}$ are derived from the correlation matrix to guarantee FDR control at the target level $\alpha$. Specifically, they introduced two shifted BH procedures, Shifted BH1 (SBH1) and  Shifted BH2 (SBH2). In both, each $P_i$ is shifted to
\begin {eqnarray*} P_i^{-} = \left \{
\begin{array}{cc}
\bar{\Psi}_1 \left( \frac{1}{\delta_i} \bar{\Psi}_1^{-1}\left( P_i \right) \right ), & \mbox{Setting 1,} \\
\bar{\Psi}_{1, \nu} \left( \frac{1}{\delta_i} \bar{\Psi}_{1, \nu}^{-1}\left( P_i \right)\right ), & \mbox{Setting 2,} \\
\end{array}
\right. \end{eqnarray*}
for some $\delta_i \in (0.1)$, before deriving an upper bound on the FDR of the resulting stepup procedure and then determining $\tilde{\alpha}$ subject to controlling the FDR at $\alpha$. In SBH1, $\delta_i$ is set to $\tau_i := 1 - R_i^2$, where $R_i^2 = \boldsymbol{\sigma}_{-i,i}^{\prime} \boldsymbol{\Sigma}_{-i,-i}\boldsymbol{\sigma}_{-i,i}$, with  $\boldsymbol{\Sigma}_{-i,-i} = \textrm{Cov}(\boldsymbol{X}_{-i})$ and $\boldsymbol{\sigma}_{-i,i} = \textrm{Cov}(\boldsymbol{X}_{-i}, X_i)$, the squared multiple correlation between $X_i$ and $\boldsymbol{X}_{-i}= (X_1, \ldots, X_d)\setminus\{X_i\}$. In SBH2, $\delta_i = \tau$ is used for all $i$, for some fixed $\tau \in (0, \lambda_{\min}(\boldsymbol{\Sigma}))$, where $\lambda_{\min}(\boldsymbol{\Sigma})$ is the smallest eigenvalue of the correlation matrix $\boldsymbol{\Sigma}$. These choices of $\delta_i$ exploit the distributional properties of squared multivariate normal and squared multivariate $t$ statistics, with $\delta_i$ serving as the shifting constant for each $P_i$.

These Shifted BH procedures represent a notable methodological advance in controlling the FDR for Gaussian mean testing against two-sided alternatives. However, they were not developed within a broader theoretical framework that accounts for general forms of dependence among the $p$-values or their associated test statistics. Building on the unified framework introduced in the previous section, we move beyond the ad hoc nature of \cite{sz2025} and demonstrate that their approach arises as a special case of a more general theory. Specifically, we extend their work by considering the following general form of shifted $p$-values:
\begin {eqnarray}
\label{GSBH pvalues}
\tilde{P}_i = \left \{
\begin{array}{cc}
\bar{\Psi}_1 \left( \frac{1}{\tau} \bar{\Psi}_1^{-1}\left( P_i \right) \right ), & \mbox{Setting 1,} \\
\bar{\Psi}_{1, \nu} \left( \frac{1}{\tau} \bar{\Psi}_{1, \nu}^{-1}\left( P_i \right)\right ), & \mbox{Setting 2,} \\
\end{array}
\right. \end{eqnarray}
and then identifying values of $\tau \in (0,1)$ that ensure the PLTDN property for the resulting test statistics, or, as noted in Remark \ref{Remark1}, for suitable increasing stochastic minorants of these statistics. This construction yields a broad class of BH-type procedures with rigorously established FDR control.

To determine values of $\tau$ that guarantee the PLTDN property, we first observe the following result: For each $i \in I_0$
$$X_i^2~\mid \boldsymbol{X}_{-i} \sim \tau_i{\chi^{\prime}}_{1}^{2}(\lambda_i(\boldsymbol{X}_{-i})), \mbox{where} \; \lambda_i \left (\boldsymbol{X}_{-i} \right )  =  [\boldsymbol{\sigma}_{-i,i}^{\prime}\boldsymbol{\Sigma}_{-i,i}^{-1}(\boldsymbol{X}_{-i}- \boldsymbol{\mu}_{-i})]^2/\tau_i,$$ and ${\chi^{\prime}}_{1}^{2}(\lambda)$ denote the non-central chi-squared random variable with $1$ degree of freedom and the non-centrality parameter $\lambda$.

This leads to the following lemma, proved in \hyperref[sec:appendixa]{Appendix A}.

\begin{lemma} \label{Lemma2} Let
\begin {eqnarray*}
\ddot{P}_i := \left \{
\begin{array}{cc}
\bar{\Psi}_1 \left( \frac{1}{\tau_i} \bar{\Psi}_1^{-1}\left( P_i \right) \right ), & \mbox{Setting 1,} \\
\bar{\Psi}_{1, \nu} \left( \frac{1}{\tau_i} \bar{\Psi}_{1, \nu}^{-1}\left( P_i \right)\right ), & \mbox{Setting 2,} \\
\end{array}
\right. \end{eqnarray*} The $\ddot{P}_i$'s are LTPDN, with Condition \ref{Cond1}a satisfied in Setting 1 and Condition \ref{Cond1}b satisfied in Setting 2.
\end{lemma}

The desired LTPDN property for the $\tilde{P}_i$’s, or for suitably chosen minorants thereof (as stated in Theorem \ref{Theorem1}), can now be established using the above lemma. Let us use, for notational convenience, $\Psi$ to denote $\Psi_1$ in Setting 1 and $\Psi_{1,\nu}$ in Setting 2. We note that $$\tilde{P}_i = \varphi_i(\ddot{P}), \; \mbox{where} \; \varphi_i(u) = \Psi \left(\frac{\tau_i}{\tau}\Psi^{-1}(u) \right ):(0,1) \to(0,1).$$ The function $\varphi_i$ is increasing, satisfies $\varphi_i(0)=0$ and $\varphi_i(1)=1$, and, as shown in the Appendix, is convex when $\tau_i\ge \tau$ and concave when $\tau_i < \tau$.

\noindent {\bf Case 1: $\tau_i \ge \tau$}.

\noindent For $i \in I_0$, \begin{eqnarray*} & & \frac{1}{u}\textrm{Pr}(\tilde{P}_i \le u\mid\boldsymbol{P}_{-i}) = \frac{1}{u}\textrm{Pr}(\ddot{P}_i \le \varphi_i^{-1}(u)~\mid~\boldsymbol{P}_{-i}) = \frac{\textrm{Pr}(\ddot{P}_i \le \varphi_i^{-1}(u)~\mid~\boldsymbol{P}_{-i})} {\varphi_i^{-1}(u)}\frac{\varphi_i^{-1}(u)}{u}. \end{eqnarray*} Suppose Condition~\ref{Cond1}a ensures the PLTDN property of $\ddot{P}_i$ on $\boldsymbol{P}_{-i}$. Since $\varphi_i^{-1}(u)$ is increasing and, being concave, satisfies that $\varphi_i^{-1}(u)/u$ is decreasing on $(0,1)$, the above expression is decreasing in $u \in (0,1)$. The same conclusion holds under Condition~\ref{Cond1}b. Hence, $\tilde{P}_i$ has the PLTDN property on $\boldsymbol{P}_{-i}$ for all $i \in I_0$ when $\tau_i \ge \tau$.

\noindent {\bf Case 2: $\tau_i < \tau$}.

\noindent Here, since $\varphi_i^{-1}(u)$ is convex, $\varphi_i^{-1}(u) < u \varphi_i^{-1}(\tilde{\alpha})/\tilde{\alpha}$, for $ u < \tilde{\alpha}$. Therefore, for each $i \in I_0$, \begin{eqnarray*} \frac{1}{u}\textrm{Pr}(\tilde{P}_i \le u~\mid~\boldsymbol{P}_{-i}) \le \frac{1}{u}\textrm{Pr}(\ddot{P}_i \le \frac{u}{\tilde{\alpha}}\varphi^{-1}_i(\tilde{\alpha})~\mid~\boldsymbol{P}_{-i}), \end{eqnarray*} which is decreasing in $u \in (0,\tilde{\alpha})$ due to the PLTDN property of $\ddot{P}_i$ on $\boldsymbol{P}_{-i}$. Thus, we have $\hat{P}_i = \tilde{\alpha}\ddot{P}_i/\varphi_i^{-1}(\tilde{\alpha})$, a minorant of $\tilde{P}_i$, which possesses the PLTDN property.

Now, since  $$\tilde{P}_i = \bar{\Psi}\left (\frac{\tau_i}{\tau}  \bar{\Psi}^{-1}(\ddot{P}_i)\right ), \; \;  \hat{P}_i = \frac{\tilde{\alpha}} {\bar{\Psi}\left ( \frac{\tau}{\tau_i} \bar{\Psi}^{-1}(\tilde{\alpha})\right)} \ddot{P}_i ,$$ and $\textrm{Pr}(\ddot{P}_i \le u) = \bar{\Psi}(\tau_i \bar{\Psi}^{-1}(u)),$ for each $i \in I_0$, we have the following results:
For each $i \in I_0$,
\begin{eqnarray}
\label{GSBH probabilities}
\textrm{Pr}(\tilde{P}_i \le u) & = & \bar{\Psi}(\tau \bar{\Psi}^{-1}(u)), \; \tau_i \ge \tau \nonumber\\
\textrm{Pr}(\hat{P}_i \le u) & = & \bar{\Psi}\left (\tau_i \bar{\Psi}^{-1} \left ( \frac{u}{\tilde{\alpha}}\bar{\Psi}\left(\frac{\tau}{\tau_i}\bar{\Psi}^{-1}(\tilde{\alpha})\right ) \right) \right ), \; \tau_i < \tau.
\end{eqnarray}

These results lead us to the following definition of a broad class of shifted BH methods with theoretically valid FDR control.

\begin{definition*}
\label{Definition1}
\noindent {\bf Generalized Shifted BH (GSBH)}: A step-up test based on the $\tilde{P}_i$'s in \ref{GSBH pvalues} and critical constants ${\alpha_i} = i H^{-1}(\alpha/d)$, $i=1,\ldots, d$, where
\begin{eqnarray*}
H(u) & = & \frac{1}{d} \left \{ \sum_{i: \tau_i \ge \tau} H_{\tau}(u) +  \sum_{i: \tau_i < \tau} H_{\tau_i} \left ( \frac{1}{d}H_{\frac{\tau}{\tau_i}}(du)\right ) \right \}: (0,1) \to (0,1), \end{eqnarray*}
and $H_{\tau} = \bar{\Psi}(\tau \bar{\Psi}^{-1}(u)):(0,1)\to(0,1) $, with $\Psi$ referring to $\Psi_ 1$ in Setting 1 and to $\Psi_{i, \nu}$ in Setting 2.   
\end{definition*}

\begin{newthm*} \label{Theorem2}
The GSBH controls the FDR at $\alpha$. 
\end{newthm*}

The theorem follows upon using the results in \ref{GSBH probabilities} with $\tilde{\alpha}=dH^{-1}(\alpha/d)$ in Theorem \ref{Theorem1} and noting that FDR $\le dH(\tilde{\alpha}/d)$.

\begin{remark*} \label{Remark2} \rm The framework underpinning the development of GSBH allows us to vary $\tau \in (0,1)$ and thereby formulate a broad class of Shifted BH methods with guaranteed FDR control. Within this framework, SBH1 from \cite{sz2025} can be seen as a modified GSBH, where each shift is tailored to the corresponding test statistic and its dependence on the others. Similarly, SBH2 from \cite{sz2025} arises as a special case of GSBH with $\tau$ chosen as a fraction between $0$ and the minimum eigenvalue $\lambda_{\min}(\Sigma)$ of the correlation matrix. However, we can now fine-tune and potentially strengthen it by setting $\tau = \lambda_{\min}(\Sigma)$, which we refer to as the new SBH2.

Our analyses aim to identify a choice of $\tau$, and thus a corresponding method, which yields the most favorable results across all settings. Since $\tau$ serves as a common partition among the $\tau_i$’s, we focus our attention on a restricted set of candidates: the strongest squared multiple correlation $\tau_{\text{min}}$ (GSBH1), the weakest squared multiple correlation $\tau_{\text{max}}$ (GSBH2), median $\tau_{\text{med}}$ (GSBH3), arithmetic mean $\bar{\tau}$ (GSBH4), geometric mean $\bar{\tau}_{\text{geo}}$ (GSBH5) and harmonic mean $\bar{\tau}_{\text{har}}$ (GSBH6). \end{remark*}

\section{FDR-Controlled Variable Selection: The Bigger Picture}
\label{sec:variable selection}

This section sheds light on the relevance of our work by demonstrating the implementation of the proposed multiple testing procedures, developed in the preceding section with rigorous control of the false discovery rate (FDR), in the context of variable selection within a linear regression framework.

Consider the linear model: $\boldsymbol{Y} = \boldsymbol{X}\boldsymbol{\beta} + \boldsymbol{\epsilon}$, where $\boldsymbol{Y}$ is $n$-dimensional response vector, $\boldsymbol{X} = (\boldsymbol{X}_1, \ldots, \boldsymbol{X}_d)$ is $n \times d$ design matrix of rank $d < n$ with its columns representing the known vectors of observations on the $d$ variables (or predictors) $\bs{X}_1, \ldots, \bs{X}_d$, $\boldsymbol{\beta} = (\beta_1, \ldots, \beta_d)$ is the unknown vector of regression coefficients corresponding to these variables, and $\boldsymbol{\epsilon} \sim N_d(\boldsymbol{0}, \eta^2 \boldsymbol{I}_d)$ is the Gaussian noise with some unknown variance $\eta^2$.

We formulate variable selection as a multiple hypothesis testing problem. Specifically, for each predictor $X_i$, we test the null hypothesis $H_i: \beta_i = 0$ against the alternative $H_i: \beta_i \neq 0$, simultaneously for $i = 1, \ldots, d$. Variables corresponding to rejected null hypotheses are selected as relevant predictors. The effectiveness of such selection procedures is most commonly evaluated using the FDR, which quantifies the expected proportion of false discoveries among all selections.

The ordinary least squares (OLS) estimator of $\boldsymbol{\beta}$ is $\hat{\boldsymbol{\beta}}= (\hat{\beta}_1, \ldots, \hat{\beta}_d)^{\prime} =  \boldsymbol{A}^{-1}\boldsymbol{X}^T\boldsymbol{Y}$, where  $\boldsymbol{A} = \boldsymbol{X}^{\prime}\boldsymbol{X}$. Under the model assumptions, $\hat{\boldsymbol{\beta}} \sim N_d(\boldsymbol{\beta}, \eta^2 \boldsymbol{A}^{-1})$. Furthermore, the estimator of $\eta^2$, $\hat{\eta}^2 = (\|\boldsymbol{Y}\|^2 - \hat{\boldsymbol{\beta}}^{\prime} \boldsymbol{A} \hat{\boldsymbol{\beta}})/(n-d)$ is independent of $\hat{\boldsymbol{\beta}}$ and follows a scaled chi-squared distribution: $\hat{\eta}^2 \sim \eta^2 \chi^2_{\nu} / \nu$, $\nu = n-d$. These facts naturally suggest using $\hat{\boldsymbol{\beta}}$ when developing FDR-controlling procedures. When $\eta^2$ is known (e.g., set to 1), one may use the $p$-values $P_i = \bar{\Psi}_1(\hat{\beta}_i^2/(\boldsymbol{A}^{-1})_{ii})$, while for unknown $\eta^2$, a natural choice is $P_i = \bar{\Psi}_{1,\nu}(\hat{\beta}_i^2/(\boldsymbol{A}^{-1})_{ii}\nu\hat{\eta}^2)$. However, the direct application of the classical BH procedure to these $p$-values, as noted before, is problematic, given the arbitrariness of the correlation structure of $\hat{\boldsymbol{\beta}}$. In contrast, our newly proposed generalized shifted BH procedures, which are provably robust to general dependence structures, provide a natural and reliable choice for FDR-controlled variable selection.

When $n \ge 2d$, our proposed GSBH procedure becomes a natural competitor to the Bonferroni–BH (BBH) procedure of \cite{st2022}. The BBH was developed under a paired multiple-testing framework obtained by recasting the knockoff-assisted variable selection problem of \cite{bc2015} as a two-sample $p$-value problem. More specifically, \cite{st2022} decomposed the OLS estimator $\hat{\bs{\beta}}$ into two independent estimators,
\begin{align}
    \hat{\bs{\beta}}_1 = & (2\bs{A} -\bs{D})^{-1} (\bs{X}+\bs{\tilde{X}})^{\prime} \bs{Y} \; \mbox{and} \; \hat{\bs{\beta}}_2 = \bs{D}^{-1}(\bs{X}-\bs{\tilde{X}})^{\prime} \bs{Y}
\nonumber \end{align} by using $\tilde{\boldsymbol{X}}$, the knockoff counterpart of $\boldsymbol{X}$ (defined in \cite{bc2015}), which satisfies the equations $\tilde{\boldsymbol{X}}^{\prime} \tilde{\boldsymbol{X}} = \boldsymbol{A}$ and $\tilde{\boldsymbol{X}}^{\prime} \boldsymbol{X} = \boldsymbol{A} - \boldsymbol{D}$, with $\bs{D}=\text{diag}\{\bs{s}\}$, for some $\bs{s} \in \mathbb{R}_+^d$, being such that $2\bs{D}-\bs{D}\bs{A}^{-1}\bs{D}$ is positive definite. These two independent parts of $\hat{\boldsymbol{\beta}}$ are distributed as follows: $\hat{\bs{\beta}}_1 \sim N_{d}({\bs{\beta}}, 2\eta^2(2{\bs{A}} - {\bs{D}})^{-1})$ and  $\hat{\bs{\beta}}_2 \sim N_{d}({\bs{\beta}}, 2\eta^2{\bs{D}}^{-1})$. An alternative, data-fission perspective on such a decomposition of Gaussian vector $\hat{\boldsymbol{\beta}}$ is discussed in \cite{sarkar2025comments}.

From $\hat{\bs{\beta}}_1$ and $\hat{\bs{\beta}}_2$, two sets of $p$-values are constructed, $\bs{P}^{(1)} = (P^{(1)}_1, \ldots, P^{(1)}_d)$ and $\bs{P}^{(2)} = (P^{(2)}_1, \ldots, P^{(2)}_d)$, where       
\begin{align}
P_i^{(1)} = \bar{\Psi}_{1}\left( \frac{\hat{\beta_{1i}}^2}{2((2\bs{A}-\bs{D})^{-1})_{ii}}\right) \; \mbox{and} \; P_{i}^{(2)} = \bar{\Psi}_{1} \left ( \frac{\hat{s_{i}\beta_{2i}}^2}{2} \right ), 
\nonumber \end{align} if $\eta^2 = 1$, and 
\begin{align}
P_{i}^{(1)} = \bar{\Psi}_{1,\nu}\left( \frac{\hat{\beta_{1i}}^2}{2\nu\hat{\eta}^2((2\bs{A}-\bs{D})^{-1})_{ii}}\right) \; \mbox{and} \; P_{i}^{(2)} = \bar{\Psi}_{1} \left ( \frac{\hat{s_{i}\beta_{2i}}^2}{2\nu\hat{\eta}^2}\right ),  \nonumber \end{align} if $\eta^2$ is unknown.

With $\hat{\bs{\beta}}^{(2)}$ being internally independent, the original BH method can be directly applied to $\bs{P}^{(2)}$ with valid FDR control. However,  $\hat{\bs{\beta}}^{(1)}$ is not internally independent. So, \cite{st2022} proposed a two-step procedure: first, use $\bs{P}^{(1)}$ with a Bonferroni threshold to screen potentially important variables, and then apply BH to the corresponding $p$-values in $\bs{P}^{(2)}$. Specifically, they apply BH at level $\sqrt{\alpha}$ to the combined $p$-values, $P_i^* = \mathds{1} (P^{(1)}_i > \sqrt{\alpha}) + \mathds{1} (P^{(1)}_i \le  \sqrt{\alpha}) P^{(2)}_{i}$, $i=1, \ldots, d$. This defines the BBH procedure. \cite{st2022} also proposed an adaptive version of BBH (Adapt-BBH). It has been obtained by replacing  $P^{(2)}_i$ in BBH by $\hat{\pi}_0 P^{(2)}_i$, where $\hat{\pi}_0$ is an estimate of $\pi_0$, the proportion of unimportant variables, obtained from $\boldsymbol{P}^{(2)}$. They recommend using an estimator of $\pi_0$ (e.g., \cite{sts2004}) that is known to provide theoretically valid FDR control for adaptive BH methods (\cite{sks2008}). The BBH and Adapt-BBH control the FDR non-asymptotically when $\eta^2 = 1$, but asymptotically when $\nu$ is large making $\hat{\eta}^2 \approx \eta^2$.  

Now that we have established that our proposed class of Shifted BH procedures based on $\bs{P}^{(1)}$, which explicitly incorporate the underlying correlation structure, can provide a more powerful alternative to Bonferroni thresholding while maintaining theoretically valid FDR control, we consider a modification of the two-step framework of \cite{st2022}. Specifically, we first screen potentially important variables by thresholding the $p$-values in $\bs{P}^{(2)}$ with an appropriately chosen cutoff. We then apply GSBH with a suitably chosen $\tau$ to the corresponding $p$-values in $\bs{P}^{(1)}$. The procedure is formally defined below.

\begin{definition*} \label{Definition2}
\noindent {\bf Shifted BBH}: A step-up procedure applied to  $$\tilde{P_i}^* = \mathds{1}(P_i^{(2)} > \sqrt{\alpha}) + \mathds{1}(P_i^{(2)} \le  \sqrt{\alpha})\tilde{P}_{i}^{(1)}, \; i=1, \ldots, d,$$ using critical constants $\alpha_i= i H^{-1}(\sqrt{\alpha}/d)$, for $i=1, \ldots, d$, where $\tilde{P}^{(1)}_{i} = \bar{\Psi} \left( \bar{\Psi}^{-1} (P_i^{(1)} )/\tau \right)$, for some appropriately chosen $\tau$ depending on $\tau_i = 1 - R_i^2$, $i=1,\ldots,d$, with $R_i^2$ denoting the squared multiple correlations obtained from $(2\bs{A} - \bs{D})^{-1}$. Here, $H$ is as defined in Definition \ref{Definition1}, and $\Psi$ denotes $\Psi_1$ when $\eta^2=1$, or $\Psi_{1,\nu}$ when $\eta^2$ is unknown.
\end{definition*}

The following result, which is a consequence of Theorem~\ref{Theorem2}, establishes the FDR control property of Shifted BBH.

\begin{newthm*} \label{Theorem3}
Shifted BBH controls the FDR at level $\alpha$ non-asymptotically when $\eta^2=1$, and asymptotically when $\eta^2$ is unknown and $\nu$ is large, so that $\hat{\eta}^2 \approx \eta^2$.
\end{newthm*}

\section{Empirical Evidence} 
\label{sec:numerical}
This section empirically examines the finite-sample performance of the proposed methods under both controlled synthetic conditions and real-world data settings, thereby bridging theoretical guarantees with practical applicability. The investigation proceeds in two parts: (i) we first assess the operating characteristics of the methods under a range of simulated environments that capture complex dependence structures and varying signal configurations; and (ii) we subsequently evaluate their empirical utility on an HIV gene expression dataset extensively analyzed in \cite{bc2015}, thereby validating their performance in relevant contexts.
\subsection{Simulations}
\label{subsec:simu}
We consider two inferential settings, described formally in Section \ref{sec:framework}. The first involves detecting nonzero means in a $d$-dimensional multivariate normal distribution, allowing for arbitrary correlations among test statistics. The second casts the problem within a variable selection framework, discussed in Section \ref{sec:variable selection}. This latter formulation corresponds to the two-sided $t$-test layout (Setting \hyperlink{Setting2}{2}) introduced earlier, providing a unified treatment across simulation designs. For brevity, we report results for mean testing using $z$-tests and reserve $t$-test–based results for the regression context.

\subsubsection{Testing for Means}
\label{subsubsec:means}
\noindent In multivariate mean testing, correlations among variables can strongly influence the rate of false discoveries. Let $\bs{X} = (X_1, \ldots, X_d)'$ denote a random vector drawn from a $d$-dimensional normal distribution with mean vector $\bs{\mu}$ and correlation matrix $\bs{\Sigma}$, as specified in Section \ref{sec:methods}. Our simulations assess the proposed procedures over a range of signal strengths and null proportions, measuring both false discovery rate (FDR) control at level $\alpha$ and power (the ability to detect true signals).

We study two simulation regimes: (1) fixed null proportion (75\%), with varying nonzero mean magnitudes from (1-5), and (2) fixed signal strength ($\mu_i=2$), with varying null proportions (50–90\%).
In each regime, method performance is averaged over 500 Monte Carlo replications, with $d \in \{40, 100\}$ and $\alpha = 0.05$.
To explore robustness, we generate correlated test statistics under six dependence structures: 
\newline \noindent (i) Equicorrelated (`Equi'): all pairwise correlations set to $\rho \in \{0.3,0.7\}$;
\newline \noindent(ii) Autoregressive of order 1 (`AR'): correlations decaying as a function of distance with $\rho \in \{0.3,0.7\}$;
\newline \noindent (iii) Inverse autoregressive of order 1 (`IAR'): inverse structure with the correlation $\rho \in \{0.3,0.7\}$; 
\newline\noindent (iv) Block diagonal (`Cluster'): partitioned into $d/4$ blocks; 
\newline \noindent (v) Sparse (`Sparse'): a sparsely populated correlation matrix with 20\% non-zero entries; and 
\newline \noindent (vi) A matrix (denoted by `Prefixed Corr 1') which indicates that a random set (25\% of total) of variables is randomly correlated with all the variables.

\medskip
Our comparisons aim to identify (a) which members of the proposed family of methods show the strongest performance, and (b) how they compare with established FDR procedures.
The methods differ in how they partition the squared multiple correlations to determine shifted $p$-values, guided by the theoretical insight that certain partitions preserve the PLTDN property required for valid FDR control. If no natural partition arises, we expect strong performance from either one of SBH2, GSBH1, or GSBH2.

Overall, results reaffirm the robustness of the BH procedure, but also highlight its practical conservatism under dependence. The BY procedure remains a solid benchmark, yet our proposed GSBH methods offer reliable and often superior alternatives under correlated settings.

\begin{figure}[ht!]
  \begin{tabular*}{\textwidth}{
    @{}m{0.5cm}
    @{}m{\dimexpr0.33\textwidth-0.25cm\relax}
    @{}m{\dimexpr0.33\textwidth-0.25cm\relax}
    @{}m{\dimexpr0.33\textwidth-0.25cm\relax}}
  \rotatebox{90}{Equi(0.3)}
  & \begin{subfigure}[b]{\linewidth}
      \centering
      \includegraphics[width=\linewidth]{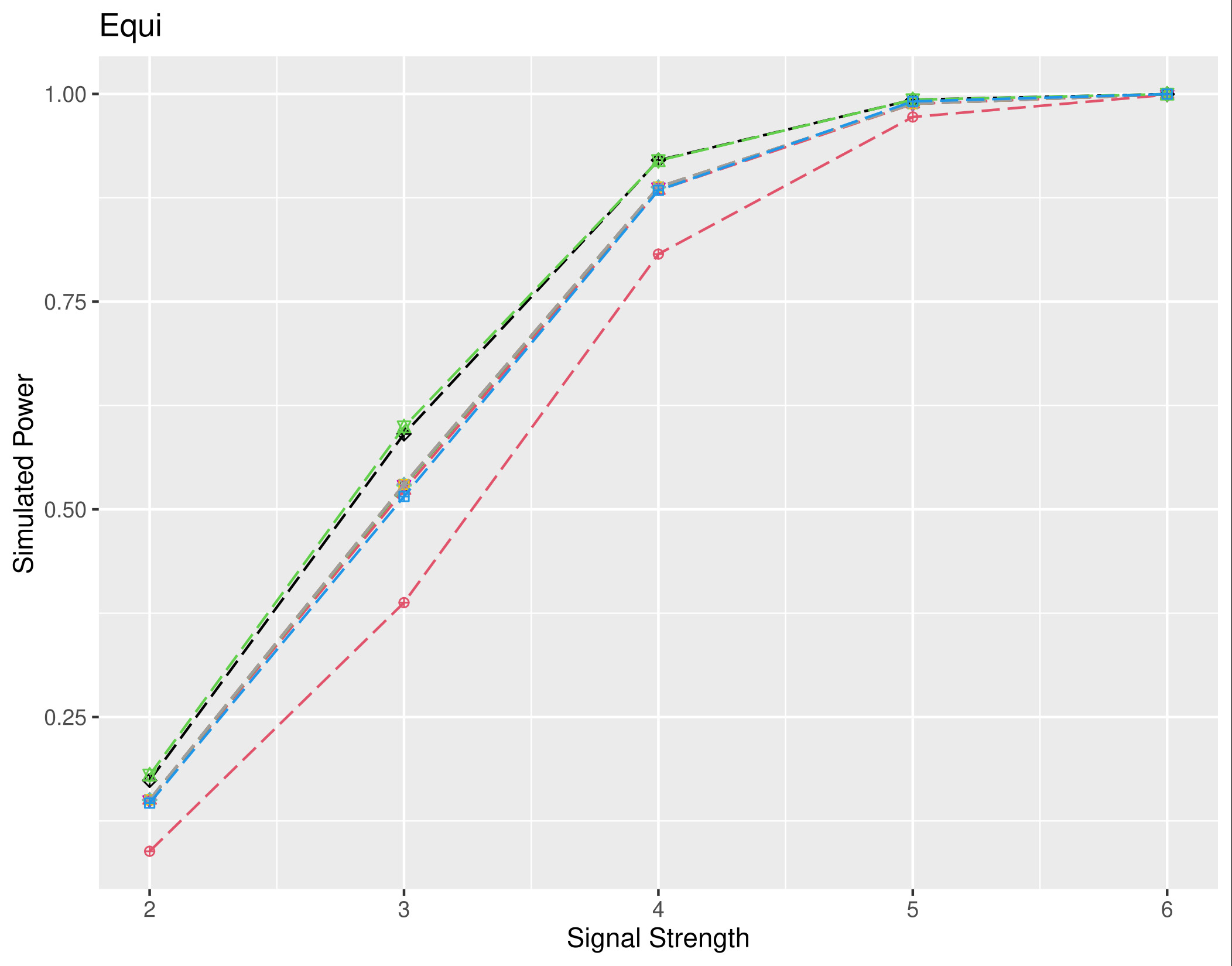} 
      \end{subfigure}  
  & \begin{subfigure}[b]{\linewidth}
      \centering
      \includegraphics[width=\linewidth]{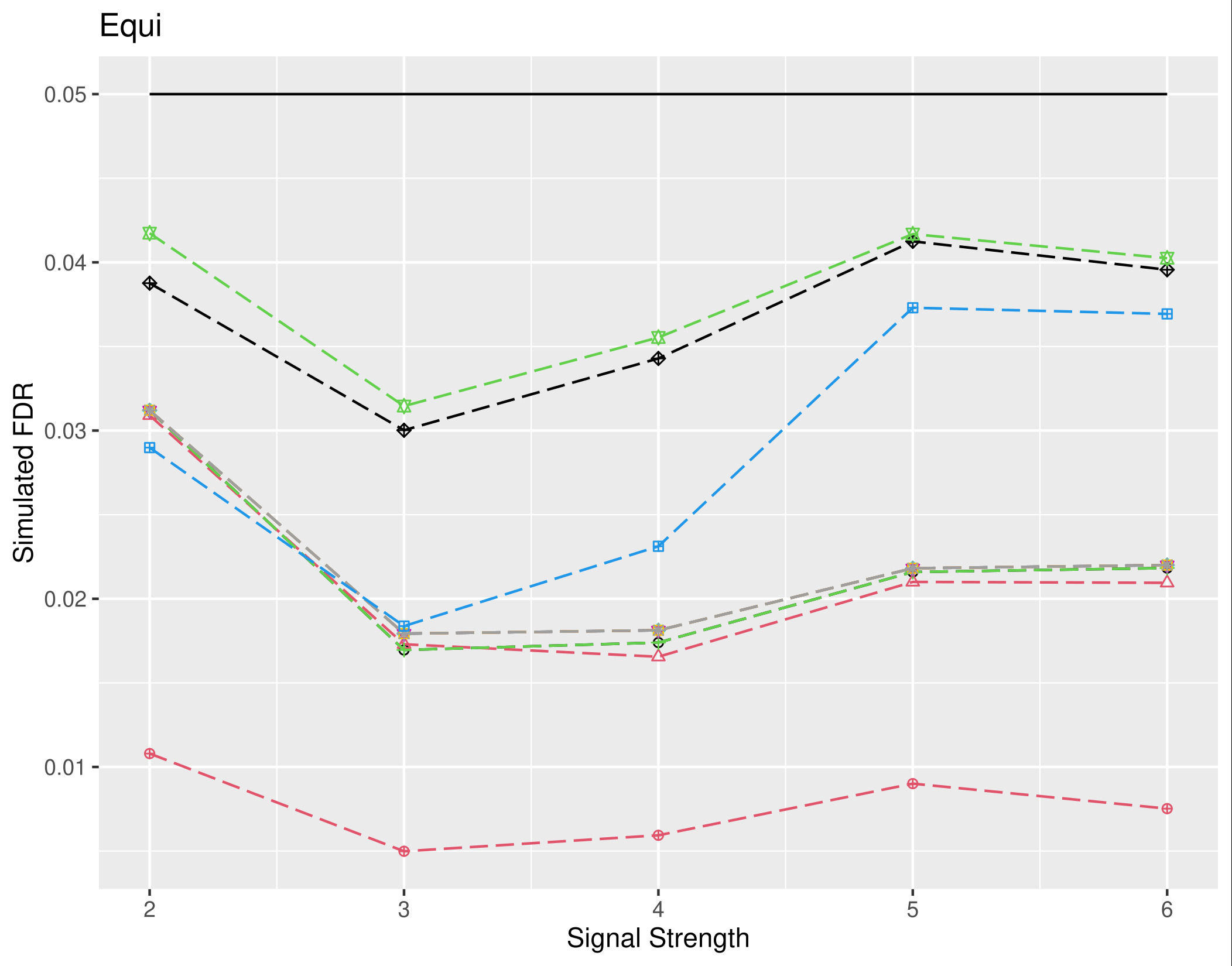} 
      \end{subfigure} 
  & \begin{subfigure}[b]{\linewidth}
      \centering
      \includegraphics[width=\linewidth]{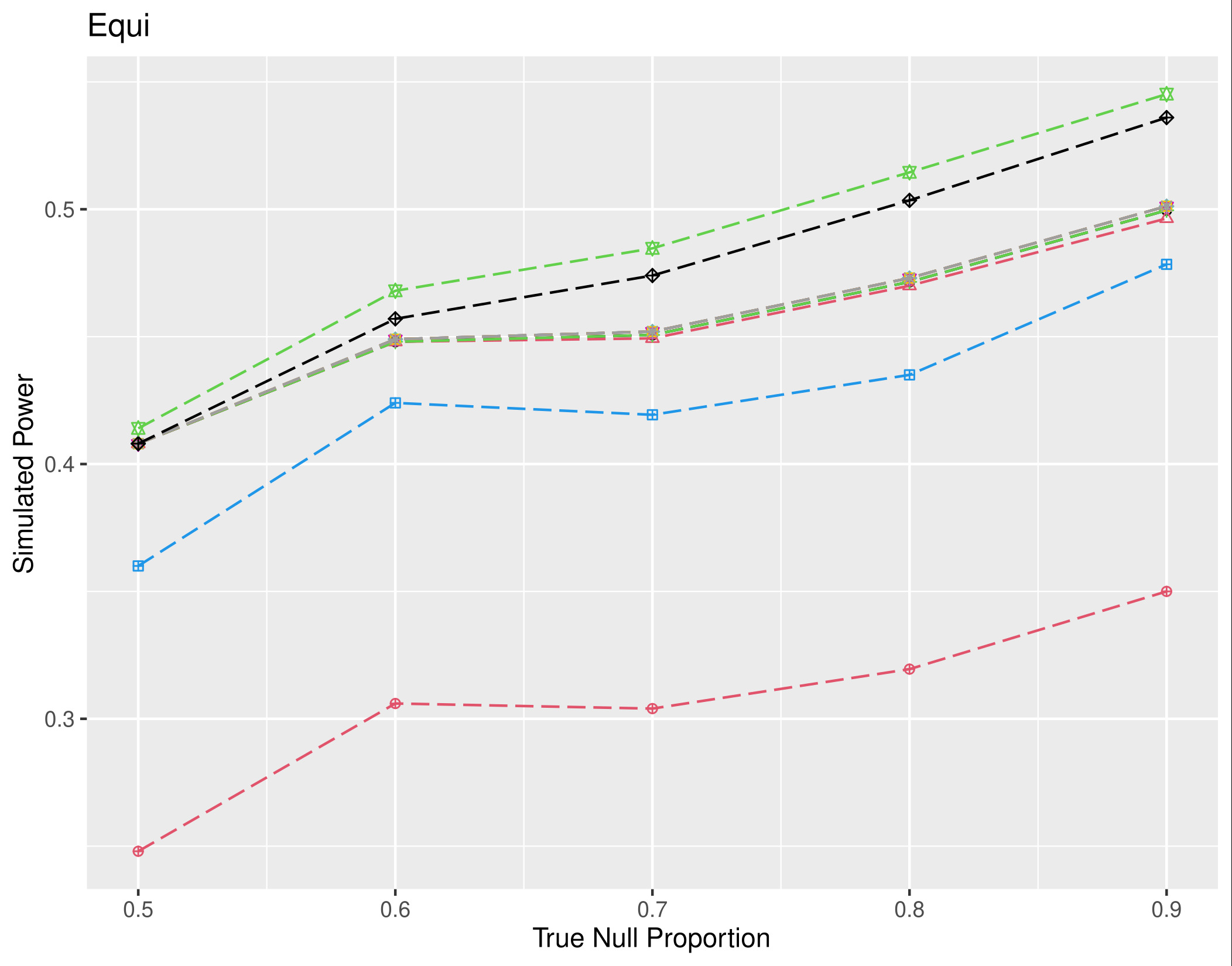} 
      \end{subfigure} \\
  \rotatebox{90}{AR(0.3)} 
  & \begin{subfigure}[b]{\linewidth}
      \centering
      \includegraphics[width=\linewidth]{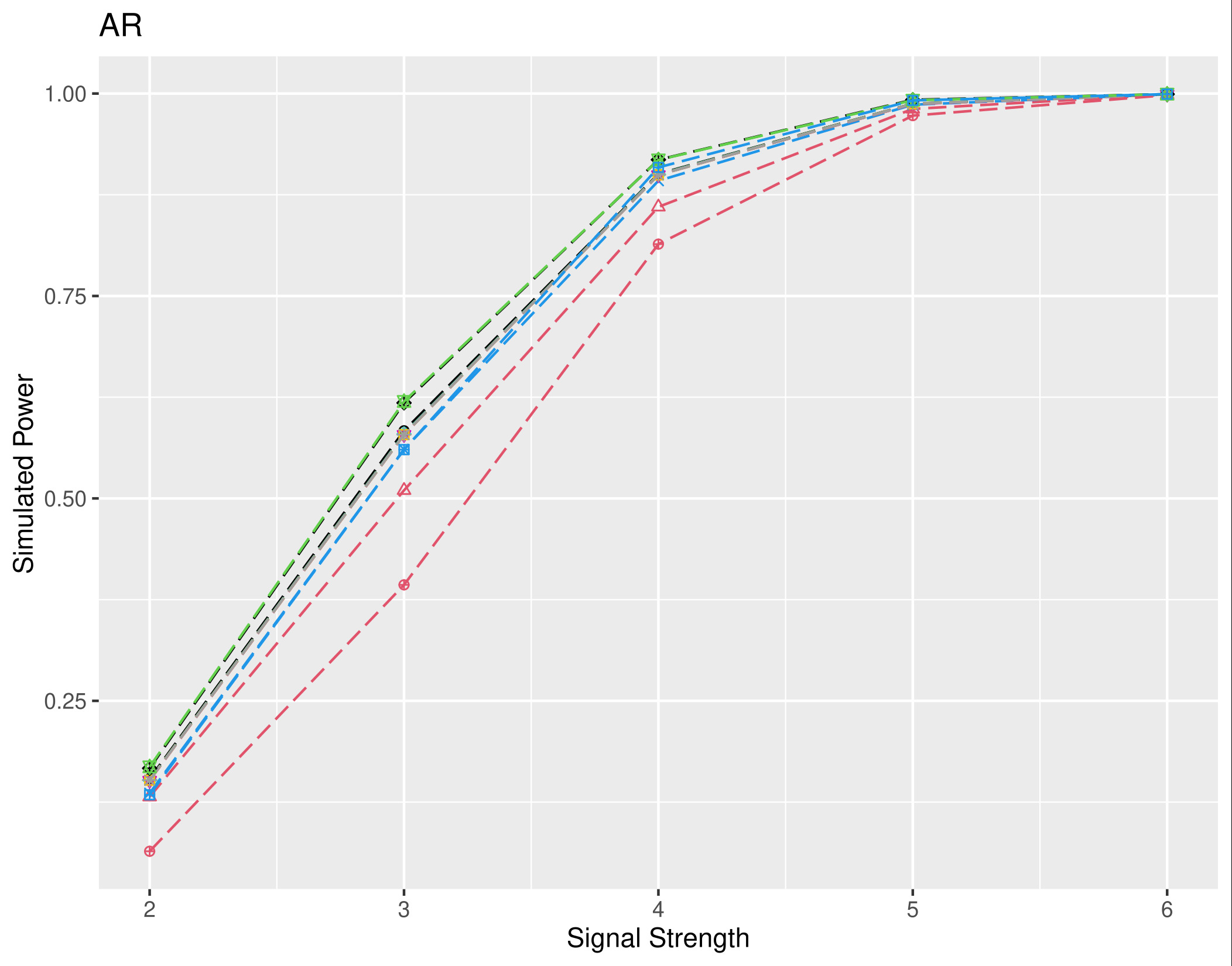} 
      \end{subfigure}  
  & \begin{subfigure}[b]{\linewidth}
      \centering
      \includegraphics[width=\linewidth]{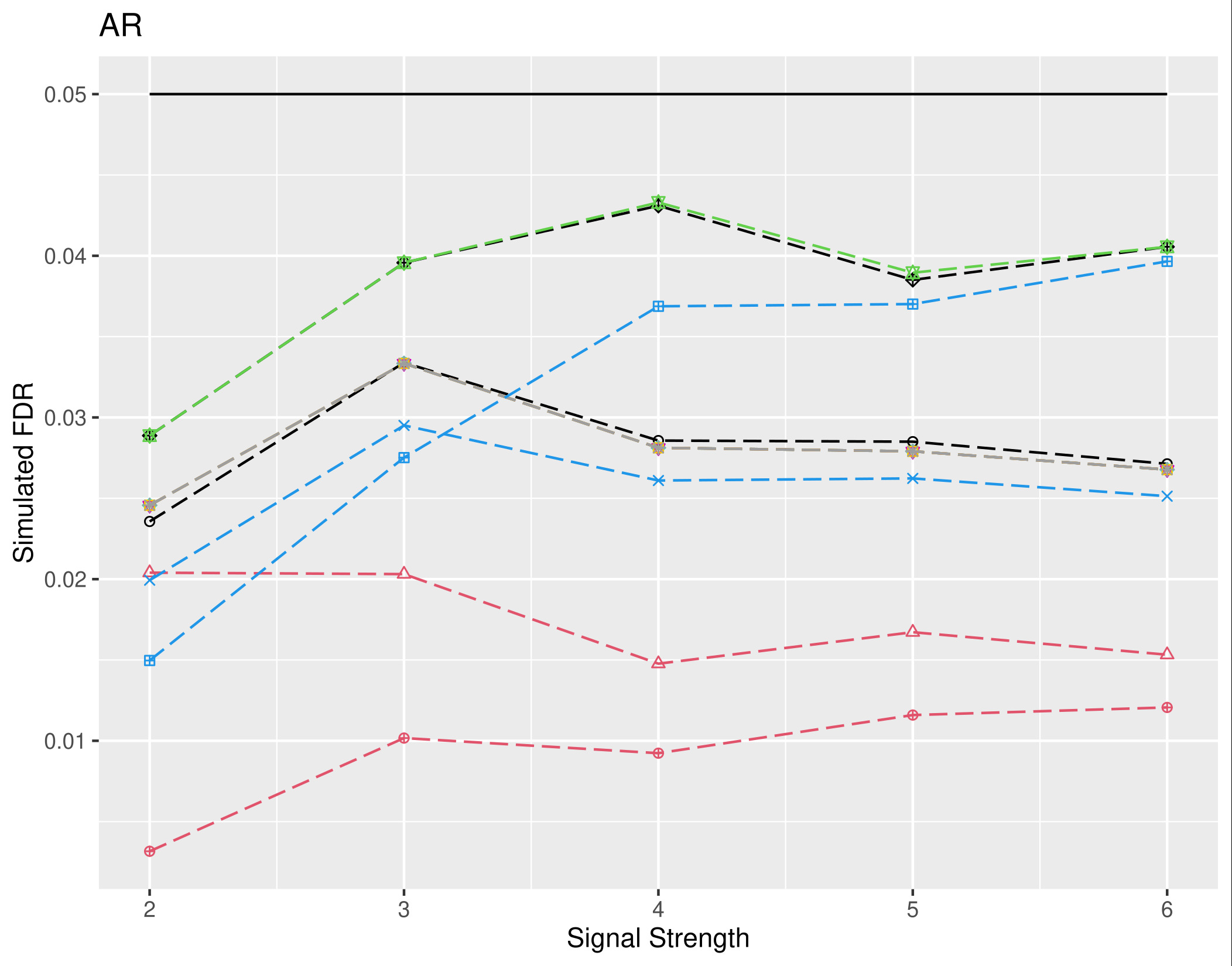} 
      \end{subfigure} 
  & \begin{subfigure}[b]{\linewidth}
      \centering
      \includegraphics[width=\linewidth]{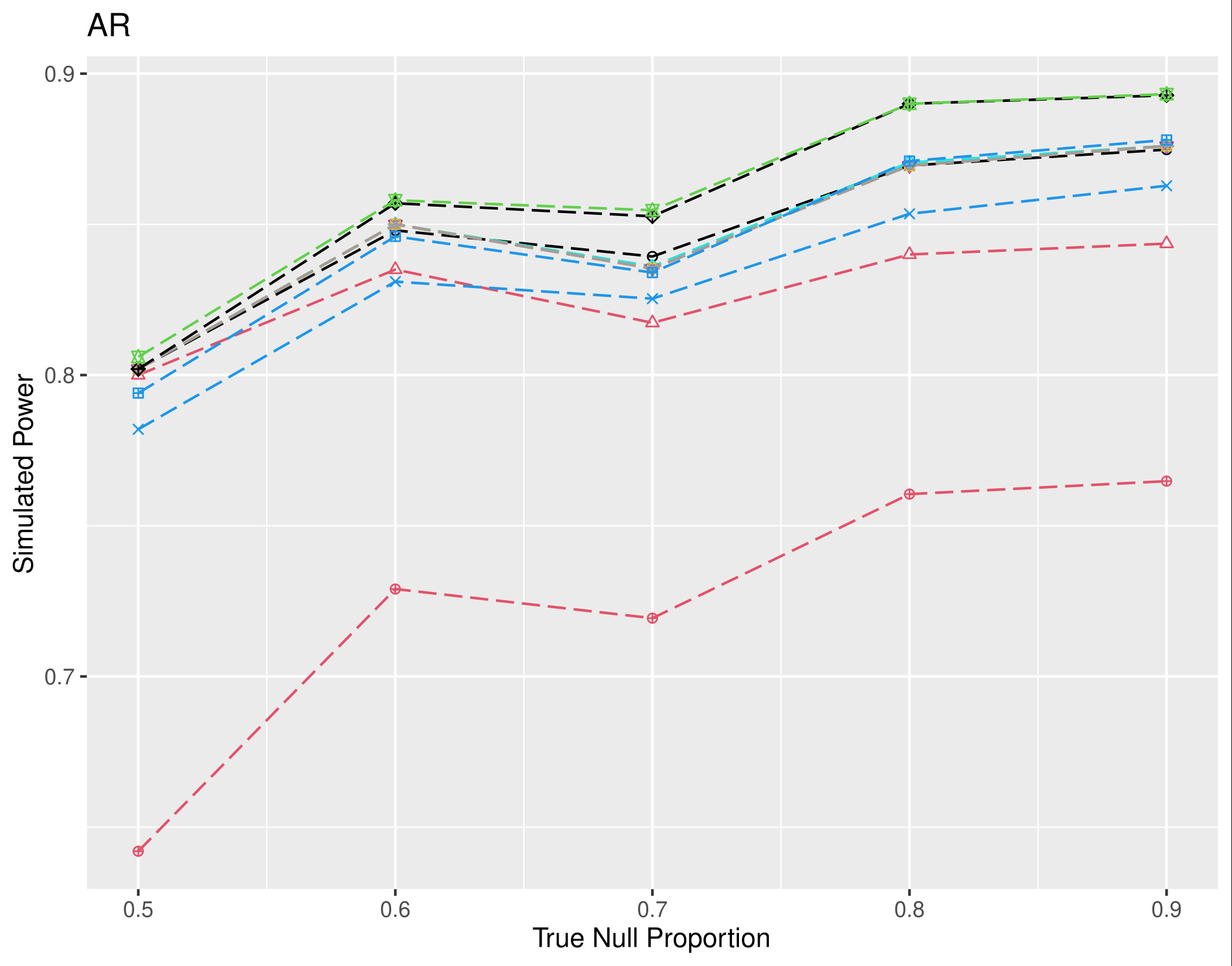} 
      \end{subfigure} \\
  \rotatebox{90}{IAR(0.3)} 
  & \begin{subfigure}[b]{\linewidth}
      \centering
      \includegraphics[width=\linewidth]{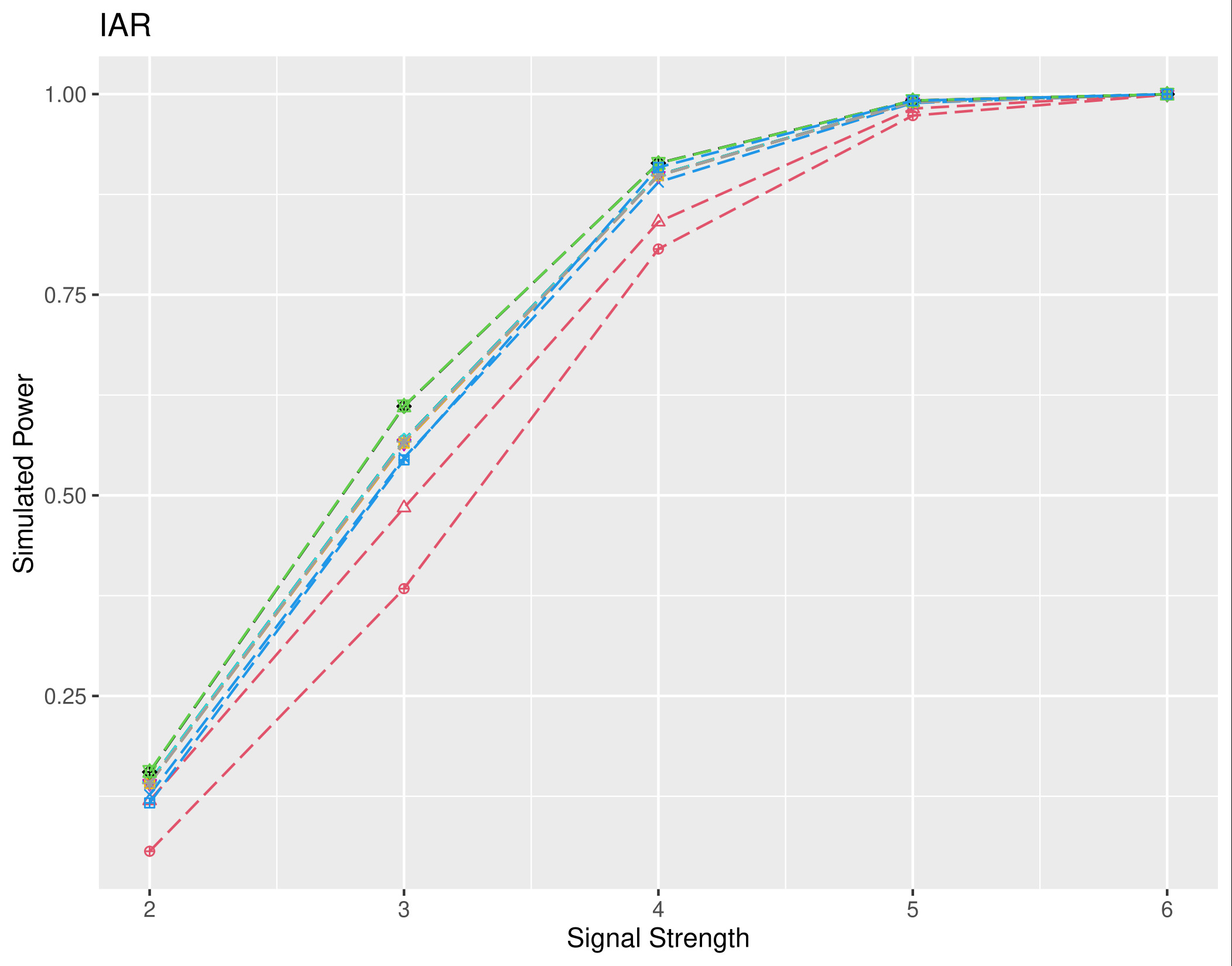} 
      \end{subfigure}  
  & \begin{subfigure}[b]{\linewidth}
      \centering
      \includegraphics[width=\linewidth]{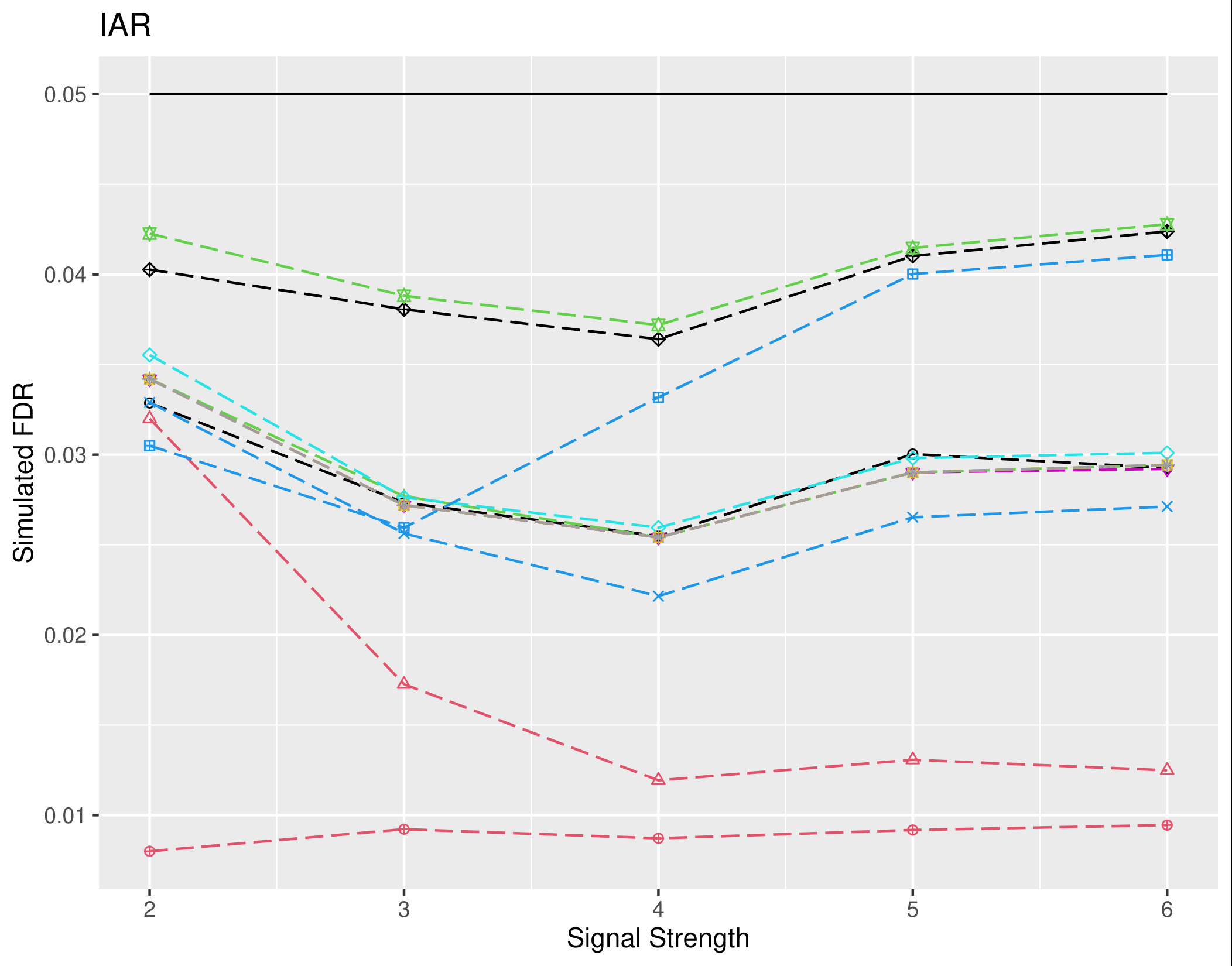} 
      \end{subfigure} 
  & \begin{subfigure}[b]{\linewidth}
      \centering
      \includegraphics[width=\linewidth]{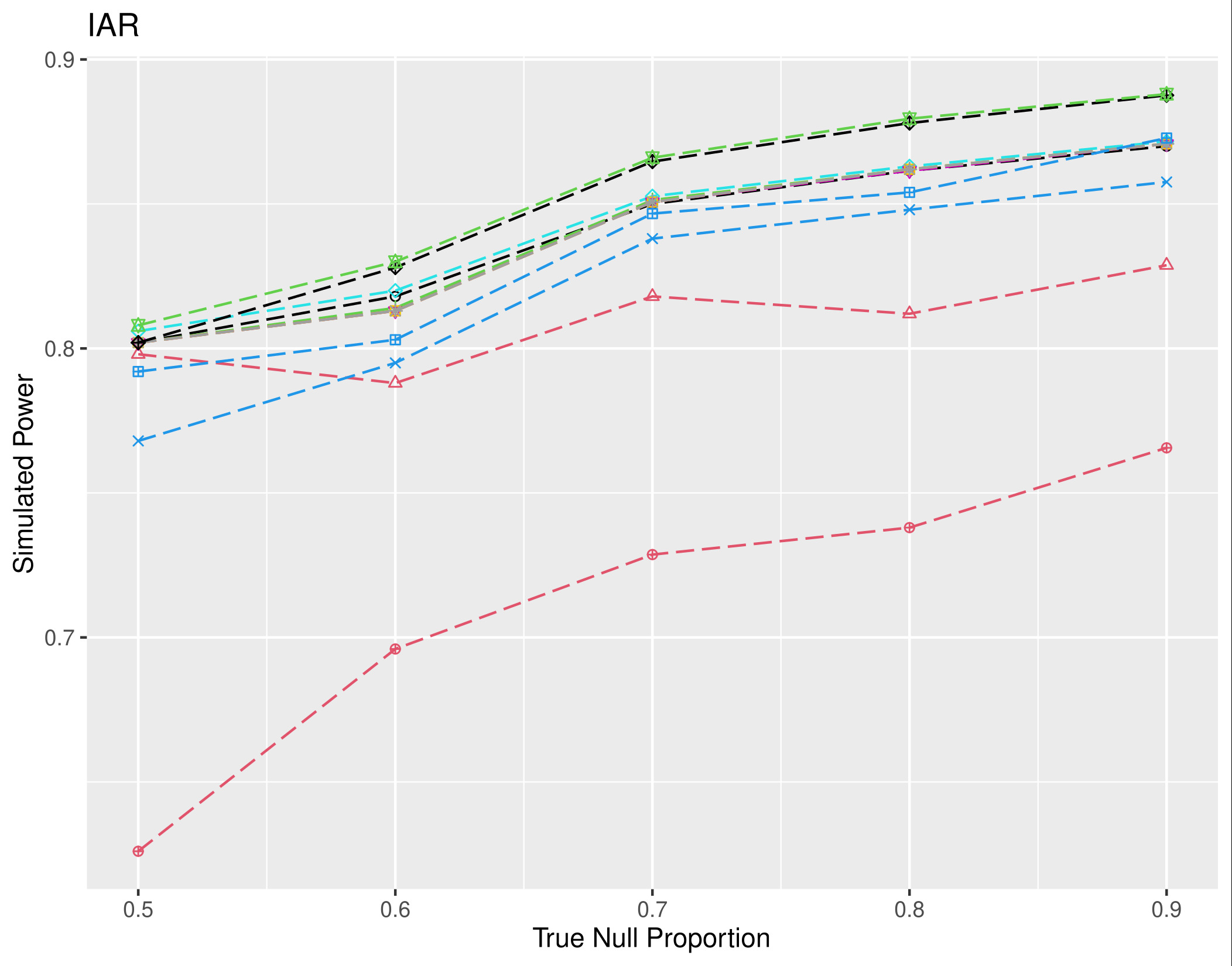} 
      \end{subfigure}    
  \end{tabular*} 
  \caption{Simulated Power (left column), simulated FDR (middle column) for fixed null proportion and simulated power (right column) for fixed signal strength, displayed for mean testing of $d=40$ parameters. Methods compared are SBH1 method (Circle and black), SBH2 method (Triangle point up and red), GSBH1 method (Plus and green), GSBH2 (Cross and blue), GSBH3 (Diamond and light blue), GSBH4 (Triangle point down and purple), GSBH5 (Square cross and yellow), GSBH6 (Star and grey), BH (Diamond plus and black), BY (Circle plus and red), dBH (Triangles up and down and green) and dBY (Square plus and blue)}
  \label{means:figure1} 
\end{figure}

For conciseness, results for 
($d,k$) = ($40,8$) are shown in the main text, with extended results in \hyperref[sec:appendixb]{Appendix B}.

\hyperref[means:figure1]{Figure 1} summarizes key outcomes. Across all dependence structures, GSBH procedures achieve higher power than both BH and dependence-adjusted BY (dBY), while maintaining valid FDR control. Under moderate correlations, GSBH methods show superior FDR stability compared to dBY. Although stronger dependence tends to make FDR control slightly conservative, the power gains remain substantial, representing a favorable trade-off.

\begin{figure}[ht!]
  \begin{tabular*}{\textwidth}{
    @{}m{0.5cm}
    @{}m{\dimexpr0.33\textwidth-0.25cm\relax}
    @{}m{\dimexpr0.33\textwidth-0.25cm\relax}
    @{}m{\dimexpr0.33\textwidth-0.25cm\relax}}
  \rotatebox{90}{Block Diagonal}
  & \begin{subfigure}[b]{\linewidth}
      \centering
      \includegraphics[width=\linewidth]{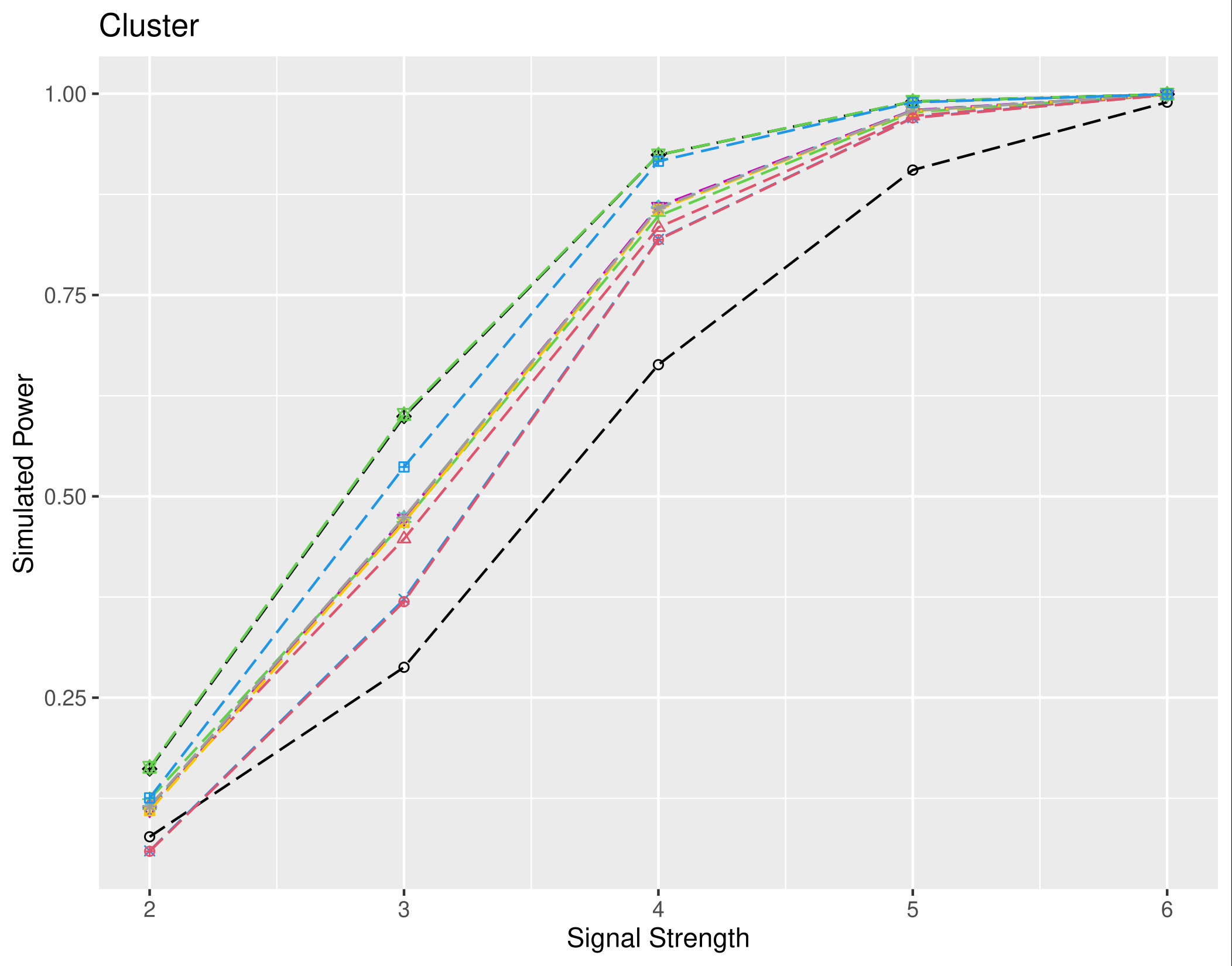} 
      \end{subfigure}  
  & \begin{subfigure}[b]{\linewidth}
      \centering
      \includegraphics[width=\linewidth]{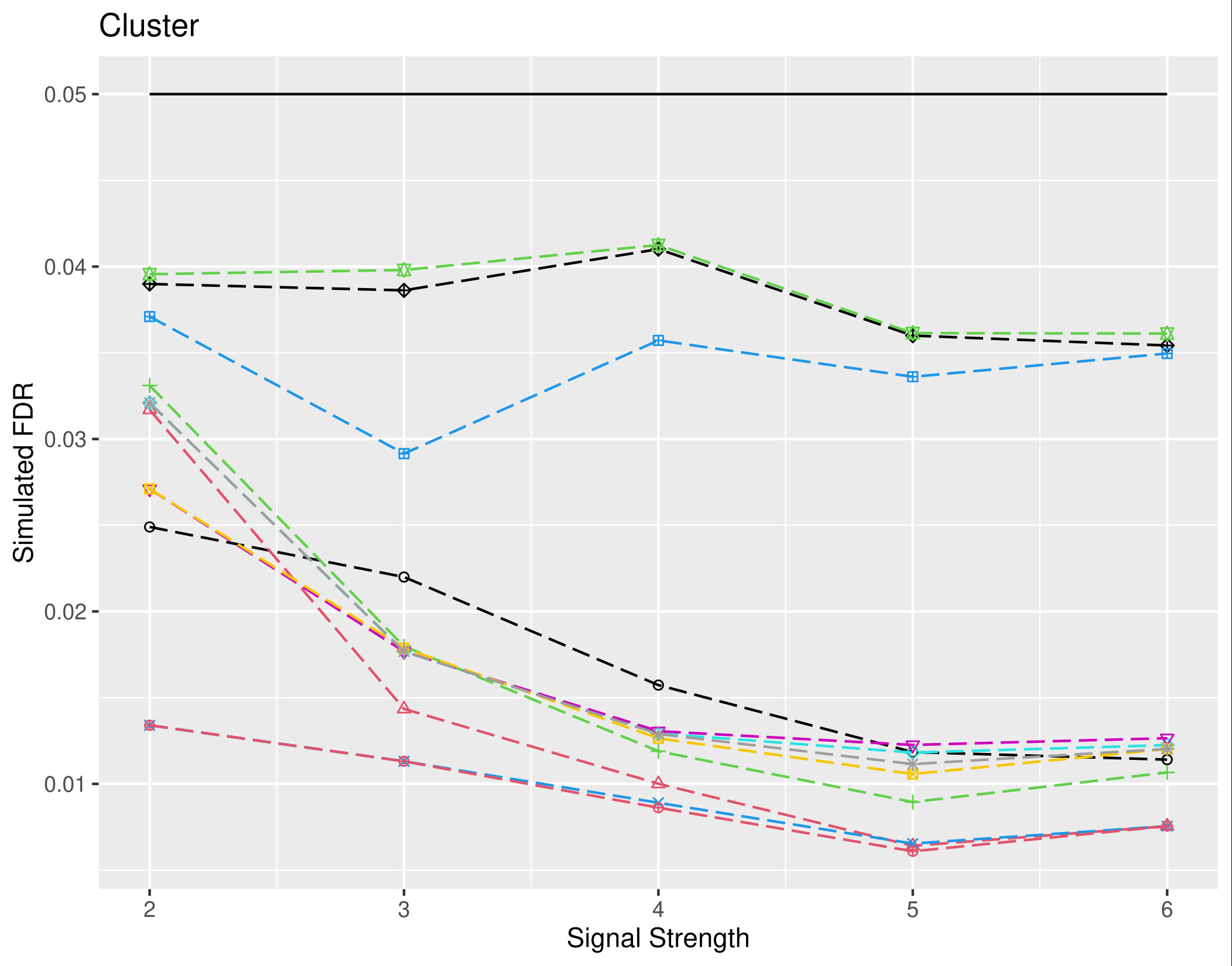} 
      \end{subfigure} 
  & \begin{subfigure}[b]{\linewidth}
      \centering
      \includegraphics[width=\linewidth]{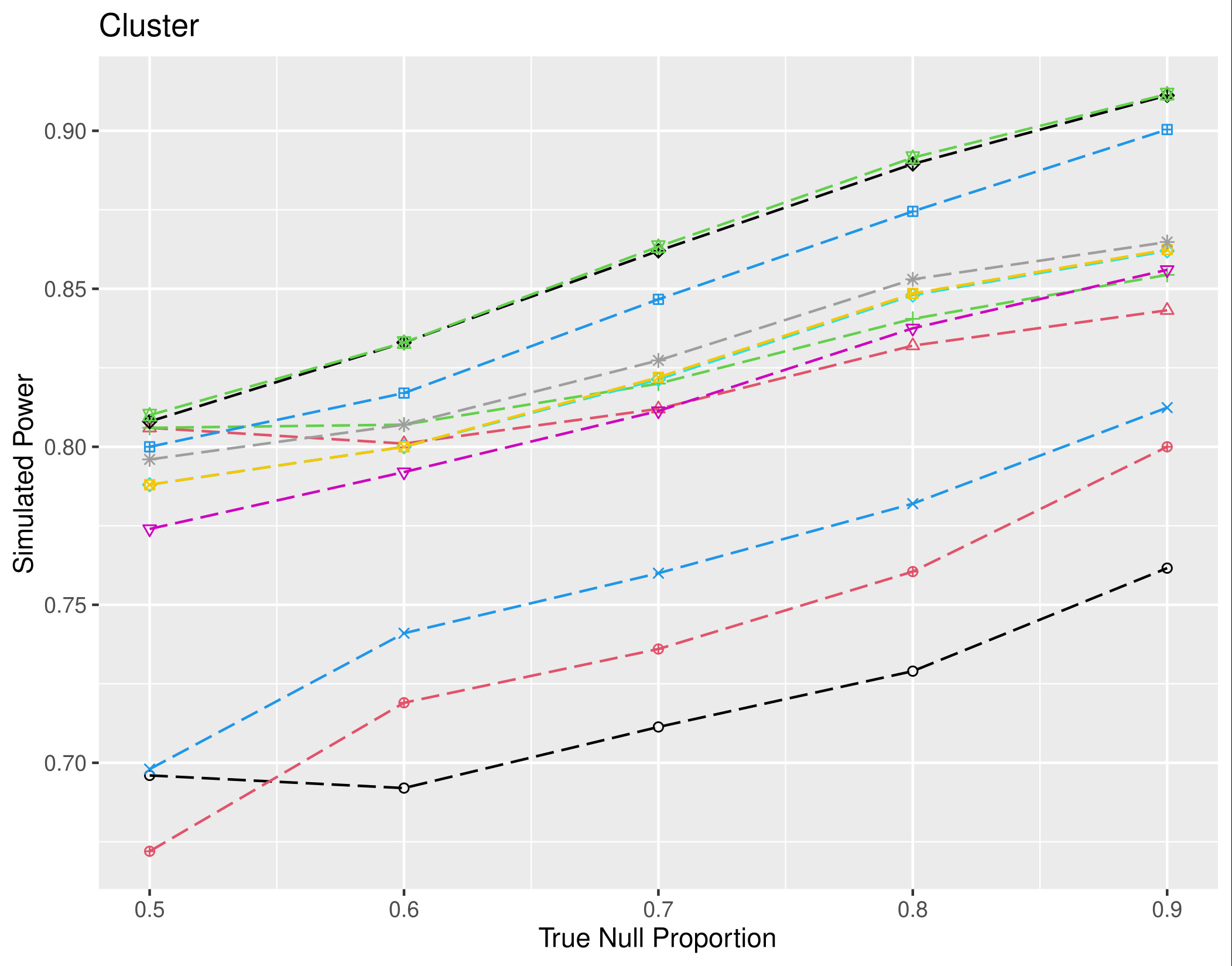} 
      \end{subfigure} \\
  \rotatebox{90}{Sparse} 
  & \begin{subfigure}[b]{\linewidth}
      \centering
      \includegraphics[width=\linewidth]{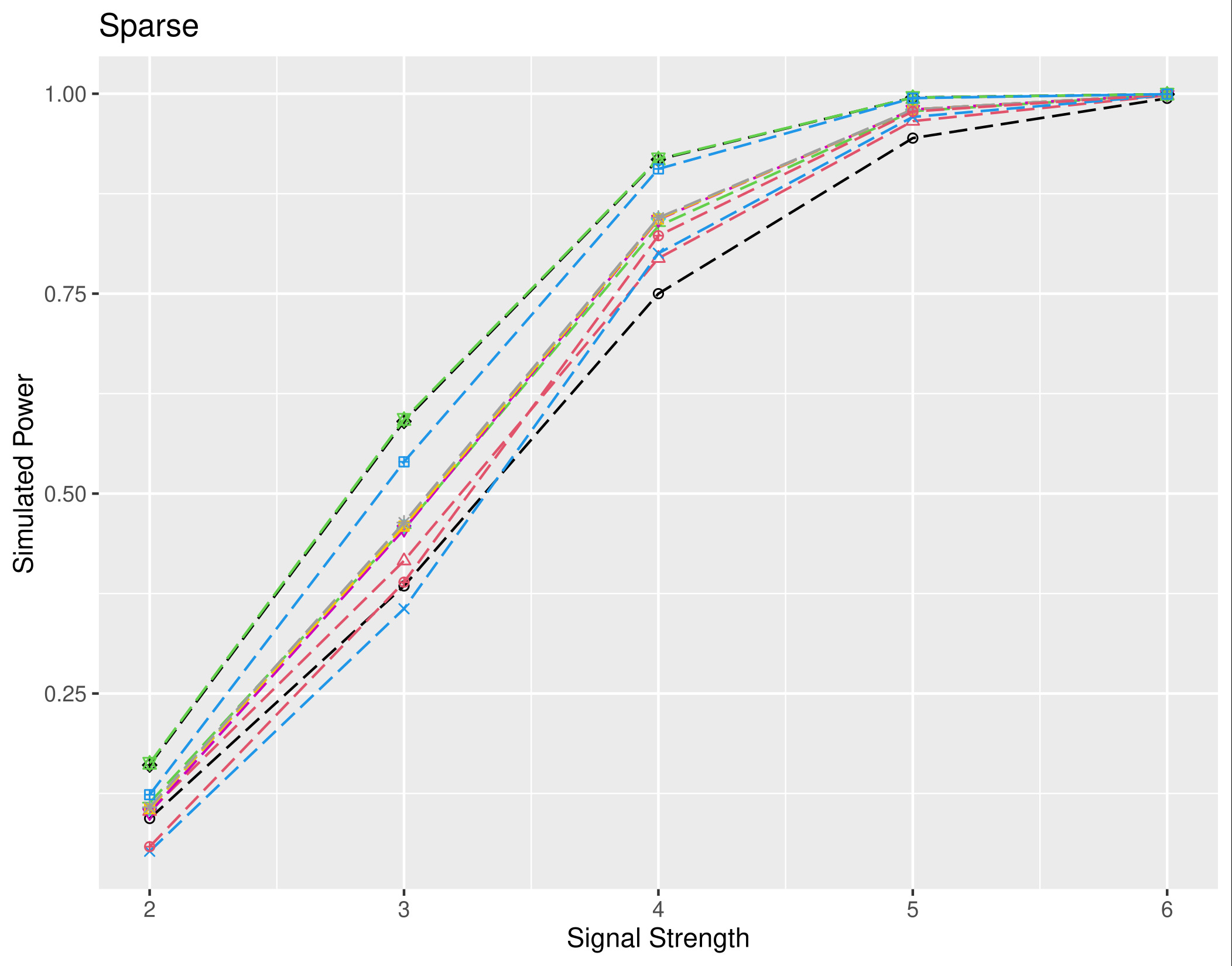} 
      \end{subfigure}  
  & \begin{subfigure}[b]{\linewidth}
      \centering
      \includegraphics[width=\linewidth]{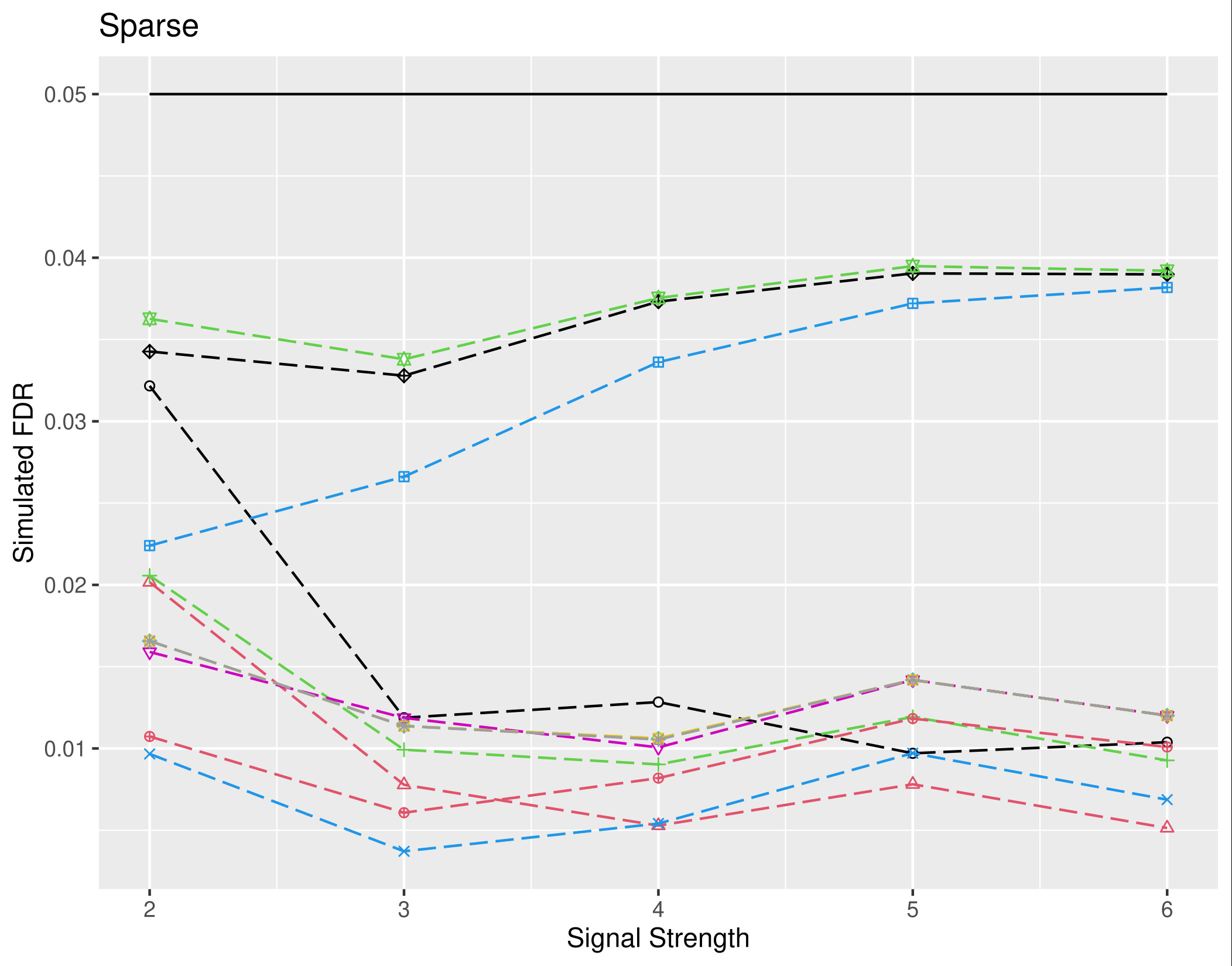} 
      \end{subfigure} 
  & \begin{subfigure}[b]{\linewidth}
      \centering
      \includegraphics[width=\linewidth]{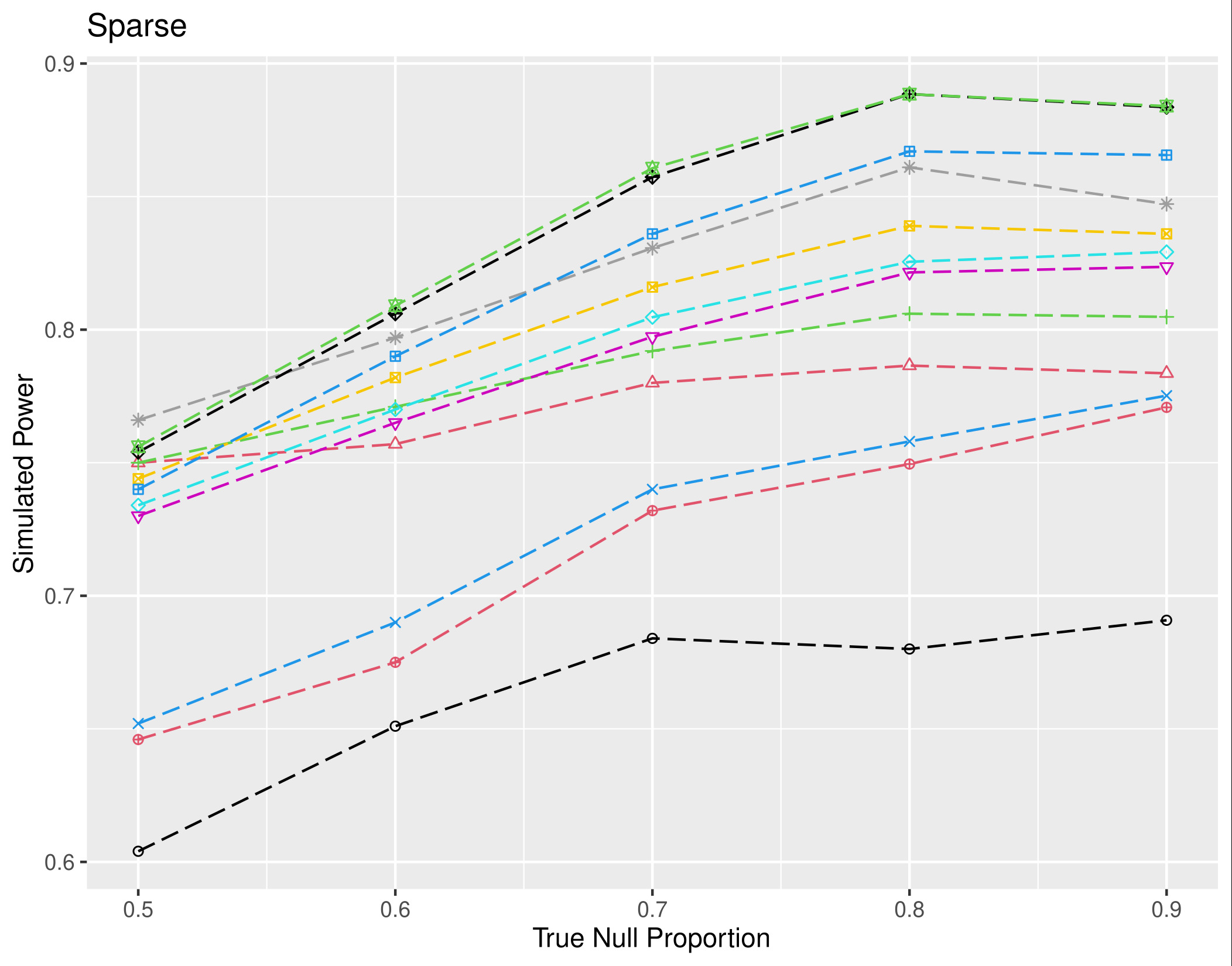} 
      \end{subfigure} \\
  \rotatebox{90}{Prefixed Corr 1} 
  & \begin{subfigure}[b]{\linewidth}
      \centering
      \includegraphics[width=\linewidth]{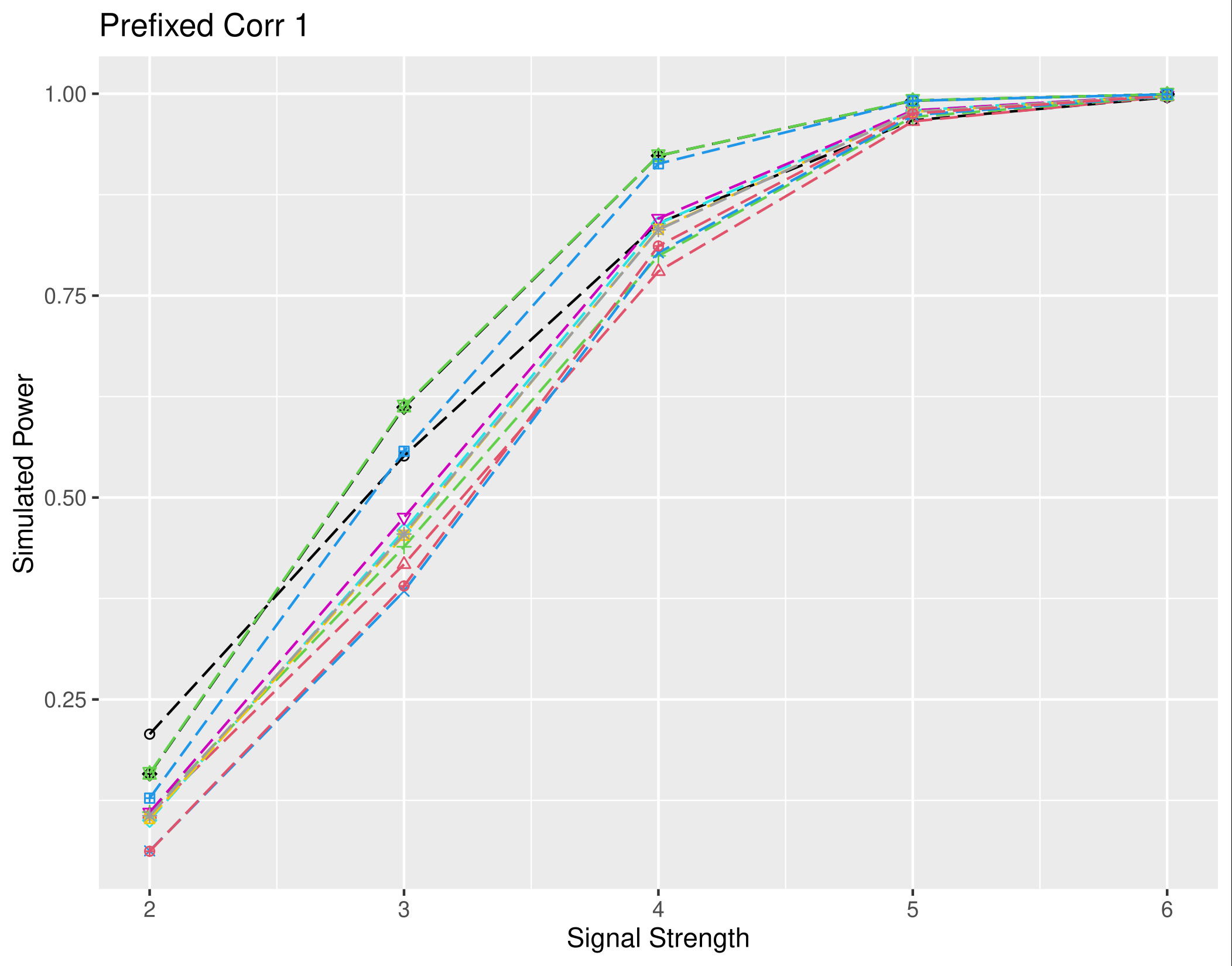} 
      \end{subfigure}  
  & \begin{subfigure}[b]{\linewidth}
      \centering
      \includegraphics[width=\linewidth]{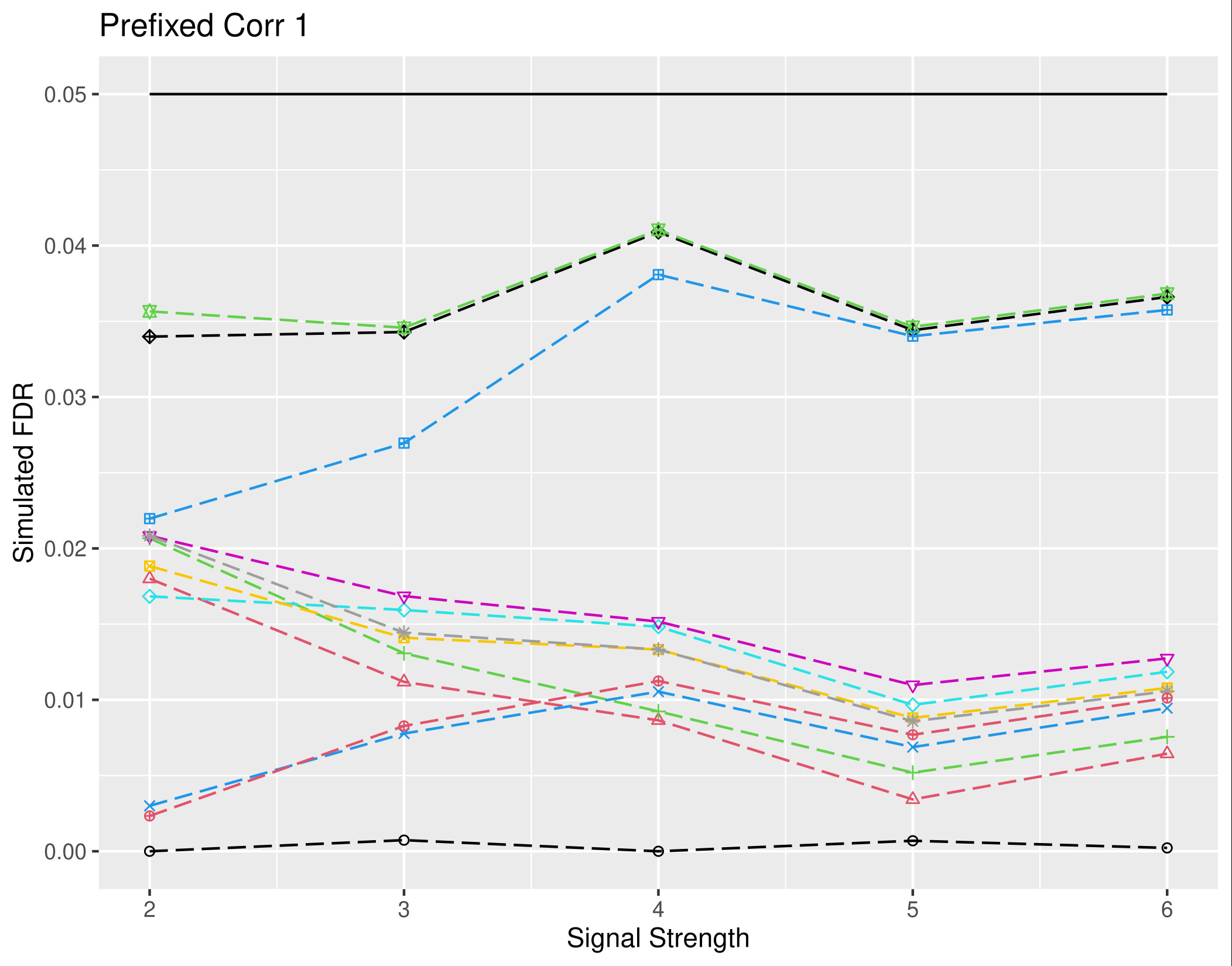} 
      \end{subfigure} 
  & \begin{subfigure}[b]{\linewidth}
      \centering
      \includegraphics[width=\linewidth]{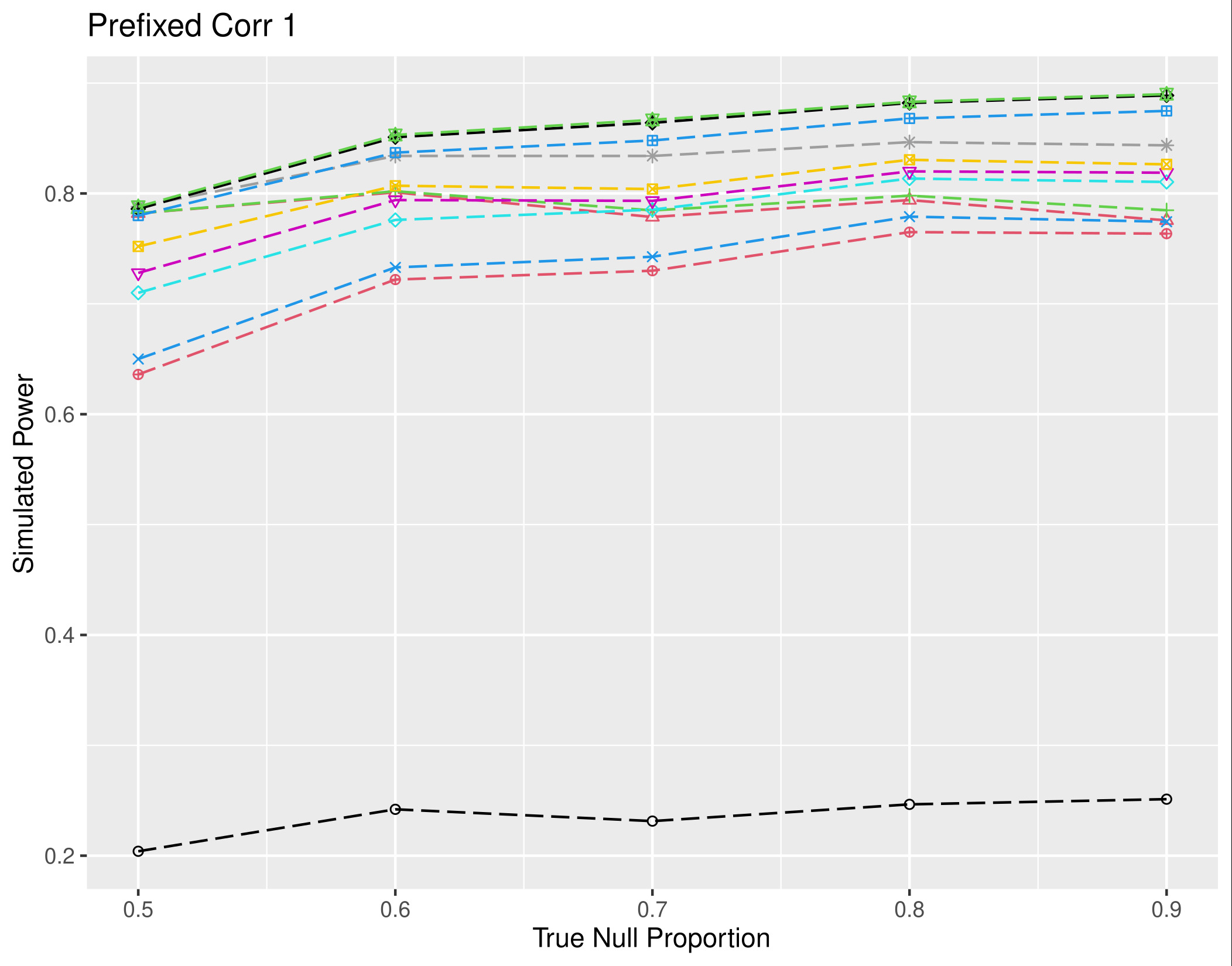} 
      \end{subfigure}    
  \end{tabular*} 
  \caption{Simulated Power (left column), simulated FDR (middle column) for fixed null proportion and simulated power (right column) for fixed signal strength, displayed for mean testing of $d=40$ parameters. Methods compared are SBH1 method (Circle and black), SBH2 method (Triangle point up and red), GSBH1 method (Plus and green), GSBH2 (Cross and blue), GSBH3 (Diamond and light blue), GSBH4 (Triangle point down and purple), GSBH5 (Square cross and yellow), GSBH6 (Star and grey), BH (Diamond plus and black), BY (Circle plus and red), dBH (Triangles up and down and green) and dBY (Square plus and blue)}
  \label{means:figure2} 
\end{figure}

Figure \ref{means:figure2} further demonstrates that GSBH procedures exhibit greater stability than SBH1 and SBH2 in complex correlation settings. Using the minimum of the squared pairwise correlations as a global shift tends to yield higher power than using the maximum. Across varying proportions of true signals, the harmonic mean–based shift (GSBH6) performs particularly well, while the median-based shift (GSBH3) provides an appealing balance of simplicity and accuracy.
Extensive simulations in \hyperref[sec:appendixb]{Appendix B} elucidates that median- and harmonic-based shifts, as well as GSBH1, match or outperform BH as signal prevalence increases.

These findings motivate extending the framework to variable selection in regression.

\subsubsection{Variable Selection}
\label{subsubsec:variable selection}
As outlined in Section \ref{sec:variable selection}, the linear regression model can be naturally viewed as a multiple testing problem, with each coefficient tested for nullity.
We generate the design matrix $\bs{X} \in \mathbb{R}^{n \times d}$ with rows independently drawn from a $d$-variate normal distribution with correlation matrix $\bs{\Sigma}$.  Columns of $\bs{X}$ are standardized to unit norm, ensuring orthonormal scaling.

Correlation structures mirror those of Section \ref{subsubsec:means}, allowing direct comparison between the two inferential settings.
Two dimensional regimes are examined: ($n,d$) = ($100,40$) and ($250,100$), with $\alpha = 0.05$.

\begin{figure}[ht!]
  \begin{tabular*}{\textwidth}{
    @{}m{0.5cm}
    @{}m{\dimexpr0.33\textwidth-0.25cm\relax}
    @{}m{\dimexpr0.33\textwidth-0.25cm\relax}
    @{}m{\dimexpr0.33\textwidth-0.25cm\relax}}
  \rotatebox{90}{Equi(0.3)}
  & \begin{subfigure}[b]{\linewidth}
      \centering
      \includegraphics[width=\linewidth]{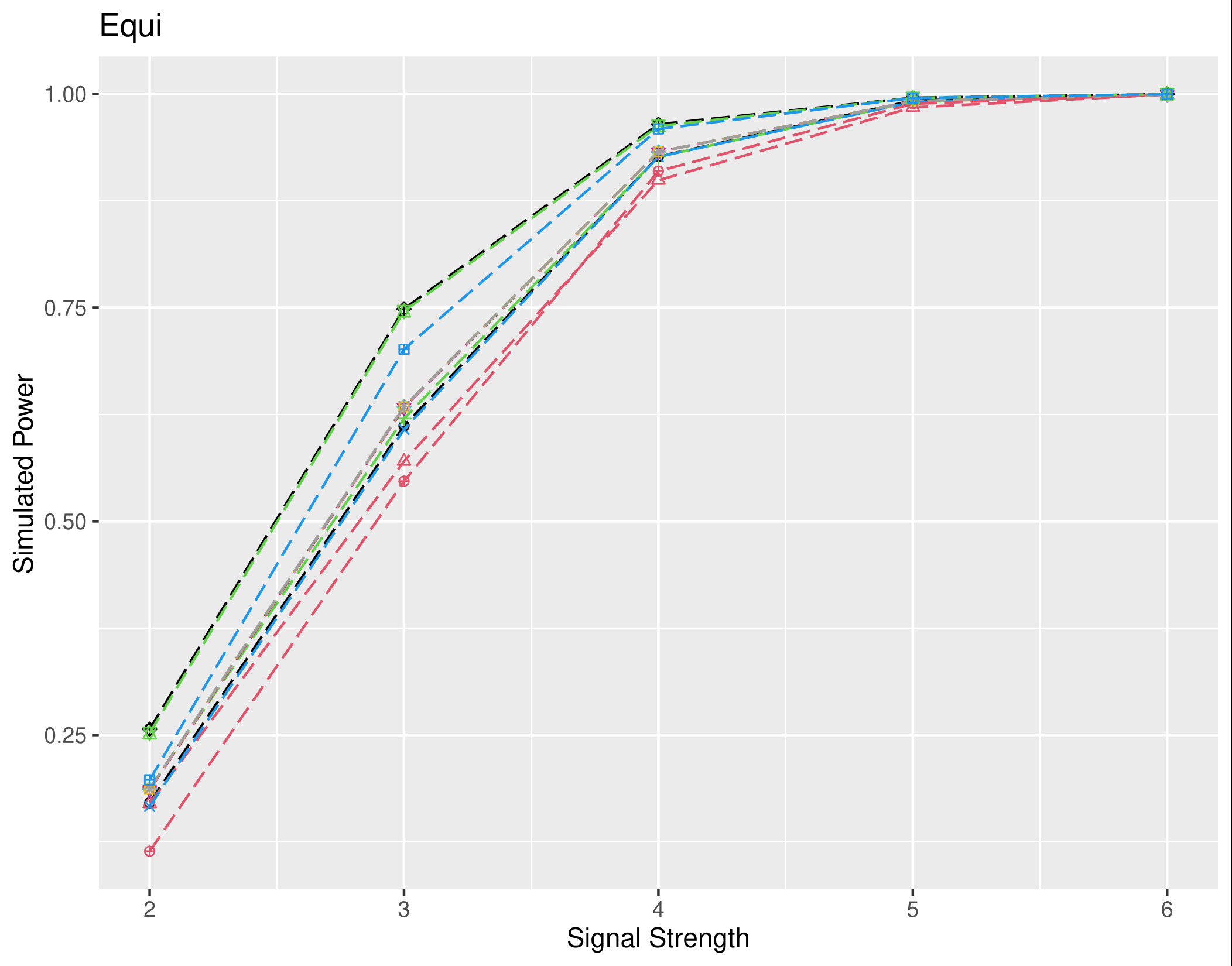} 
      \end{subfigure}  
  & \begin{subfigure}[b]{\linewidth}
      \centering
      \includegraphics[width=\linewidth]{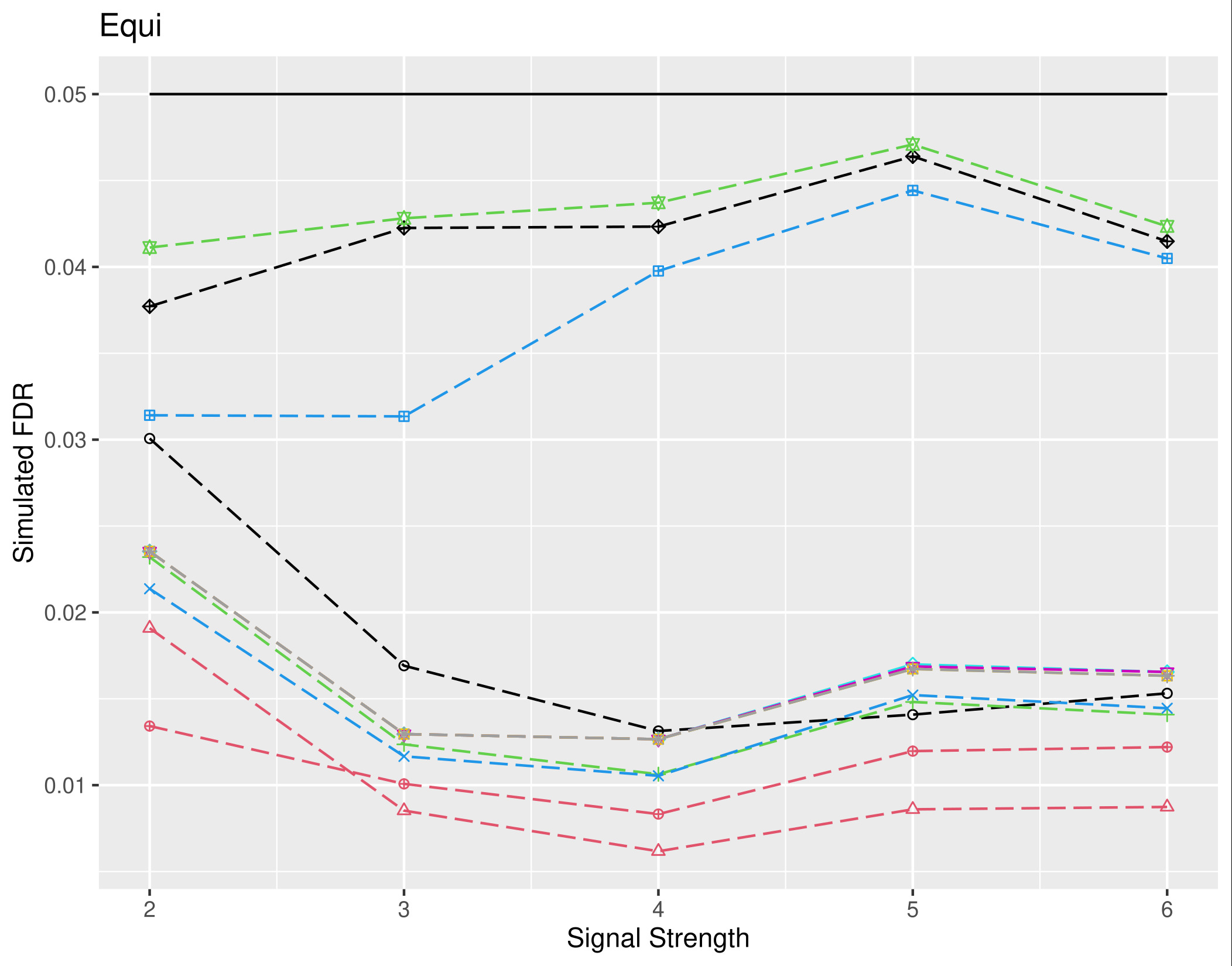} 
      \end{subfigure} 
  & \begin{subfigure}[b]{\linewidth}
      \centering
      \includegraphics[width=\linewidth]{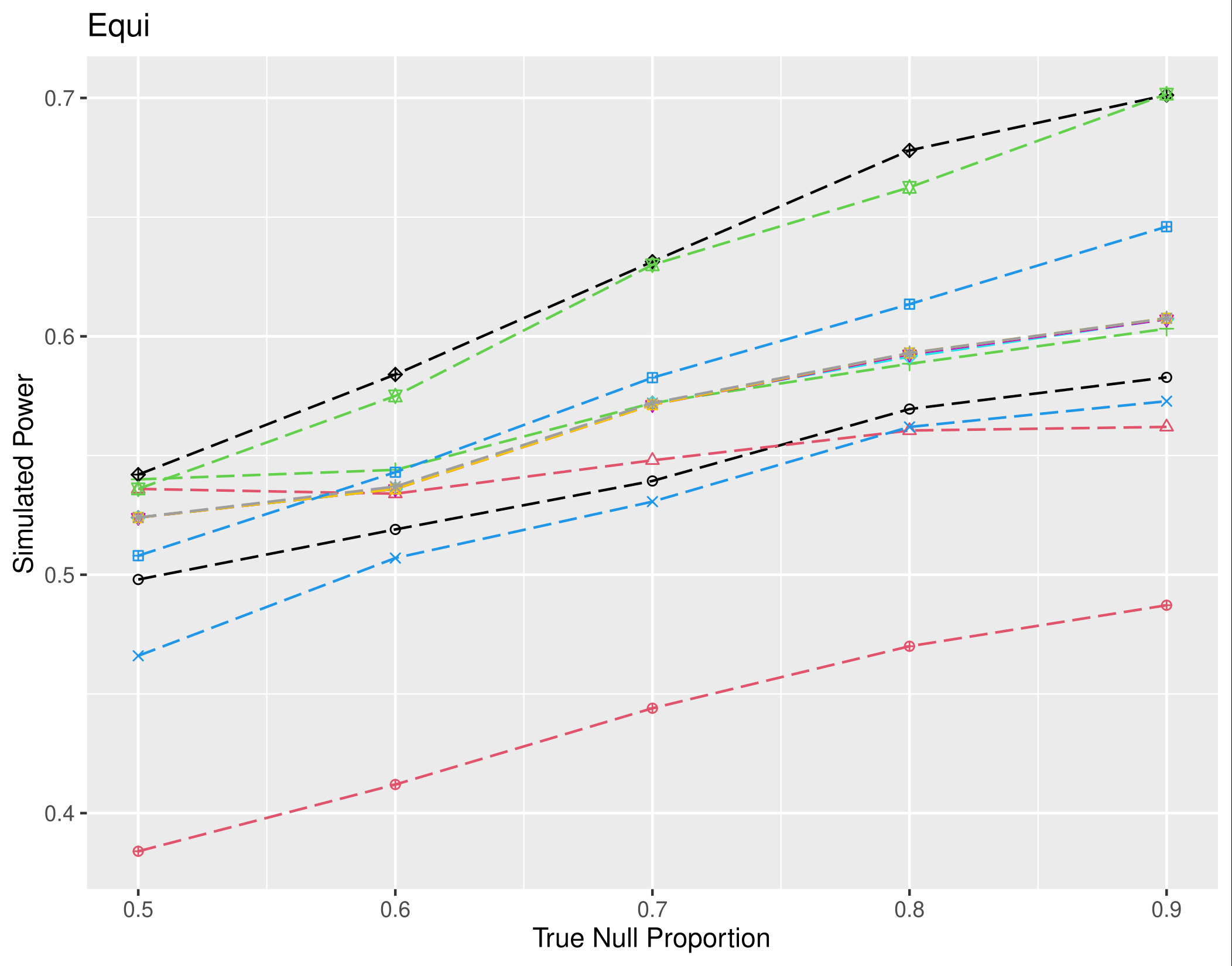} 
      \end{subfigure} \\
  \rotatebox{90}{AR(0.3)} 
  & \begin{subfigure}[b]{\linewidth}
      \centering
      \includegraphics[width=\linewidth]{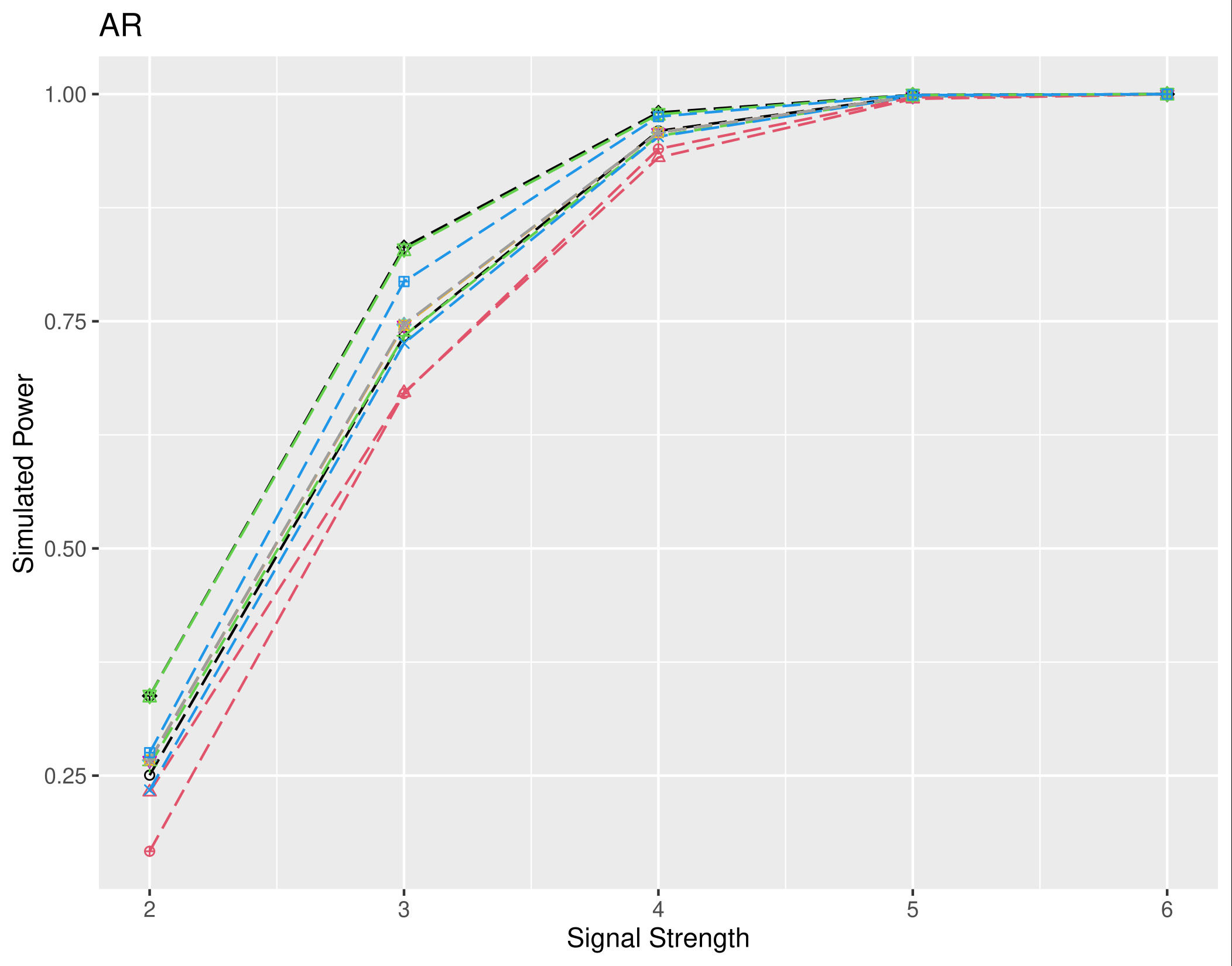} 
      \end{subfigure}  
  & \begin{subfigure}[b]{\linewidth}
      \centering
      \includegraphics[width=\linewidth]{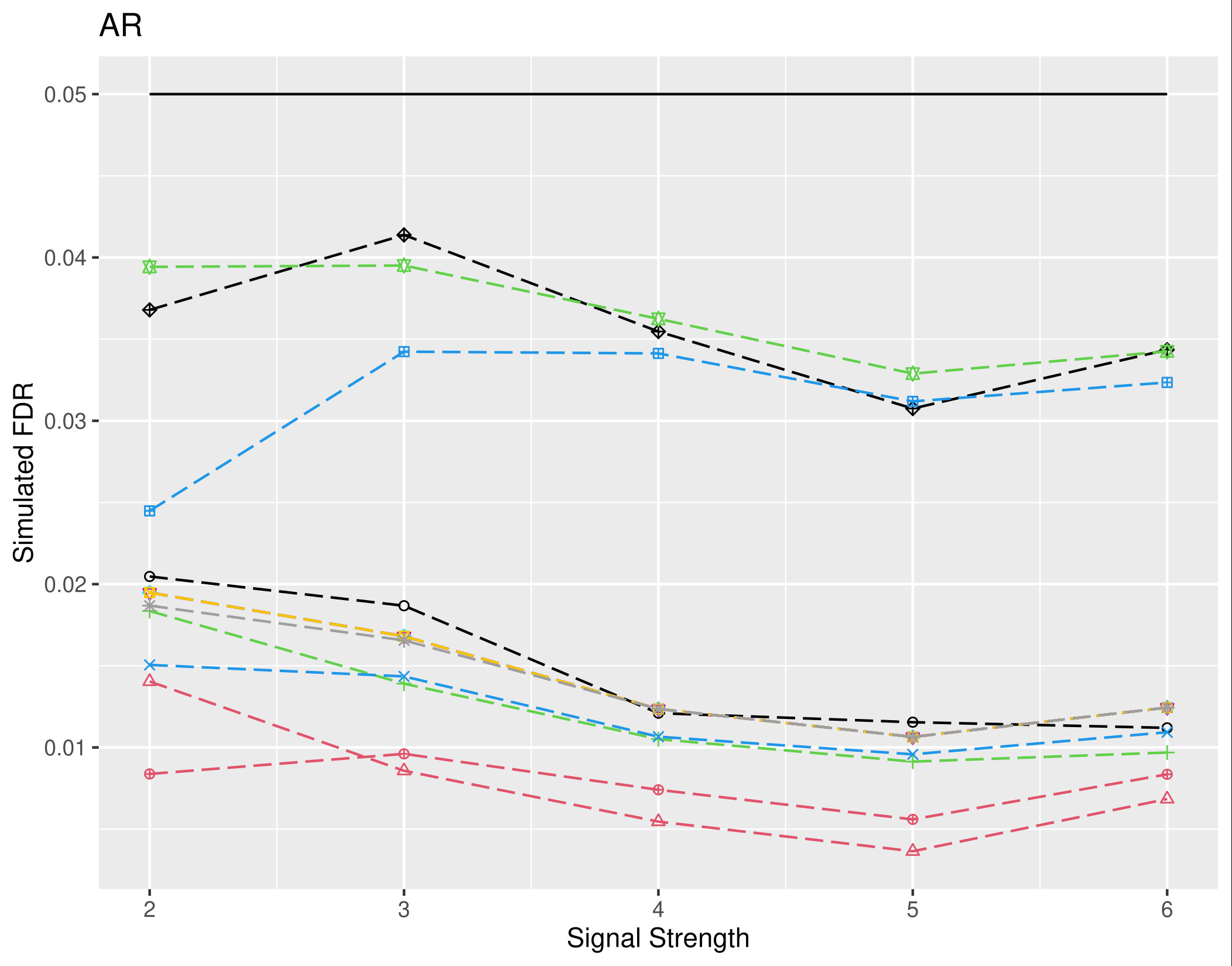} 
      \end{subfigure} 
  & \begin{subfigure}[b]{\linewidth}
      \centering
      \includegraphics[width=\linewidth]{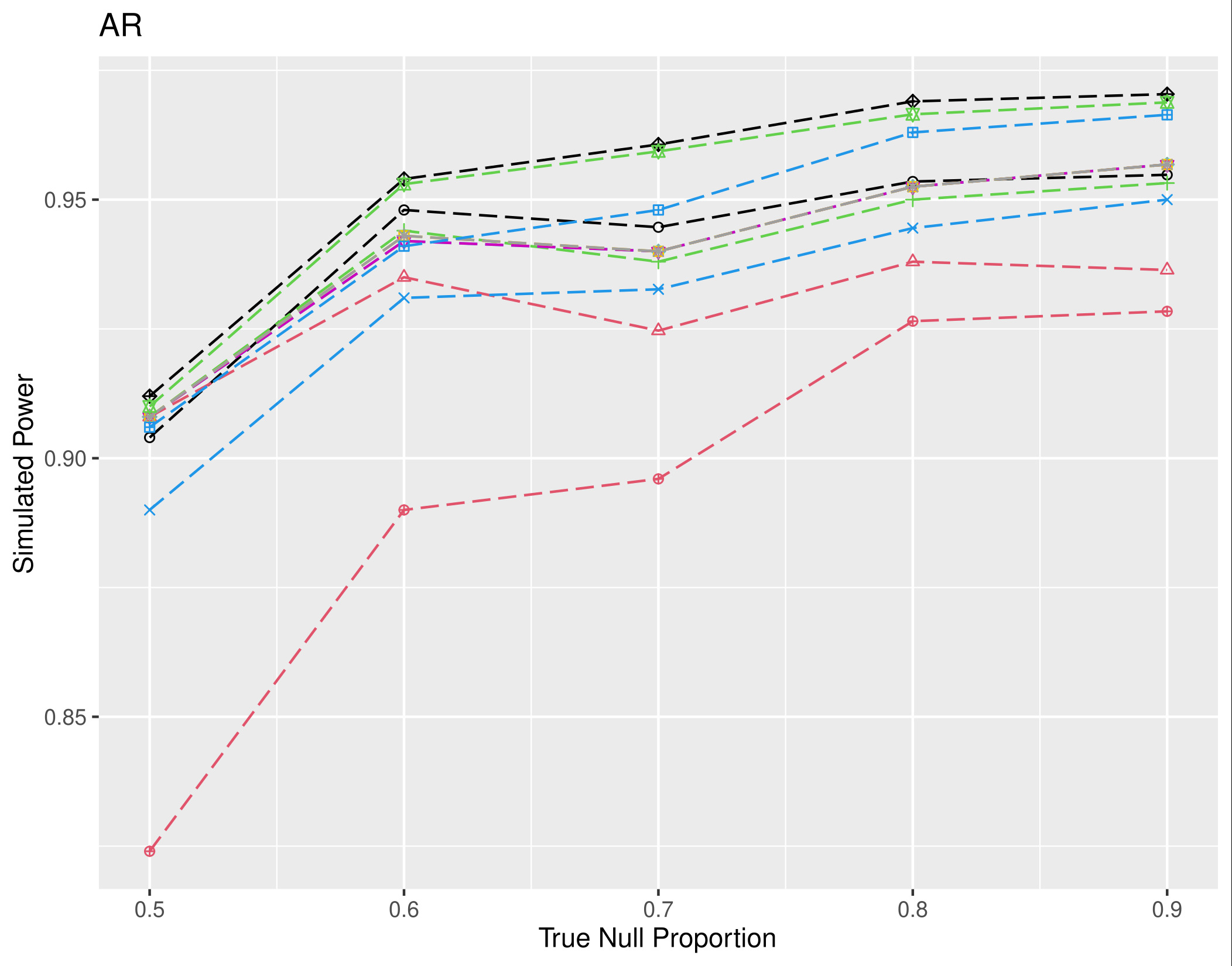} 
      \end{subfigure} \\
  \rotatebox{90}{IAR(0.3)} 
  & \begin{subfigure}[b]{\linewidth}
      \centering
      \includegraphics[width=\linewidth]{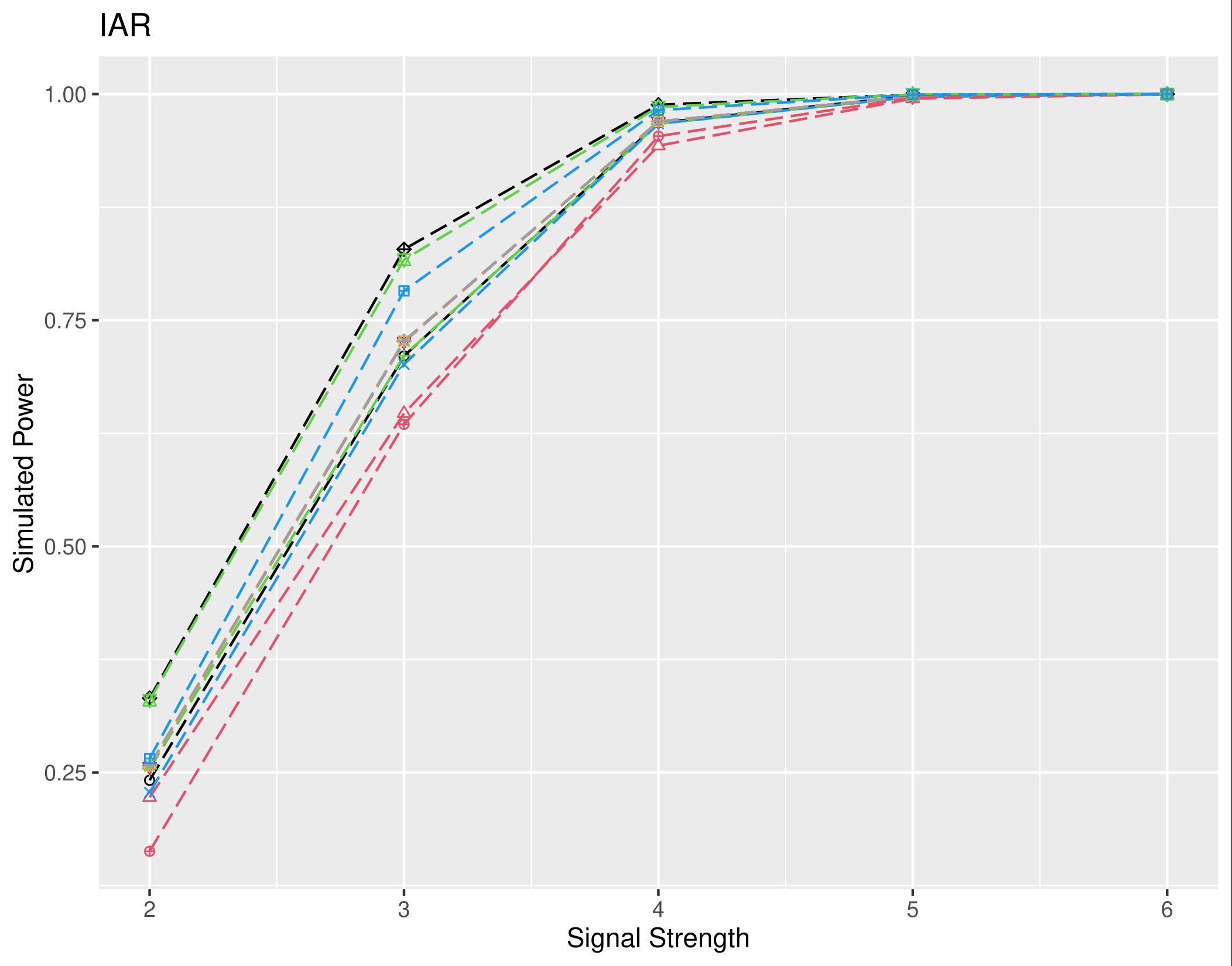} 
      \end{subfigure}  
  & \begin{subfigure}[b]{\linewidth}
      \centering
      \includegraphics[width=\linewidth]{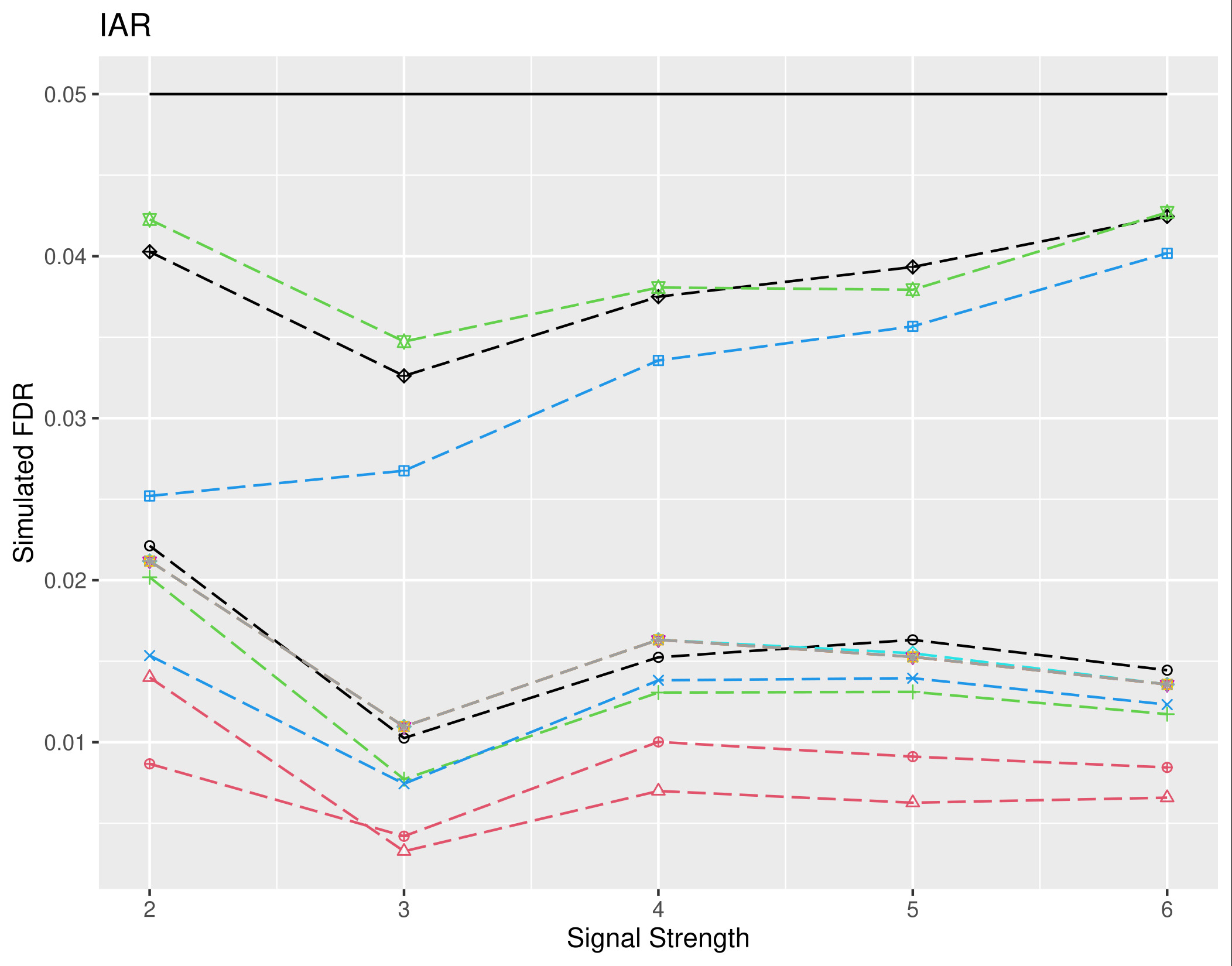} 
      \end{subfigure} 
  & \begin{subfigure}[b]{\linewidth}
      \centering
      \includegraphics[width=\linewidth]{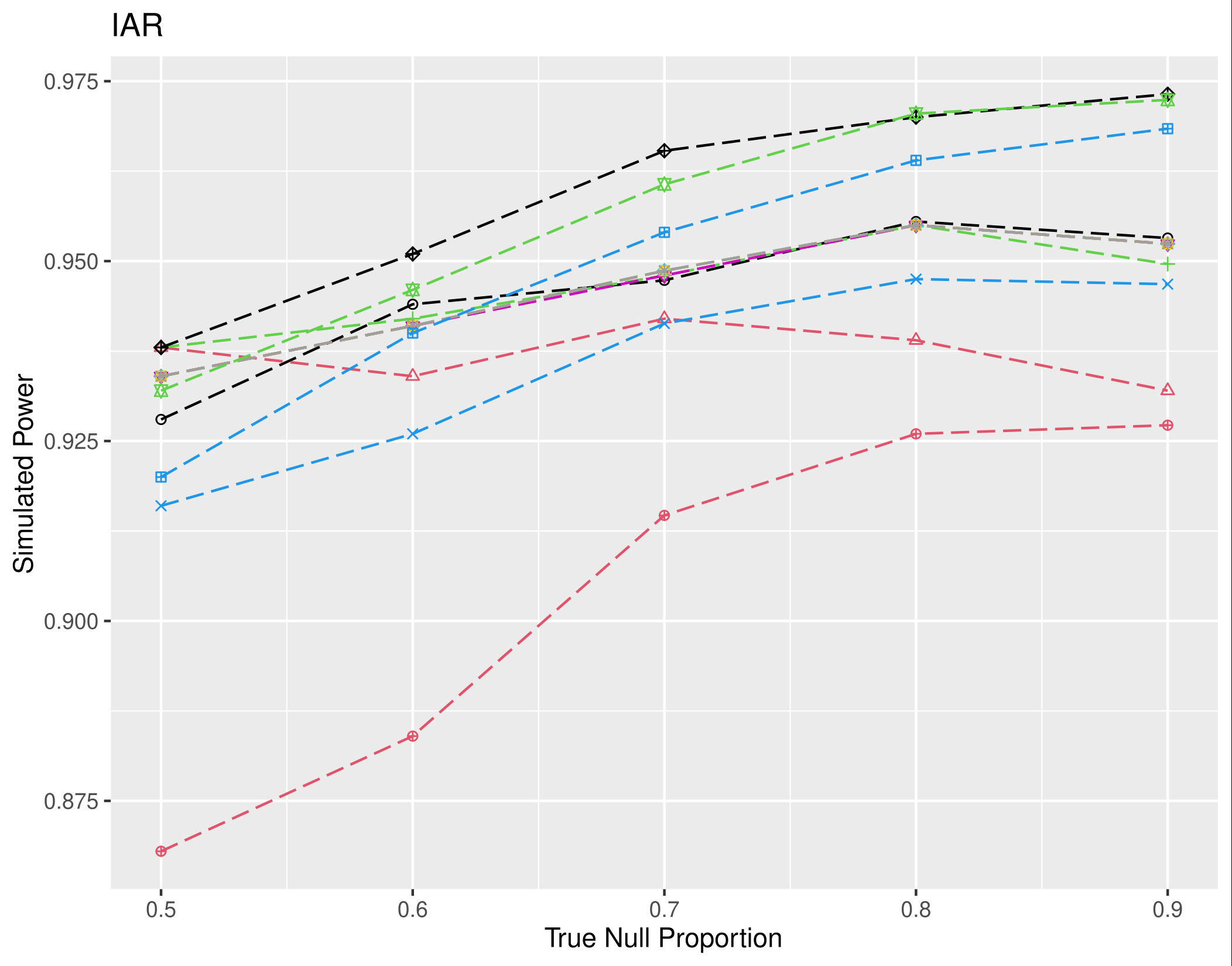} 
      \end{subfigure}    
  \end{tabular*} 
  \caption{Simulated Power (left column), simulated FDR (middle column) for fixed null proportion and simulated power (right column) for fixed signal strength, displayed for variable selection from $d=40$ parameters. Methods compared are SBH1 method (Circle and black), SBH2 method (Triangle point up and red), GSBH1 method (Plus and green), GSBH2 (Cross and blue), GSBH3 (Diamond and light blue), GSBH4 (Triangle point down and purple), GSBH5 (Square cross and yellow), GSBH6 (Star and grey), BH (Diamond plus and black), BY (Circle plus and red), dBH (Triangles up and down and green) and dBY (Square plus and blue)}
  \label{var_sel:figure1} 
\end{figure}

Figure \ref{var_sel:figure1} and supplementary figures (in \hyperref[sec:appendixb]{Appendix B}) ascertain that the proposed methods retain strong signal detection while maintaining reliable FDR control.
Across all designs, GSBH methods match or outperform BY, echoing results from the mean-testing simulations.
Among them, GSBH1 (based on the minimum shift $\tau_{\min}$) performs robustly, approaching BH’s power in some settings.
The median (GSBH3) and harmonic mean (GSBH6) variants outperform other shift-based methods across diverse correlation structures, reflecting the flexibility of our generalization framework. We therefore recommend GSBH3 and GSBH6 as practical default choices, and GSBH1 as a theoretically grounded option under the PLTDN condition.

\bigskip
\noindent\textbf{Knockoff-Assisted Framework:} To test the versatility of our approach, we embed the proposed procedures in a knockoff-assisted framework. We generate design matrices 
$\bs{X}$ with rows drawn independently from $d$-dimensional standard multivariate normal with three different structured correlation matrices -- equicorrelated, AR(1), and inverse AR(1) -- with $\rho \in \{0.3, 0.7\}$. Knockoff variables $\tilde{\bs{X}}$ are generated from the equation $\bs{\tilde{X}} = \bs{XA^{-1}(A-D)} + \bs{\tilde{U}C}$ where the columns of $\bs{\tilde{U}} \in \mathbb{R}^d$ are orthonormal and orthogonal to column space of $\bs{X}$ (cf. \cite{bc2015}).

We evaluate the performance of our methods against several existing procedures, including the Bonferroni-BH (BBH) and its adaptive variant (Adapt-BBH) proposed in \cite{st2022}, as well as the original knockoff filter of \cite{bc2015}. Additionally, we consider three hybrid procedures: 

\noindent $\bullet$ Rev-BBH: Initial selection based on the independent $p$-values followed by BH adjustment on the dependent set; 

\noindent $\bullet$ BBY: Initial Bonferroni selection followed by BY adjustment; 

\noindent $\bullet$ BH on the independent set of $p$-values.

Simulation settings considered include $(n, d) = (100, 40)$ with $d/5$ non-null effects and target FDR levels $\alpha \in \{0.05, 0.1, 0.2\}$. Incorporating the Shifted BH framework into the Bonferroni-BH paired one, we had already defined the SBBH class of methods. Members of this group would be SBBH1 and SBBH2 which correspond to SBH1 in the modified GSBH class and SBH2 in the GSBH class. Among others, we consider the promising ones from \ref{subsubsec:means}: SBBH3 ($\tau=\tau_{min}$), SBBH4 ($\tau=\tau_{med}$) and SBBH5 ($\tau=\tau_{har}$) which respectively correspond to GSBH1, GSBH3 and GSBH6 from the GSBH class.

\begin{figure}[ht!]
  \begin{tabular*}{\textwidth}{
    @{}m{0.5cm}
    @{}m{\dimexpr0.50\textwidth-0.25cm\relax}
    @{}m{\dimexpr0.50\textwidth-0.25cm\relax}}
  \rotatebox{90}{Equi(0.3)}
  & \begin{subfigure}[b]{\linewidth}
      \centering
      \includegraphics[width=0.8\linewidth]{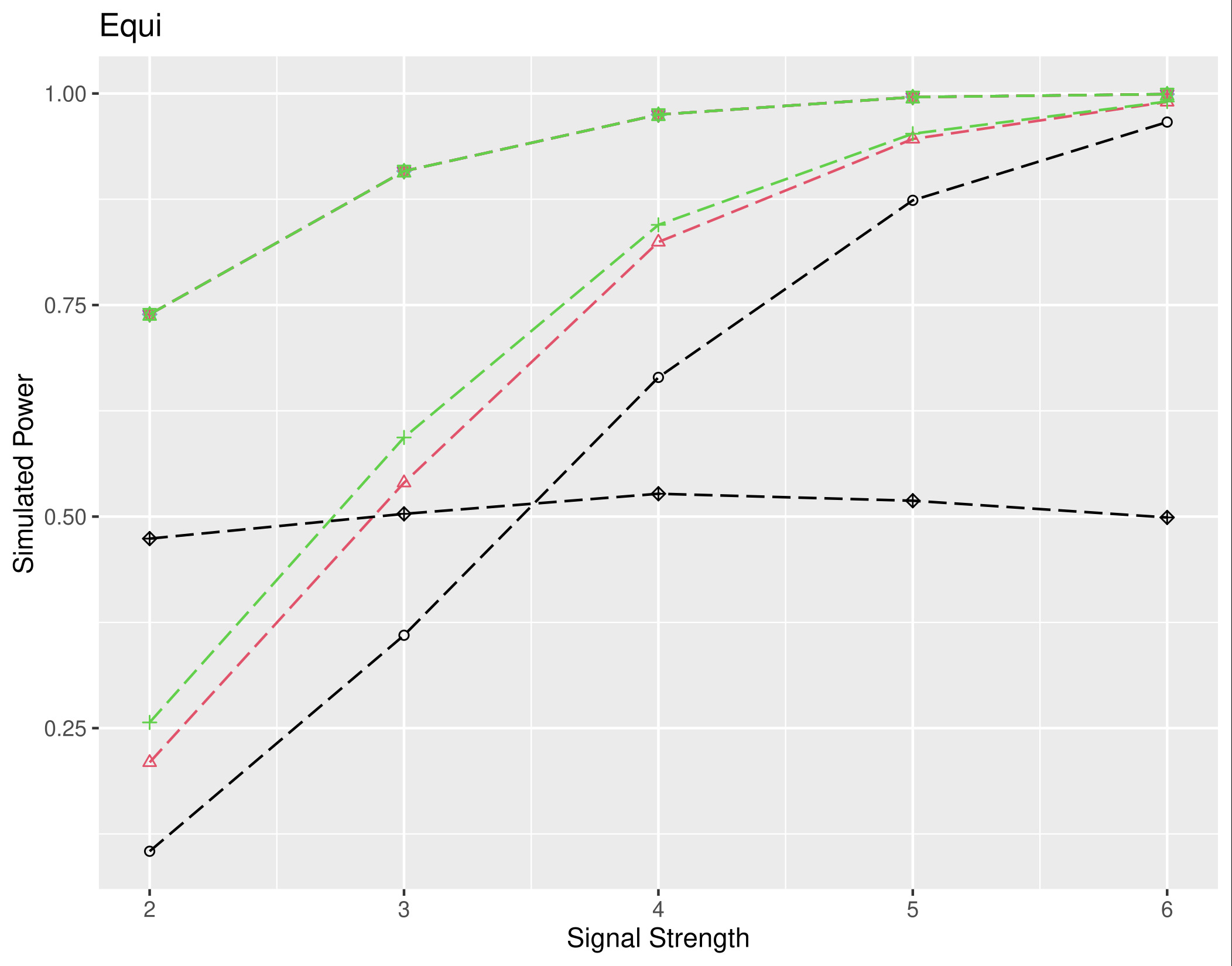} 
      \end{subfigure}  
  & \begin{subfigure}[b]{\linewidth}
      \centering
      \includegraphics[width=0.8\linewidth]{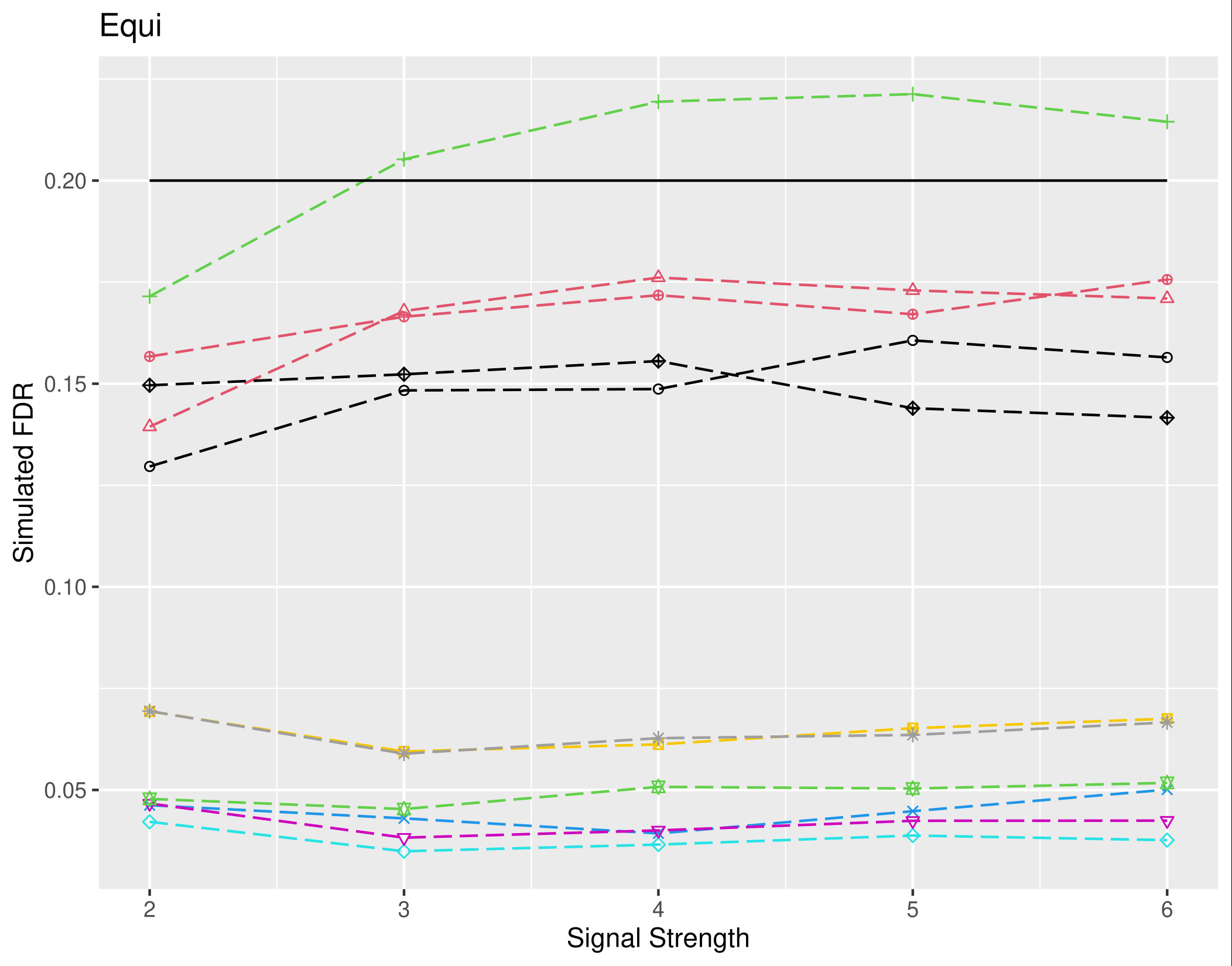} 
      \end{subfigure} \\
  \rotatebox{90}{AR(0.3)} 
  & \begin{subfigure}[b]{\linewidth}
      \centering
      \includegraphics[width=0.8\linewidth]{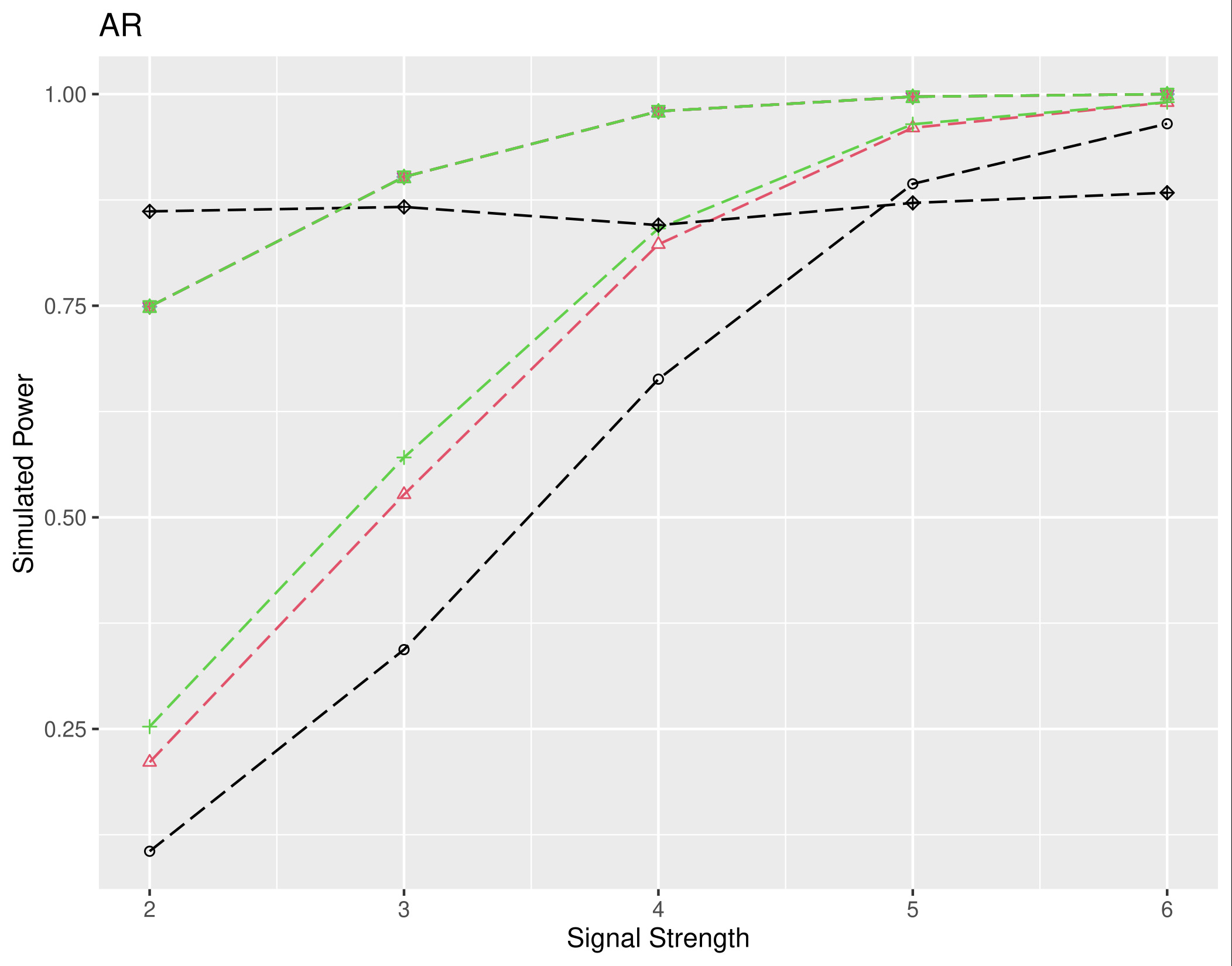} 
      \end{subfigure}  
  & \begin{subfigure}[b]{\linewidth}
      \centering
      \includegraphics[width=0.8\linewidth]{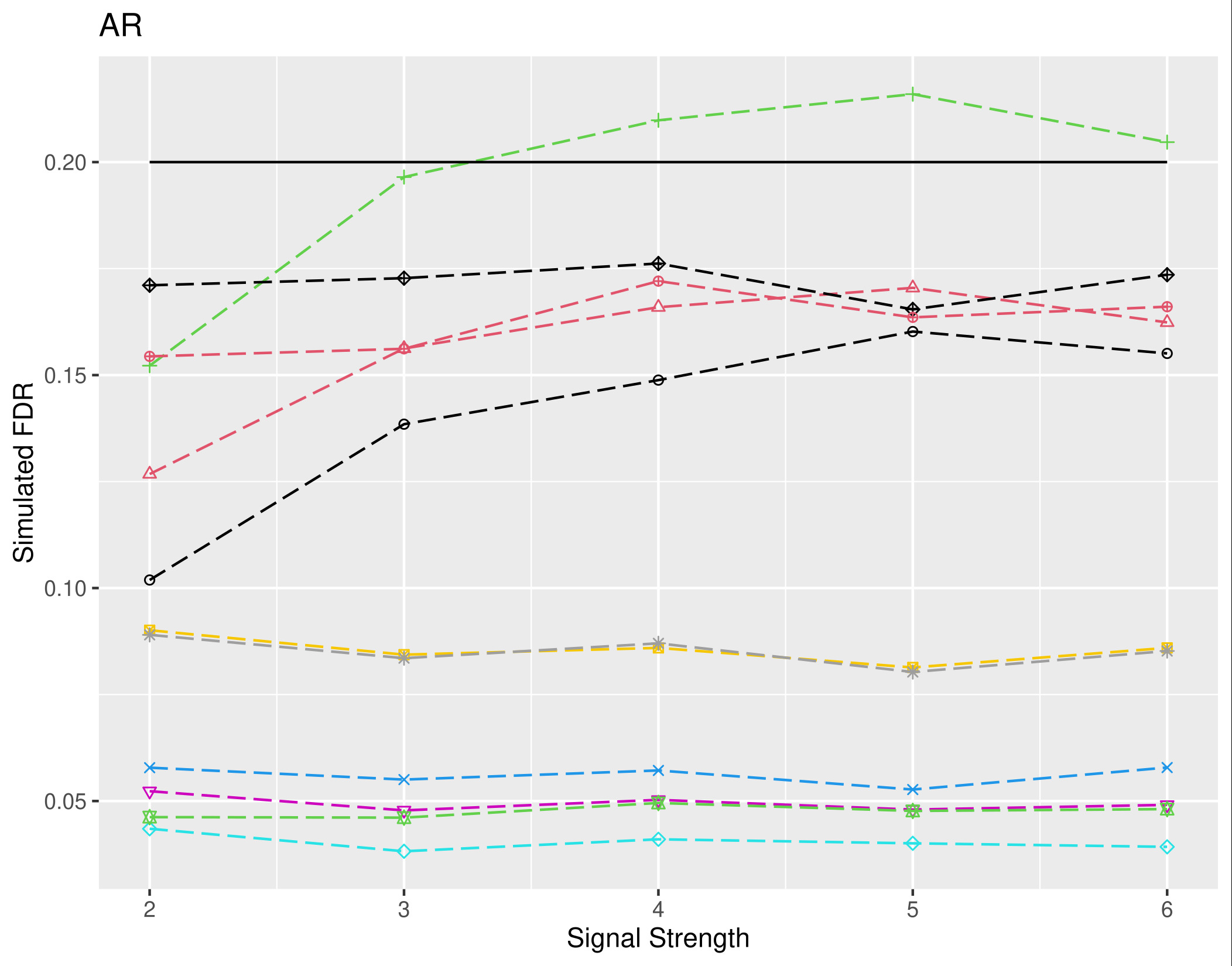} 
      \end{subfigure} \\
  \rotatebox{90}{IAR(0.3)} 
  & \begin{subfigure}[b]{\linewidth}
      \centering
      \includegraphics[width=0.8\linewidth]{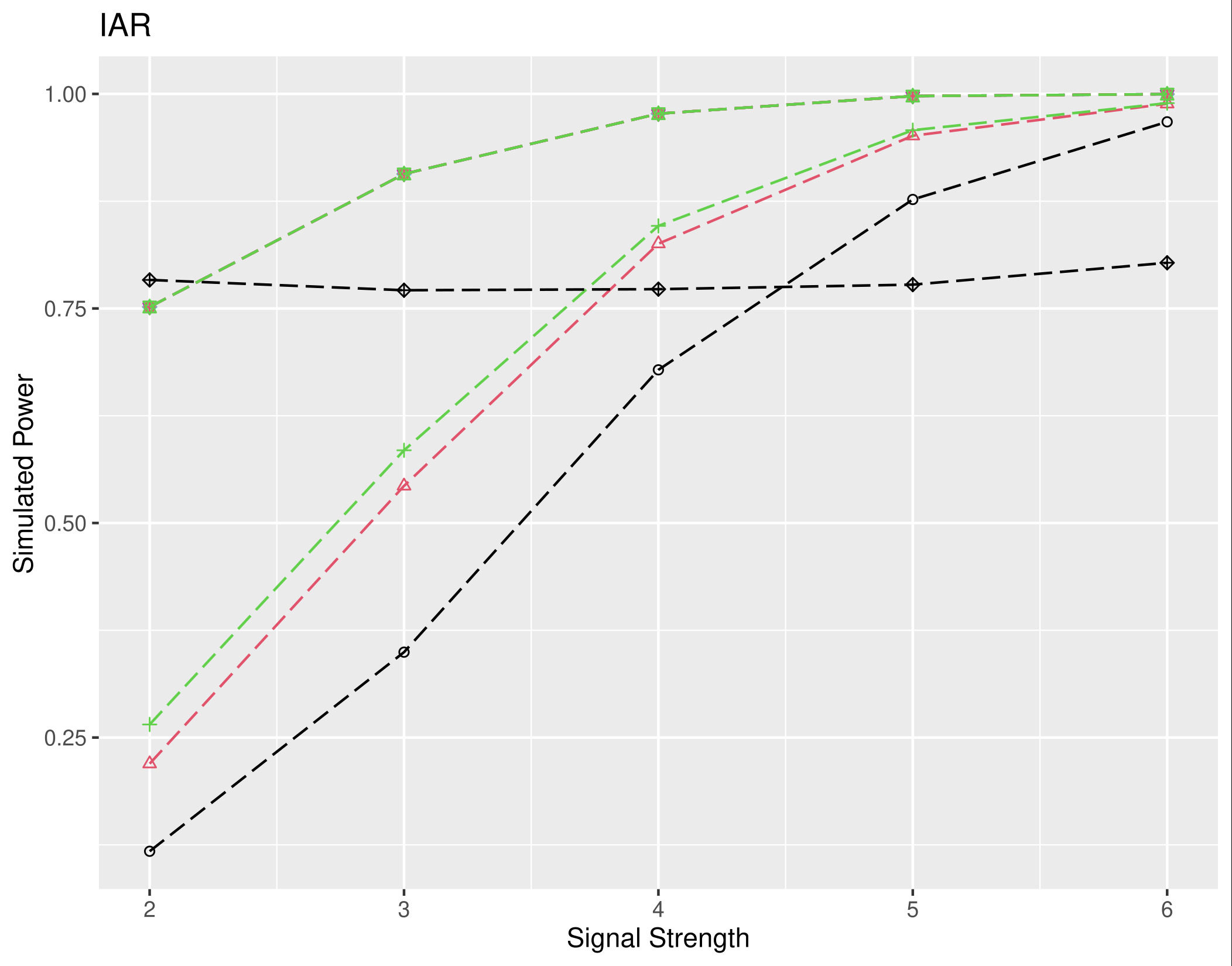} 
      \end{subfigure}  
  & \begin{subfigure}[b]{\linewidth}
      \centering
      \includegraphics[width=0.8\linewidth]{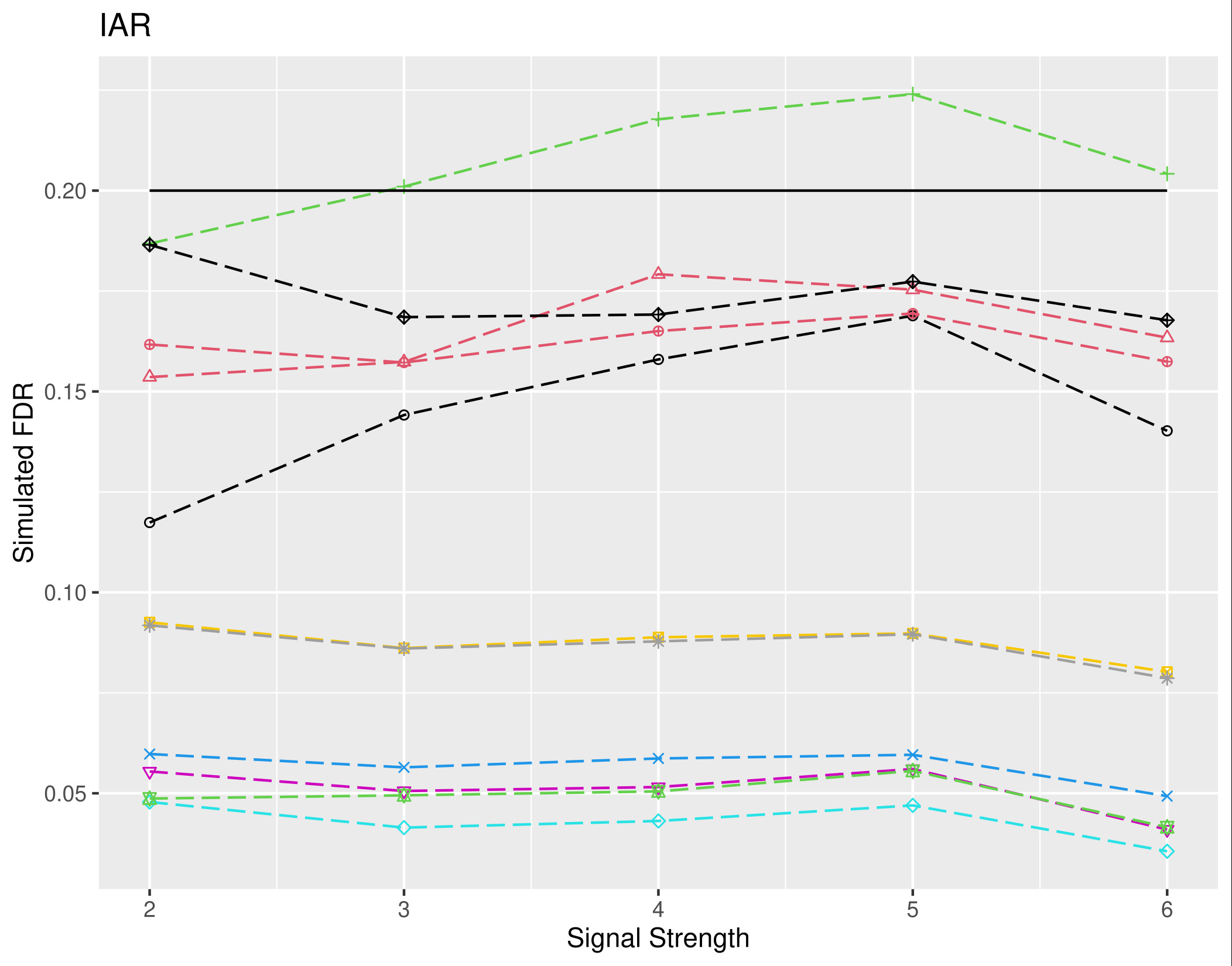} 
      \end{subfigure}    
  \end{tabular*} 
  \caption{Simulated Power (left column) and simulated FDR (right column) displayed for knockoff-assisted variable selection from $d=40$ parameters at a level of $\alpha=0.2$. Methods compared are BH method (Circle and black), BBH method (Triangle point up and red), Adapt-BBH method (Plus and green), SBBH1 (Cross and blue), SBBH2 (Diamond and light blue), SBBH3 (Triangle point down and purple), SBBH4 (Square cross and yellow), SBBH5 (Star and grey), Knockoff (Diamond plus and black), Rev-BBH (Circle plus and red) and BBY (Triangles up and down and green)}
  \label{knockoff:figure1} 
\end{figure}

The comparative performance of the proposed SBBH methods against existing procedures, including BBH and its adaptive variant (ABBH), is presented in Figure \ref{knockoff:figure1}, with additional results provided in \hyperref[sec:appendixb]{Appendix B}. While SBBH1 shows some loss of power under strong inter-variable correlations, often performing slightly below BBH, the other SBBH variants achieve more balanced performance, maintaining conservative FDR control while exhibiting markedly higher power. Notably, although the original knockoff filter of \cite{bc2015} begins to surpass BBH at higher significance levels (e.g., $\alpha = 0.2$), it still falls short of the proposed methods in both detection power and stability.

The Rev-BBH procedure, which combines independent and dependent 
p-value sets within a paired testing framework, demonstrates improved power and can be effective in identifying true signals. However, this approach lacks formal theoretical guarantees to prevent the inclusion of spurious discoveries, thereby limiting its interpretability in rigorous inferential settings and sharing the same fundamental shortcomings as the original BH method. In contrast, the BBY procedure provides a conservative reference point, while the proposed SBBH methods represent principled extensions that offer formal FDR control at arbitrary target levels.

The power curves for SBBH2–SBBH5 generally fall between, or partially overlap with, those of Rev-BBH and BBY, suggesting a form of near-optimal performance within the knockoff-assisted multiple testing framework. These results highlight the flexibility and robustness of the generalized shift-based BH methodology in handling complex dependence structures.

\subsection{Real Data Analysis}
\label{subsec:real data}

To assess the practical performance of the proposed methods, we apply them to a real-world HIV drug resistance dataset originally analyzed in \cite{rhee2006}. The study investigates mutations in the protease gene associated with resistance to protease inhibitors (PIs) used in antiretroviral therapy. The dataset contains binary genotype information indicating the presence or absence of specific mutations, along with corresponding phenotypic resistance measurements. Following the preprocessing steps of \cite{bc2015}, resistance values were log-transformed to better approximate normality across the seven PIs examined. The design matrix $\bs{X}$ comprises binary entries, where 
$X_{ij} = 1$ denotes the presence of mutation $j$ in sample $i$. The goal is to identify mutations significantly associated with resistance phenotypes, using the treatment-selected mutation (TSM) panels from \cite{rhee2006} as a biological benchmark.

In this analysis, we considered the leading procedures from Sections \ref{subsubsec:means} and \ref{subsubsec:variable selection}, specifically GSBH1 (renamed as SBH3), GSBH3 (renamed as SBH4), and GSBH6 (renamed as SBH5), along with their corresponding BBH-type analogues. These were compared against SBH1 and SBH2 from \cite{sz2025}, the classical BH and BY procedures, their dependence-adjusted versions from \cite{fl2022}, the knockoff filter of \cite{bc2015}, and both the original and adaptive BBH methods from \cite{st2022}. All procedures were evaluated under FDR control at nominal levels $\alpha \in \{0.05, 0.1, 0.2\}$.

\begin{figure}[ht!]
\begin{tabular*}{\textwidth}{
    @{}m{0.5cm}
    @{}m{\dimexpr0.33\textwidth-0.25cm\relax}
    @{}m{\dimexpr0.33\textwidth-0.25cm\relax}
    @{}m{\dimexpr0.33\textwidth-0.25cm\relax}}
  & \begin{subfigure}[b]{\linewidth}
      \centering
      \includegraphics[width=\linewidth]{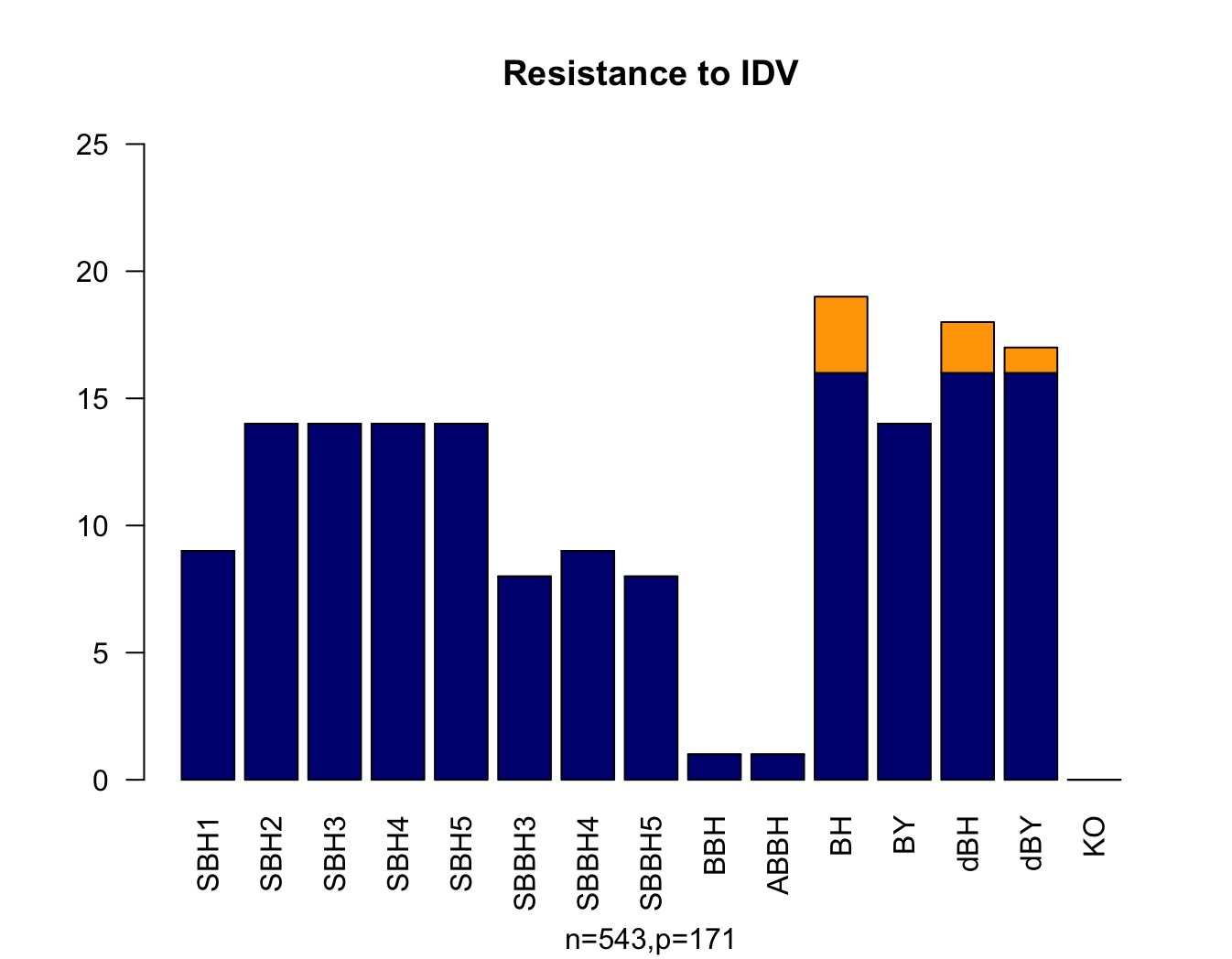}
      \caption{IDV}
      \end{subfigure}  
  & \begin{subfigure}[b]{\linewidth}
      \centering
      \includegraphics[width=\linewidth]{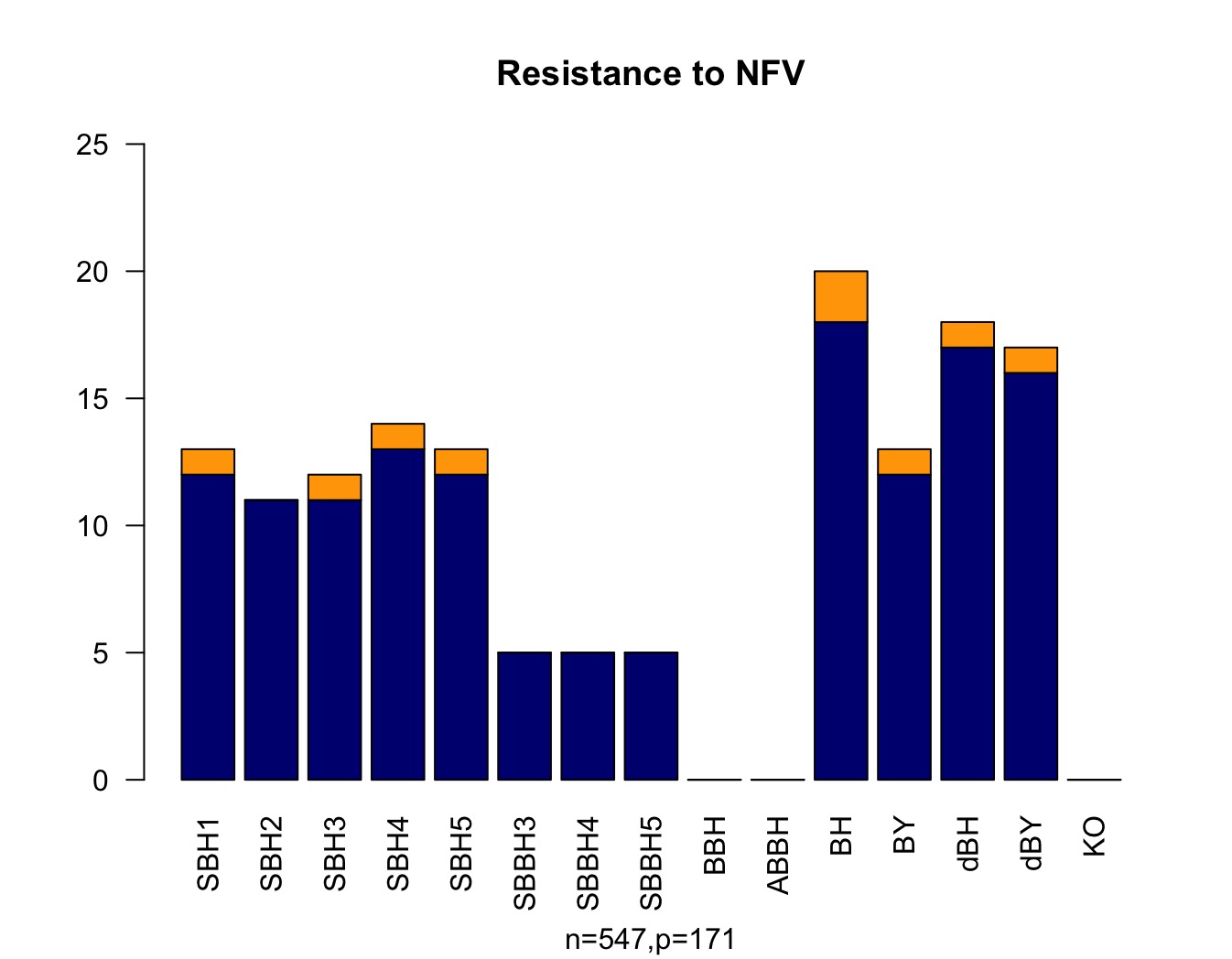}
      \caption{NFV}
      \end{subfigure} 
  & \begin{subfigure}[b]{\linewidth}
      \centering
      \includegraphics[width=\linewidth]{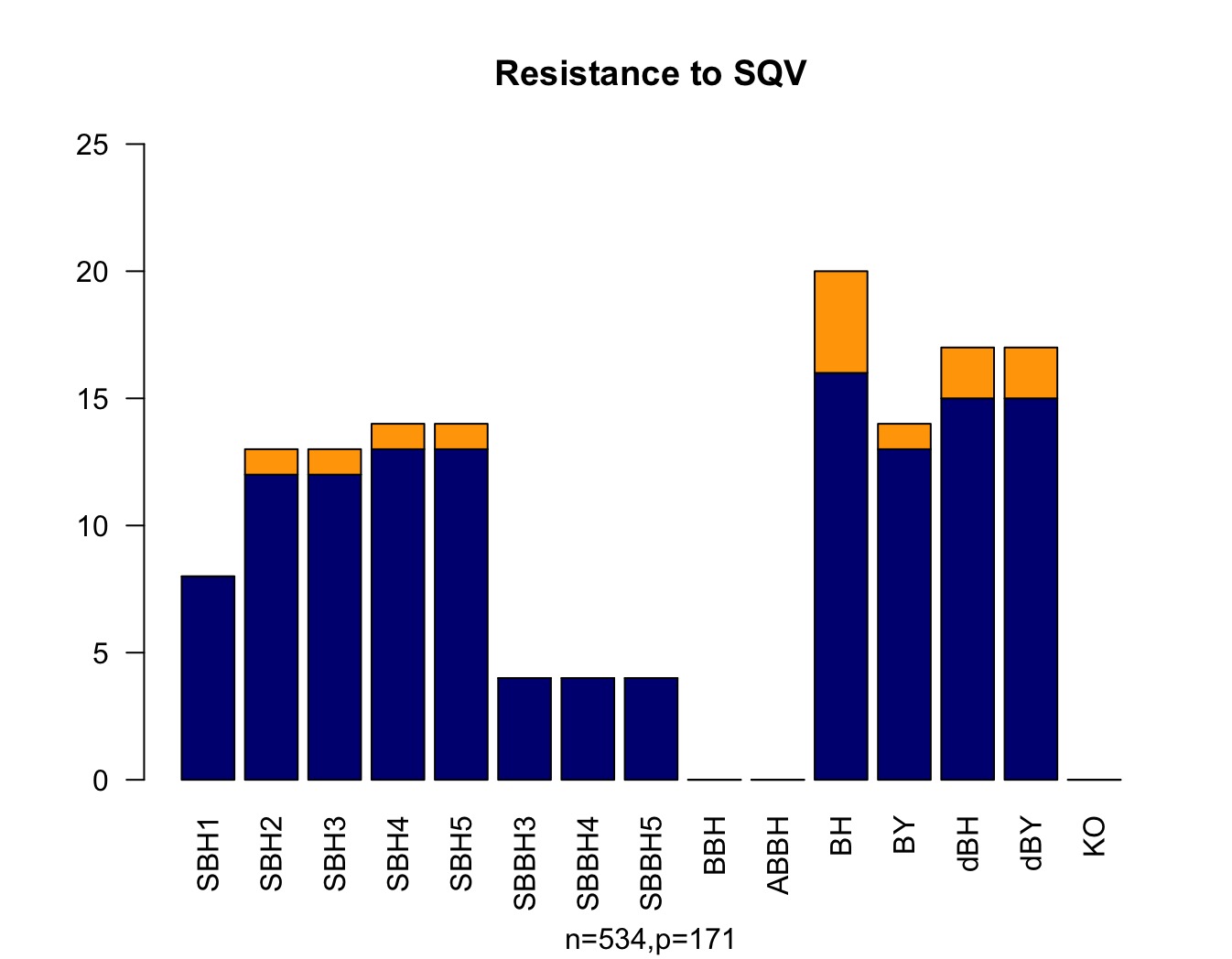}
      \caption{SQV}
      \end{subfigure} \\
\end{tabular*}
\caption{Selected results of drug resistance for $\alpha = 0.05$: (a) resistance to IDV; (b) resistance to NFV; (c) resistance to SQV. Dark blue indicates protease positions that appear in the treatment-selected mutation (TSM) panel for the PI class of treatments, while orange indicates positions selected by the method that do not appear in the TSM list.}
    \label{real:figure1}
    \end{figure}

The proposed methods produced a reasonable number of true discoveries while consistently maintaining control of the false discovery proportion (FDP). Although the total number of rejections was occasionally smaller than that obtained by conventional procedures such as BH or dBH, this difference reflected a principled adherence to the nominal error level $\alpha$, rather than excessive conservatism. In contrast, both BH and dBH failed to control the FDP across all drugs examined, often identifying mutations with no established relevance to drug resistance.

This distinction is particularly important in biomedical contexts, where false associations may lead to misleading or potentially harmful conclusions. In this regard, the proposed methods achieved reliable control of false discoveries while retaining substantial detection power, yielding stable and interpretable results across the panel of protease inhibitors (PIs). The observed tradeoff between discovery rate and rigorous error control underscores the practical robustness and inferential credibility of the GSBH-based procedures in high-stakes scientific applications.

\section{Concluding Remarks}
\label{sec:conclusion}
Multiplicity remains a central challenge in modern statistical inference, particularly under dependence. In this paper, we provide a rigorous framework for FDR control in two-sided Gaussian mean testing, introducing positive left-tail dependence under the null (PLTDN) to extend classical PRDS conditions beyond one-sided tests. Building on this insight, we develop generalized shifted BH (GSBH) and shifted Bonferroni–BH (SBBH) procedures that unify and extend recent dependence-adjusted and shift-based approaches, achieving exact FDR control while remaining computationally efficient.

Our simulations demonstrate that these procedures reliably maintain nominal FDR control across diverse correlation structures and deliver substantial power gains relative to existing methods. Median- and harmonic-mean–based variants, in particular, balance theoretical rigor with empirical performance. In real-data applications, such as HIV drug resistance, the methods identify scientifically meaningful signals while avoiding spurious discoveries, highlighting their practical relevance in dependent testing scenarios.

These results underscore the broader significance of the PLTDN framework. It provides a principled approach to understanding dependence structures under two-sided testing framework that guarantee valid FDR control, with natural extensions to variable selection in regression and other structured inference problems. By accommodating scenarios where the covariance matrix is fully known or known up to an unknown scalar, this work lays the foundation for future extensions to fully unknown covariance structures, further advancing the theory and practice of FDR controlling methodologies.

\newpage
\bibliographystyle{agsm_1}
\bibliography{biblio_1}

\section{Appendix A}
\label{sec:appendixa}
 
\subsection{Proof of Lemma 2 }
\label{proof of Lemma2}

The following result on chi-squared and $F$-distributions will facilitate our proof of this lemma.

\begin{result*}
\label{Result1}
\begin{itemize}
\item [(i).] For any fixed $\theta > 0$ and $h \ge  0$, $\bar{\Psi}_{m+h}(\theta\bar{\Psi}_{m}^{-1}(u))$ is concave (or convex) in $u \in (0,1)$ when $\theta \le 1$ (or $\theta >  1$, $h = 0$). \ 

\item [(ii).] For any fixed $\theta \in (0,1)$ and $h \ge 0$, $\bar{\Psi}_{m+h,n}(\theta\bar{\Psi}_{m,n}^{-1} (u))$ is concave (or convex) in $u \in (0,1)$ when $\theta \le 1$ (or  $\theta > 1$ and $h=0$).\ 
 
\item[(iii).] For any fixed $0<u<u^{\prime}< \infty$, $\frac{\bar{\Psi}_{m}(x \bar{\Psi}_{m}^{-1}(u))}{\bar{\Psi}_{m}(x\bar{\Psi}_{m}^{-1}(u^{\prime}))}\; \uparrow x > 0$.
\end{itemize}
\end{result*} 

Proof of (i). Letting $g(u) = \bar{\Psi}_{m+h}(\theta \bar{\Psi}_m^{-1}(u))$, we note that 
$$ \frac{d}{du}g(u) \propto \exp\{ - \frac{1}{2} (\theta- 1) \bar{\Psi}_m^{-1}(u) \}[\bar{\Psi}_m^{-1}(u)]^{\frac{h}{2}},$$ which is decreasing (or increasing) in $u \in (0,1)$ if $\theta \le 1$ (or if $\theta > 1$ and $h=0$), proving the desired result. 

Proof of (ii). The proof is very similar to that of part (i).  Let $h(u) = \bar{\Psi}_{m+h,n}(\theta \bar{\Psi}_{m,n}^{-1}(u))$. Then, we note that, 
\[ \frac{d}{du}h(u) \propto \left ( \frac{1+\bar{\Psi}_{m,n}^{-1}(u)}{1+\theta \bar{\Psi}_{m,n}^{-1}(u)}\right )^{\frac{m+n}{2}}\left (\frac{\bar{\Psi}_{m,n}^{-1}(u)}{1+ \theta \bar{\Psi}_{m,n}^{-1}(u)}\right )^{\frac{h}{2}} \] ,
which is decreasing (or increasing) in $u \in (0,1)$ if $\theta \le 1$ (or if $\theta > 1$ and $h=0$), proving the desired result.

Proof of (iii). Let $\psi_{m}(x)$  be the density of $\Psi_m $. Since the density of $\theta^{-1} \chi_{m}^2$, which is $ \theta \psi_m(x\theta)$, is totally positive of order 2 (TP$_2$) in $(x, 1/\theta)$, its survival function $\bar{\Psi}_m(x\theta)$ is also TP$_2$ in $(x, 1/\theta)$ (see, e.g., \cite{Karlin1968}). In other words, for any fixed $0 < \theta < \theta^{\prime} < \infty$, the ratio $\textrm{Pr}(\theta^{-1} \chi_{m}^2 \ge x)/\textrm{Pr}({\theta^{\prime}}^{-1} \chi_{m}^2 \ge x) = \bar{\Psi}_m(x\theta^{\prime})/\bar{\Psi}_m(x\theta)$ is decreasing in $x>0$, from which the desired result follows. 

Now, we are ready to prove Lemma \ref{Lemma2}.

Proof of Lemma \ref{Lemma2}: 
Noting that $$X_i^2 \mid \boldsymbol{X}_{-i} \stackrel{d} = \tau_i E_{J_i}(\tau_i\chi_{1+2J_i}^2),$$ 
with the expectation taken with respect to $J_i\mid \boldsymbol{X}_{-i} \sim \textrm{Poisson}(\lambda(\boldsymbol{X}_{-i})/2)$, we have  

\begin {eqnarray} & & \frac{\textrm{Pr}(\ddot{P}_i < u \mid \tilde{\boldsymbol{X}}_{-i})}{u} = \frac { \textrm {Pr}( \frac{1}{\tau_i}X_i^2 \ge \bar{\Psi}_1^{-1}(u) \mid \boldsymbol{X}_{-i})}{u} \nonumber \\ & = & E_{J_i\mid \boldsymbol{X}_{-i}} \left \{\frac{\textrm {Pr}(\chi_{1+2J_i}^2 \ge \bar{\Psi}^{-1}(u))}{u} \right \} = E_{J_i\mid \boldsymbol{X}_{-i}} \left \{\frac{\bar{\Psi}_{1+2J_i}( \bar{\Psi}^{-1}(u))}{u} \right \}. \end{eqnarray} This is decreasing in $u \in (0,1)$, since $\bar{\Psi}_{1+2J_i}(\bar{\Psi}^{-1}(u))$ is concave in (0,1) (from Result \ref{Result1} (i)), for any fixed $\boldsymbol{X}_{-i}$ and hence for any fixed $\boldsymbol{P}_{-i}$. Thus the fact that $\ddot{P}_i$ satisfies Condition \ref{Cond1}a, for each $i$, in Setting \hyperlink{Setting1}{1} is proved.

The fact that the $\ddot{P}_i$'s defined in Setting \hyperlink{Setting2}{2} satisfies Condition \ref{Cond1}b is given below. 

For any $0<u < u^{\prime} < 1$, we have the following, having  expressed $(\ddot{P}_i, \boldsymbol{P}_{-i)}$ in terms of $(X_i, \boldsymbol{X}_{-i})$, 
\begin{eqnarray}
& & \textrm{Pr}(\ddot{P}_i \le u^{\prime}, g(\boldsymbol{P}_{-i}) \ge t) \nonumber \\
& = & E \left \{ \textrm{Pr}(\ddot{P}_i \le u^{\prime} \mid \boldsymbol{X}_{-i}, V) \mathbbm{1} (g(\boldsymbol{X}_{-i}, V )\ge t) \right \} \nonumber \\
& = & E \left [ \left \{ \textrm{Pr}\left (\chi_{1+2J_i}^2 \ge V\bar{\Psi}_{1,\nu}^{-1}(u^{\prime})\right ) \right \} \mathbbm{1}(g(\boldsymbol{X}_{-i},V) \ge t ) \right ] \nonumber \\
& = & E \left \{ \bar{\Psi}_{1+2J_i} \left ( V{\Psi}_{1,\nu}^{-1}(u^{\prime})\right )  \mathbbm{1}(g(\boldsymbol{X}_{-i},V) \ge t ), \right \}\end{eqnarray} with the expectation taken with respect to $(V, J_i, \bs{X}_i)$. 
This expectation, when taken with respect to $V$, conditionally given $(J_i,\boldsymbol{X}_{-i})$, equals
\begin{eqnarray}
\label{proof for Cond1b}
& & E_{V} \left \{ \frac{\bar{\Psi}_{1+2J_i} \left ( V\bar{\Psi}_{1,\nu}^{-1}(u^{\prime})\right )}{\bar{\Psi}_{1+2J_i} \left ( V\bar{\Psi}_{1,\nu}^{-1}(u)\right )} \bar{\Psi}_{1+2J_i} \left (V\bar{\Psi}_{1,\nu}^{-1}(u)\right ) \mathbbm{1}(g(\boldsymbol{X}_{-i},V) \ge t ) \right \}\nonumber \\ & \le & \frac {E_V \left [ \bar{\Psi}_{1+2J_i} \left (V\bar{\Psi}_{1,\nu}^{-1}(u^{\prime})\right ) \right ]}{E_V \left [\bar{\Psi}_{1+2J_i} \left ( V\bar{\Psi}_{1,\nu}^{-1}(u)\right )\right ]} E_V \left \{ \bar{\Psi}_{1+2J_i} \left ( V\bar{\Psi}_{1,\nu}^{-1}(u)\right )\mathbbm{1}((g( \boldsymbol{X}_{-i},V) \ge t ) \right \} \nonumber \\
& = & \frac{\bar{\Psi}_{1+2J_i, \nu}(\bar{\Psi}_{1,\nu}^{-1}(u^{\prime}))}{\bar{\Psi}_{1+2J_i, \nu}(\bar{\Psi}_{1,\nu}^{-1}(u))} E_V \left \{ \bar{\Psi}_{1+2J_i} \left ( V\bar{\Psi}_{1,\nu}^{-1}(u)\right )\mathbbm{1}((g(\boldsymbol{X}_{-i},V) \ge t ) \right \}. \nonumber \\
& \le & \frac{u^{\prime}}{u} E_V\left \{\bar{\Psi}_{1+2J_i} \left ( V\bar{\Psi}_{1,\nu}^{-1}(u)\right )\mathbbm{1}(g(\boldsymbol{X}_{-i},V) \ge t ) \right \} \end{eqnarray}
The first inequality in \ref{proof for Cond1b} follows first noting that $\bar{\Psi}_{1+2J_i} \left ( V\bar{\Psi}_{1,\nu}^{-1}(u^{\prime})\right )/\bar{\Psi}_{1+2J_i} \left ( V\bar{\Psi}_{1,\nu}^{-1}(u)\right )$ and   $\mathbbm{1}(g((\boldsymbol{X}_{-i},V) \ge t )$ are, respectively, increasing (from Result \ref{Result1} (iii)) and decreasing in $V$, and then applying Kimball's inequality. The last inequality follows from Result \ref{Result1} (ii), since $\bar{\Psi}_{1+2J_i, \nu} (\bar{\Psi}_{1,\nu}^{-1}(u))$ is concave in $(0,1)$.

Taking expectation with respect to $(J_i, \boldsymbol{X}_{-i})$ on both sides of \ref{proof for Cond1b}, we have \[\textrm{Pr}(P_i \le u^{\prime}, g(\boldsymbol{P}_{-i}) \ge t) \le \frac{u^{\prime}}{u} \textrm{Pr}(P_i \le u, g(\boldsymbol{P}_{-i}) \ge t), \] the desired decreasing property of $\textrm{Pr}(P_i \le u, g(\boldsymbol{P}_{-i}) \ge t)/u$ in $u \in (0,1)$.


\section{Appendix B}
\label{sec:appendixb}

This section includes further figures visualized in simulations settings and the real data analysis that have been omitted in the main paper.

\begin{figure}[ht!]
  \begin{tabular*}{\textwidth}{
    @{}m{0.5cm}
    @{}m{\dimexpr0.33\textwidth-0.25cm\relax}
    @{}m{\dimexpr0.33\textwidth-0.25cm\relax}
    @{}m{\dimexpr0.33\textwidth-0.25cm\relax}}
  \rotatebox{90}{Equi(0.7)}
  & \begin{subfigure}[b]{\linewidth}
      \centering
      \includegraphics[width=\linewidth]{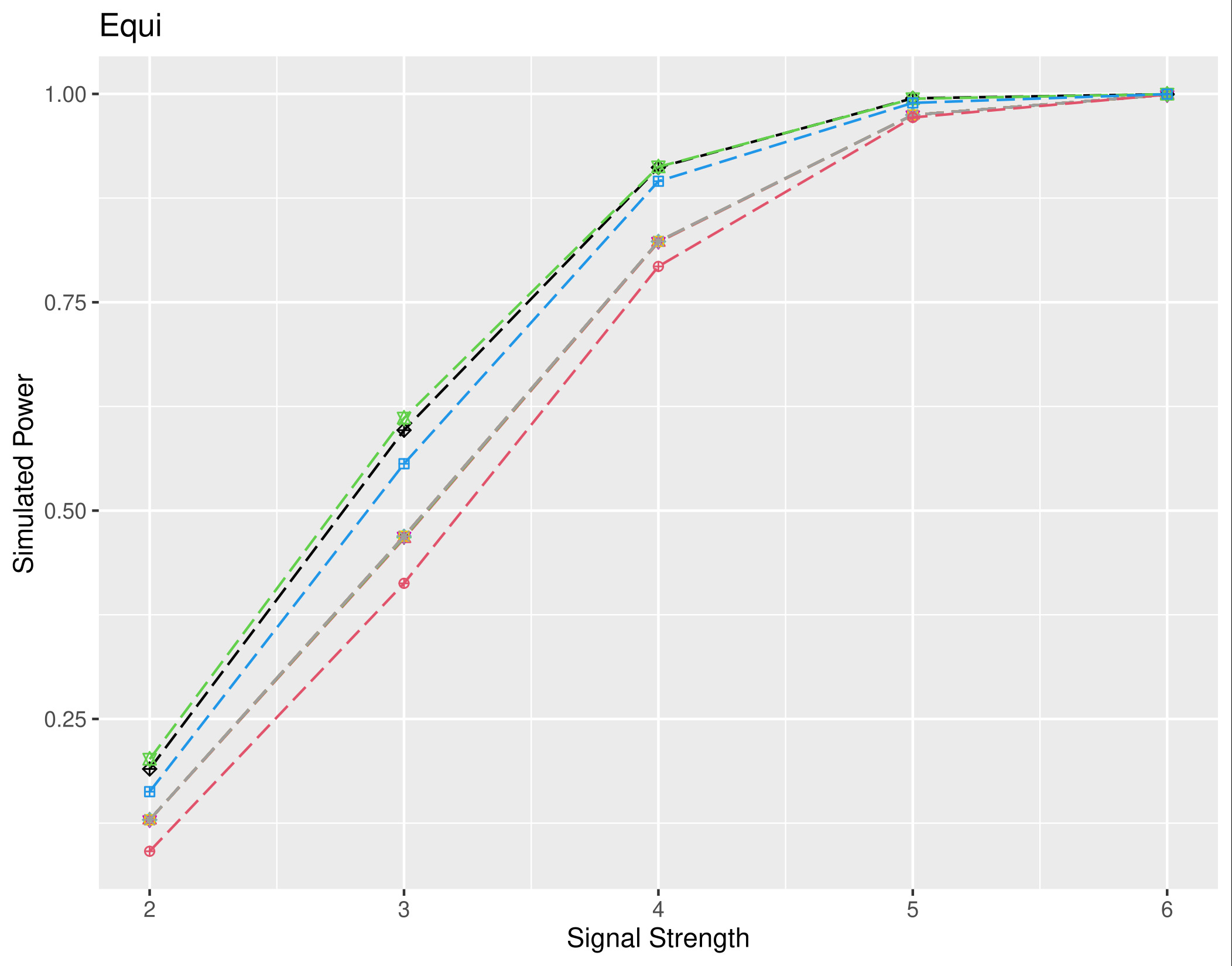} 
      \end{subfigure}  
  & \begin{subfigure}[b]{\linewidth}
      \centering
      \includegraphics[width=\linewidth]{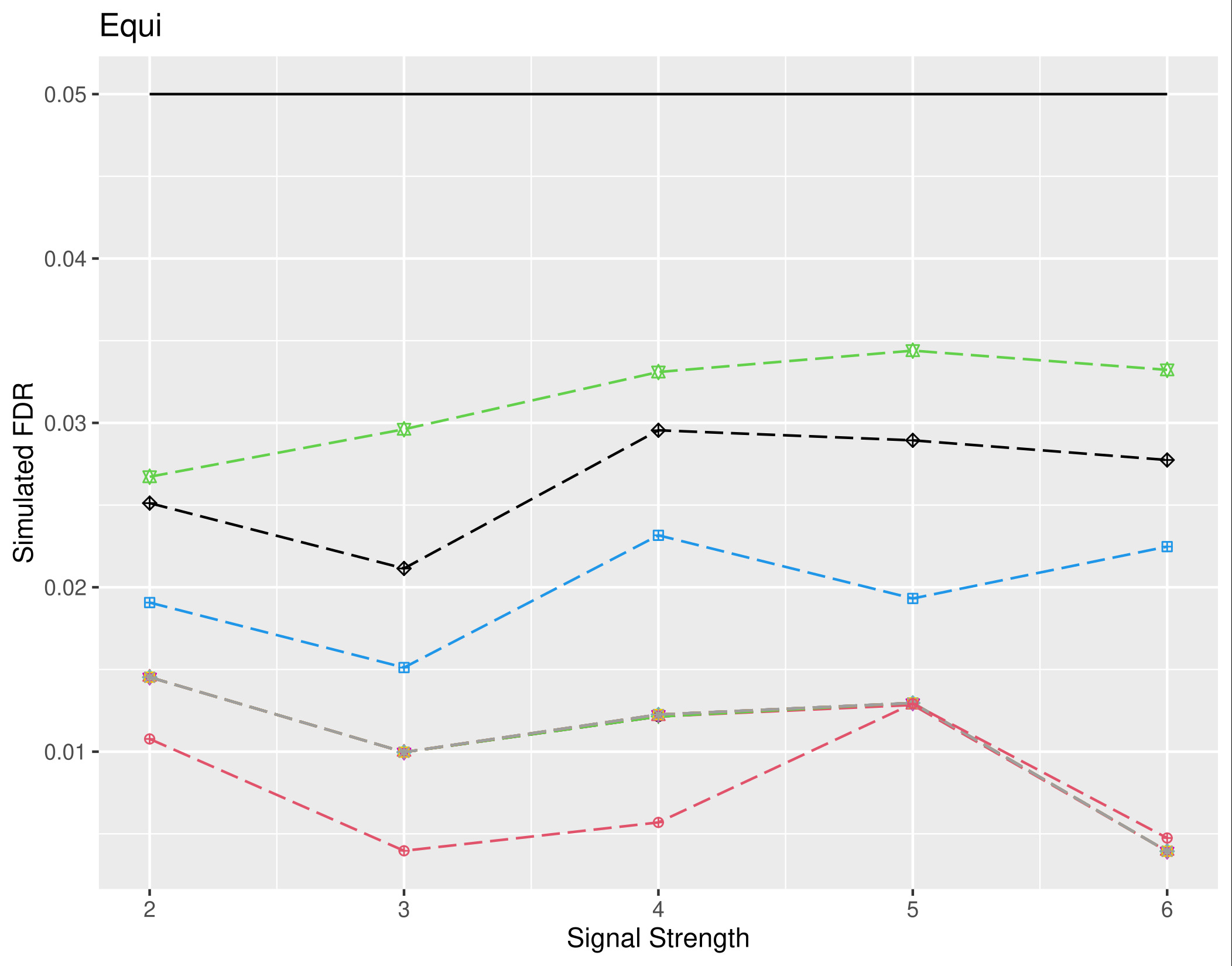} 
      \end{subfigure} 
  & \begin{subfigure}[b]{\linewidth}
      \centering
      \includegraphics[width=\linewidth]{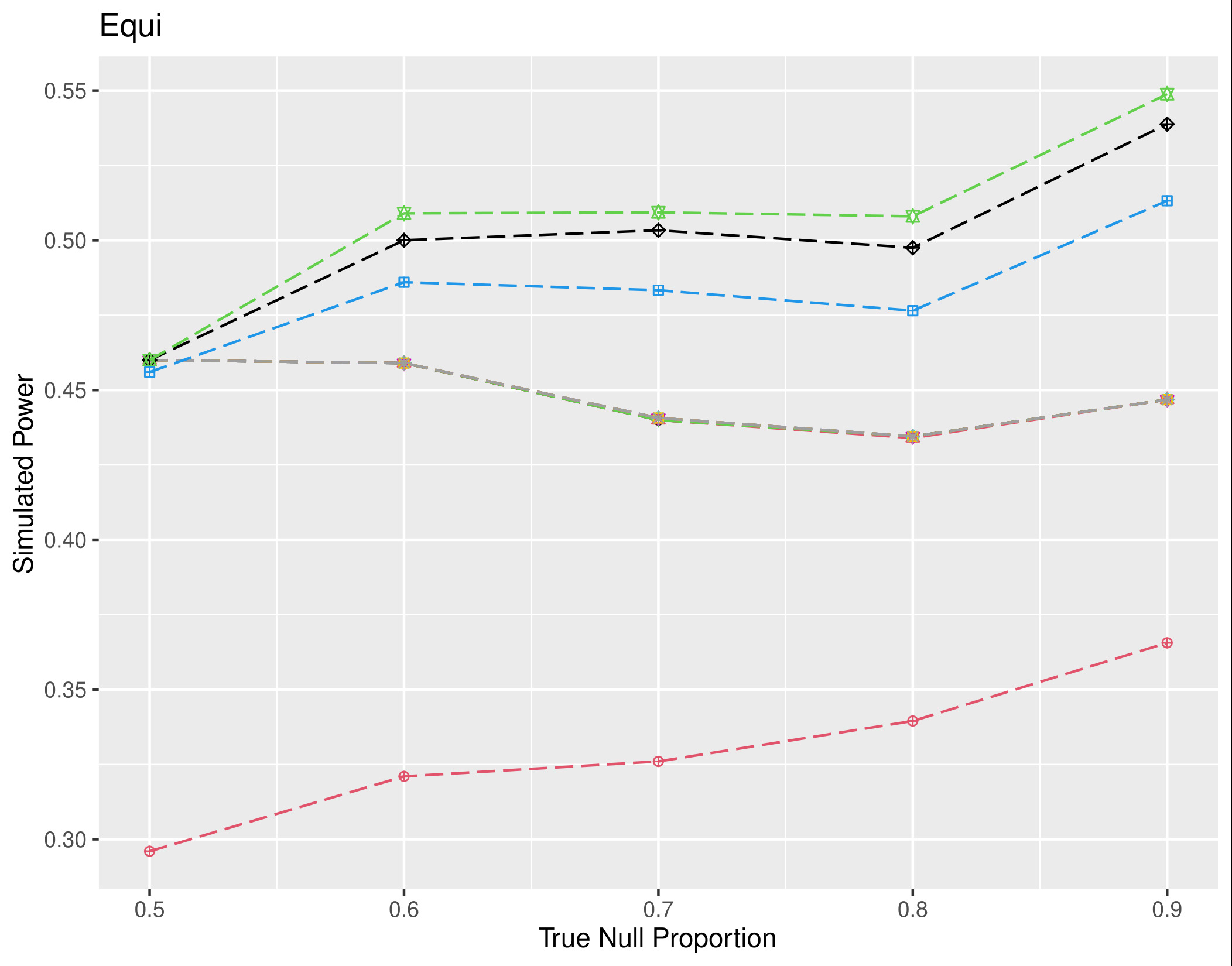} 
      \end{subfigure} \\
  \rotatebox{90}{AR(0.7)} 
  & \begin{subfigure}[b]{\linewidth}
      \centering
      \includegraphics[width=\linewidth]{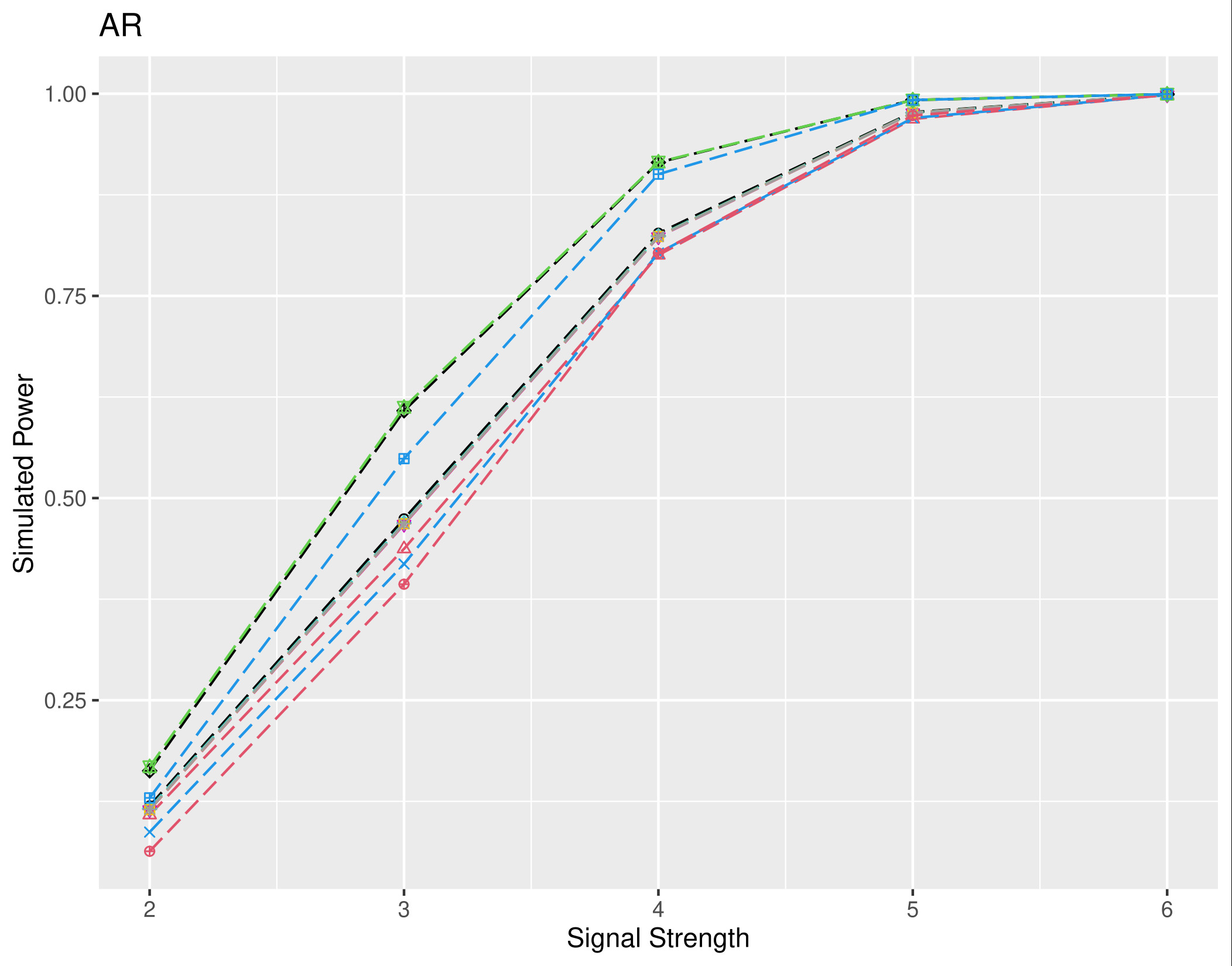} 
      \end{subfigure}  
  & \begin{subfigure}[b]{\linewidth}
      \centering
      \includegraphics[width=\linewidth]{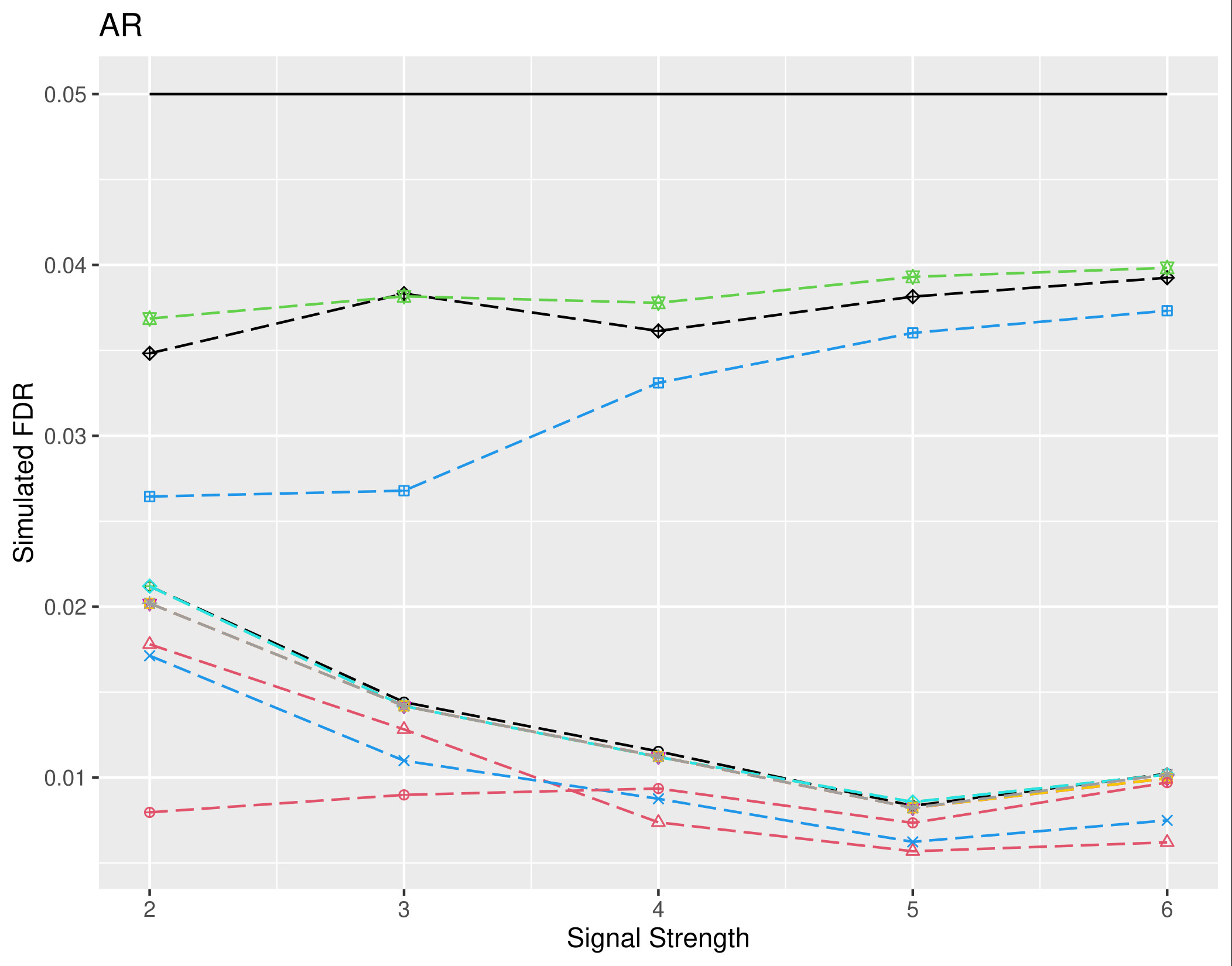} 
      \end{subfigure} 
  & \begin{subfigure}[b]{\linewidth}
      \centering
      \includegraphics[width=\linewidth]{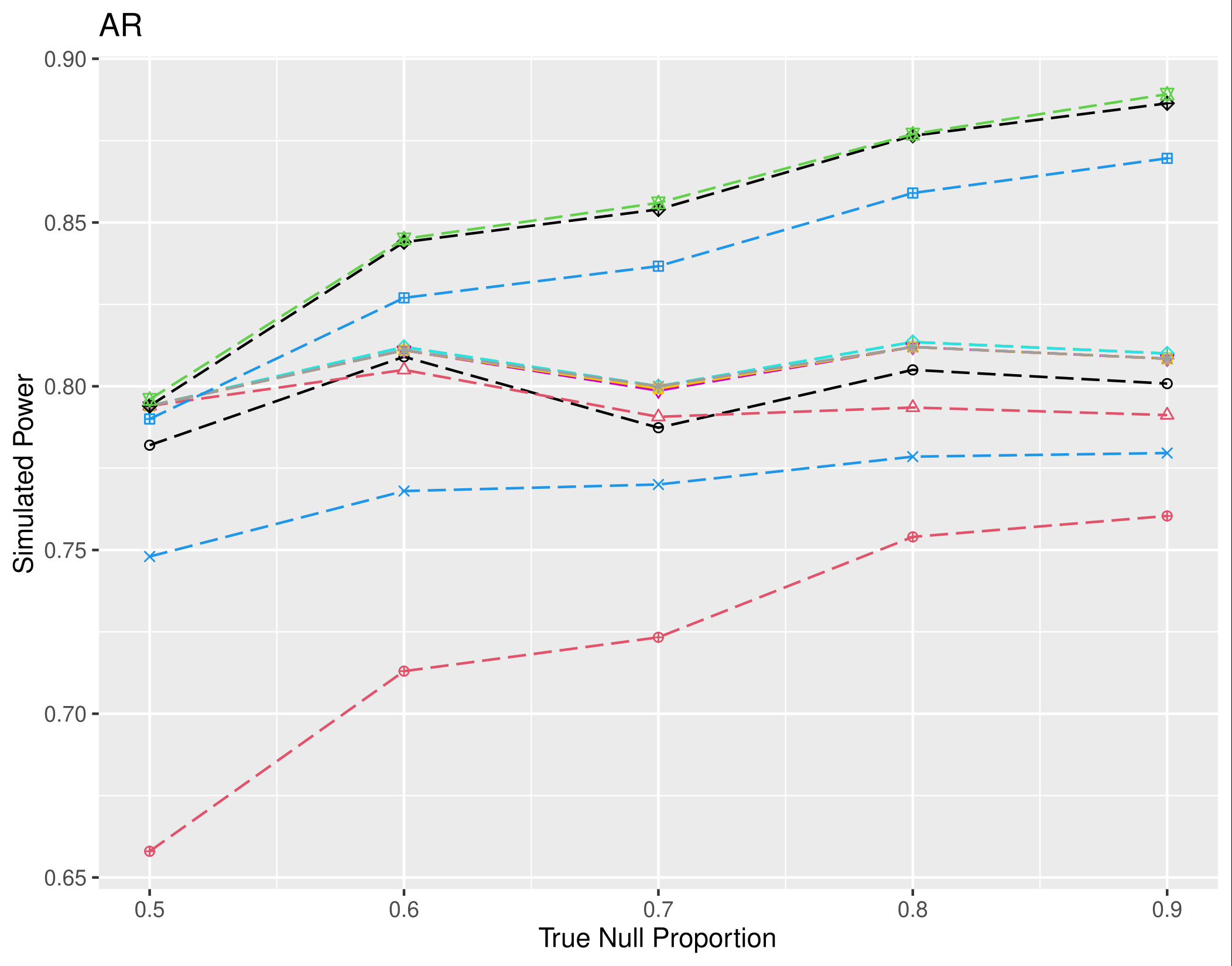} 
      \end{subfigure} \\
  \rotatebox{90}{IAR(0.7)} 
  & \begin{subfigure}[b]{\linewidth}
      \centering
      \includegraphics[width=\linewidth]{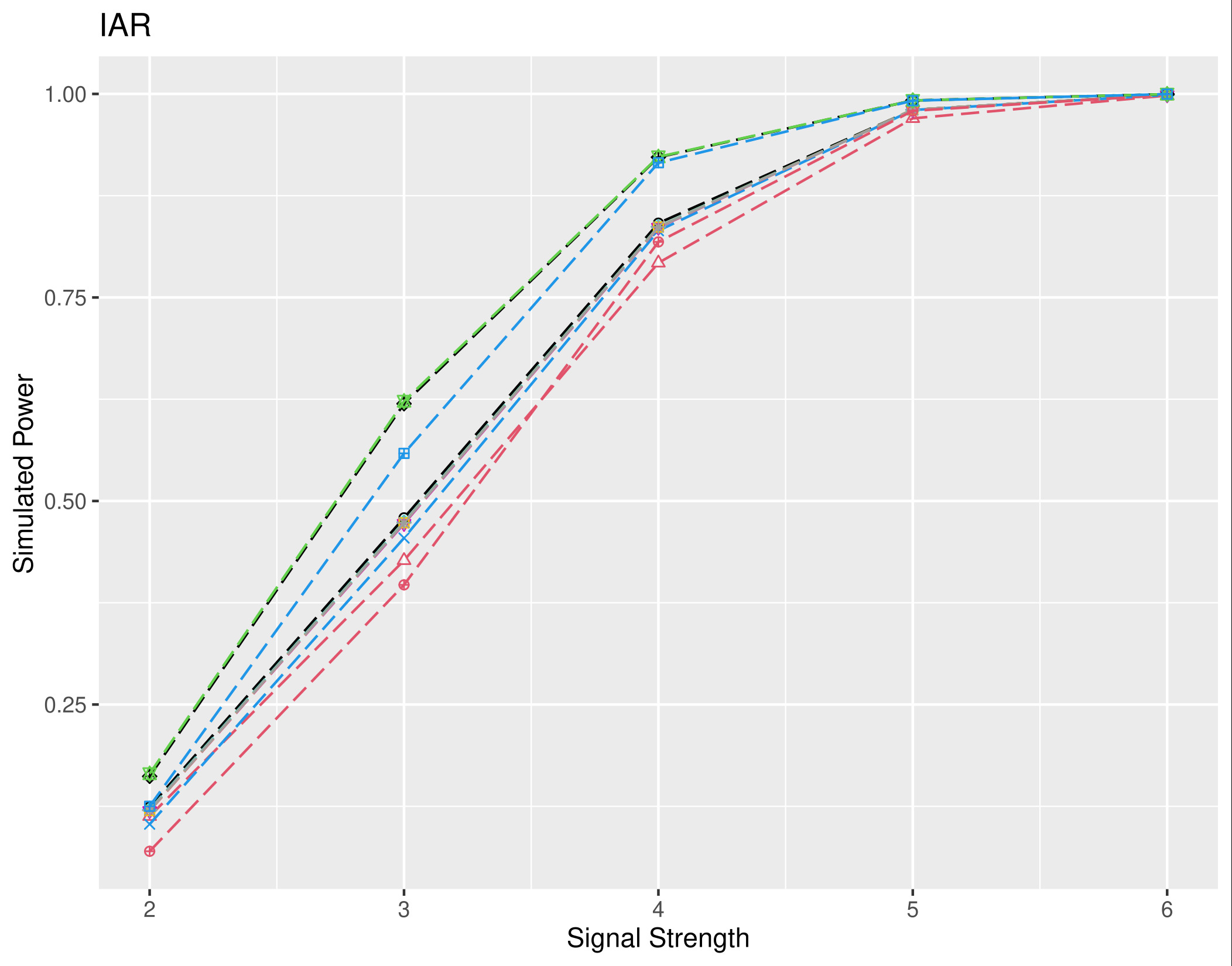} 
      \end{subfigure}  
  & \begin{subfigure}[b]{\linewidth}
      \centering
      \includegraphics[width=\linewidth]{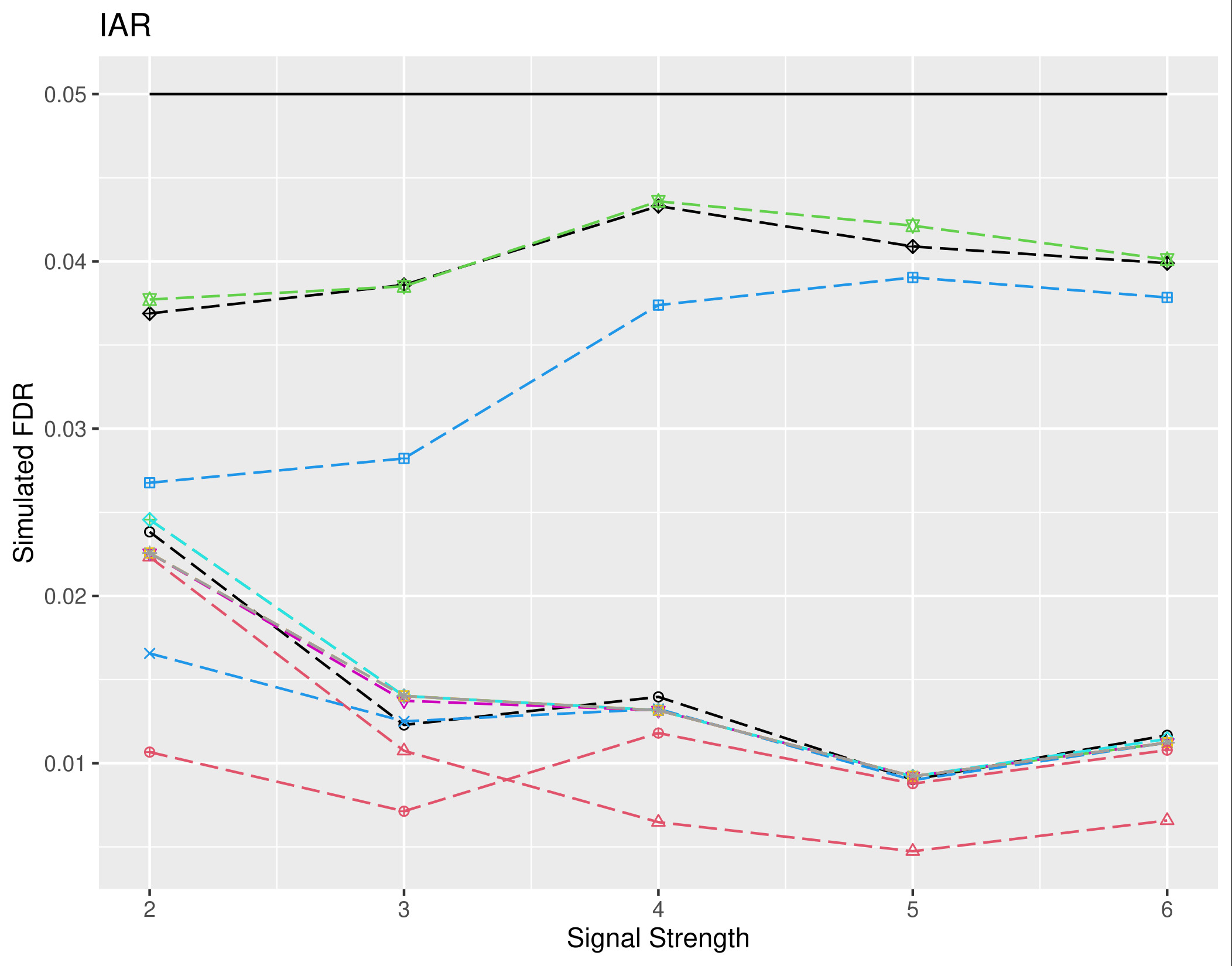} 
      \end{subfigure} 
  & \begin{subfigure}[b]{\linewidth}
      \centering
      \includegraphics[width=\linewidth]{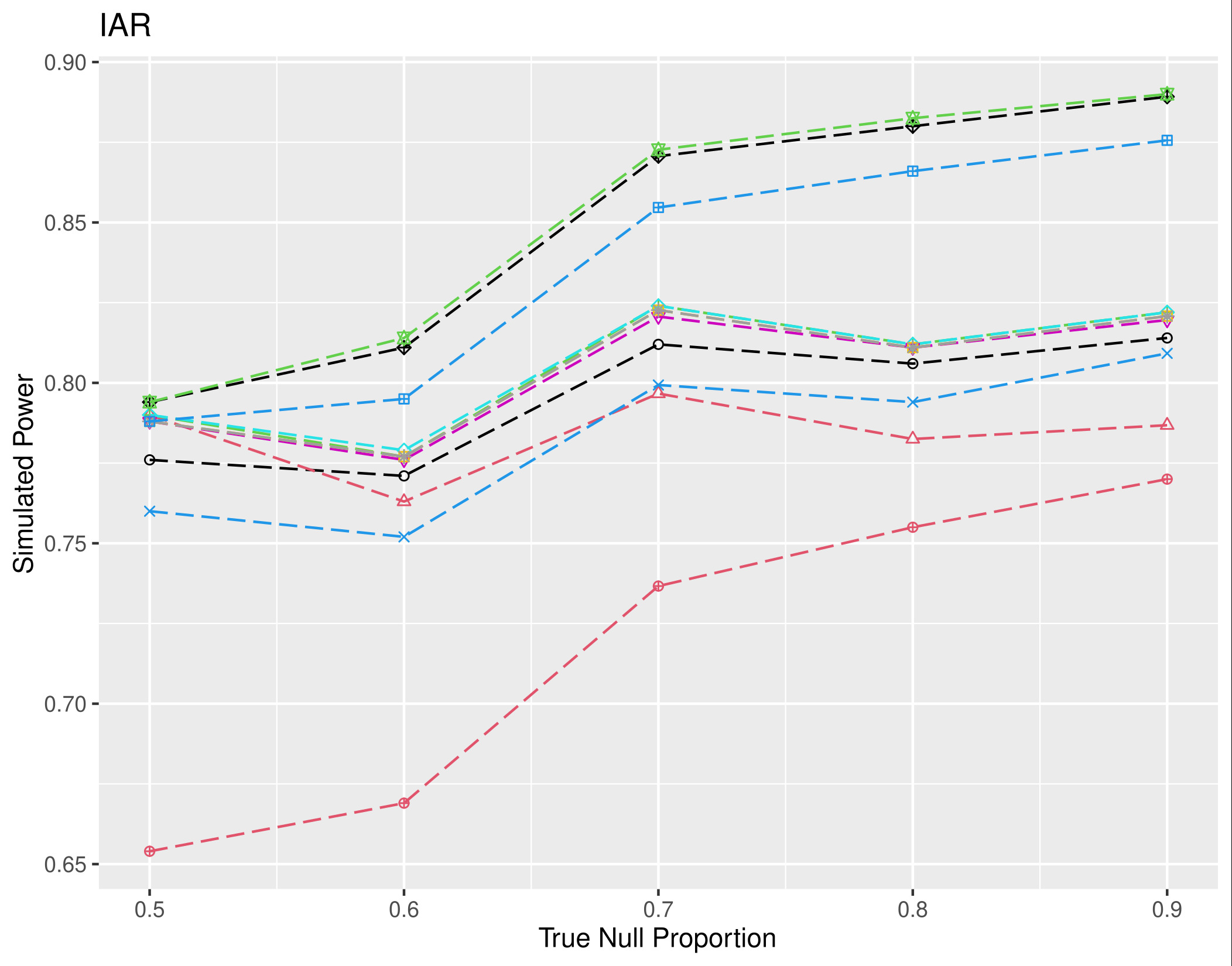} 
      \end{subfigure}    
  \end{tabular*} 
  \caption{Simulated Power (left column), simulated FDR (middle column) for fixed null proportion and simulated power (right column) for fixed signal strength, displayed for mean testing of $d=40$ parameters. Methods compared are SBH1 method (Circle and black), SBH2 method (Triangle point up and red), GSBH1 method (Plus and green), GSBH2 (Cross and blue), GSBH3 (Diamond and light blue), GSBH4 (Triangle point down and purple), GSBH5 (Square cross and yellow), GSBH6 (Star and grey), BH (Diamond plus and black), BY (Circle plus and red), dBH (Triangles up and down and green) and dBY (Square plus and blue)}
  \label{means:figure3} 
\end{figure}

\begin{figure}[ht]
  \begin{tabular*}{\textwidth}{
    @{}m{0.5cm}
    @{}m{\dimexpr0.33\textwidth-0.25cm\relax}
    @{}m{\dimexpr0.33\textwidth-0.25cm\relax}
    @{}m{\dimexpr0.33\textwidth-0.25cm\relax}}
  \rotatebox{90}{Equi(0.3)}
  & \begin{subfigure}[b]{\linewidth}
      \centering
      \includegraphics[width=\linewidth]{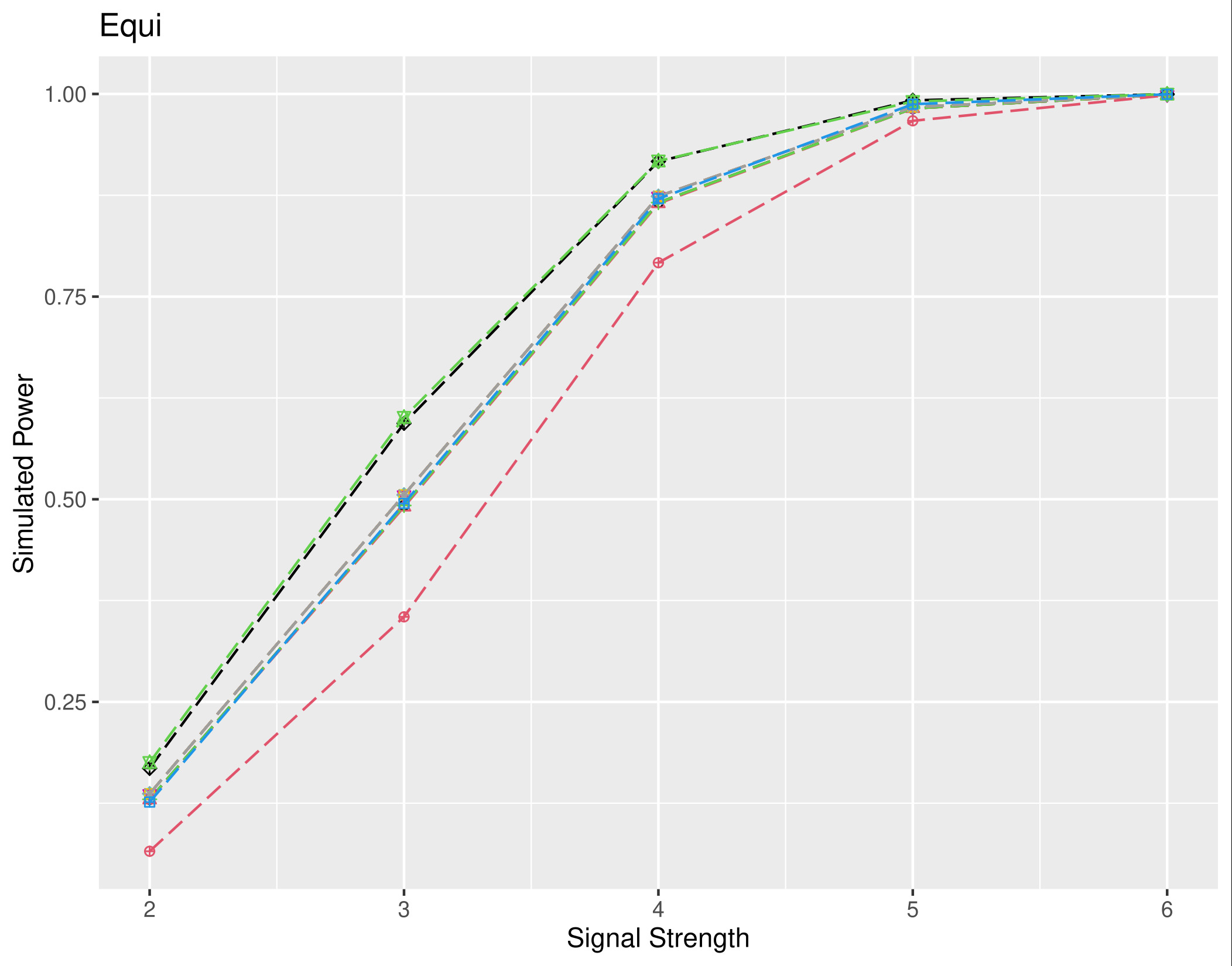} 
      \end{subfigure}  
  & \begin{subfigure}[b]{\linewidth}
      \centering
      \includegraphics[width=\linewidth]{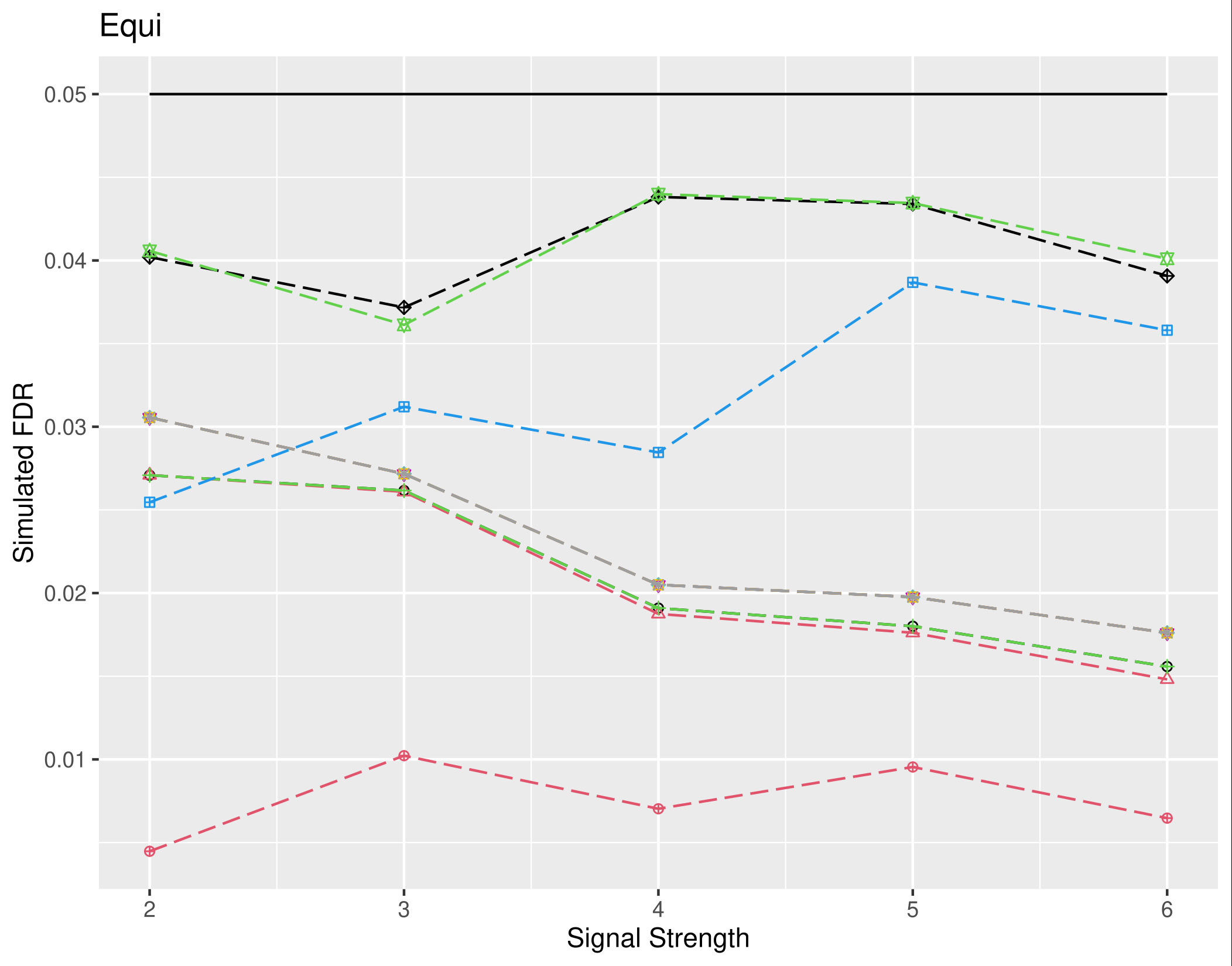} 
      \end{subfigure} 
  & \begin{subfigure}[b]{\linewidth}
      \centering
      \includegraphics[width=\linewidth]{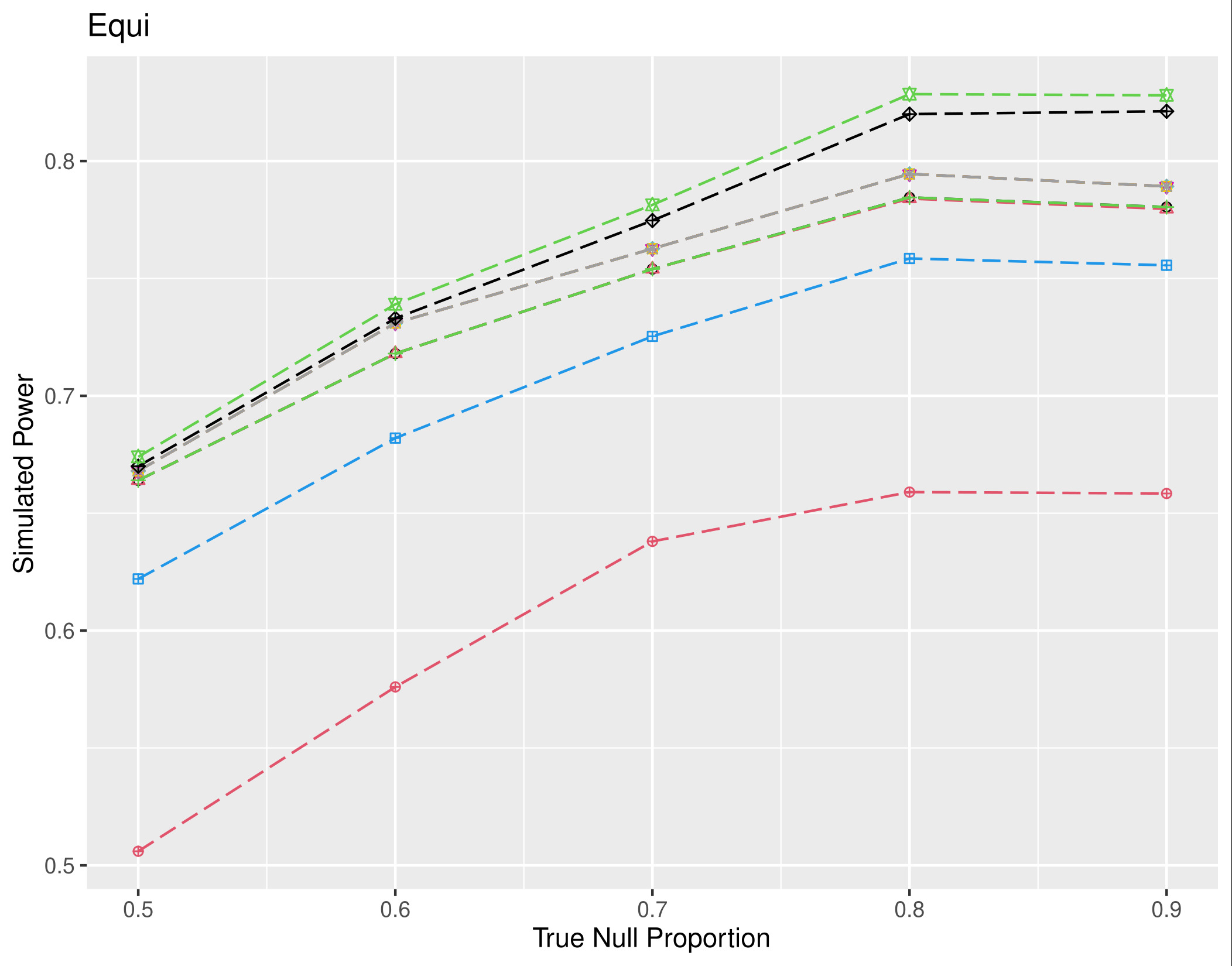} 
      \end{subfigure} \\
  \rotatebox{90}{AR(0.3)} 
  & \begin{subfigure}[b]{\linewidth}
      \centering
      \includegraphics[width=\linewidth]{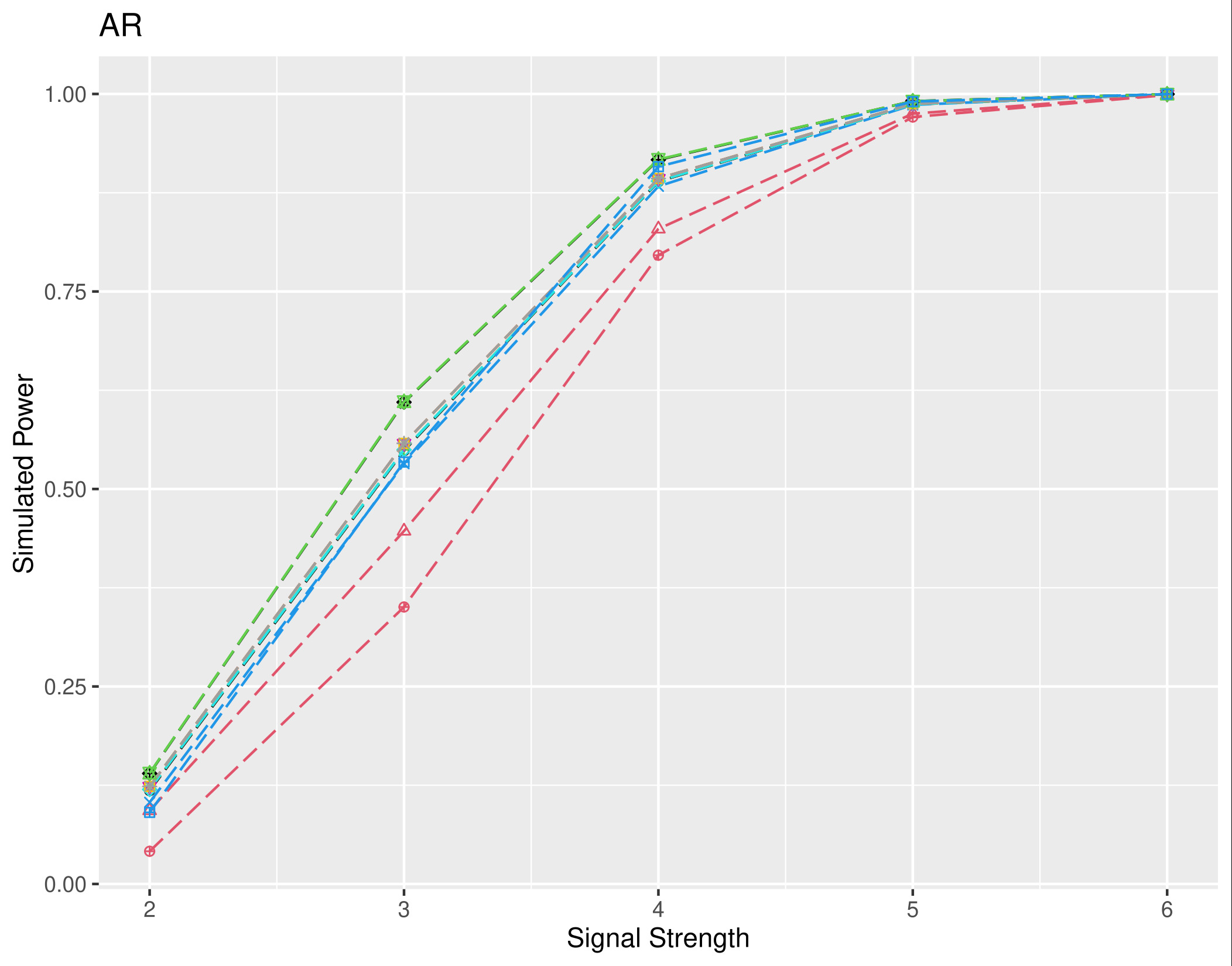} 
      \end{subfigure}  
  & \begin{subfigure}[b]{\linewidth}
      \centering
      \includegraphics[width=\linewidth]{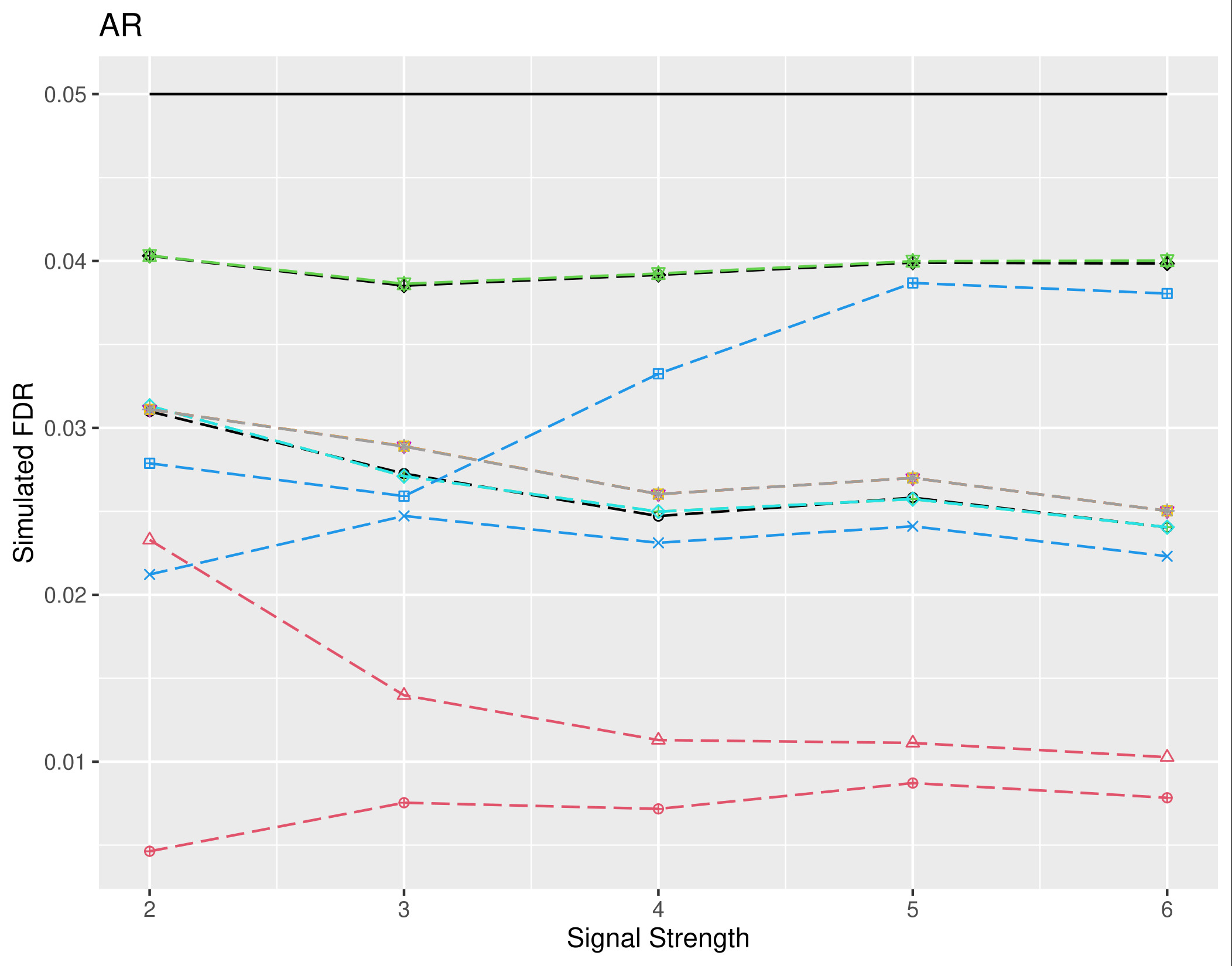} 
      \end{subfigure} 
  & \begin{subfigure}[b]{\linewidth}
      \centering
      \includegraphics[width=\linewidth]{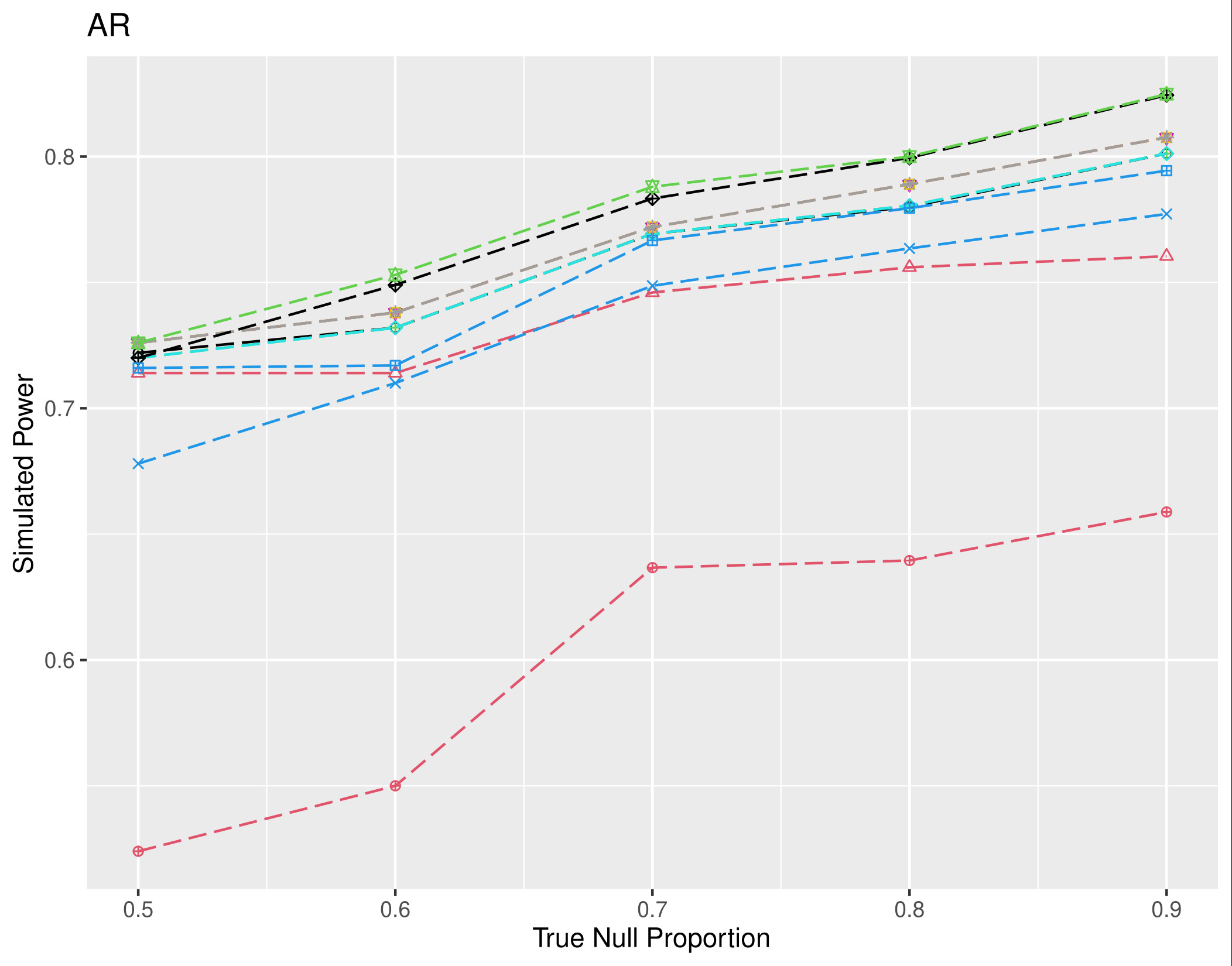} 
      \end{subfigure} \\
  \rotatebox{90}{IAR(0.3)} 
  & \begin{subfigure}[b]{\linewidth}
      \centering
      \includegraphics[width=\linewidth]{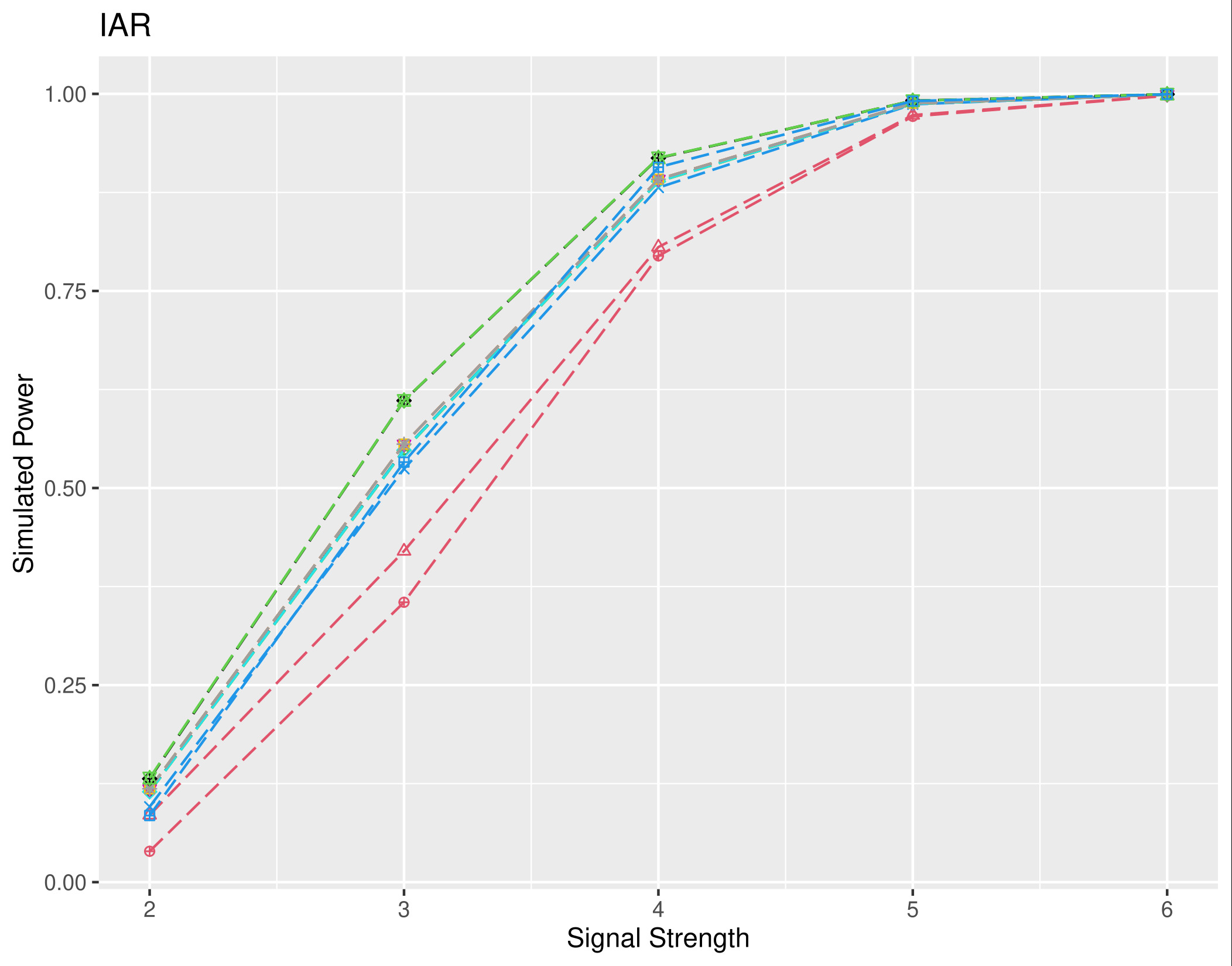} 
      \end{subfigure}  
  & \begin{subfigure}[b]{\linewidth}
      \centering
      \includegraphics[width=\linewidth]{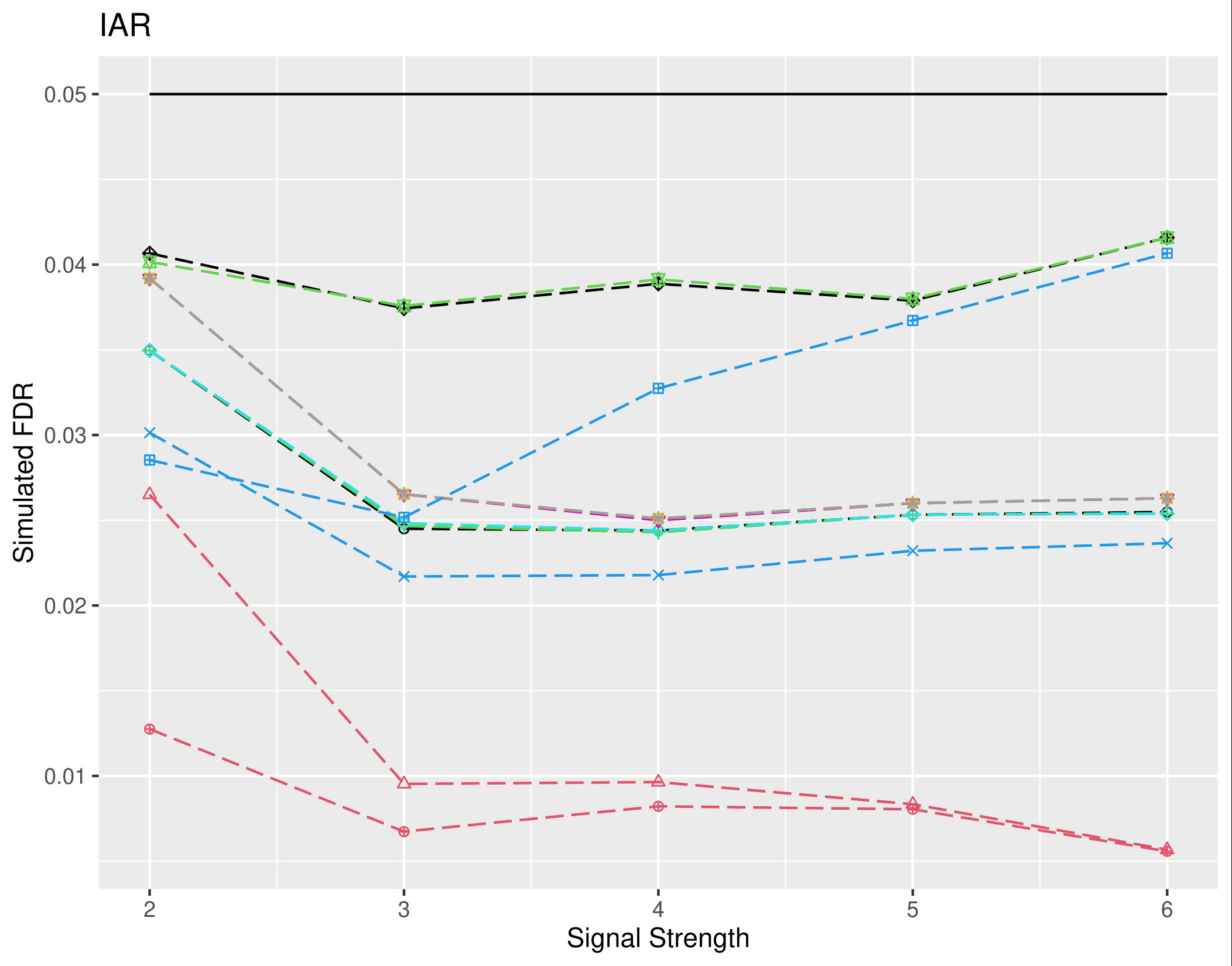} 
      \end{subfigure} 
  & \begin{subfigure}[b]{\linewidth}
      \centering
      \includegraphics[width=\linewidth]{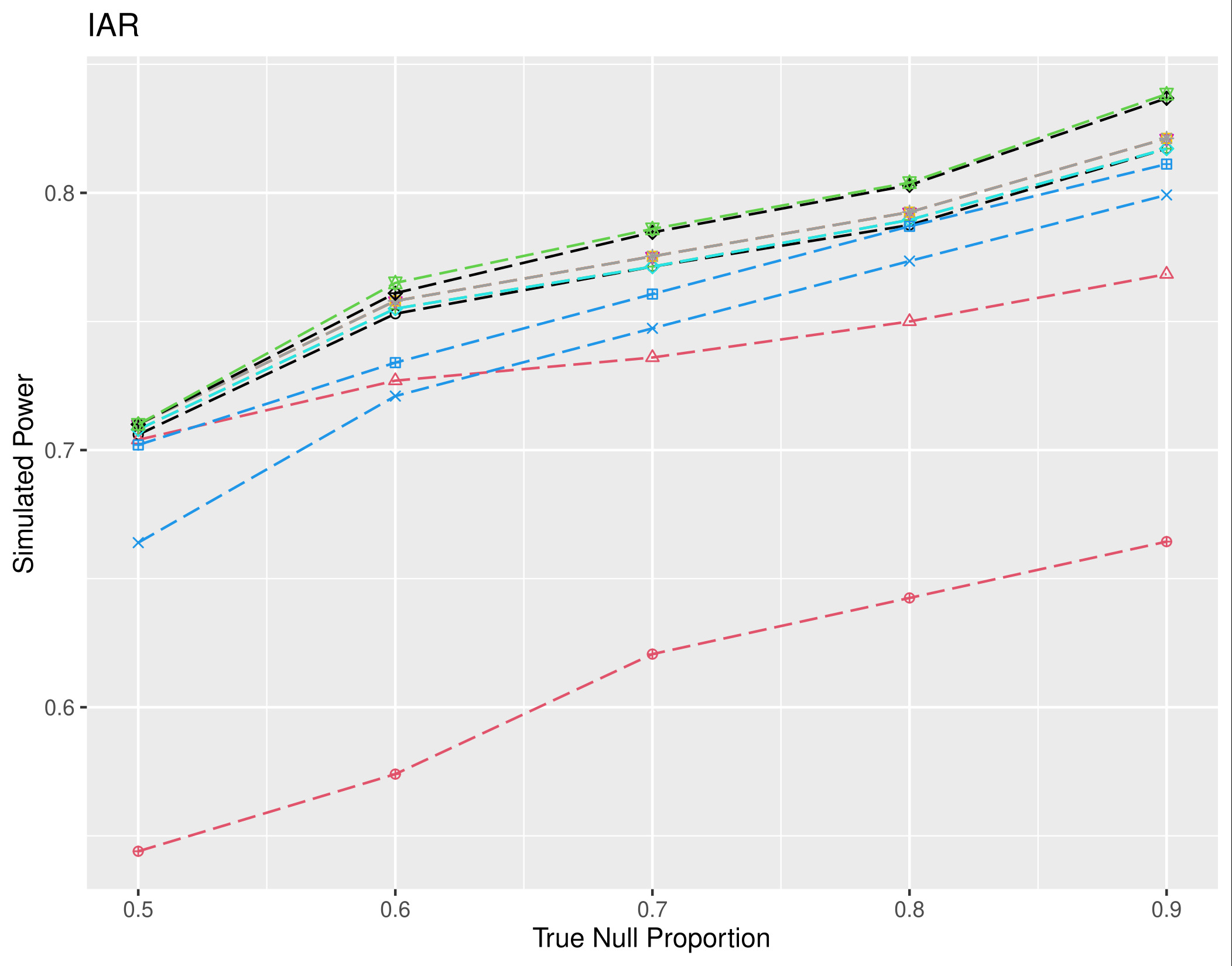} 
      \end{subfigure}    
  \end{tabular*} 
  \caption{Simulated Power (left column), simulated FDR (middle column) for fixed null proportion and simulated power (right column) for fixed signal strength, displayed for mean testing of $d=100$ parameters. Methods compared are SBH1 method (Circle and black), SBH2 method (Triangle point up and red), GSBH1 method (Plus and green), GSBH2 (Cross and blue), GSBH3 (Diamond and light blue), GSBH4 (Triangle point down and purple), GSBH5 (Square cross and yellow), GSBH6 (Star and grey), BH (Diamond plus and black), BY (Circle plus and red), dBH (Triangles up and down and green) and dBY (Square plus and blue)}
  \label{means:figure4} 
\end{figure}

\begin{figure}[ht]
  \begin{tabular*}{\textwidth}{
    @{}m{0.5cm}
    @{}m{\dimexpr0.33\textwidth-0.25cm\relax}
    @{}m{\dimexpr0.33\textwidth-0.25cm\relax}
    @{}m{\dimexpr0.33\textwidth-0.25cm\relax}}
  \rotatebox{90}{Equi(0.7)}
  & \begin{subfigure}[b]{\linewidth}
      \centering
      \includegraphics[width=\linewidth]{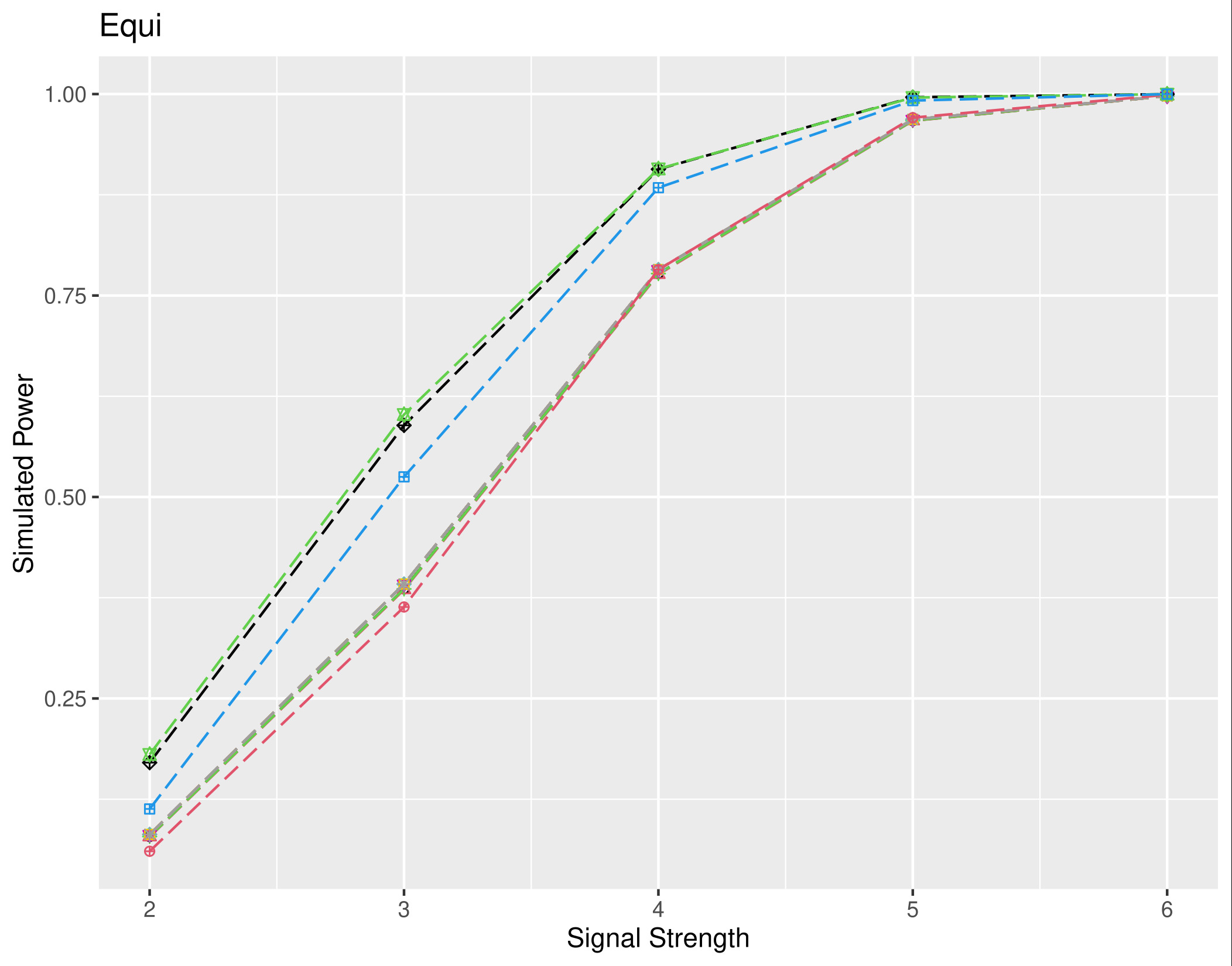} 
      \end{subfigure}  
  & \begin{subfigure}[b]{\linewidth}
      \centering
      \includegraphics[width=\linewidth]{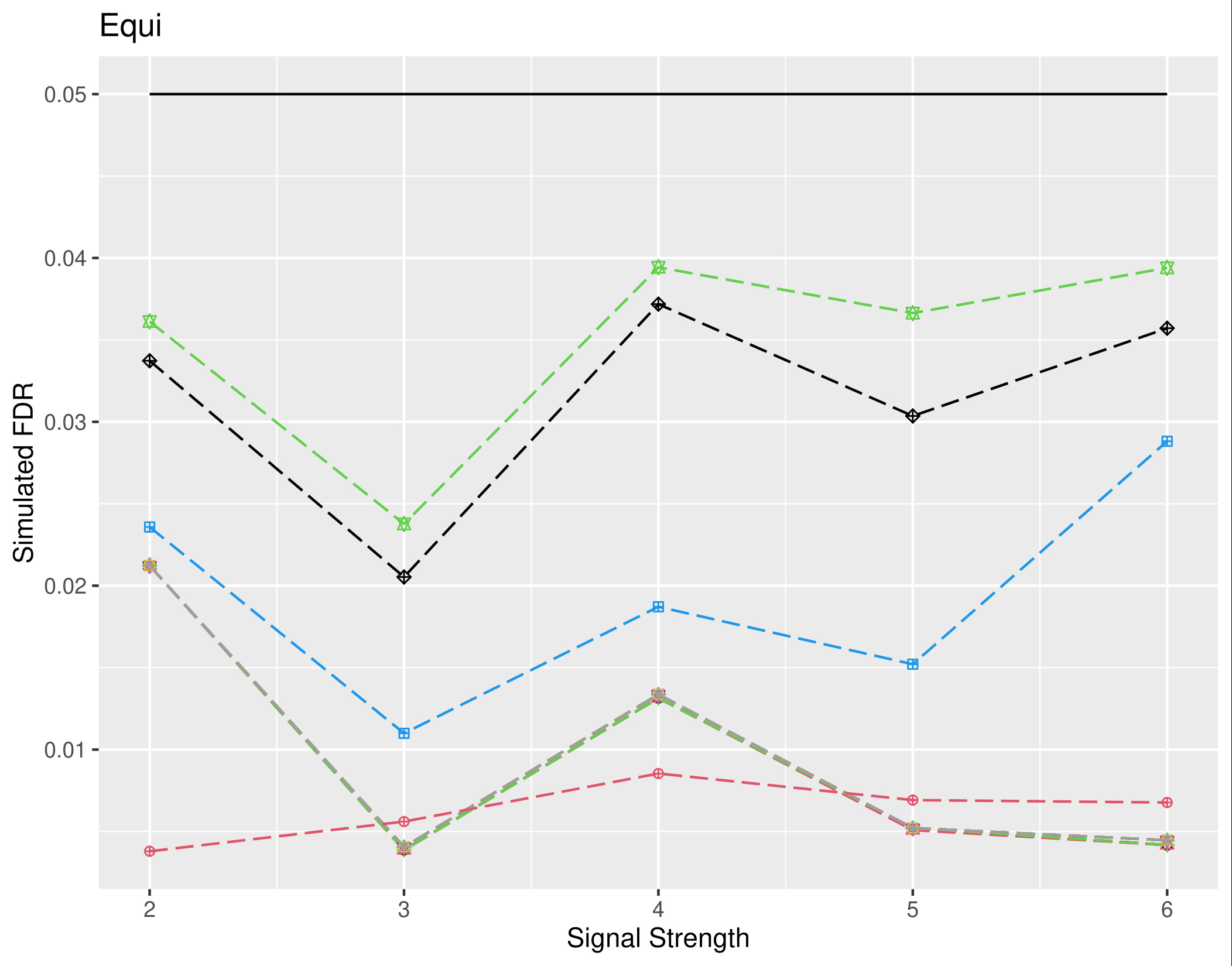} 
      \end{subfigure} 
  & \begin{subfigure}[b]{\linewidth}
      \centering
      \includegraphics[width=\linewidth]{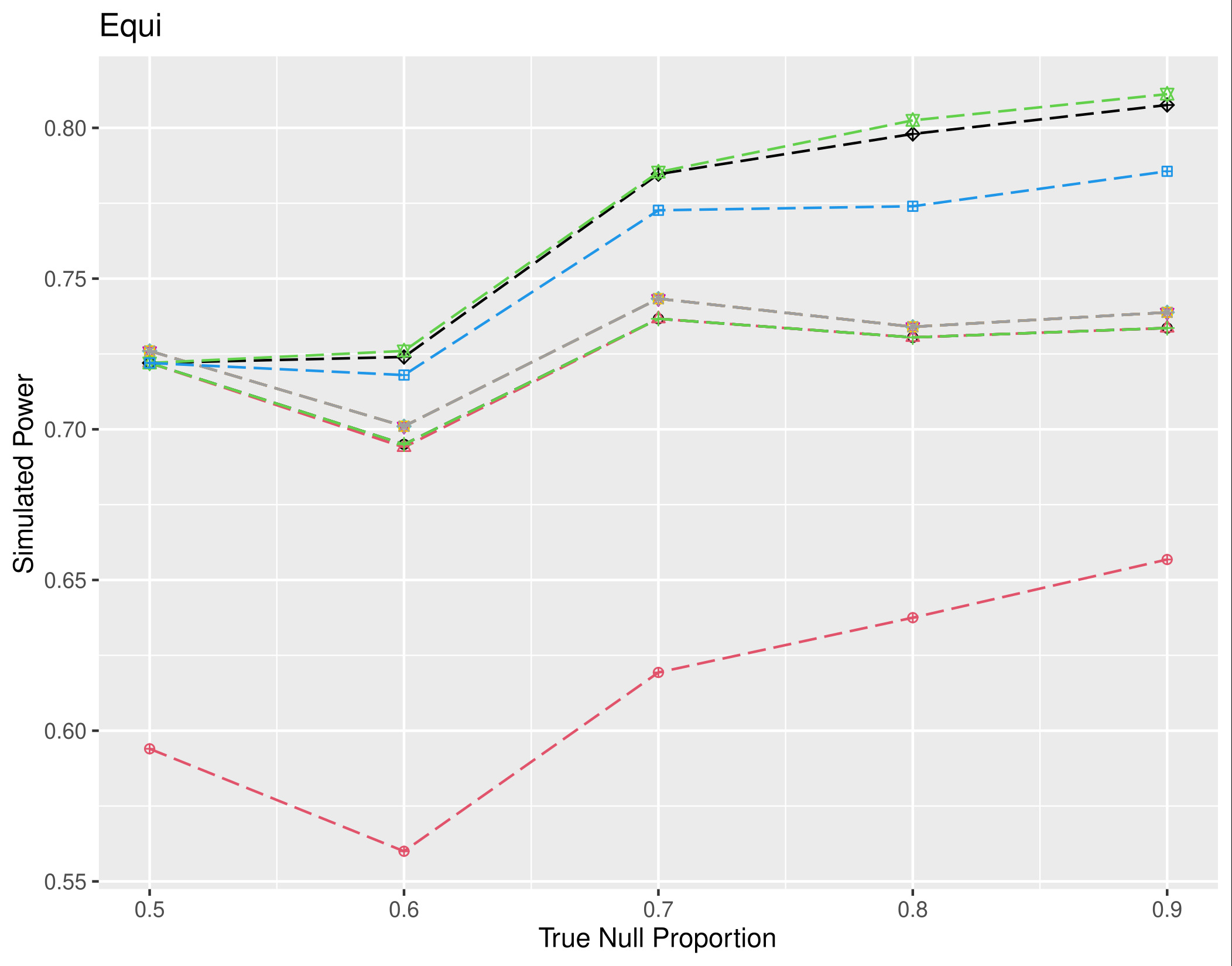} 
      \end{subfigure} \\
  \rotatebox{90}{AR(0.7)} 
  & \begin{subfigure}[b]{\linewidth}
      \centering
      \includegraphics[width=\linewidth]{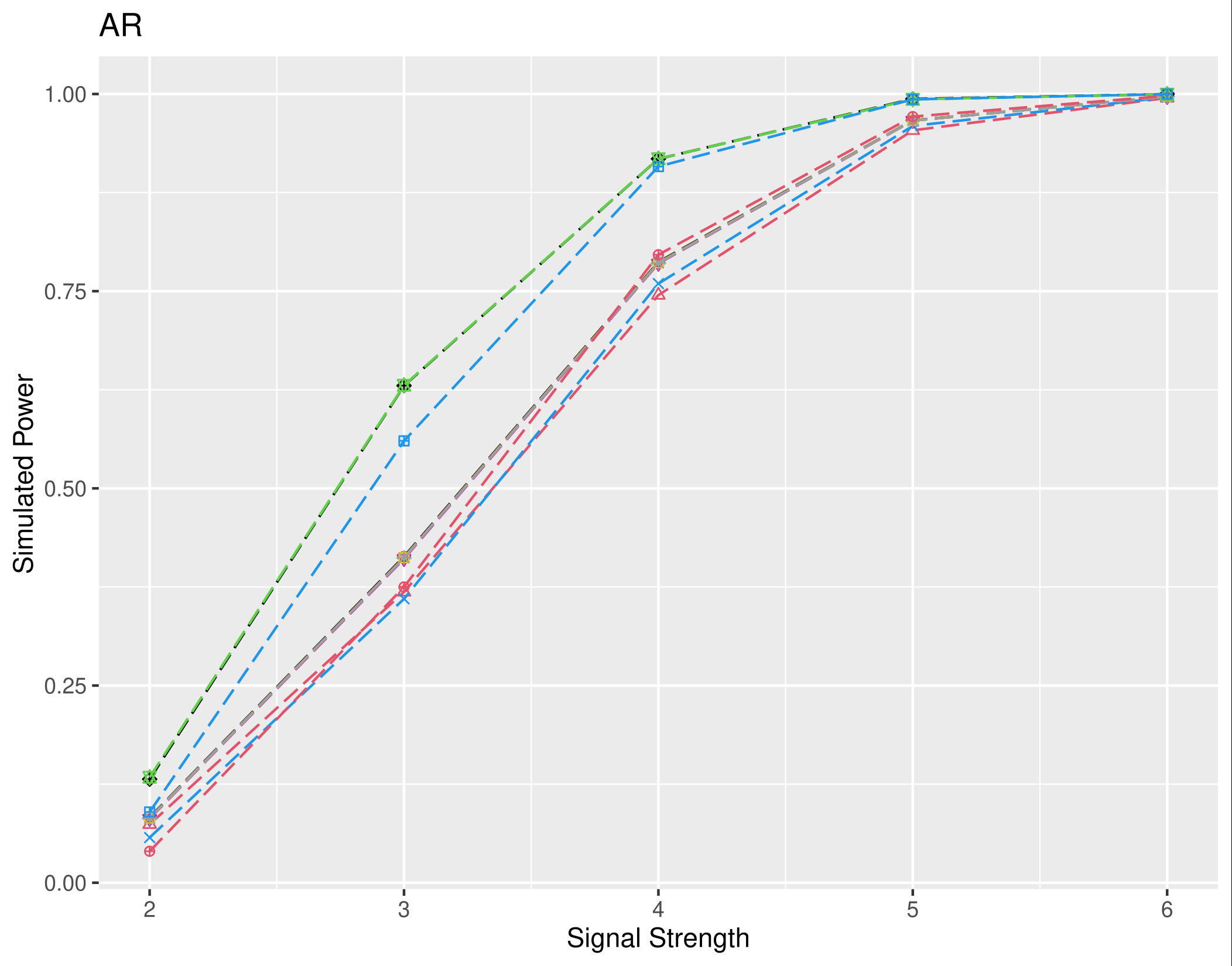} 
      \end{subfigure}  
  & \begin{subfigure}[b]{\linewidth}
      \centering
      \includegraphics[width=\linewidth]{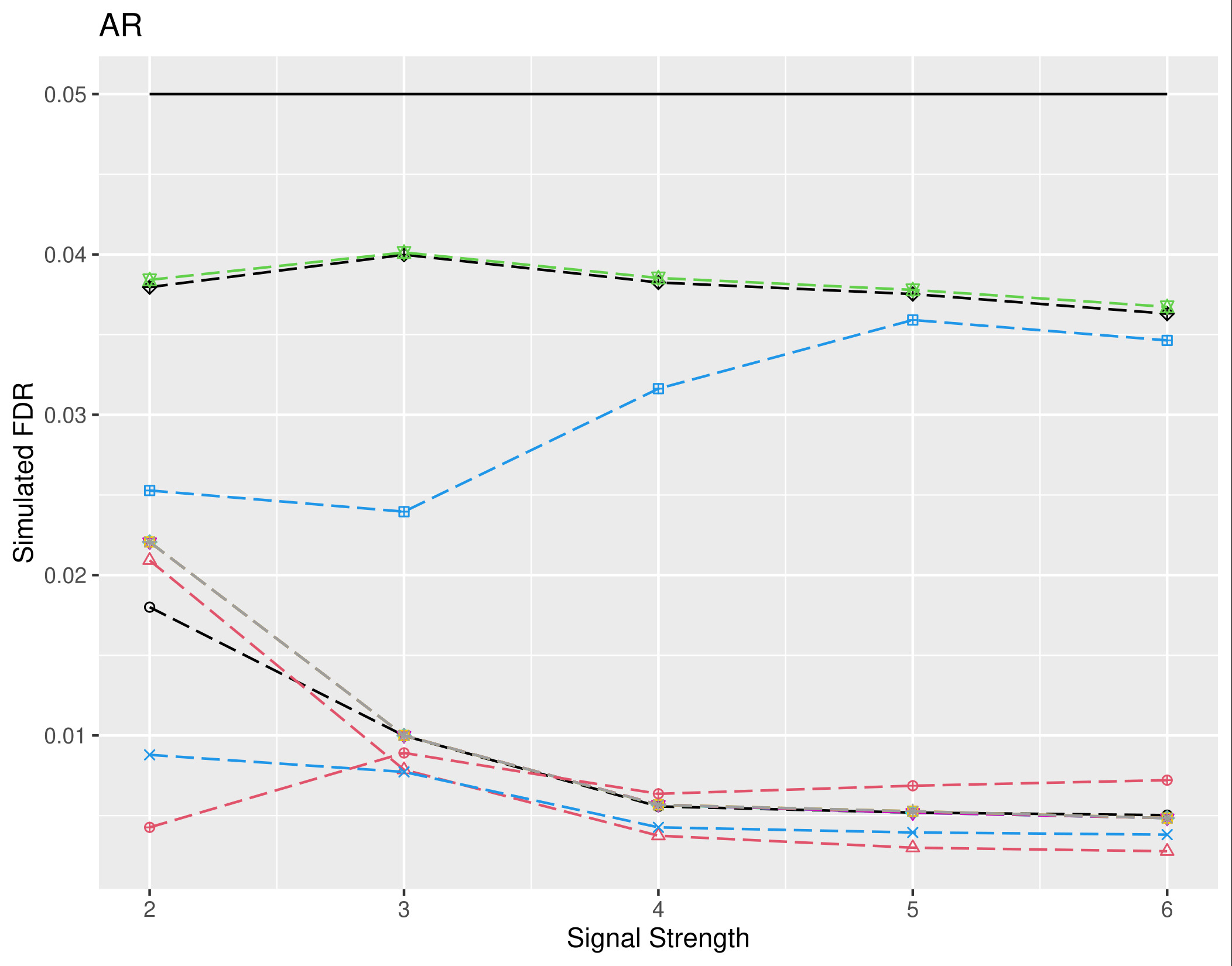} 
      \end{subfigure} 
  & \begin{subfigure}[b]{\linewidth}
      \centering
      \includegraphics[width=\linewidth]{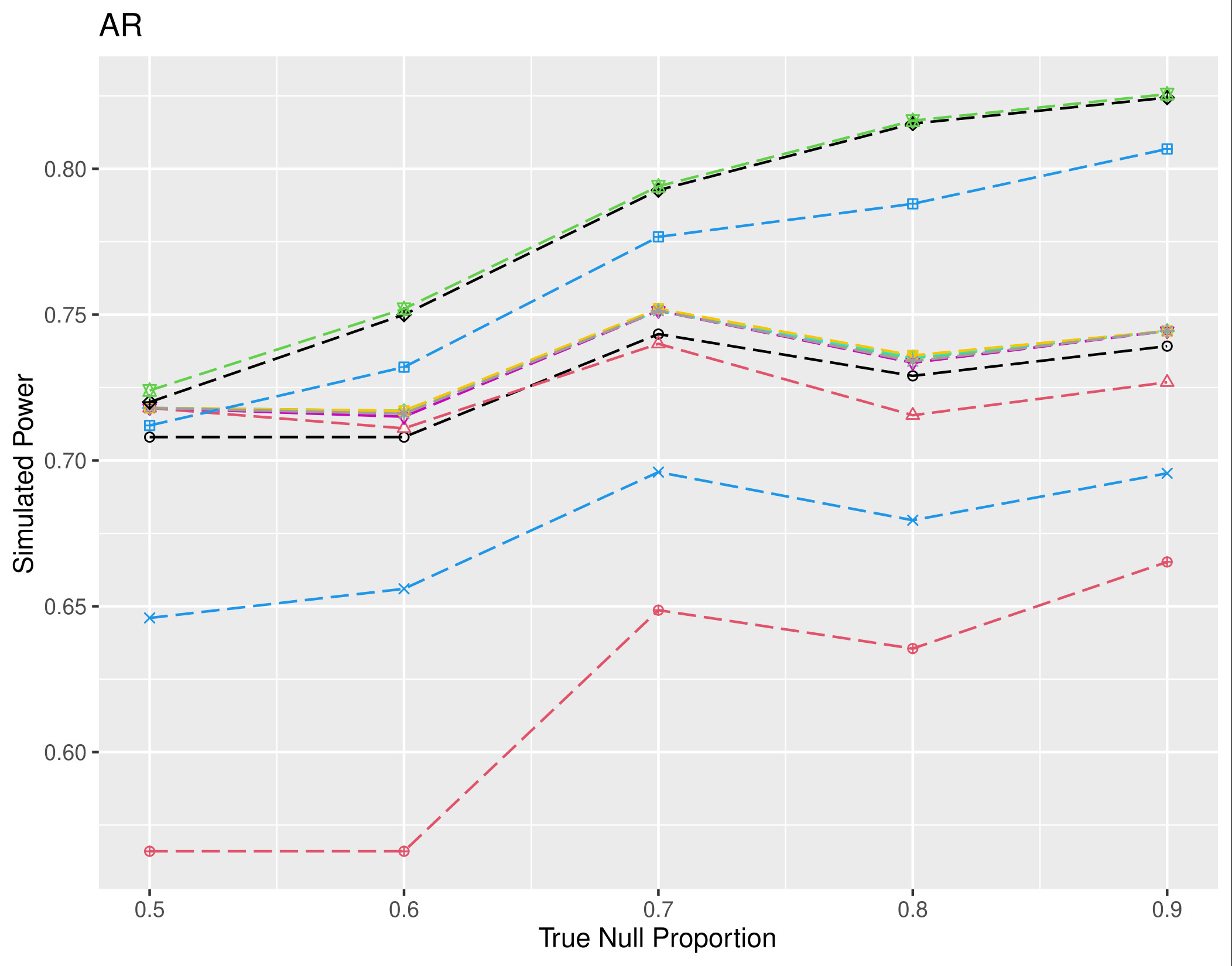} 
      \end{subfigure} \\
  \rotatebox{90}{IAR(0.7)} 
  & \begin{subfigure}[b]{\linewidth}
      \centering
      \includegraphics[width=\linewidth]{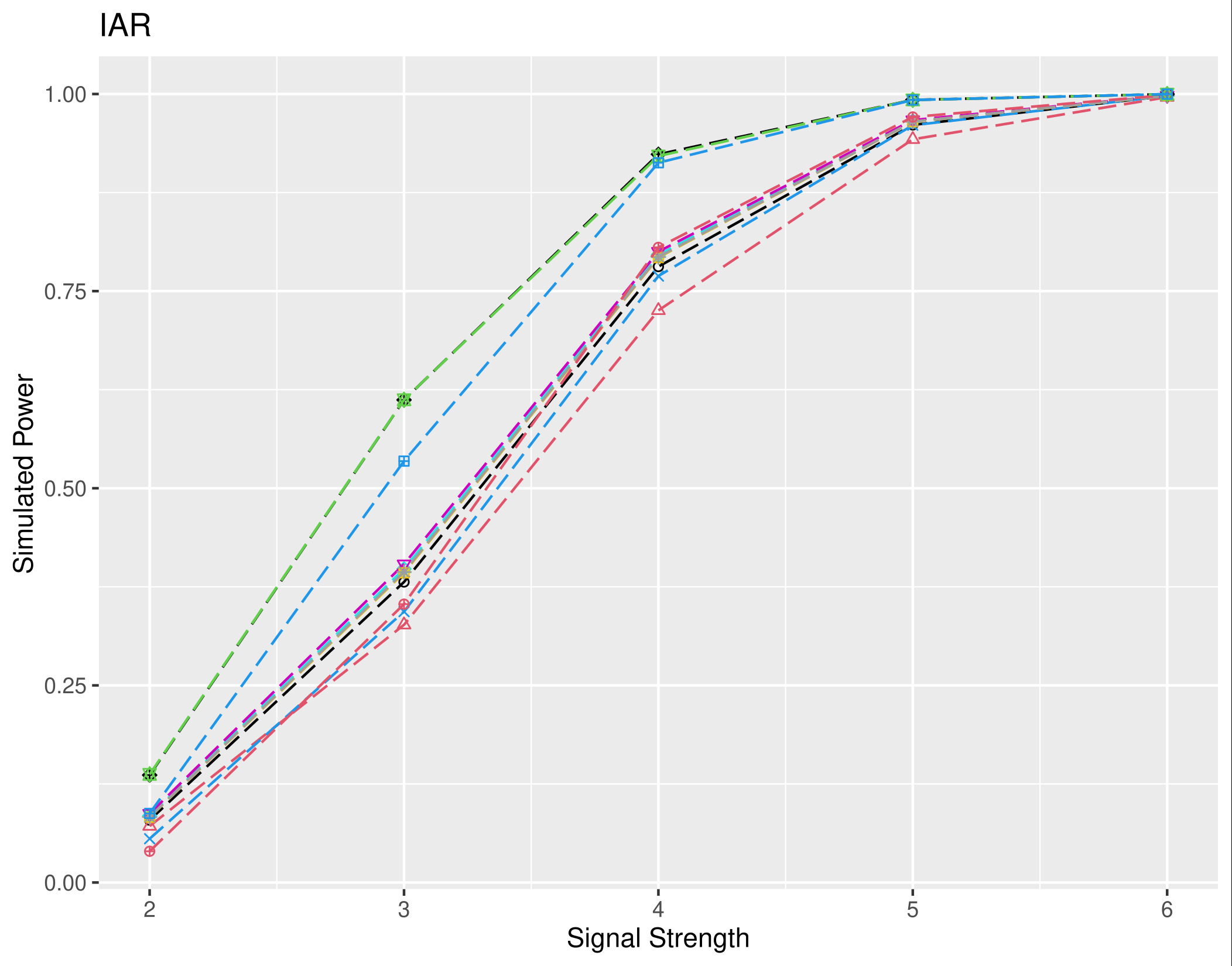} 
      \end{subfigure}  
  & \begin{subfigure}[b]{\linewidth}
      \centering
      \includegraphics[width=\linewidth]{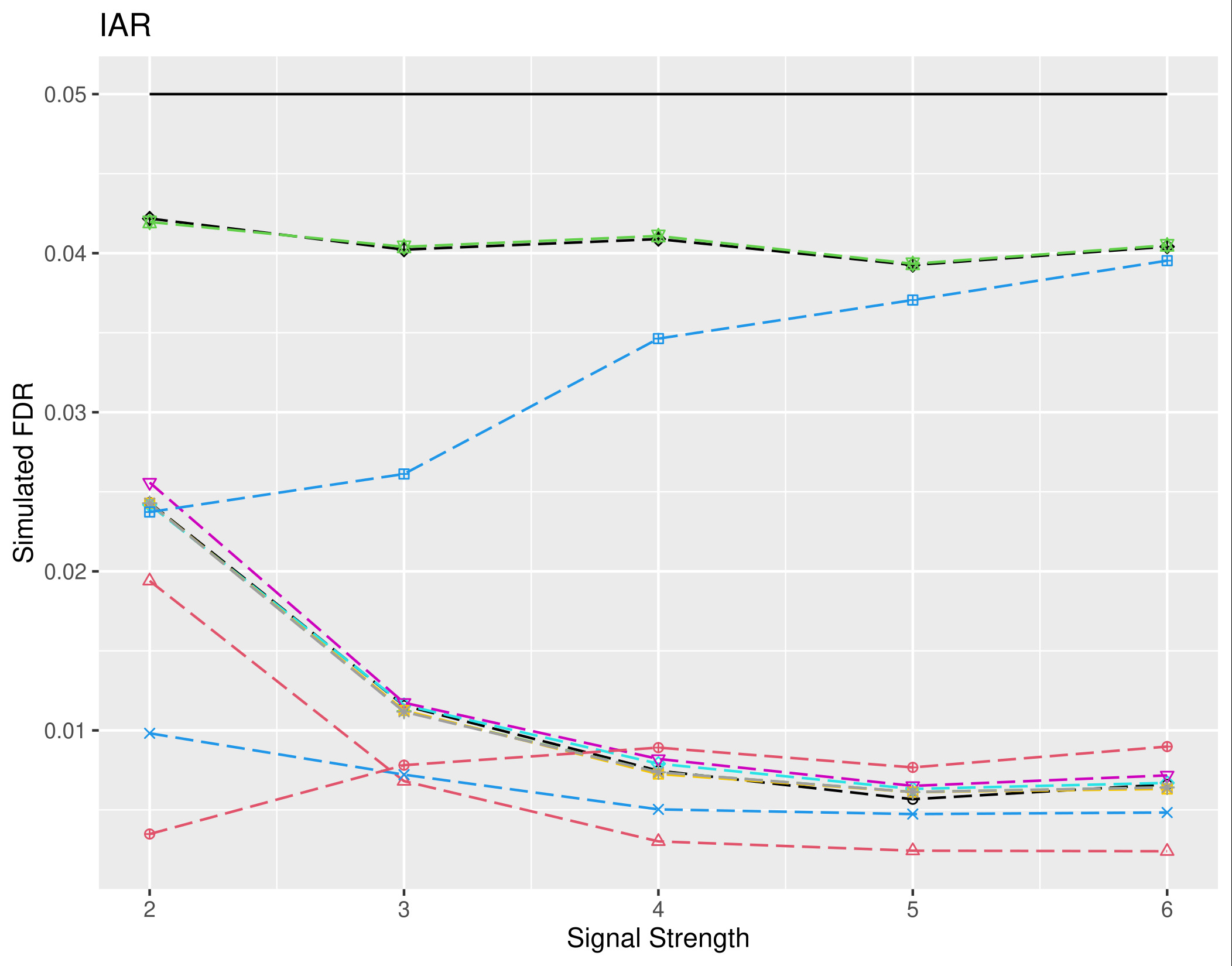} 
      \end{subfigure} 
  & \begin{subfigure}[b]{\linewidth}
      \centering
      \includegraphics[width=\linewidth]{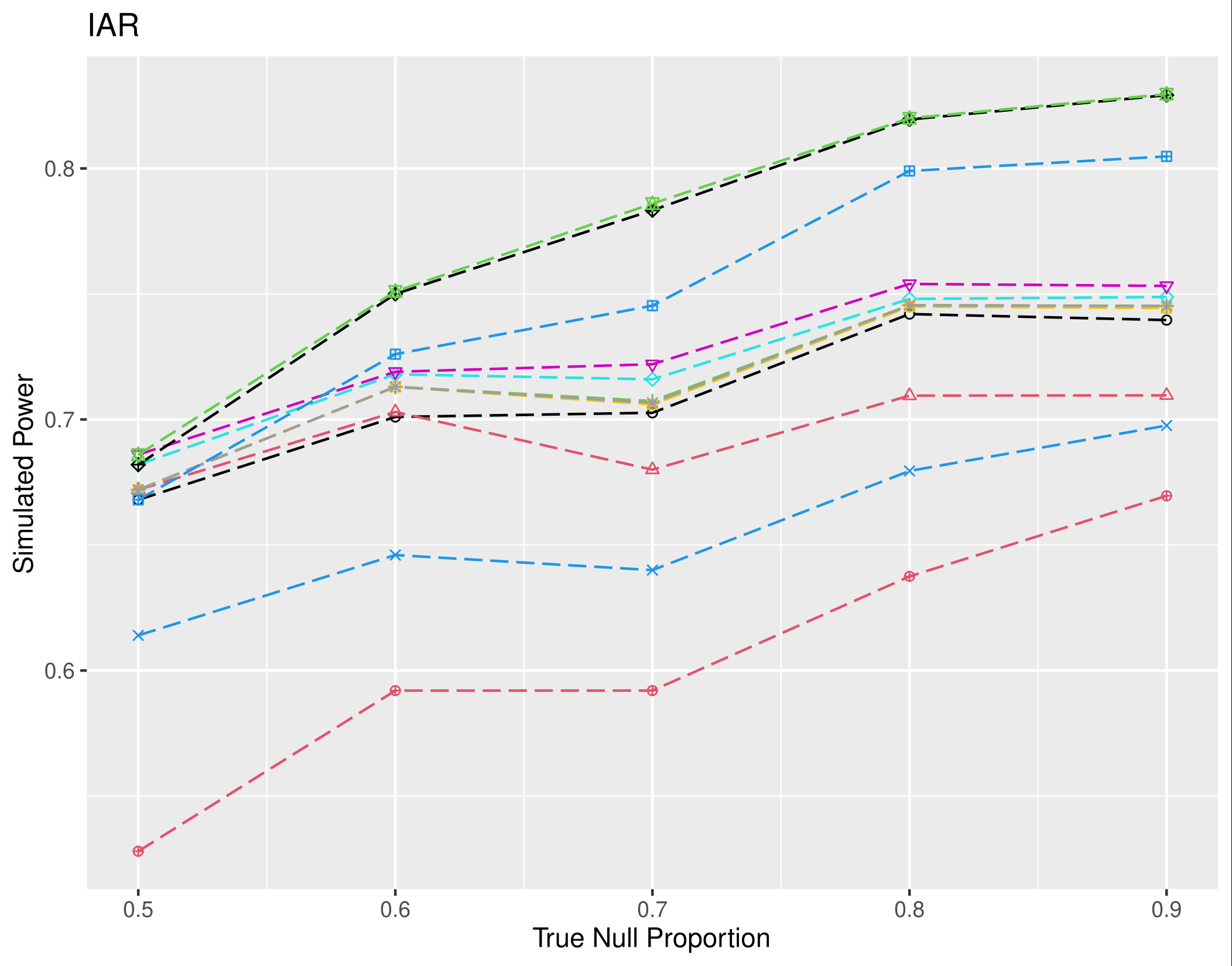} 
      \end{subfigure}    
  \end{tabular*} 
  \caption{Simulated Power (left column), simulated FDR (middle column) for fixed null proportion and simulated power (right column) for fixed signal strength, displayed for mean testing of $d=100$ parameters. Methods compared are SBH1 method (Circle and black), SBH2 method (Triangle point up and red), GSBH1 method (Plus and green), GSBH2 (Cross and blue), GSBH3 (Diamond and light blue), GSBH4 (Triangle point down and purple), GSBH5 (Square cross and yellow), GSBH6 (Star and grey), BH (Diamond plus and black), BY (Circle plus and red), dBH (Triangles up and down and green) and dBY (Square plus and blue)}
  \label{means:figure5} 
\end{figure}

\begin{figure}[ht]
  \begin{tabular*}{\textwidth}{
    @{}m{0.5cm}
    @{}m{\dimexpr0.33\textwidth-0.25cm\relax}
    @{}m{\dimexpr0.33\textwidth-0.25cm\relax}
    @{}m{\dimexpr0.33\textwidth-0.25cm\relax}}
  \rotatebox{90}{Block Diagonal}
  & \begin{subfigure}[b]{\linewidth}
      \centering
      \includegraphics[width=\linewidth]{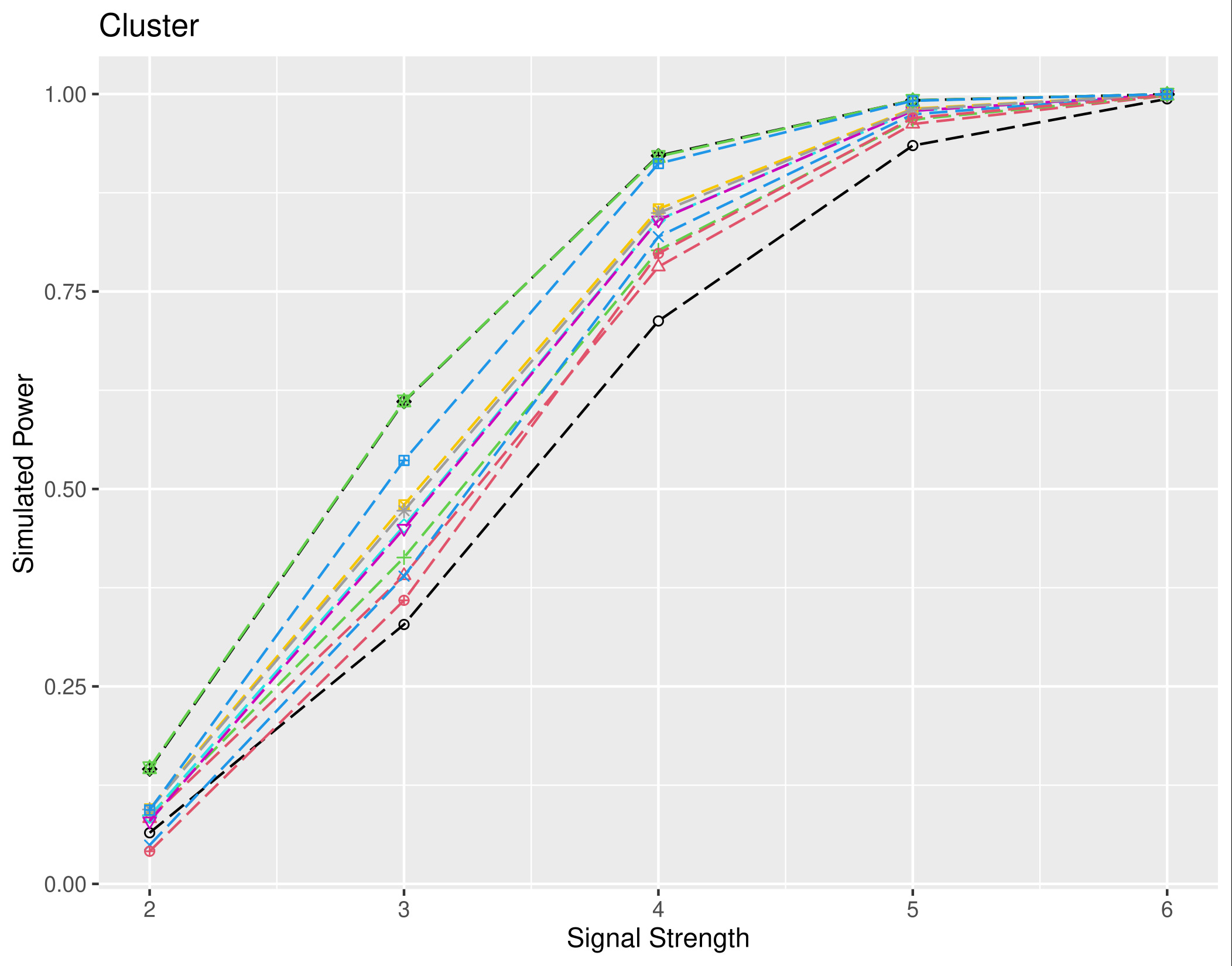} 
      \end{subfigure}  
  & \begin{subfigure}[b]{\linewidth}
      \centering
      \includegraphics[width=\linewidth]{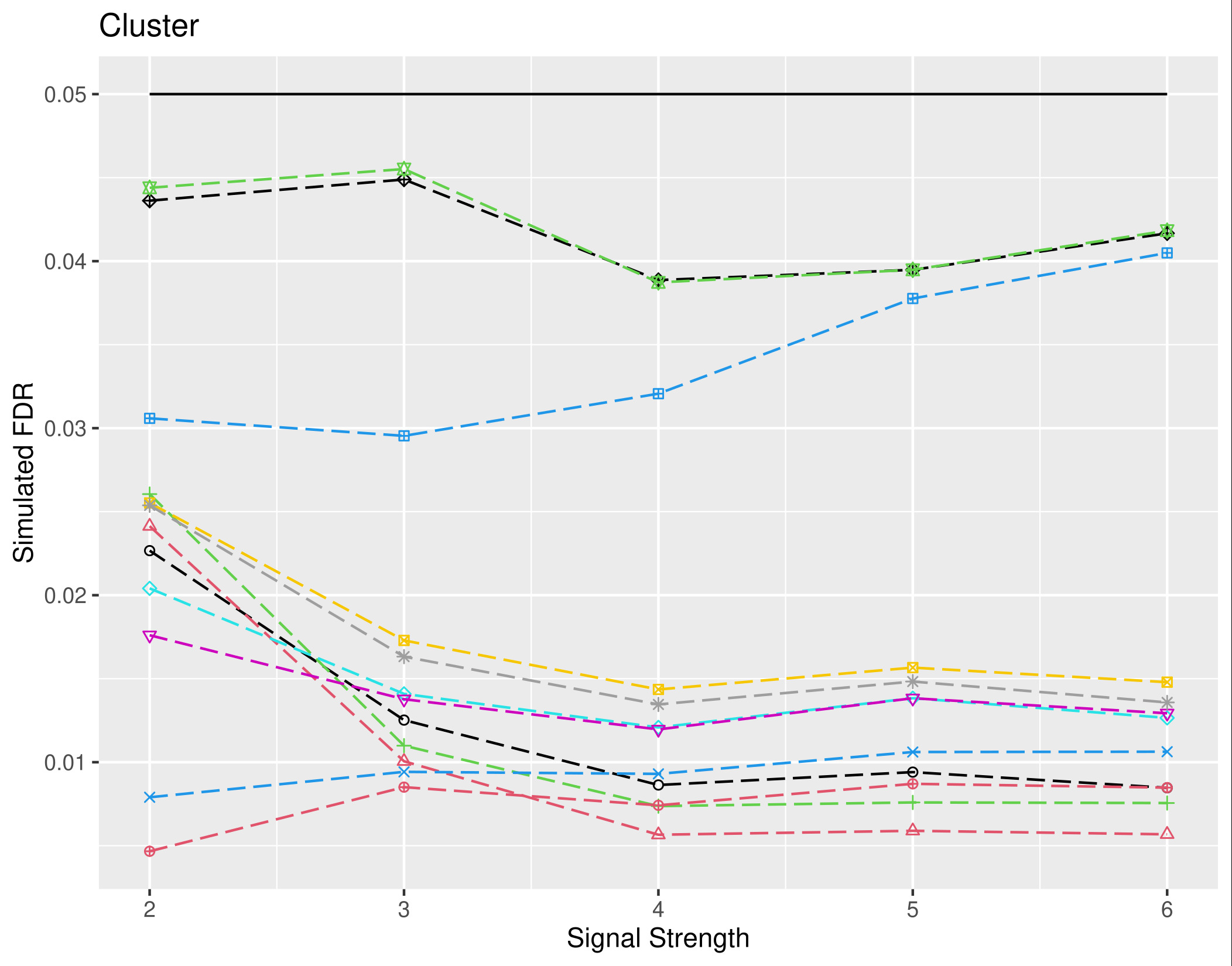} 
      \end{subfigure} 
  & \begin{subfigure}[b]{\linewidth}
      \centering
      \includegraphics[width=\linewidth]{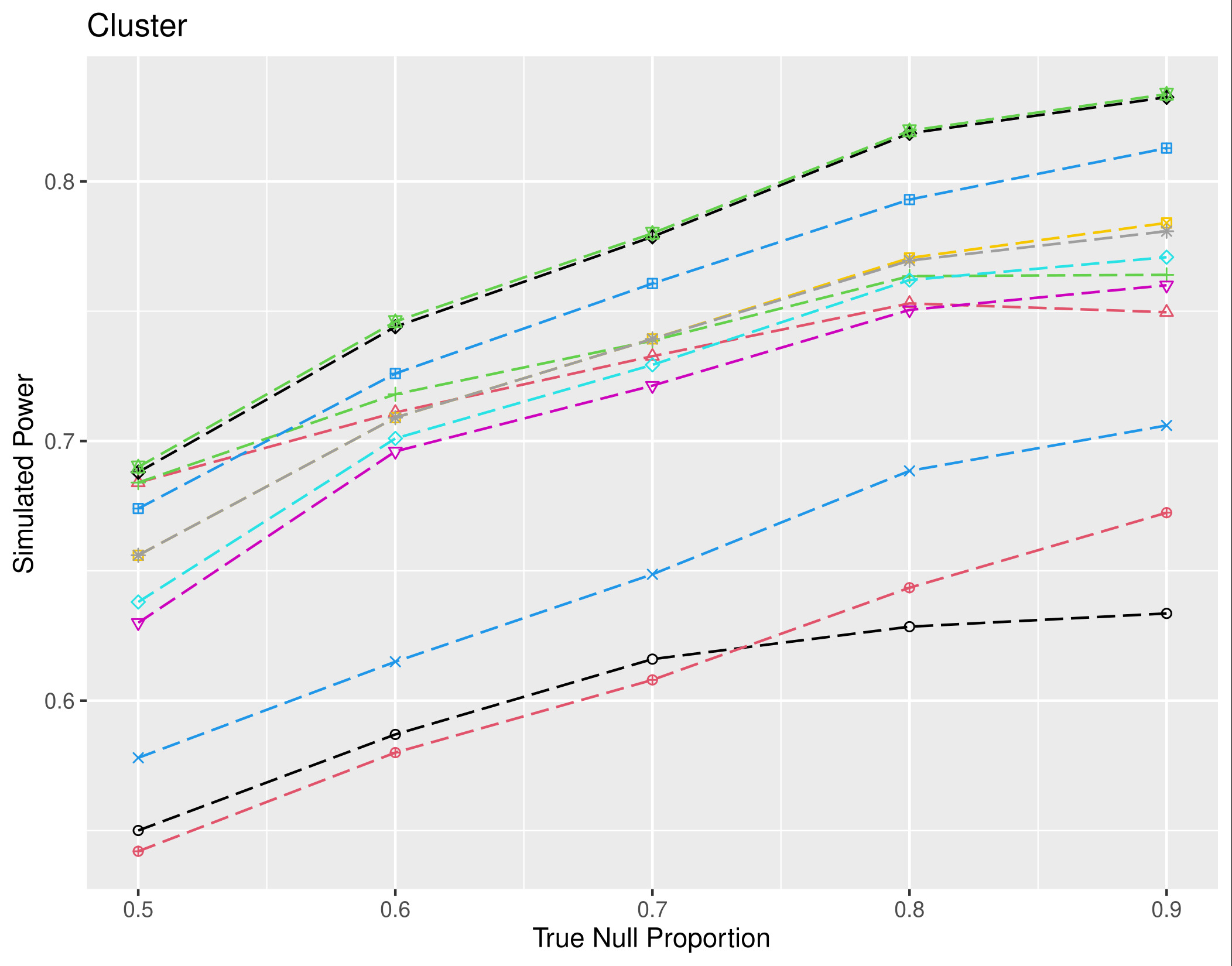} 
      \end{subfigure} \\
  \rotatebox{90}{Sparse} 
  & \begin{subfigure}[b]{\linewidth}
      \centering
      \includegraphics[width=\linewidth]{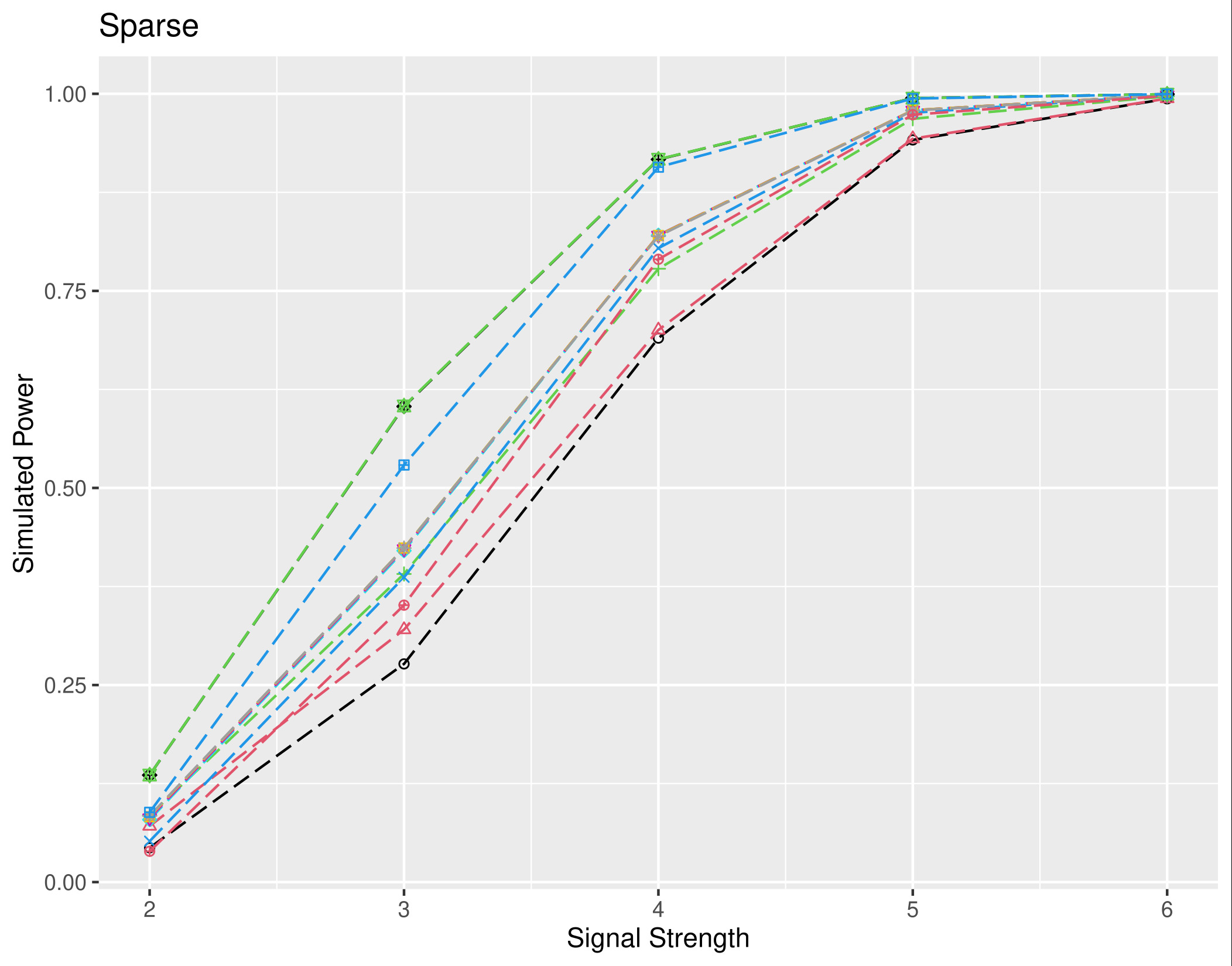} 
      \end{subfigure}  
  & \begin{subfigure}[b]{\linewidth}
      \centering
      \includegraphics[width=\linewidth]{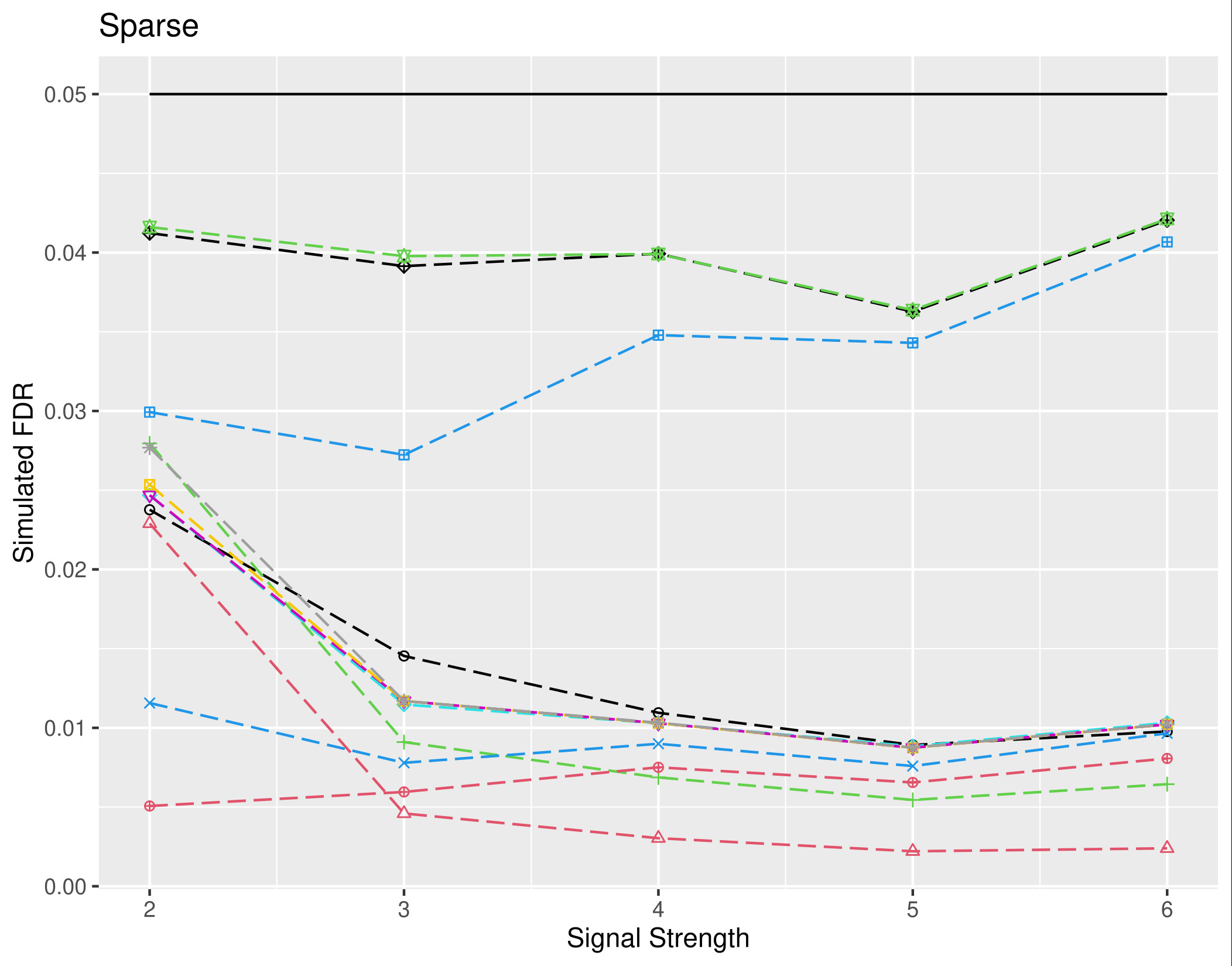} 
      \end{subfigure} 
  & \begin{subfigure}[b]{\linewidth}
      \centering
      \includegraphics[width=\linewidth]{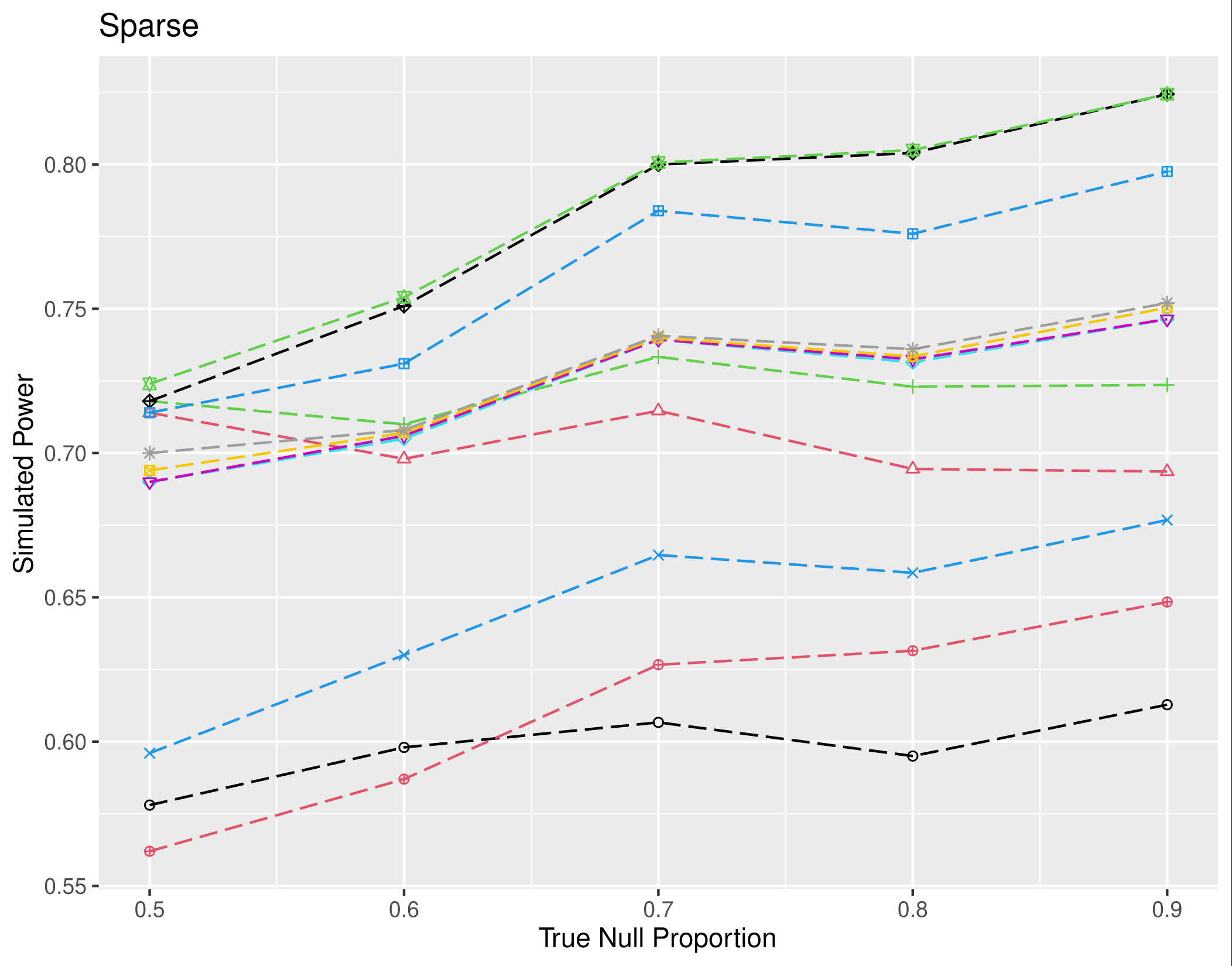} 
      \end{subfigure} \\
  \rotatebox{90}{Prefixed Corr 1} 
  & \begin{subfigure}[b]{\linewidth}
      \centering
      \includegraphics[width=\linewidth]{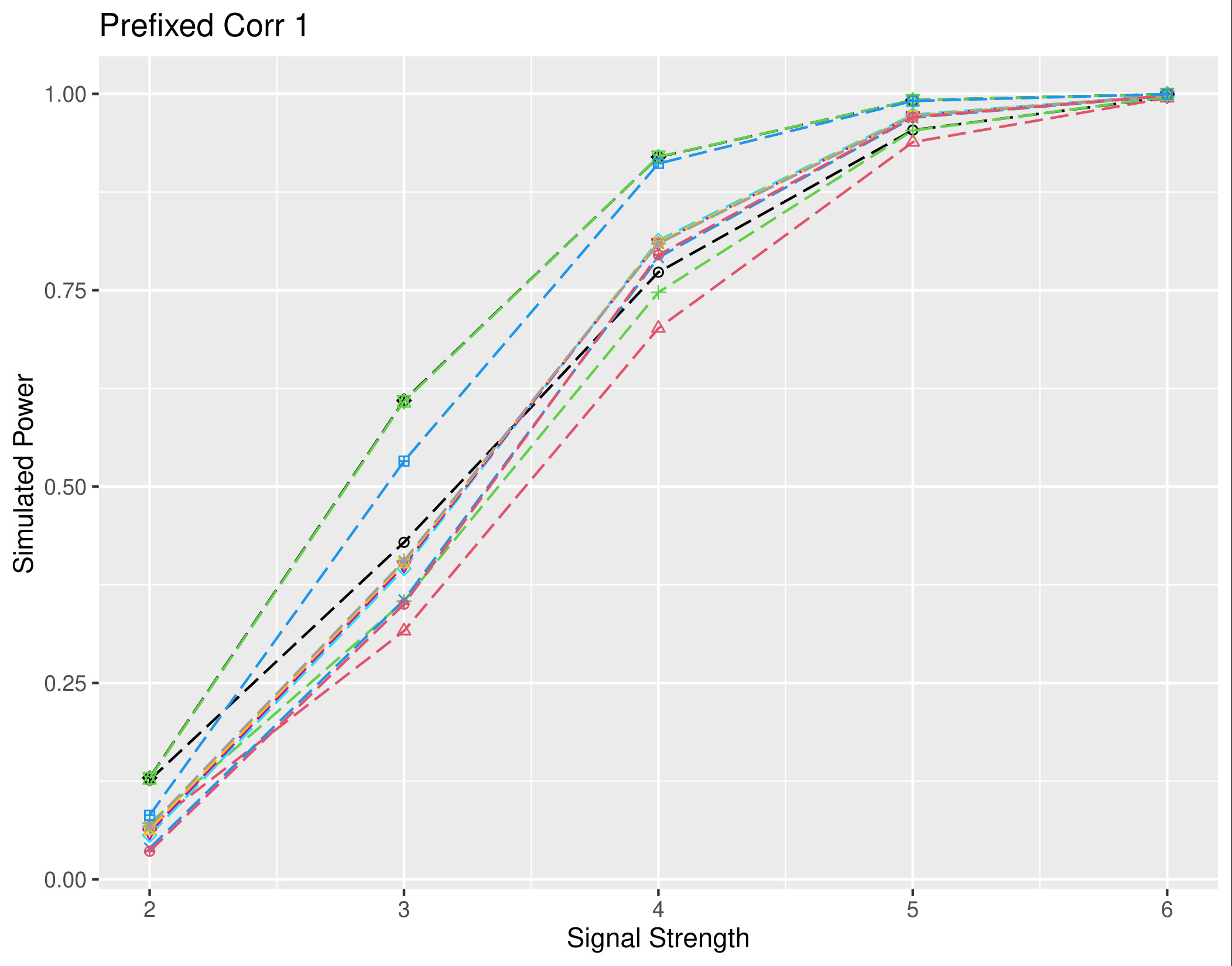} 
      \end{subfigure}  
  & \begin{subfigure}[b]{\linewidth}
      \centering
      \includegraphics[width=\linewidth]{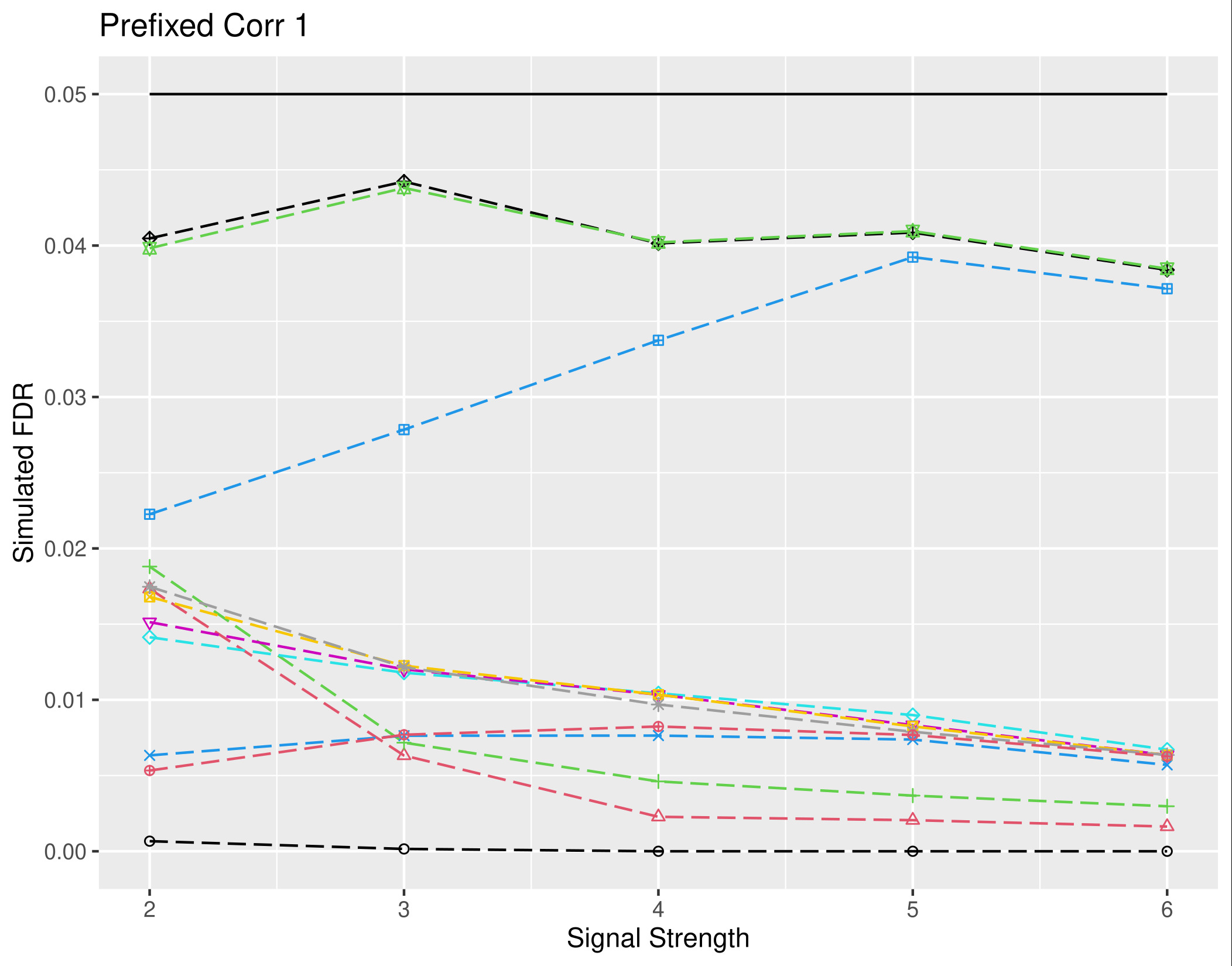} 
      \end{subfigure} 
  & \begin{subfigure}[b]{\linewidth}
      \centering
      \includegraphics[width=\linewidth]{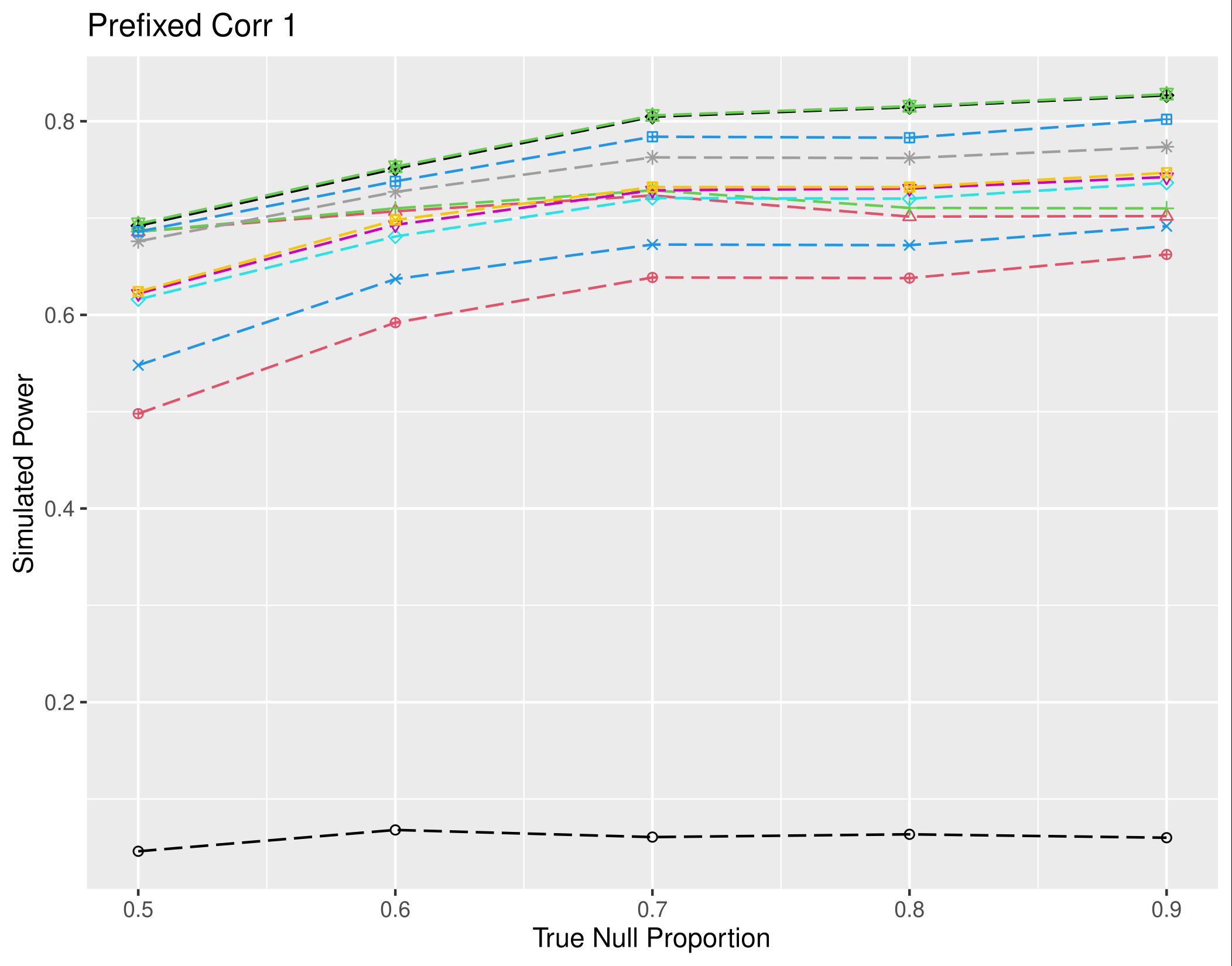} 
      \end{subfigure}    
  \end{tabular*} 
  \caption{Simulated Power (left column), simulated FDR (middle column) for fixed null proportion and simulated power (right column) for fixed signal strength, displayed for mean testing of $d=100$ parameters. Methods compared are SBH1 method (Circle and black), SBH2 method (Triangle point up and red), GSBH1 method (Plus and green), GSBH2 (Cross and blue), GSBH3 (Diamond and light blue), GSBH4 (Triangle point down and purple), GSBH5 (Square cross and yellow), GSBH6 (Star and grey), BH (Diamond plus and black), BY (Circle plus and red), dBH (Triangles up and down and green) and dBY (Square plus and blue)}
  \label{means:figure6} 
\end{figure}

\begin{figure}[ht]
  \begin{tabular*}{\textwidth}{
    @{}m{0.5cm}
    @{}m{\dimexpr0.33\textwidth-0.25cm\relax}
    @{}m{\dimexpr0.33\textwidth-0.25cm\relax}
    @{}m{\dimexpr0.33\textwidth-0.25cm\relax}}
  \rotatebox{90}{Equi(0.7)}
  & \begin{subfigure}[b]{\linewidth}
      \centering
      \includegraphics[width=\linewidth]{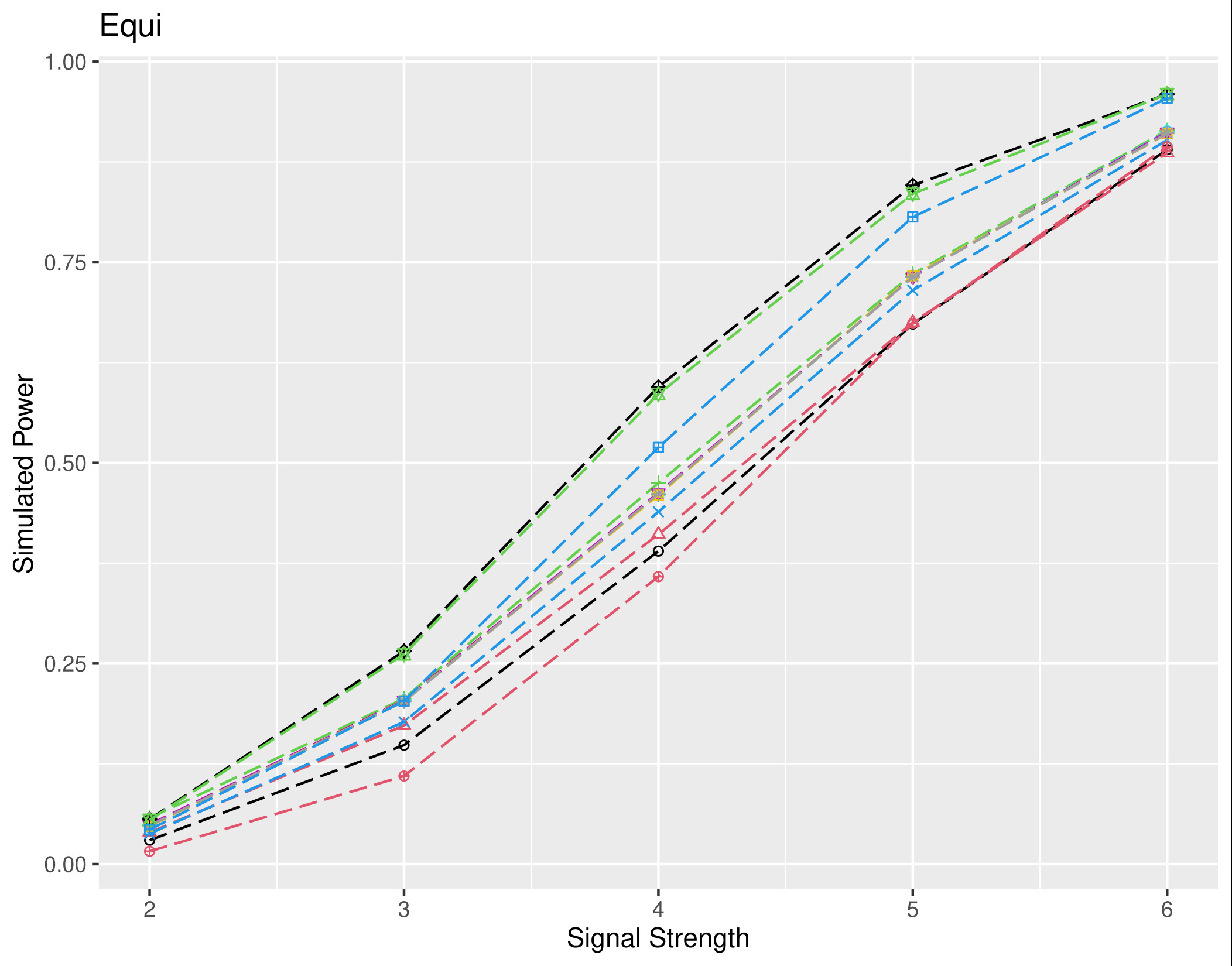} 
      \end{subfigure}  
  & \begin{subfigure}[b]{\linewidth}
      \centering
      \includegraphics[width=\linewidth]{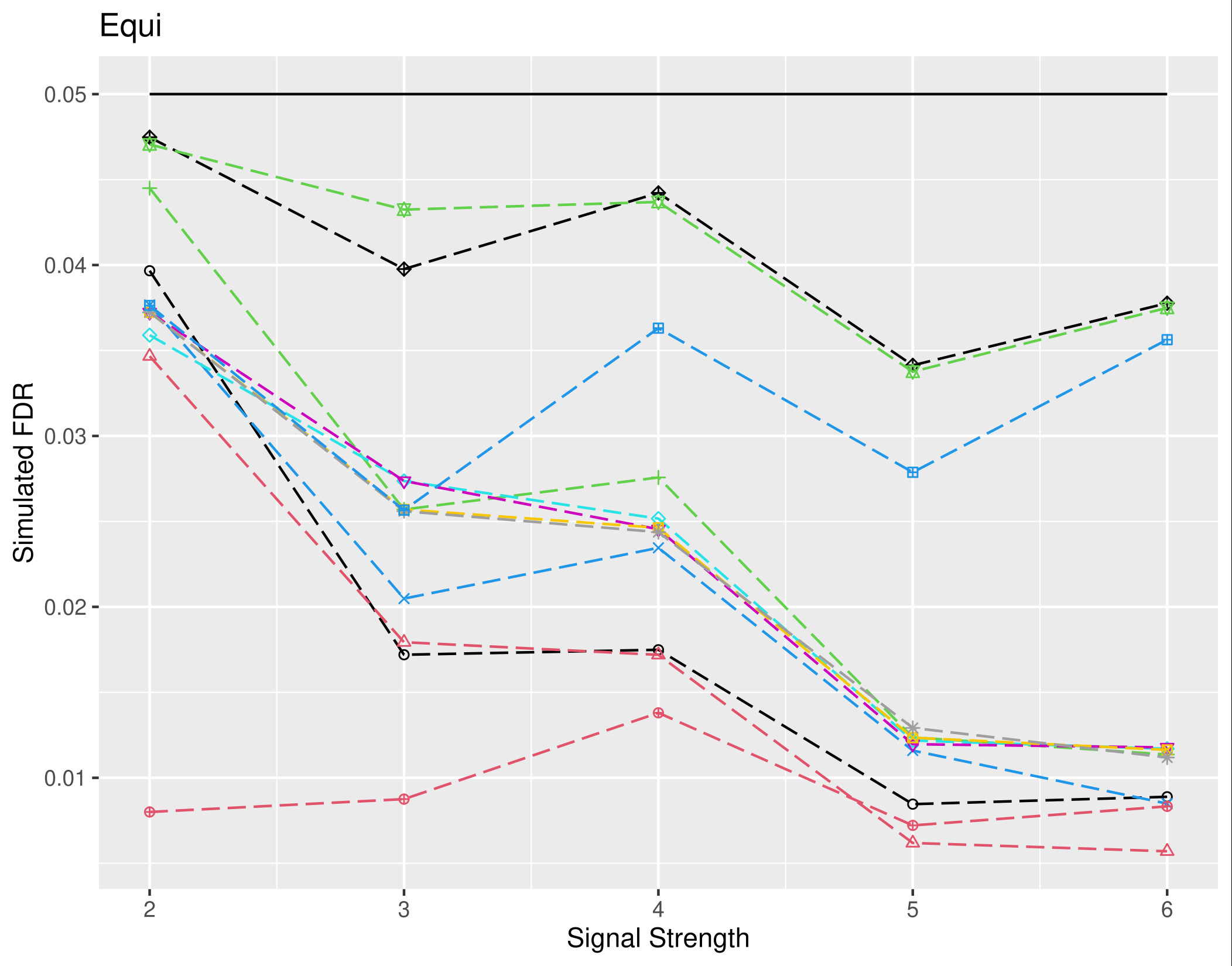} 
      \end{subfigure} 
  & \begin{subfigure}[b]{\linewidth}
      \centering
      \includegraphics[width=\linewidth]{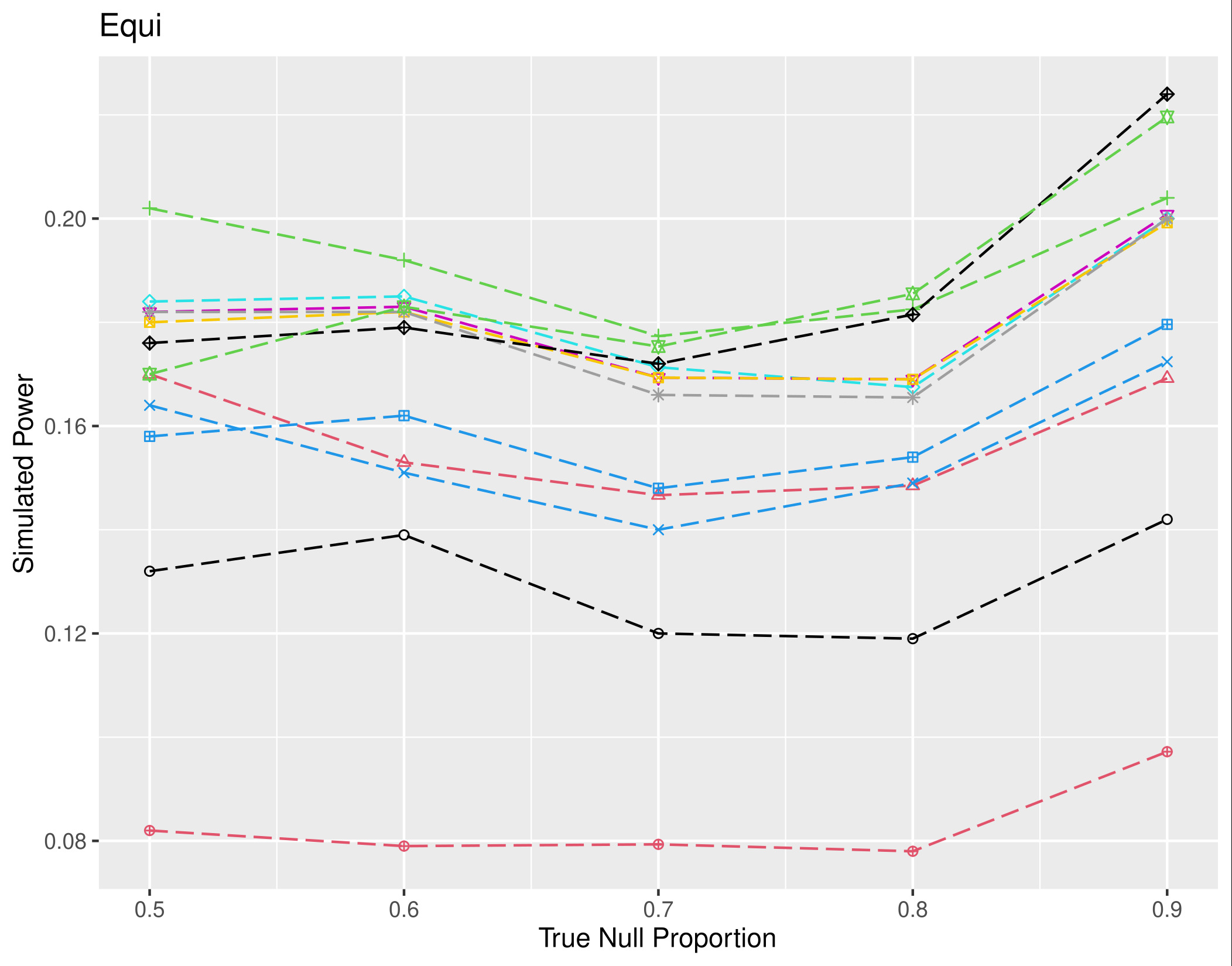} 
      \end{subfigure} \\
  \rotatebox{90}{AR(0.7)} 
  & \begin{subfigure}[b]{\linewidth}
      \centering
      \includegraphics[width=\linewidth]{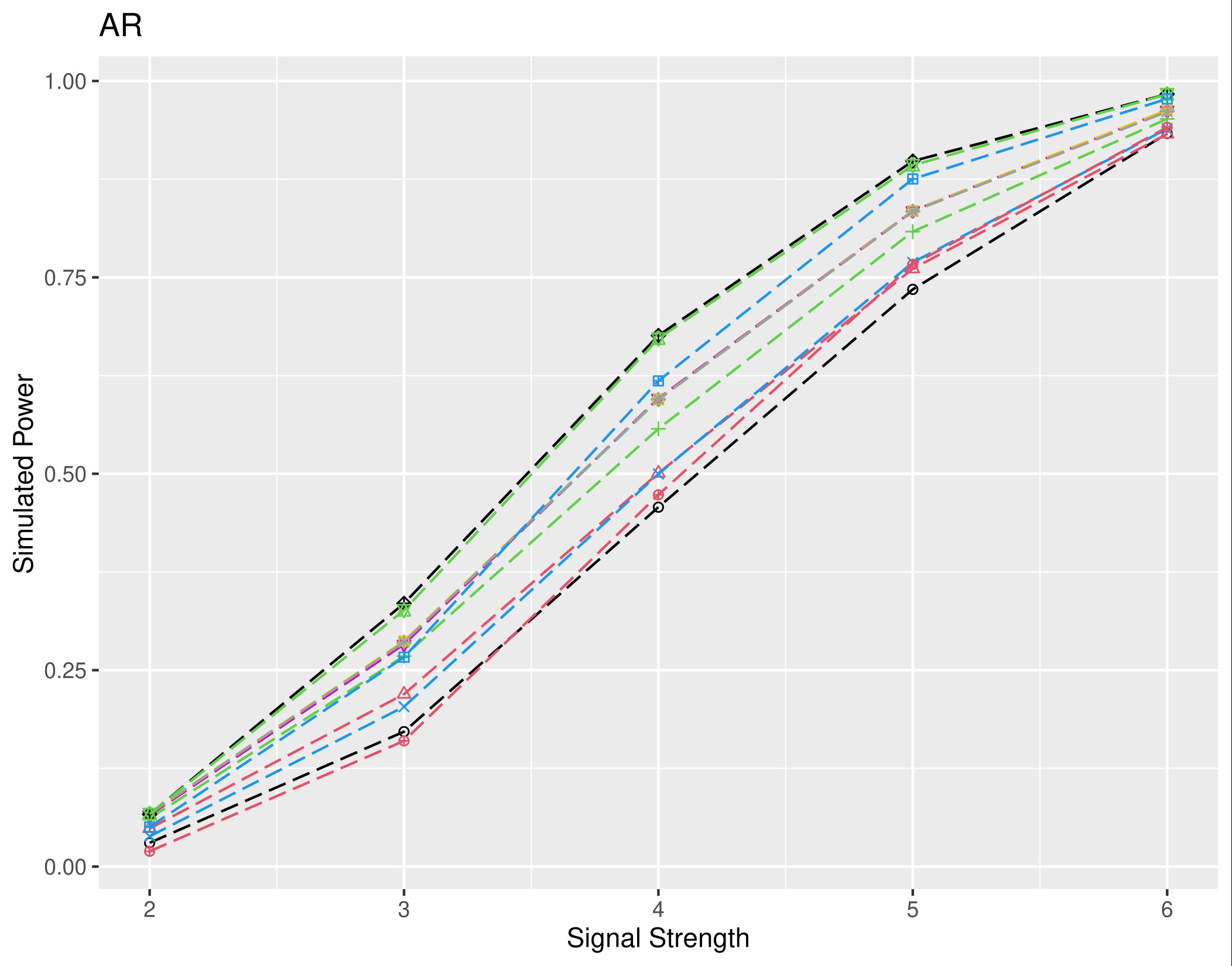} 
      \end{subfigure}  
  & \begin{subfigure}[b]{\linewidth}
      \centering
      \includegraphics[width=\linewidth]{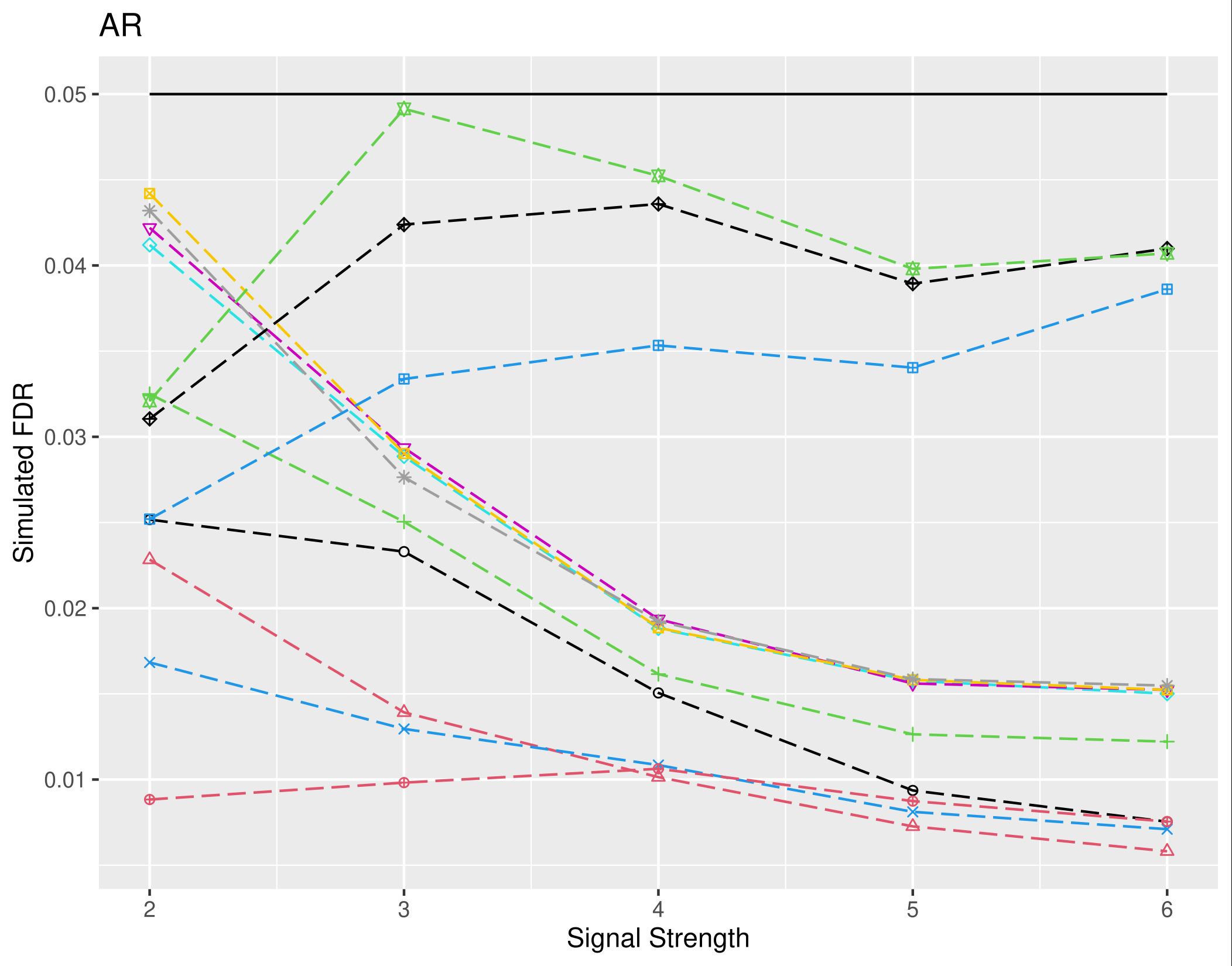} 
      \end{subfigure} 
  & \begin{subfigure}[b]{\linewidth}
      \centering
      \includegraphics[width=\linewidth]{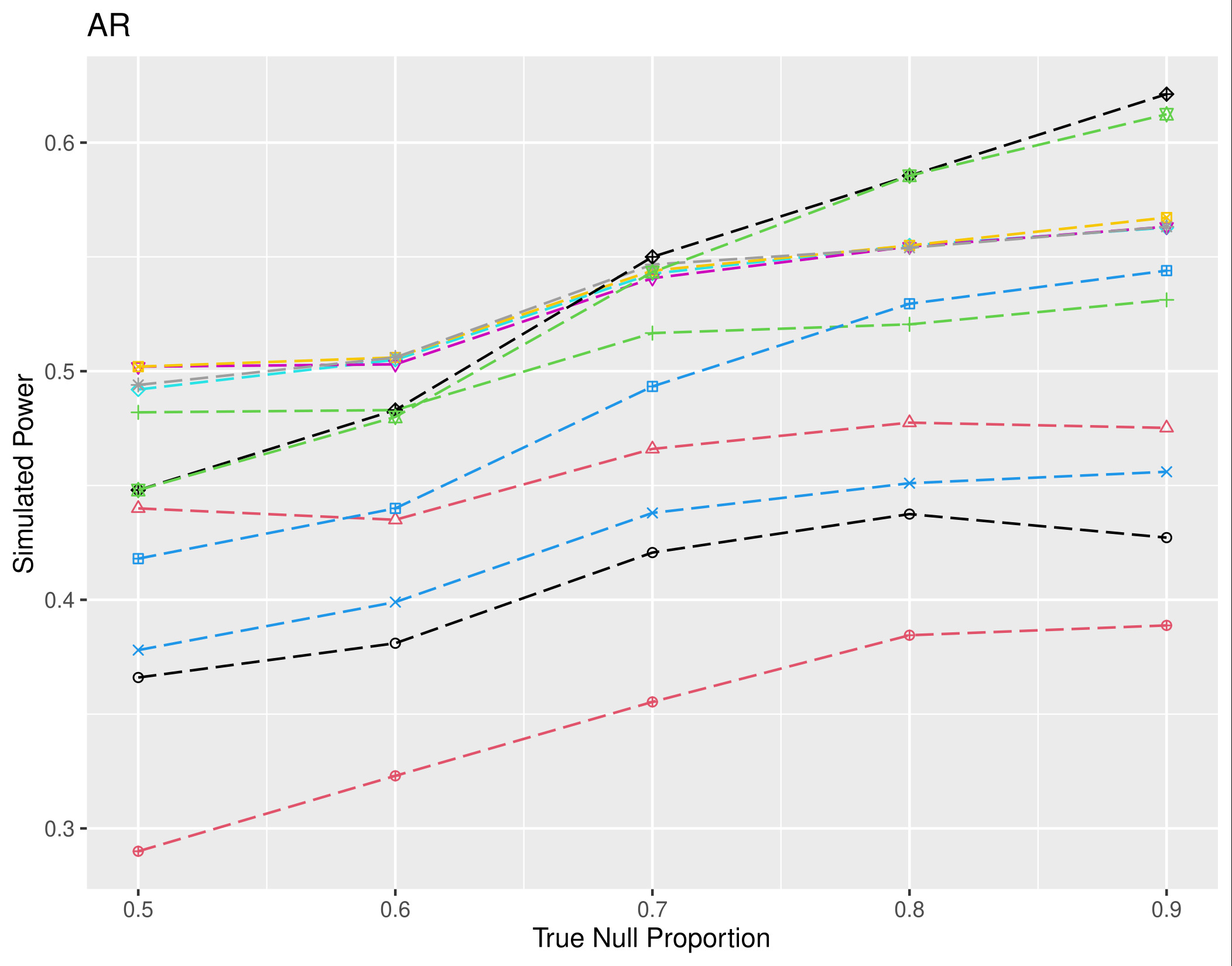} 
      \end{subfigure} \\
  \rotatebox{90}{IAR(0.7)} 
  & \begin{subfigure}[b]{\linewidth}
      \centering
      \includegraphics[width=\linewidth]{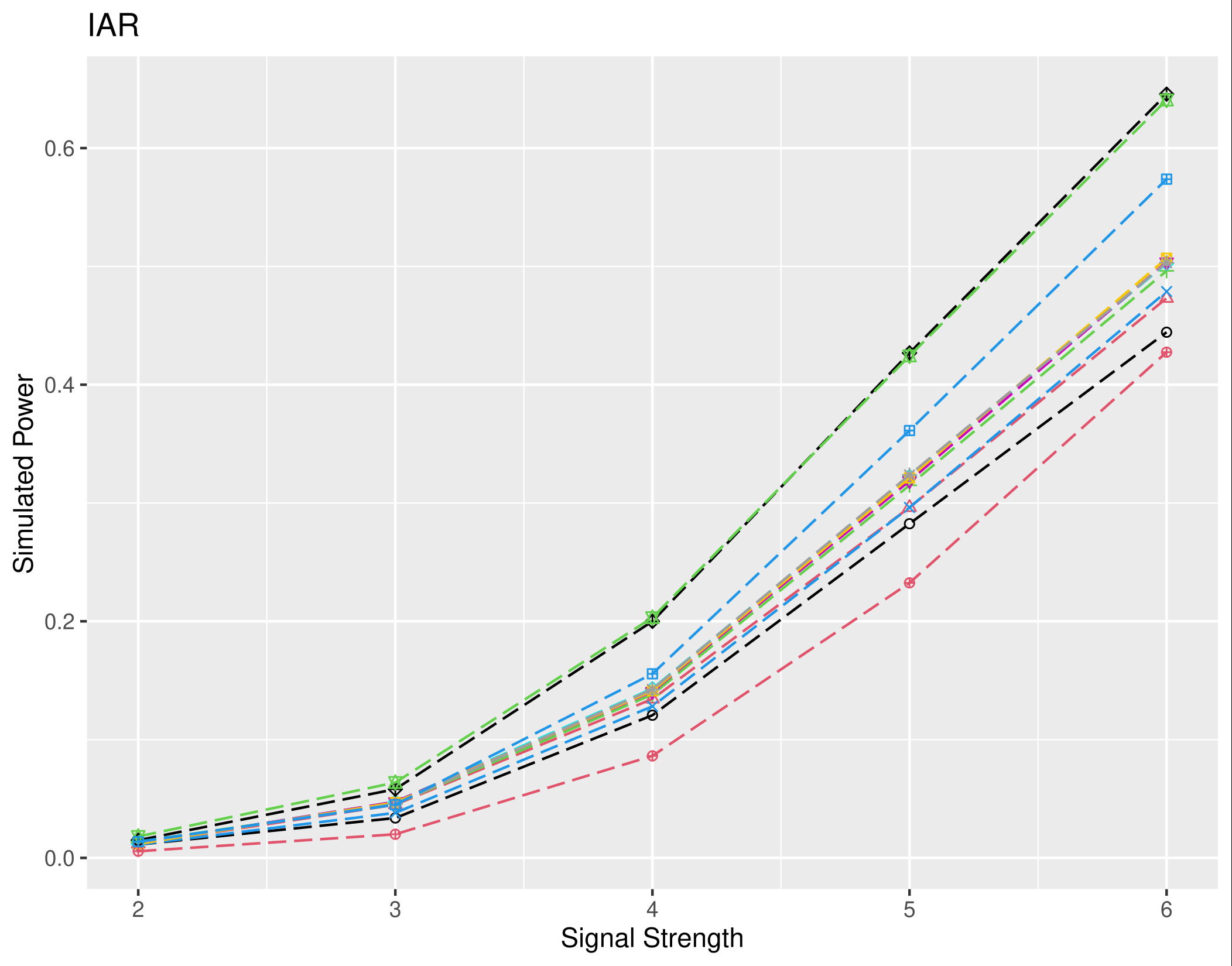} 
      \end{subfigure}  
  & \begin{subfigure}[b]{\linewidth}
      \centering
      \includegraphics[width=\linewidth]{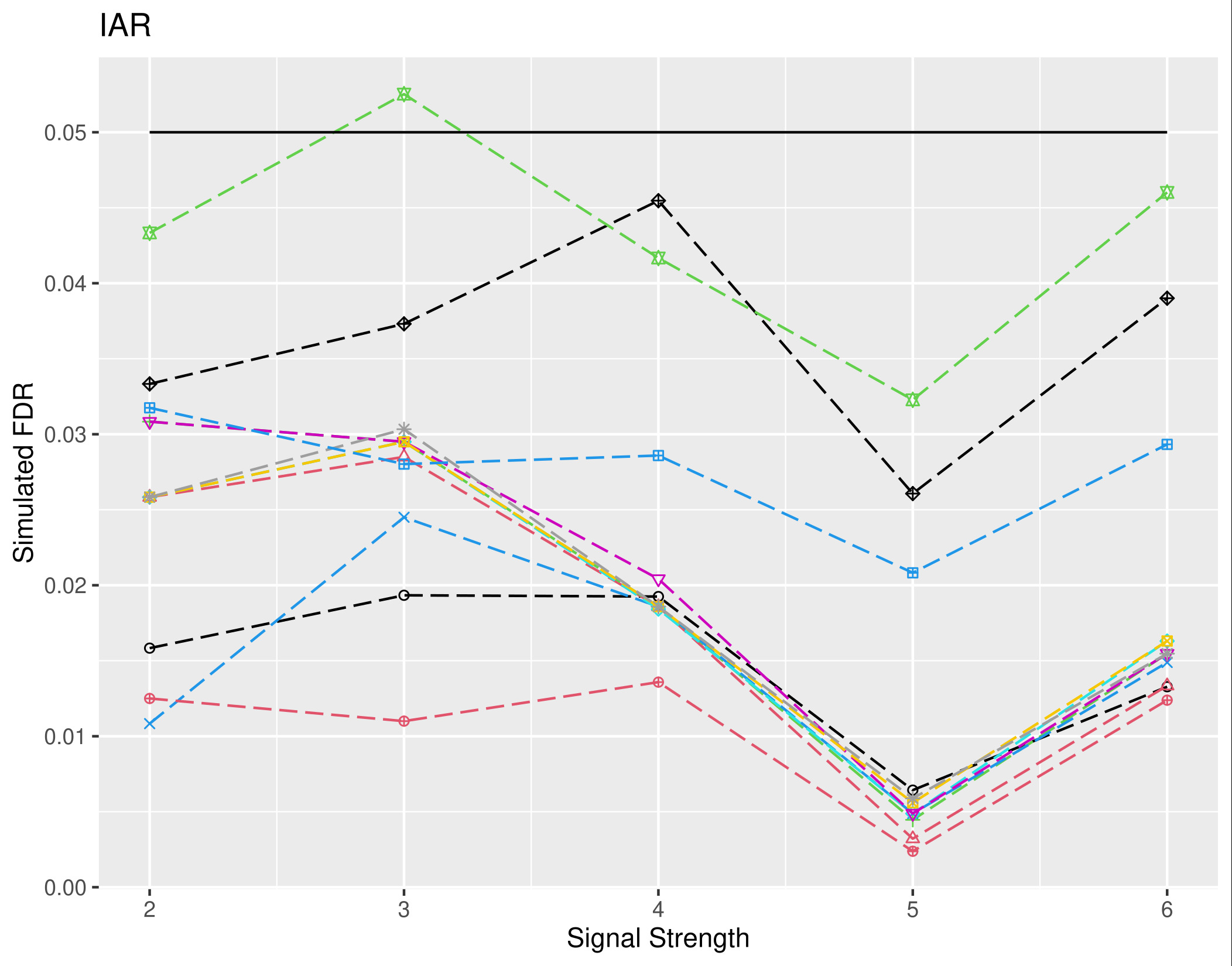} 
      \end{subfigure} 
  & \begin{subfigure}[b]{\linewidth}
      \centering
      \includegraphics[width=\linewidth]{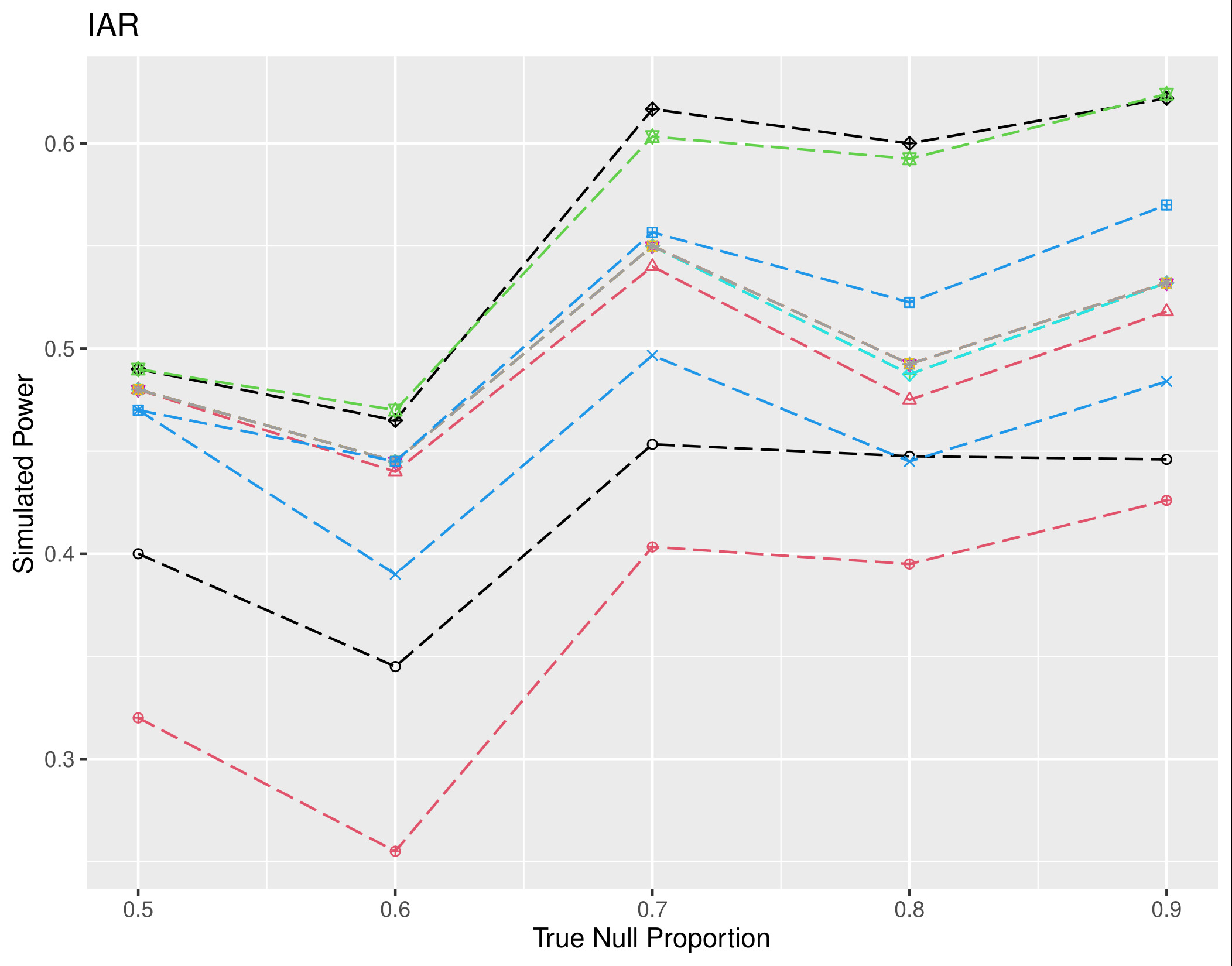} 
      \end{subfigure}    
  \end{tabular*} 
  \caption{Simulated Power (left column), simulated FDR (middle column) for fixed null proportion and simulated power (right column) for fixed signal strength, displayed for variable selection from $d=40$ parameters. Methods compared are SBH1 method (Circle and black), SBH2 method (Triangle point up and red), GSBH1 method (Plus and green), GSBH2 (Cross and blue), GSBH3 (Diamond and light blue), GSBH4 (Triangle point down and purple), GSBH5 (Square cross and yellow), GSBH6 (Star and grey), BH (Diamond plus and black), BY (Circle plus and red), dBH (Triangles up and down and green) and dBY (Square plus and blue)}
  \label{var_sel:figure2} 
\end{figure}

\begin{figure}[ht]
  \begin{tabular*}{\textwidth}{
    @{}m{0.5cm}
    @{}m{\dimexpr0.33\textwidth-0.25cm\relax}
    @{}m{\dimexpr0.33\textwidth-0.25cm\relax}
    @{}m{\dimexpr0.33\textwidth-0.25cm\relax}}
  \rotatebox{90}{Block Diagonal}
  & \begin{subfigure}[b]{\linewidth}
      \centering
      \includegraphics[width=\linewidth]{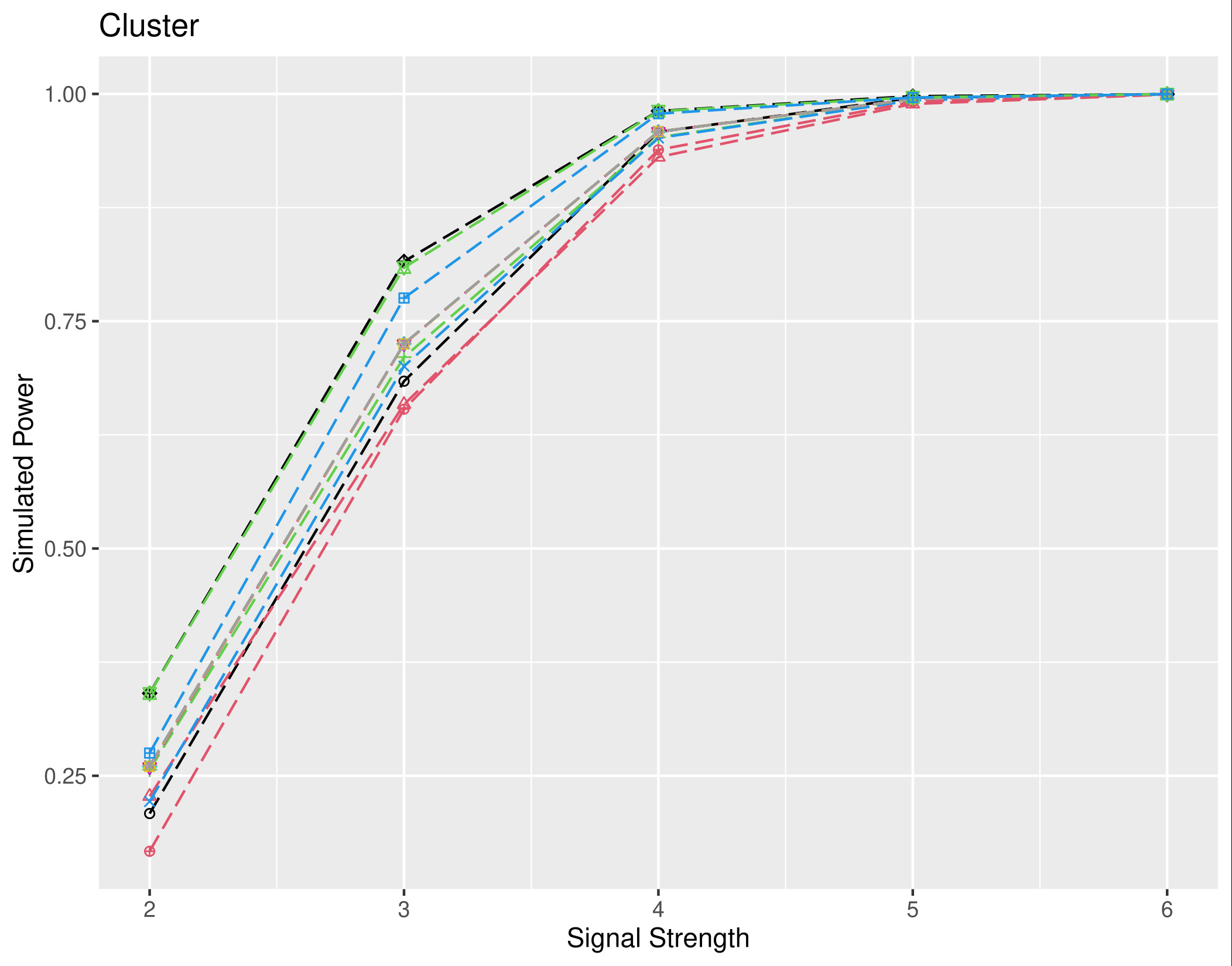} 
      \end{subfigure}  
  & \begin{subfigure}[b]{\linewidth}
      \centering
      \includegraphics[width=\linewidth]{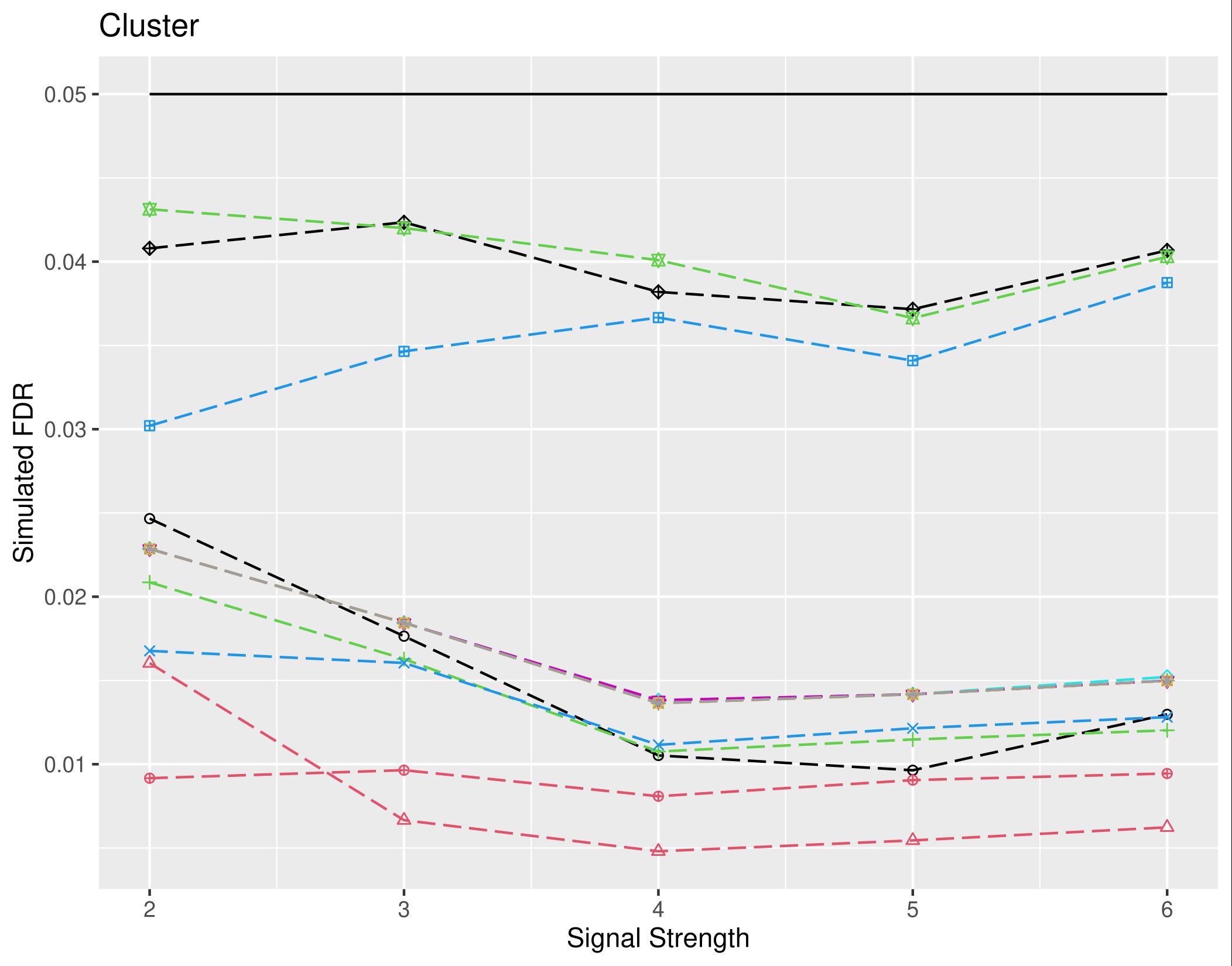} 
      \end{subfigure} 
  & \begin{subfigure}[b]{\linewidth}
      \centering
      \includegraphics[width=\linewidth]{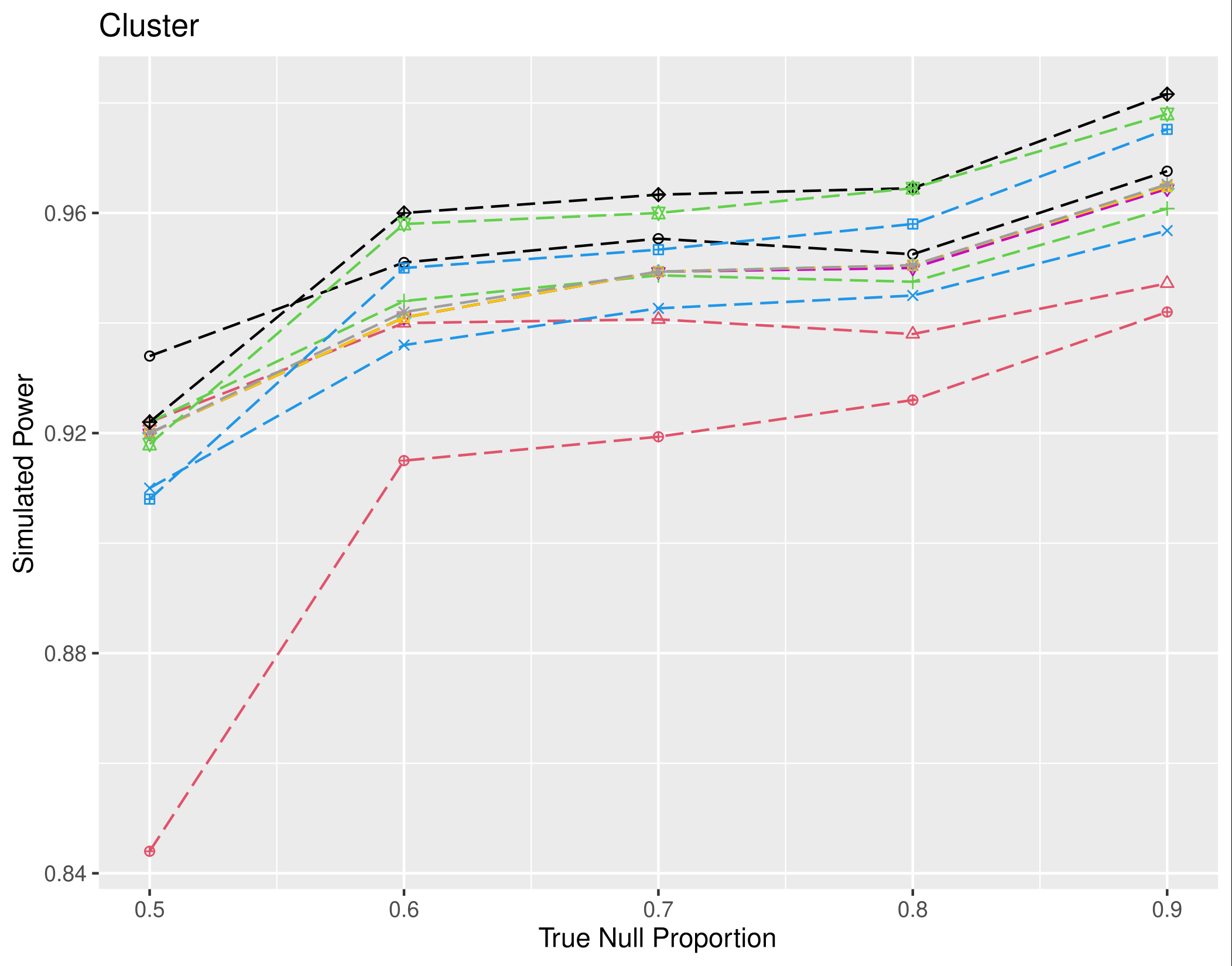} 
      \end{subfigure} \\
  \rotatebox{90}{Sparse} 
  & \begin{subfigure}[b]{\linewidth}
      \centering
      \includegraphics[width=\linewidth]{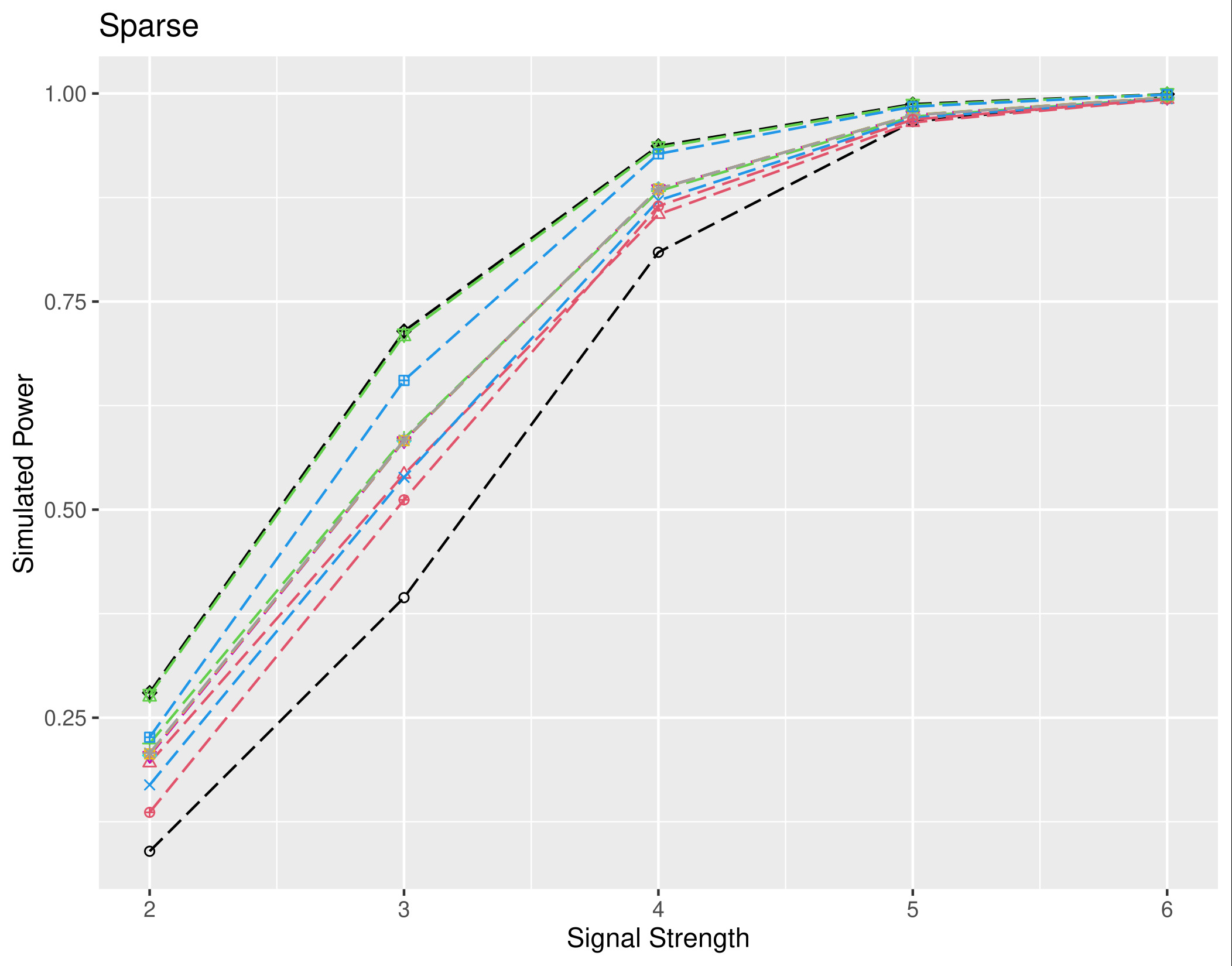} 
      \end{subfigure}  
  & \begin{subfigure}[b]{\linewidth}
      \centering
      \includegraphics[width=\linewidth]{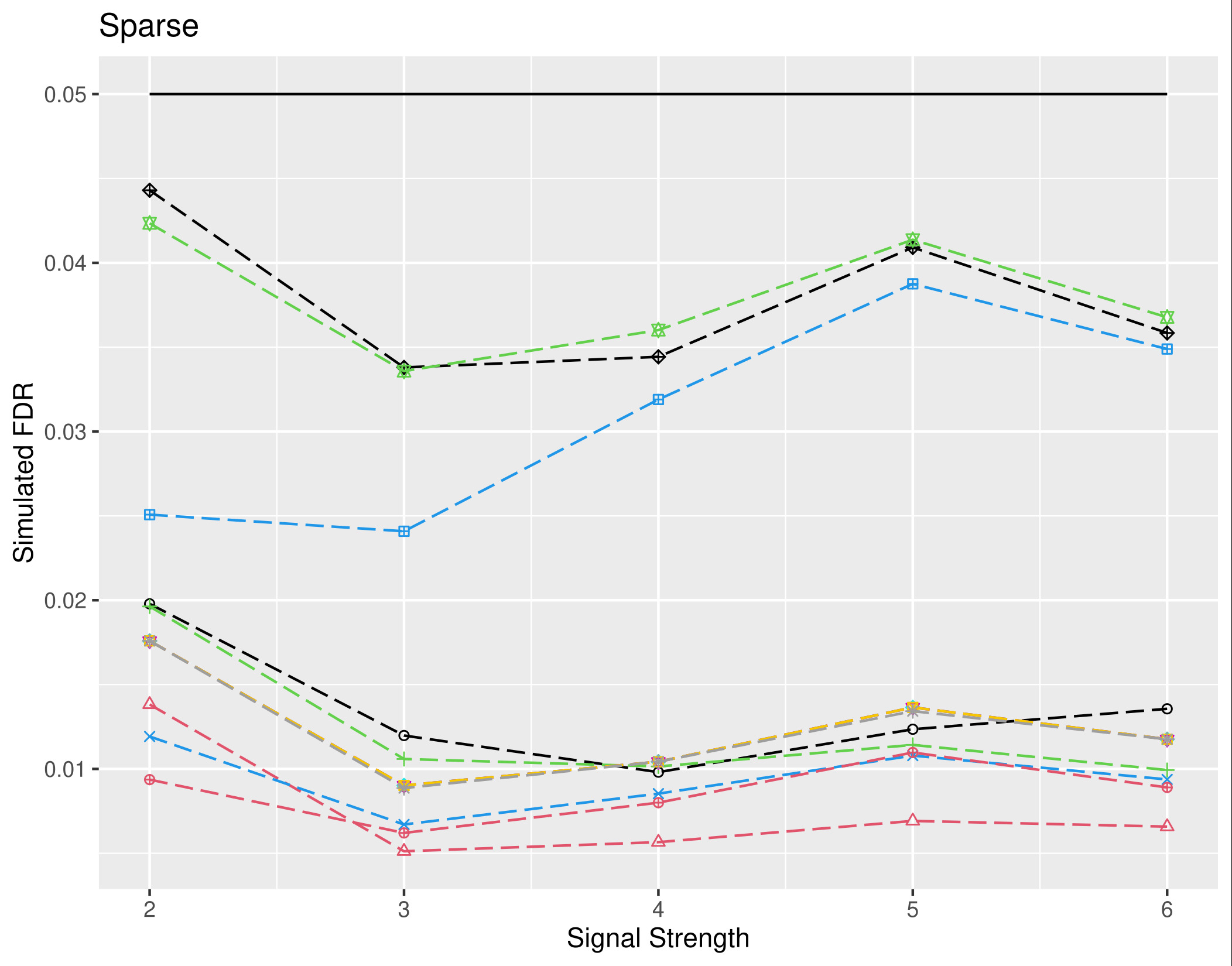} 
      \end{subfigure} 
  & \begin{subfigure}[b]{\linewidth}
      \centering
      \includegraphics[width=\linewidth]{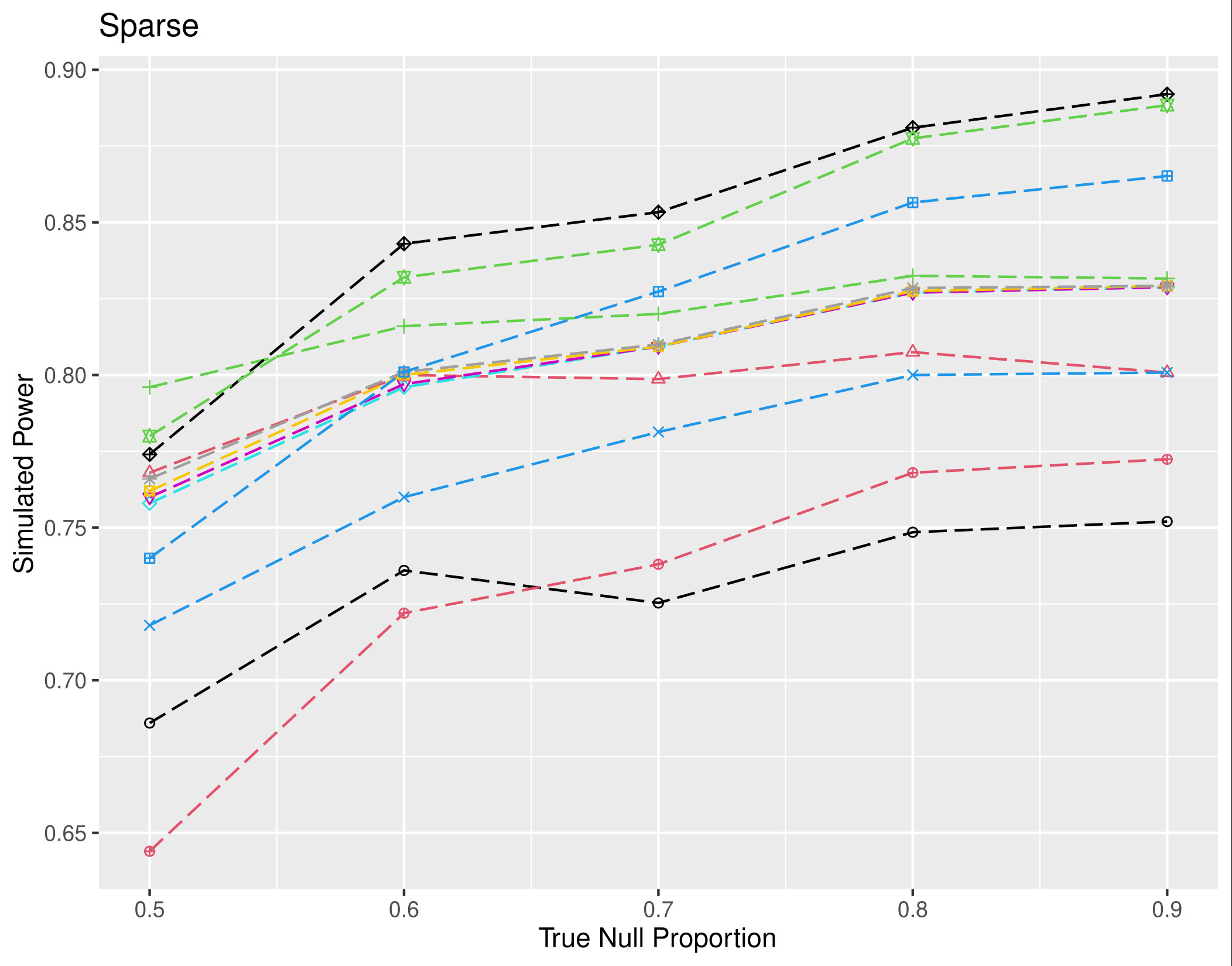} 
      \end{subfigure} \\
  \rotatebox{90}{Prefixed Corr 1} 
  & \begin{subfigure}[b]{\linewidth}
      \centering
      \includegraphics[width=\linewidth]{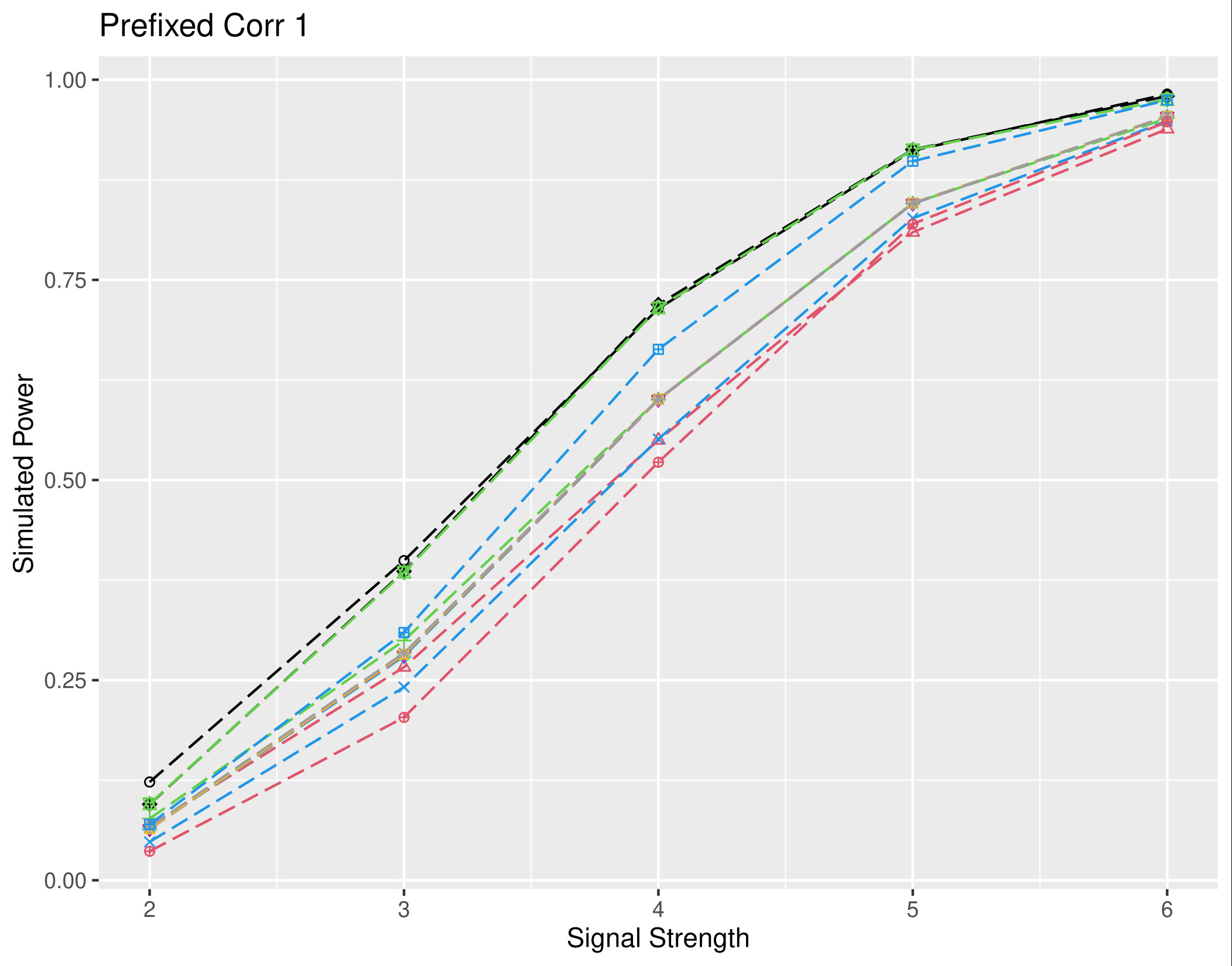} 
      \end{subfigure}  
  & \begin{subfigure}[b]{\linewidth}
      \centering
      \includegraphics[width=\linewidth]{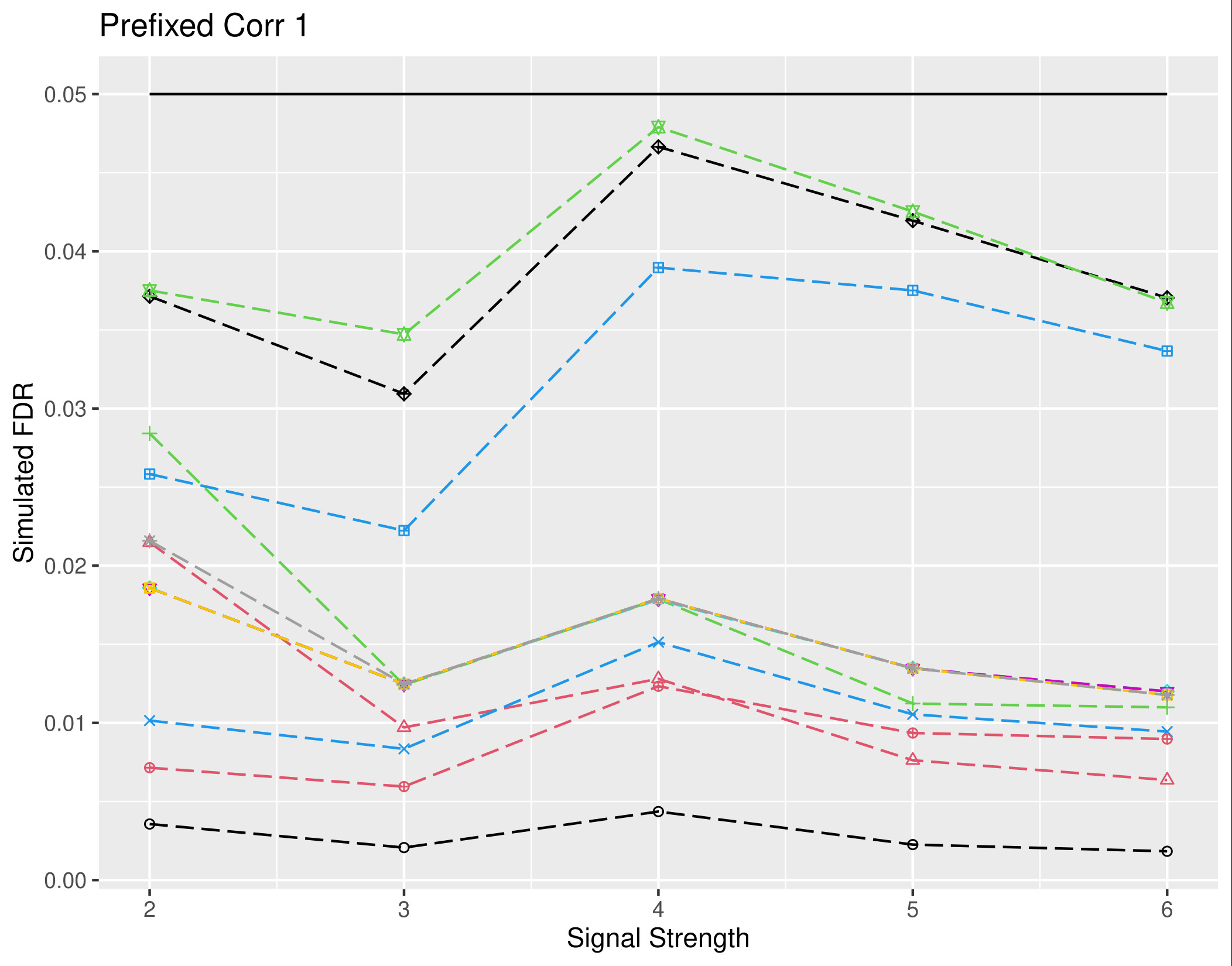} 
      \end{subfigure} 
  & \begin{subfigure}[b]{\linewidth}
      \centering
      \includegraphics[width=\linewidth]{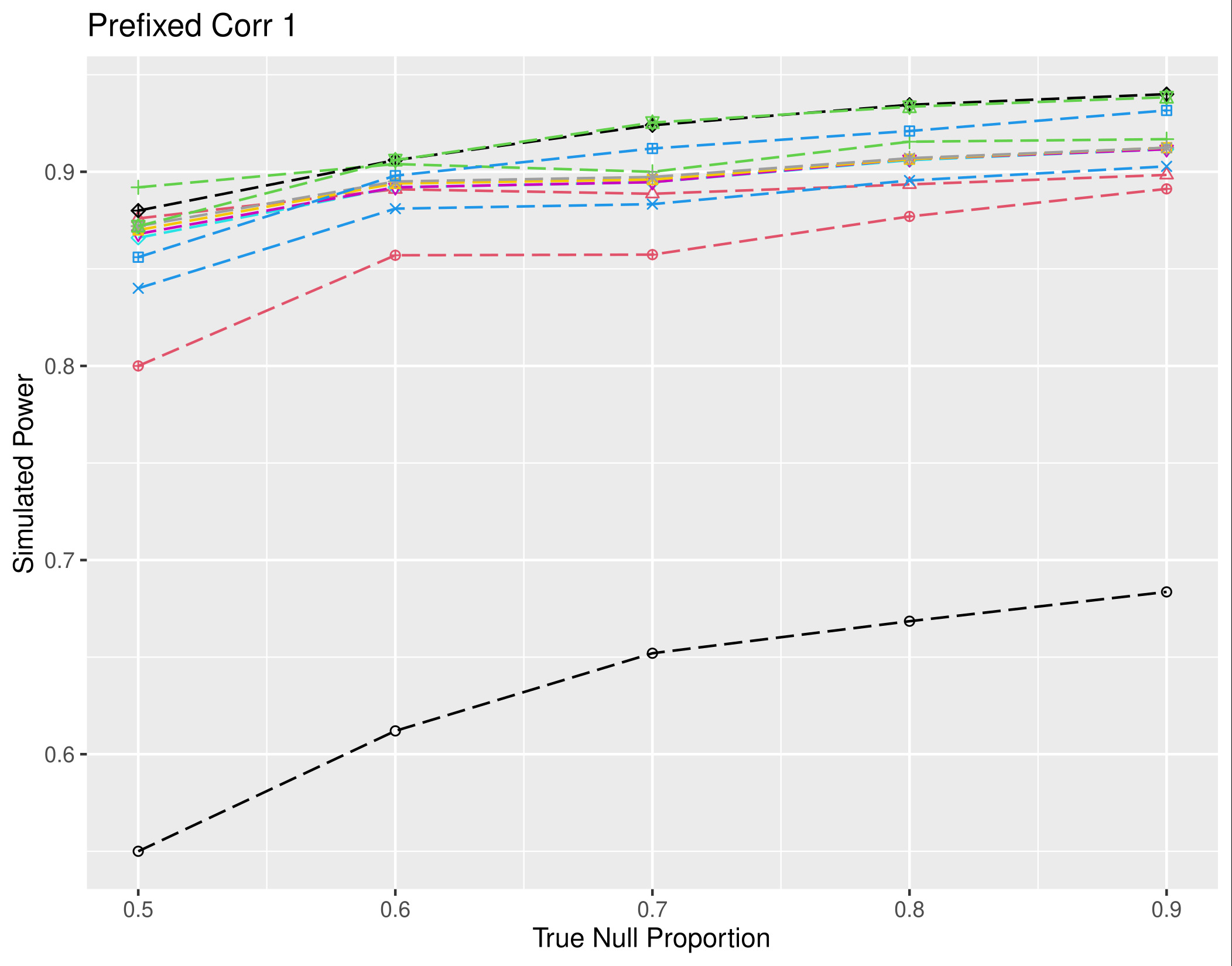} 
      \end{subfigure}    
  \end{tabular*} 
  \caption{Simulated Power (left column), simulated FDR (middle column) for fixed null proportion and simulated power (right column) for fixed signal strength, displayed for variable selection from $d=40$ parameters. Methods compared are SBH1 method (Circle and black), SBH2 method (Triangle point up and red), GSBH1 method (Plus and green), GSBH2 (Cross and blue), GSBH3 (Diamond and light blue), GSBH4 (Triangle point down and purple), GSBH5 (Square cross and yellow), GSBH6 (Star and grey), BH (Diamond plus and black), BY (Circle plus and red), dBH (Triangles up and down and green) and dBY (Square plus and blue)}
  \label{var_sel:figure3} 
\end{figure}

\begin{figure}[ht]
  \begin{tabular*}{\textwidth}{
    @{}m{0.5cm}
    @{}m{\dimexpr0.33\textwidth-0.25cm\relax}
    @{}m{\dimexpr0.33\textwidth-0.25cm\relax}
    @{}m{\dimexpr0.33\textwidth-0.25cm\relax}}
  \rotatebox{90}{Equi(0.3)}
  & \begin{subfigure}[b]{\linewidth}
      \centering
      \includegraphics[width=\linewidth]{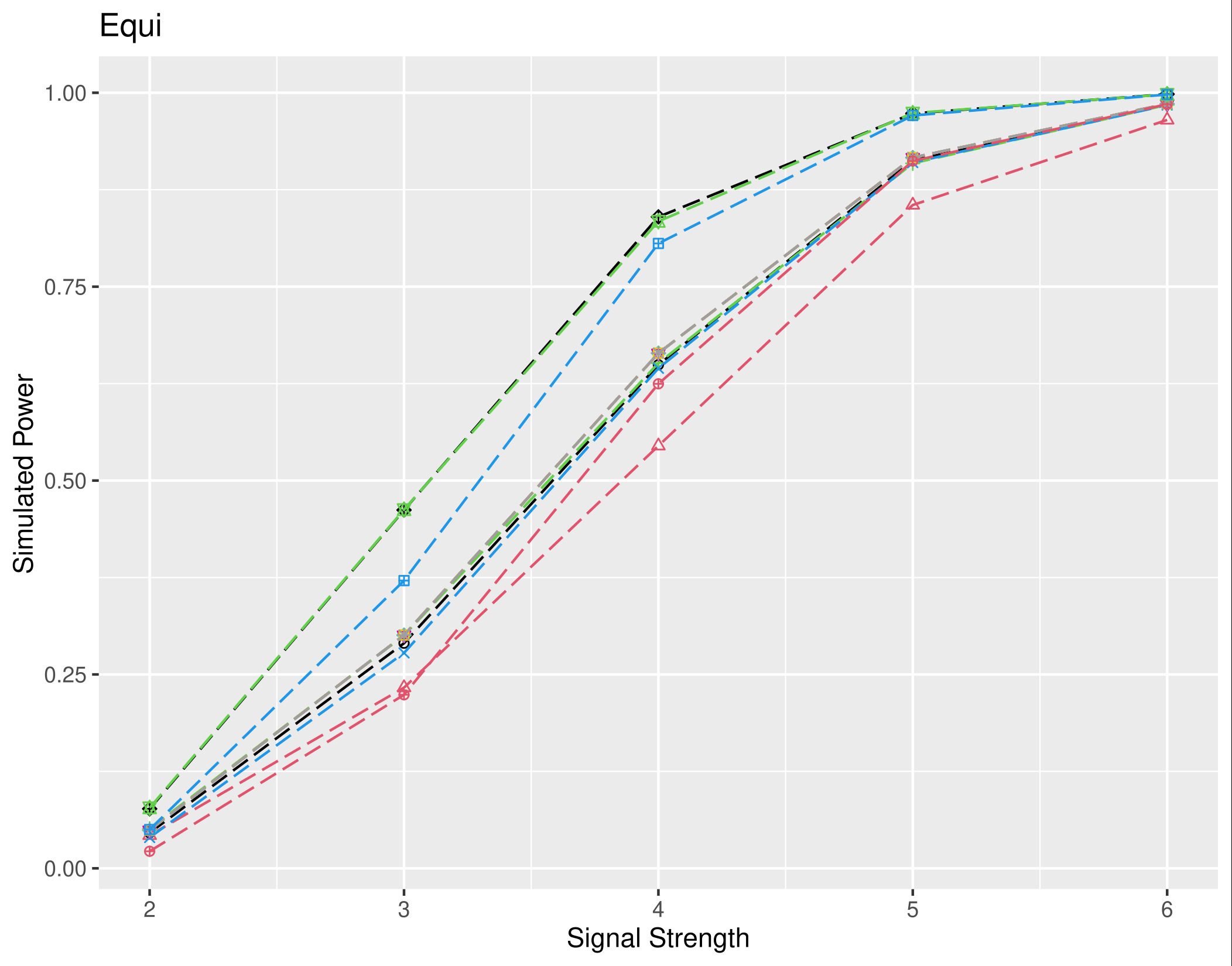} 
      \end{subfigure}  
  & \begin{subfigure}[b]{\linewidth}
      \centering
      \includegraphics[width=\linewidth]{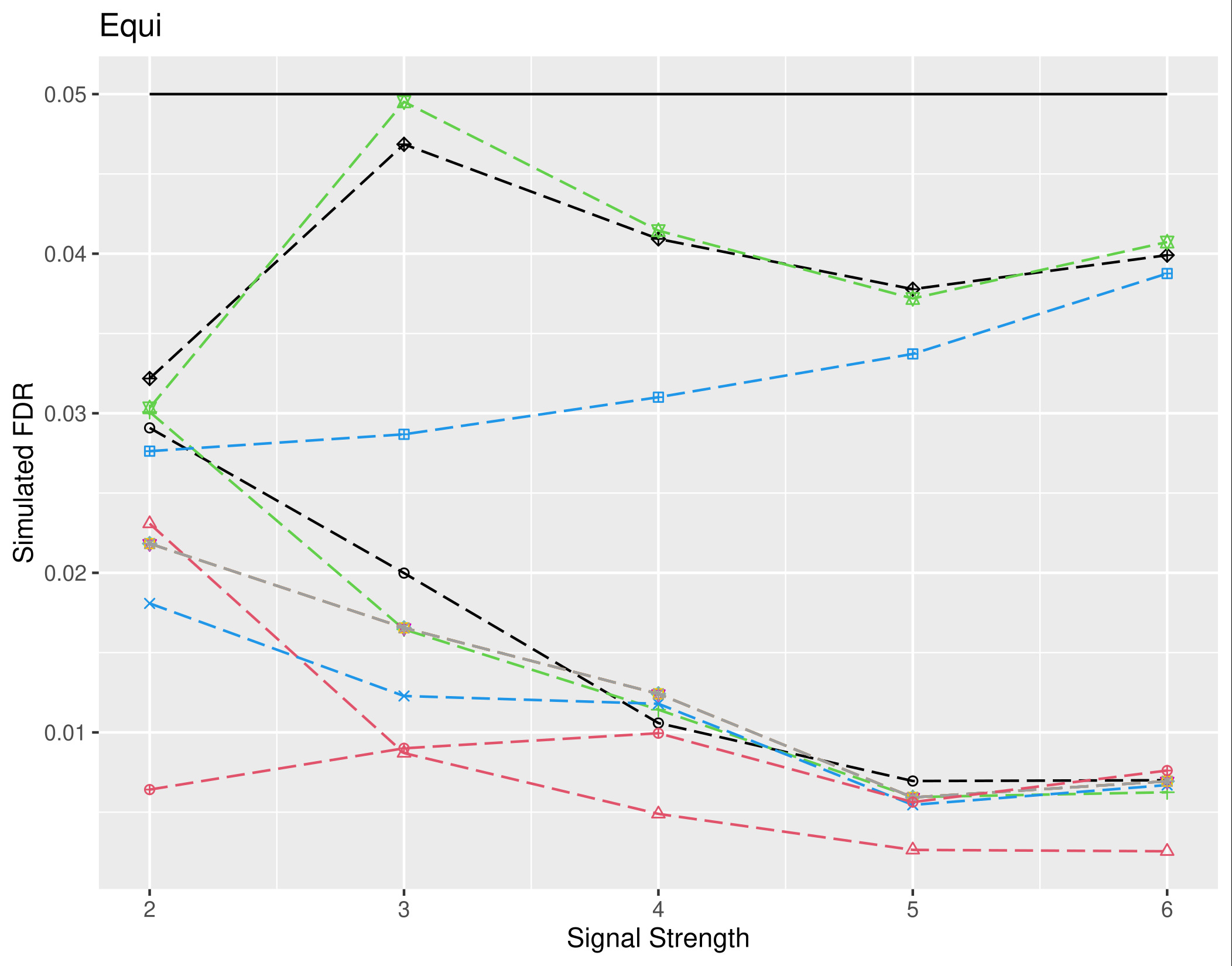} 
      \end{subfigure} 
  & \begin{subfigure}[b]{\linewidth}
      \centering
      \includegraphics[width=\linewidth]{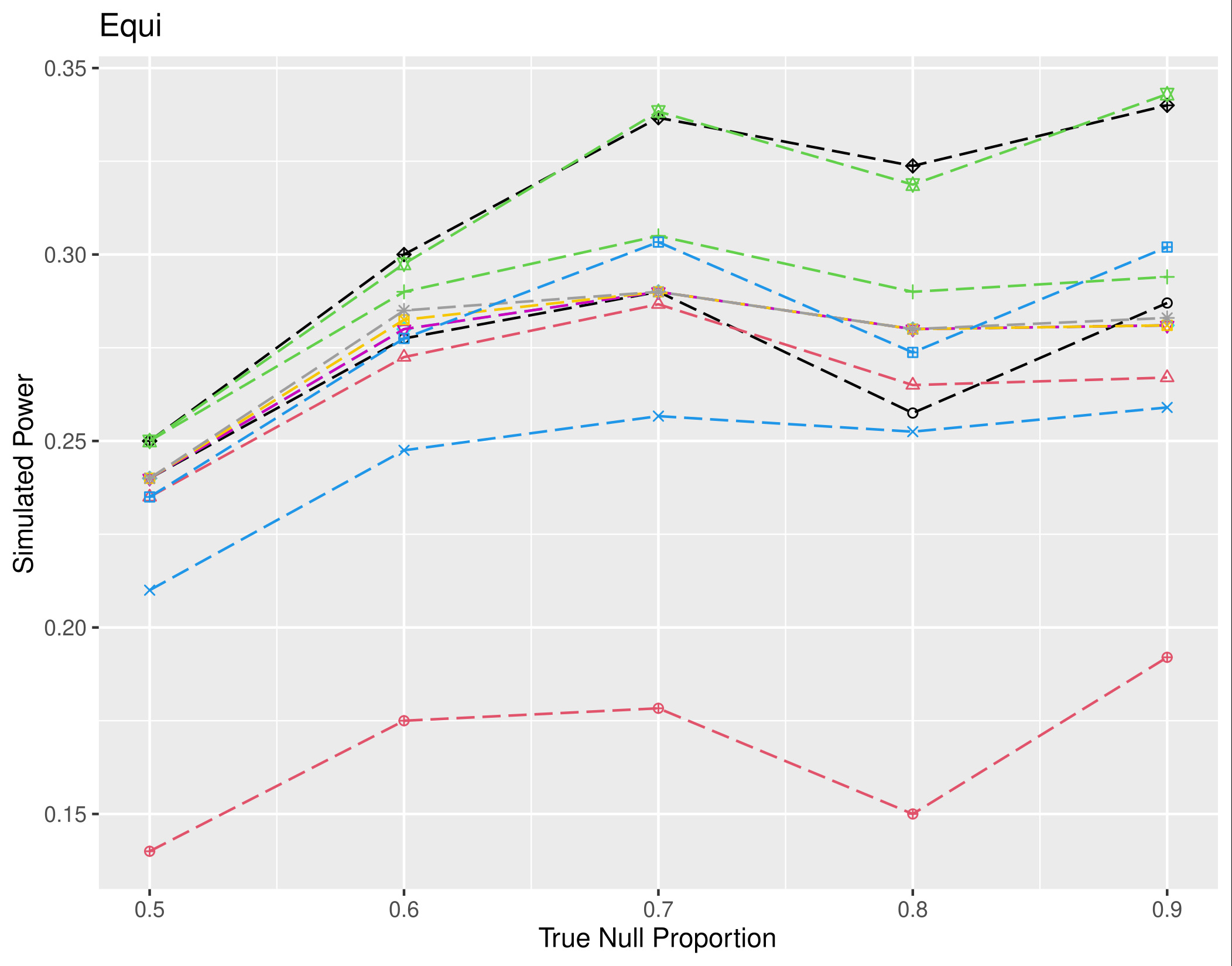} 
      \end{subfigure} \\
  \rotatebox{90}{AR(0.3)} 
  & \begin{subfigure}[b]{\linewidth}
      \centering
      \includegraphics[width=\linewidth]{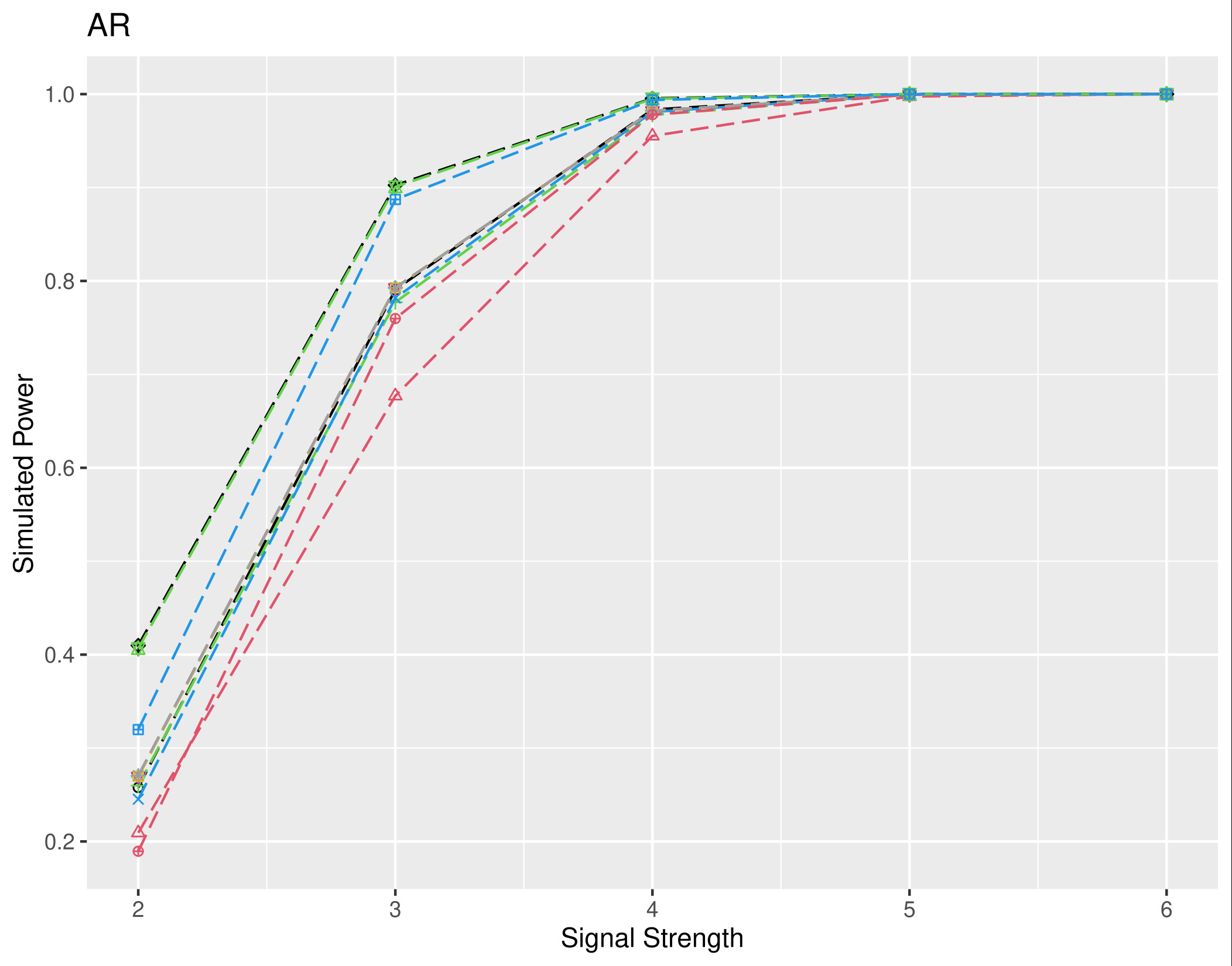} 
      \end{subfigure}  
  & \begin{subfigure}[b]{\linewidth}
      \centering
      \includegraphics[width=\linewidth]{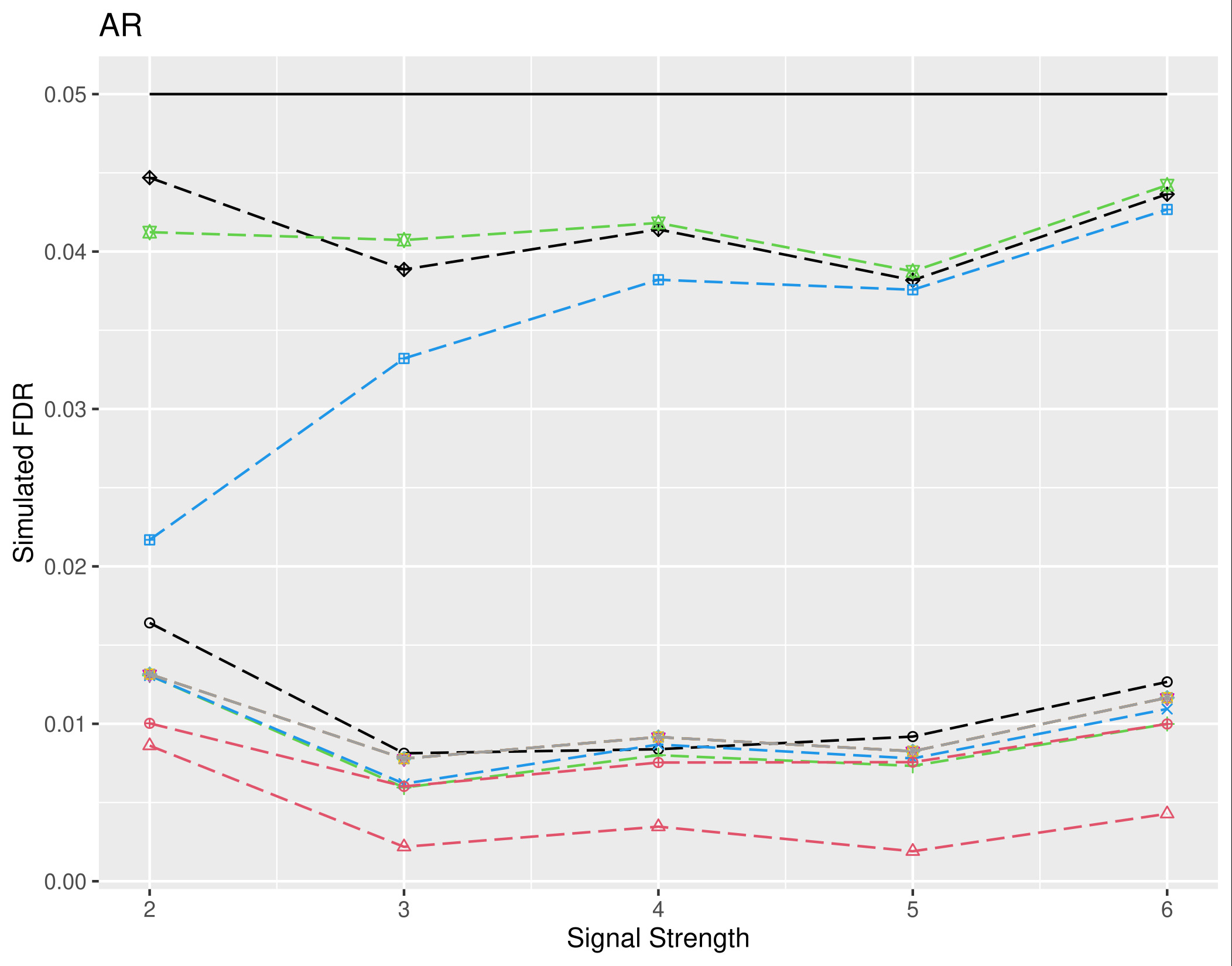} 
      \end{subfigure} 
  & \begin{subfigure}[b]{\linewidth}
      \centering
      \includegraphics[width=\linewidth]{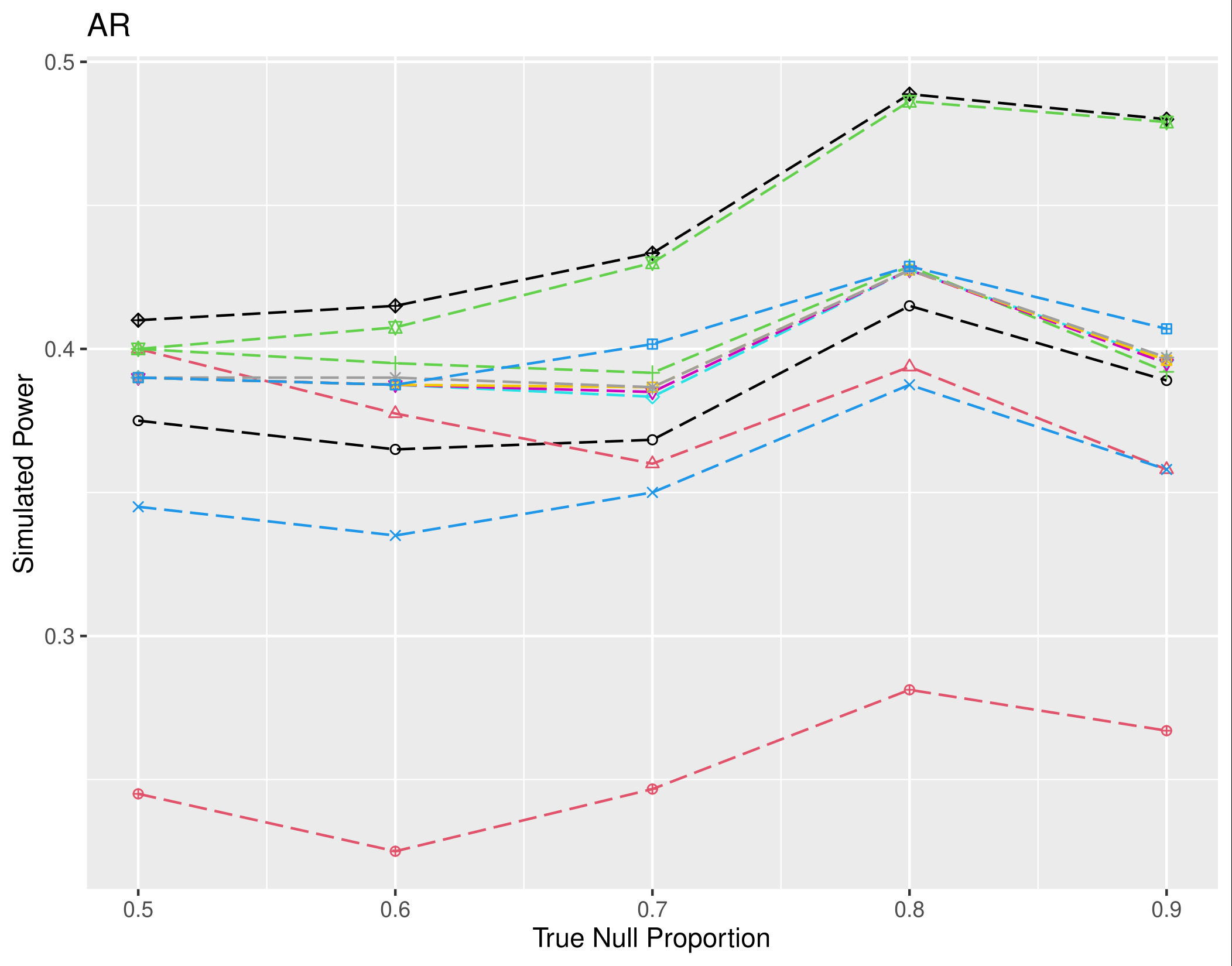} 
      \end{subfigure} \\
  \rotatebox{90}{IAR(0.3)} 
  & \begin{subfigure}[b]{\linewidth}
      \centering
      \includegraphics[width=\linewidth]{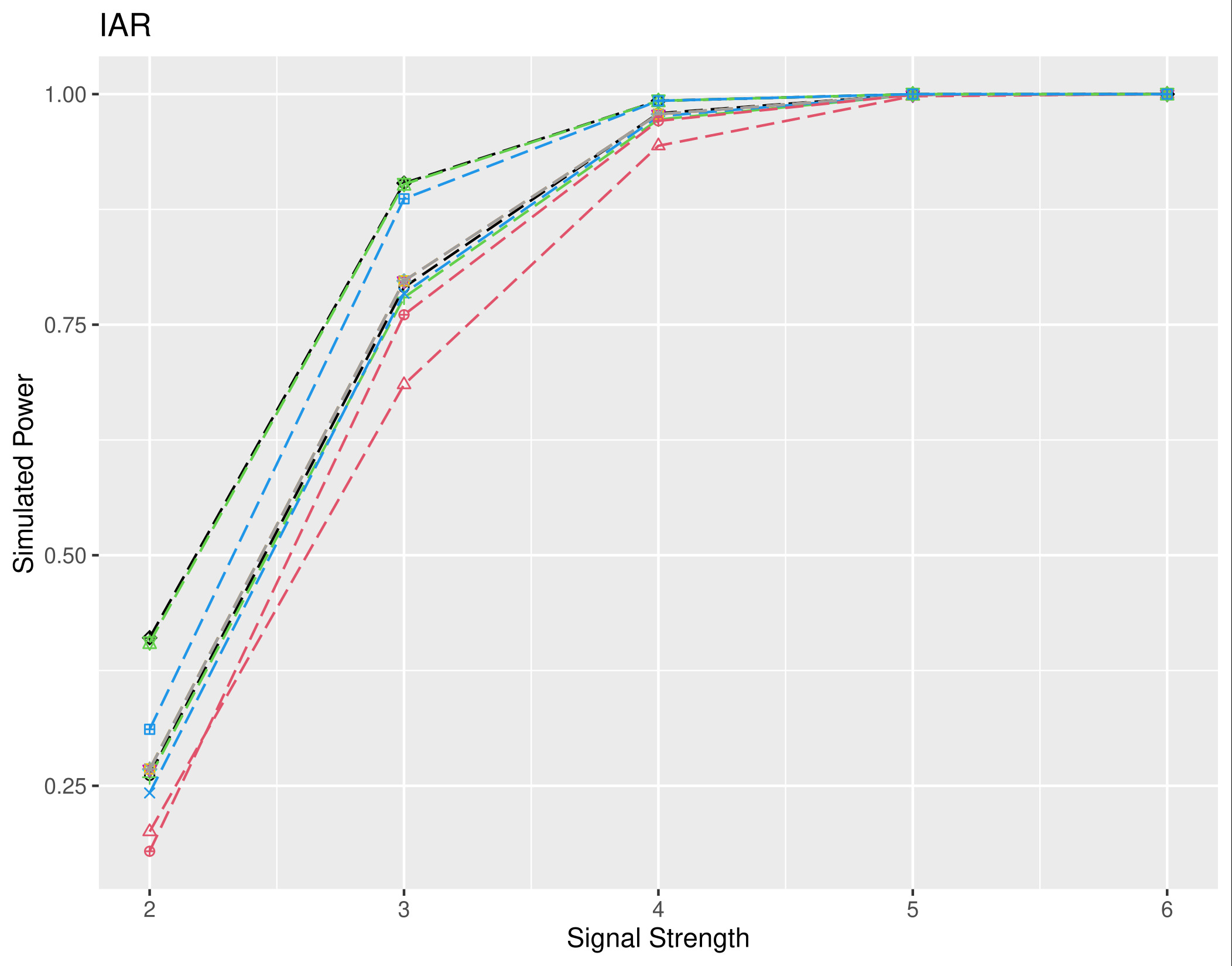} 
      \end{subfigure}  
  & \begin{subfigure}[b]{\linewidth}
      \centering
      \includegraphics[width=\linewidth]{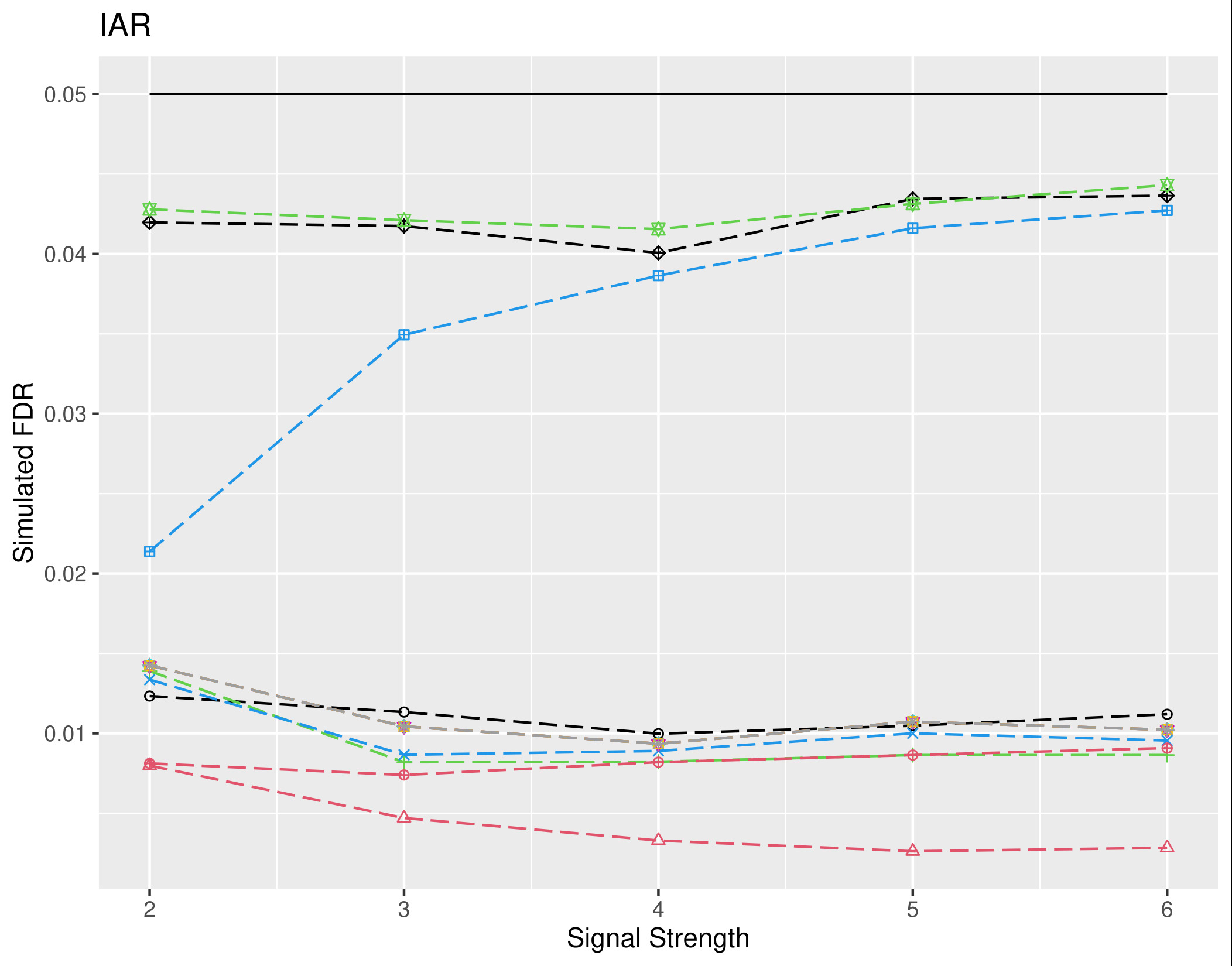} 
      \end{subfigure} 
  & \begin{subfigure}[b]{\linewidth}
      \centering
      \includegraphics[width=\linewidth]{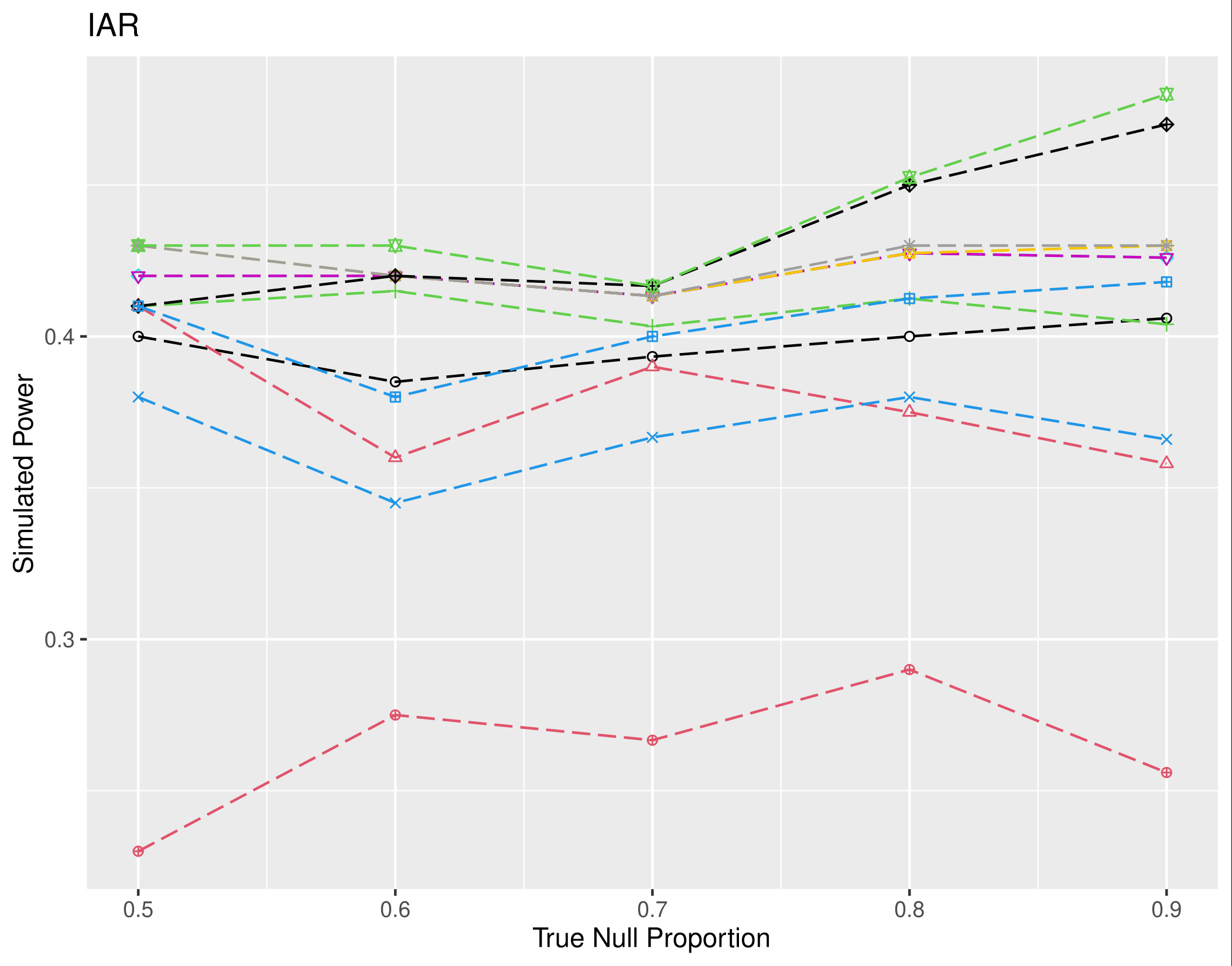} 
      \end{subfigure}    
  \end{tabular*} 
  \caption{Simulated Power (left column), simulated FDR (middle column) for fixed null proportion and simulated power (right column) for fixed signal strength, displayed for variable selection from $d=100$ parameters. Methods compared are SBH1 method (Circle and black), SBH2 method (Triangle point up and red), GSBH1 method (Plus and green), GSBH2 (Cross and blue), GSBH3 (Diamond and light blue), GSBH4 (Triangle point down and purple), GSBH5 (Square cross and yellow), GSBH6 (Star and grey), BH (Diamond plus and black), BY (Circle plus and red), dBH (Triangles up and down and green) and dBY (Square plus and blue)}
  \label{var_sel:figure4} 
\end{figure}

\begin{figure}[ht]
  \begin{tabular*}{\textwidth}{
    @{}m{0.5cm}
    @{}m{\dimexpr0.33\textwidth-0.25cm\relax}
    @{}m{\dimexpr0.33\textwidth-0.25cm\relax}
    @{}m{\dimexpr0.33\textwidth-0.25cm\relax}}
  \rotatebox{90}{Equi(0.7)}
  & \begin{subfigure}[b]{\linewidth}
      \centering
      \includegraphics[width=\linewidth]{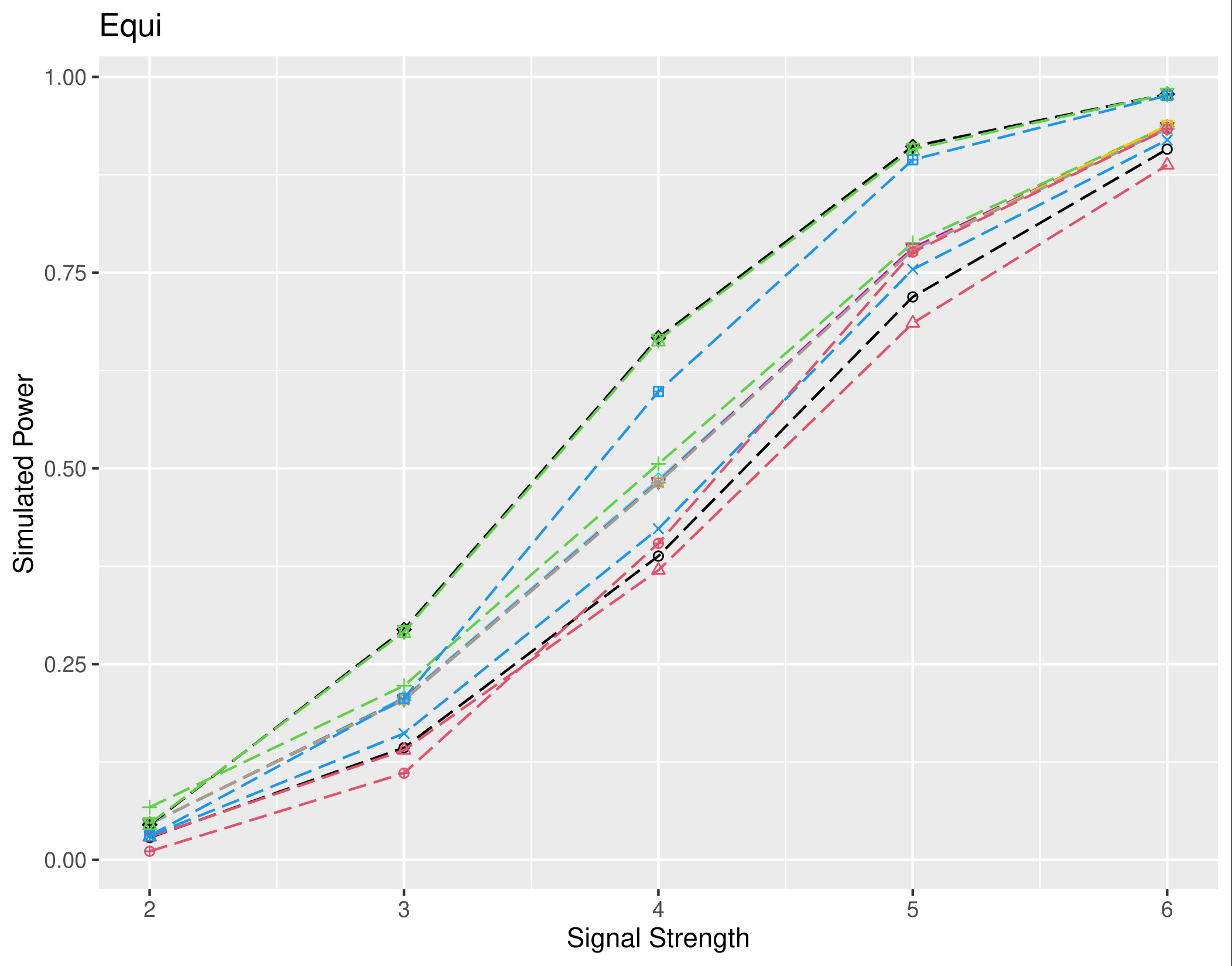} 
      \end{subfigure}  
  & \begin{subfigure}[b]{\linewidth}
      \centering
      \includegraphics[width=\linewidth]{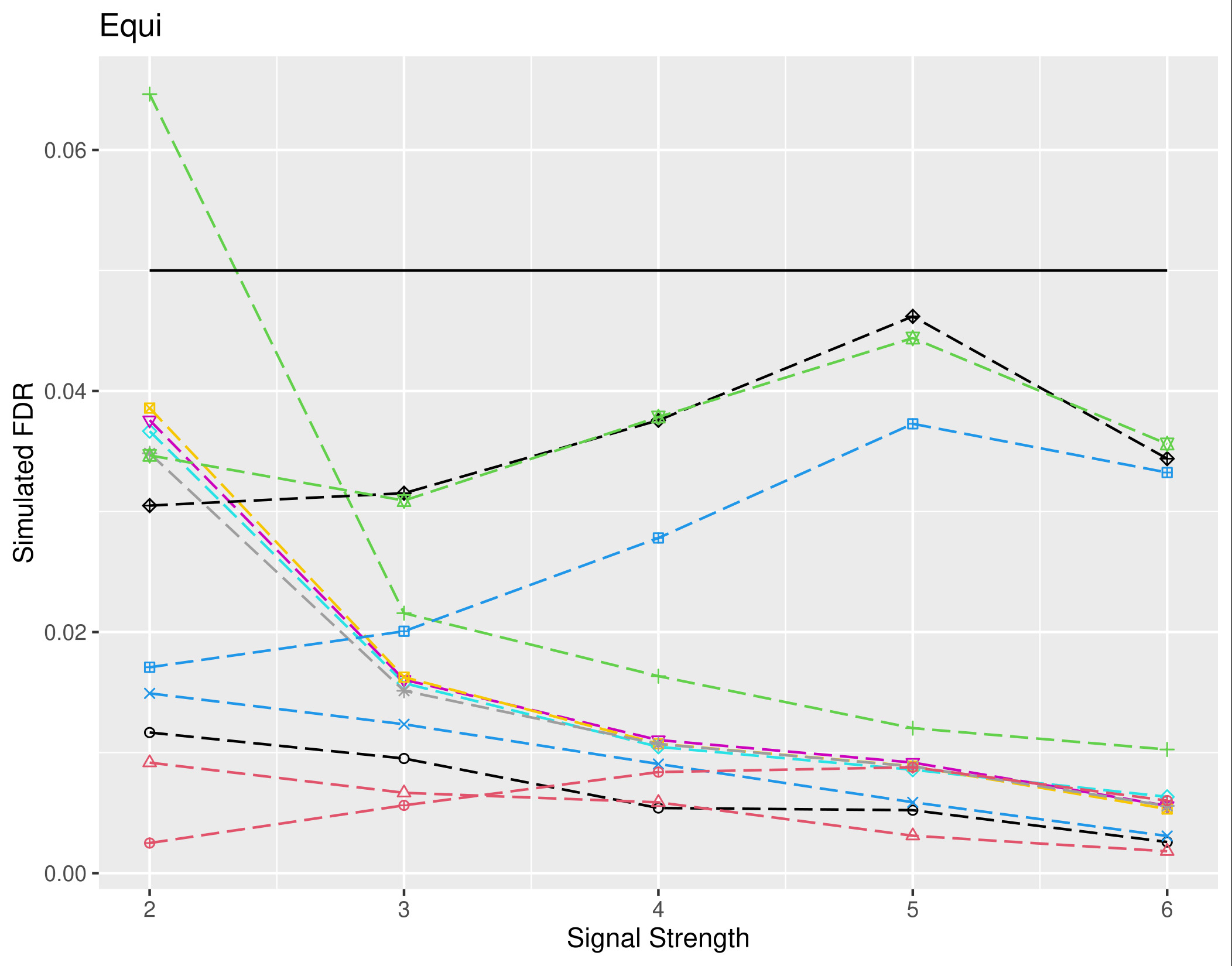} 
      \end{subfigure} 
  & \begin{subfigure}[b]{\linewidth}
      \centering
      \includegraphics[width=\linewidth]{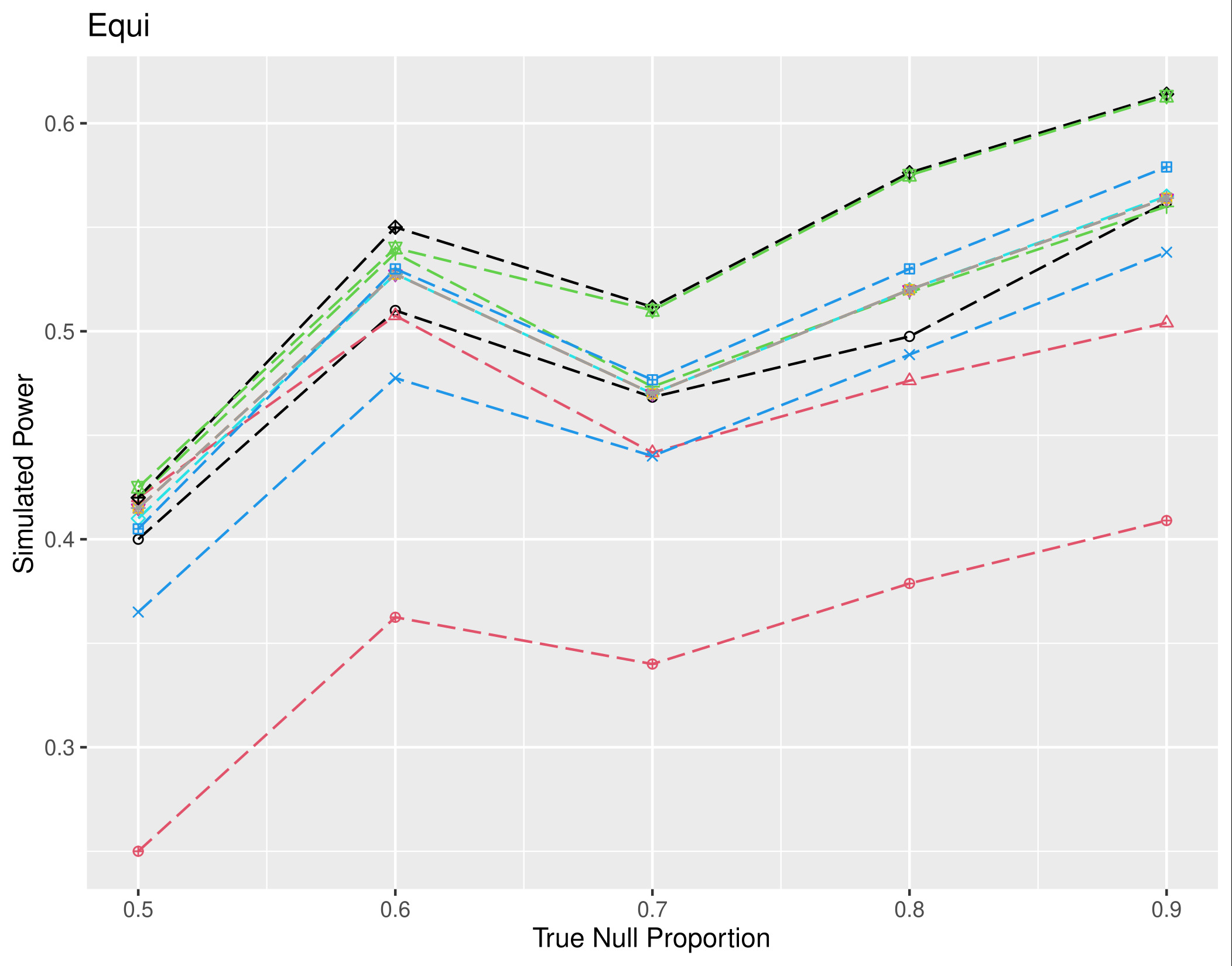} 
      \end{subfigure} \\
  \rotatebox{90}{AR(0.7)} 
  & \begin{subfigure}[b]{\linewidth}
      \centering
      \includegraphics[width=\linewidth]{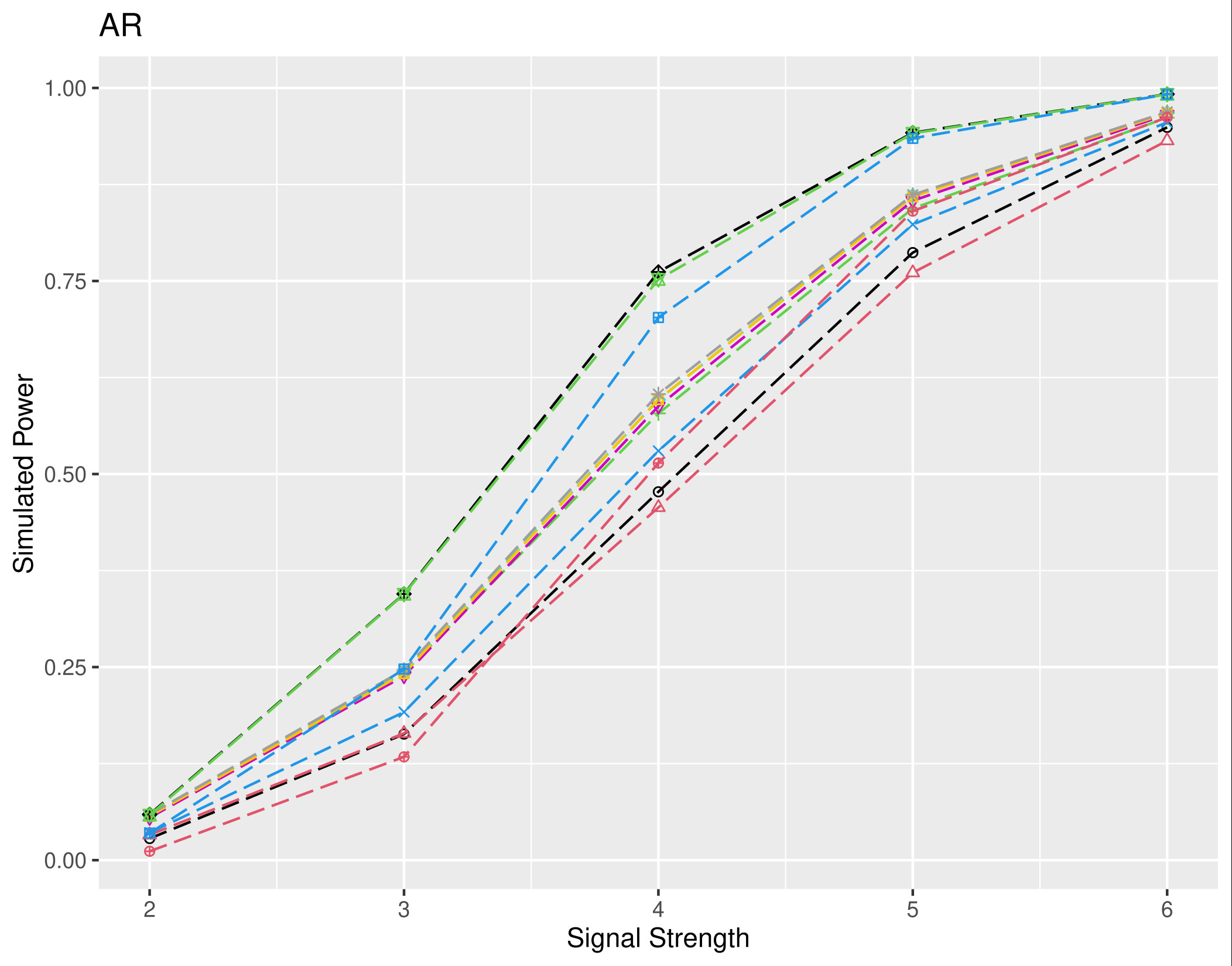} 
      \end{subfigure}  
  & \begin{subfigure}[b]{\linewidth}
      \centering
      \includegraphics[width=\linewidth]{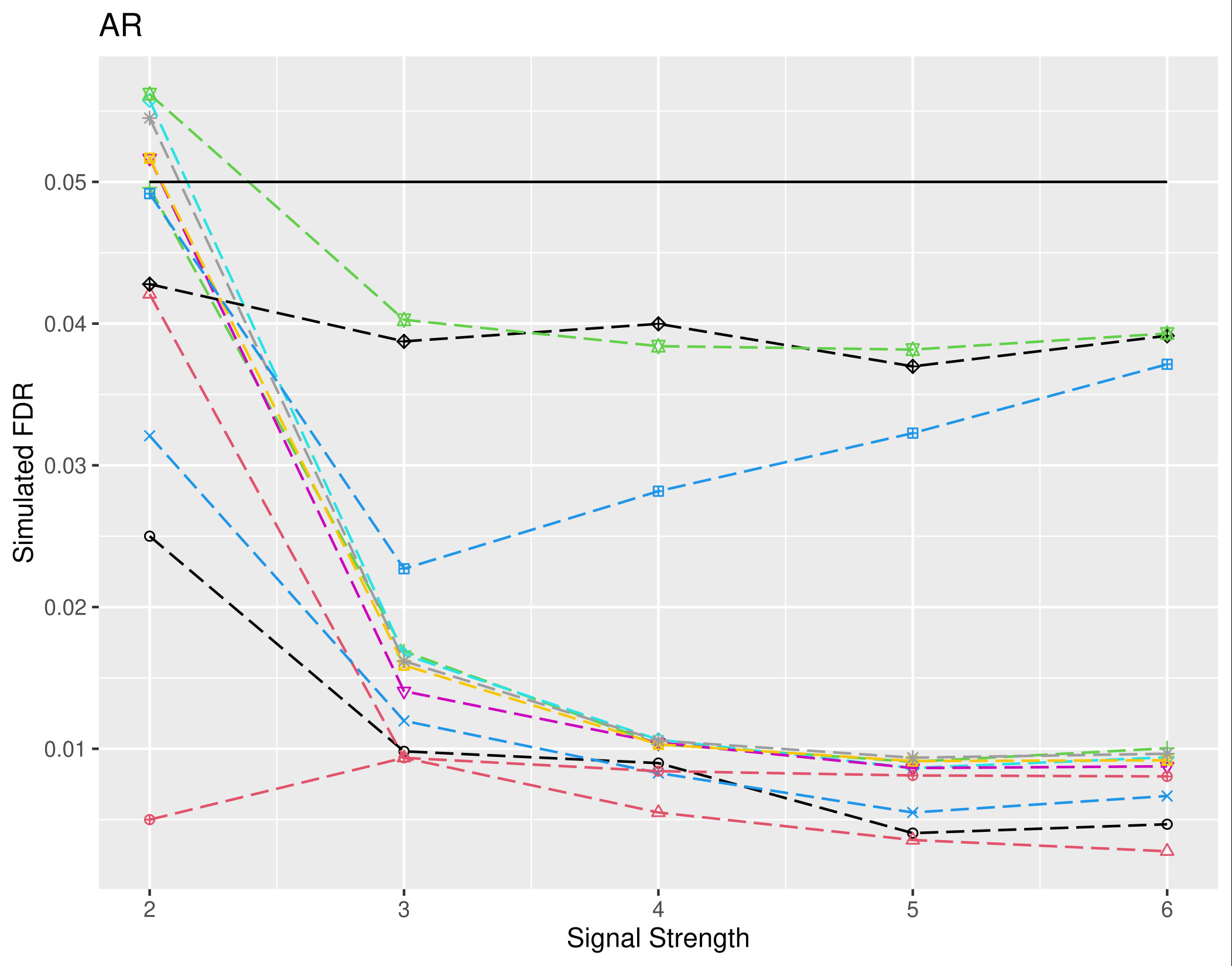} 
      \end{subfigure} 
  & \begin{subfigure}[b]{\linewidth}
      \centering
      \includegraphics[width=\linewidth]{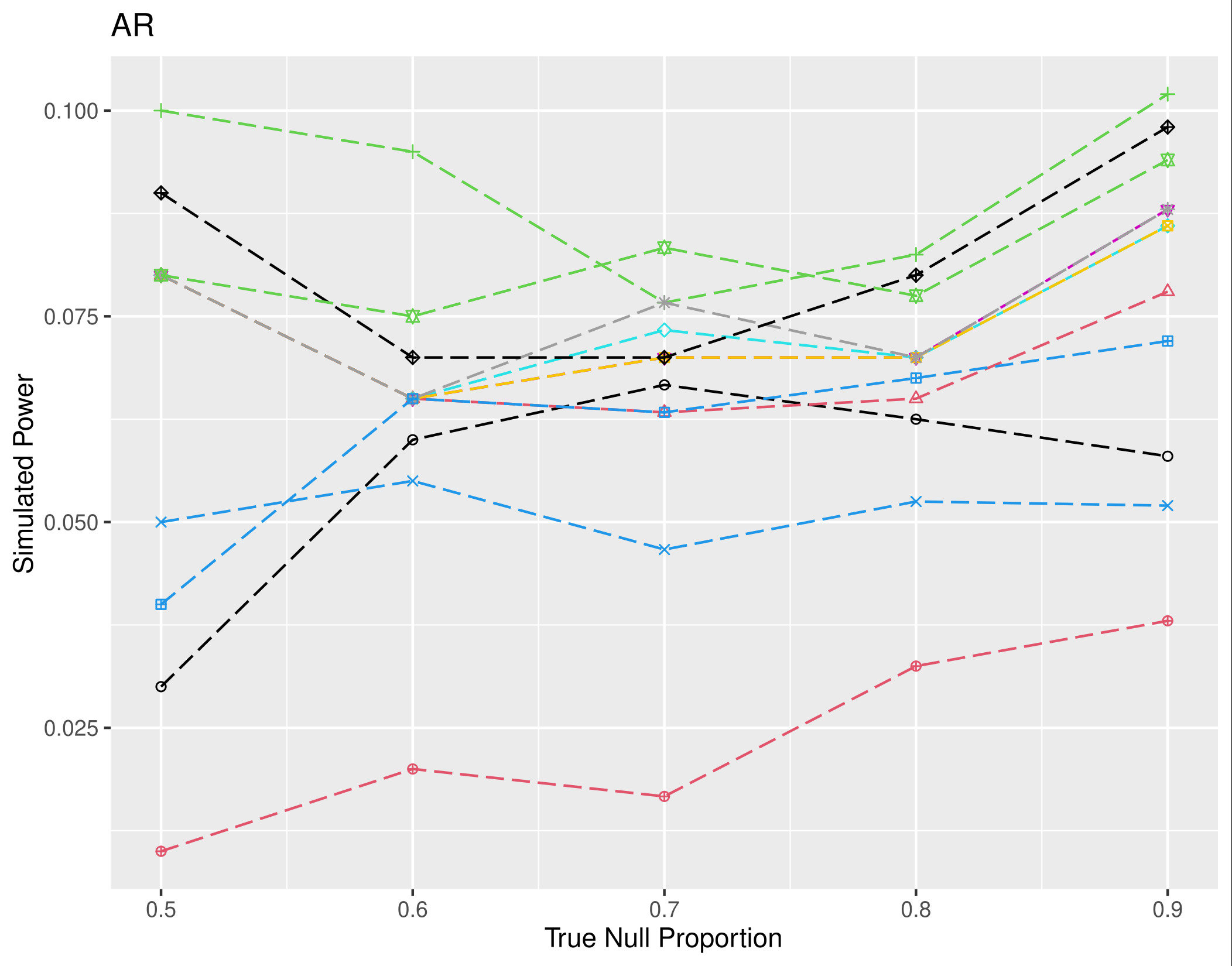} 
      \end{subfigure} \\
  \rotatebox{90}{IAR(0.7)} 
  & \begin{subfigure}[b]{\linewidth}
      \centering
      \includegraphics[width=\linewidth]{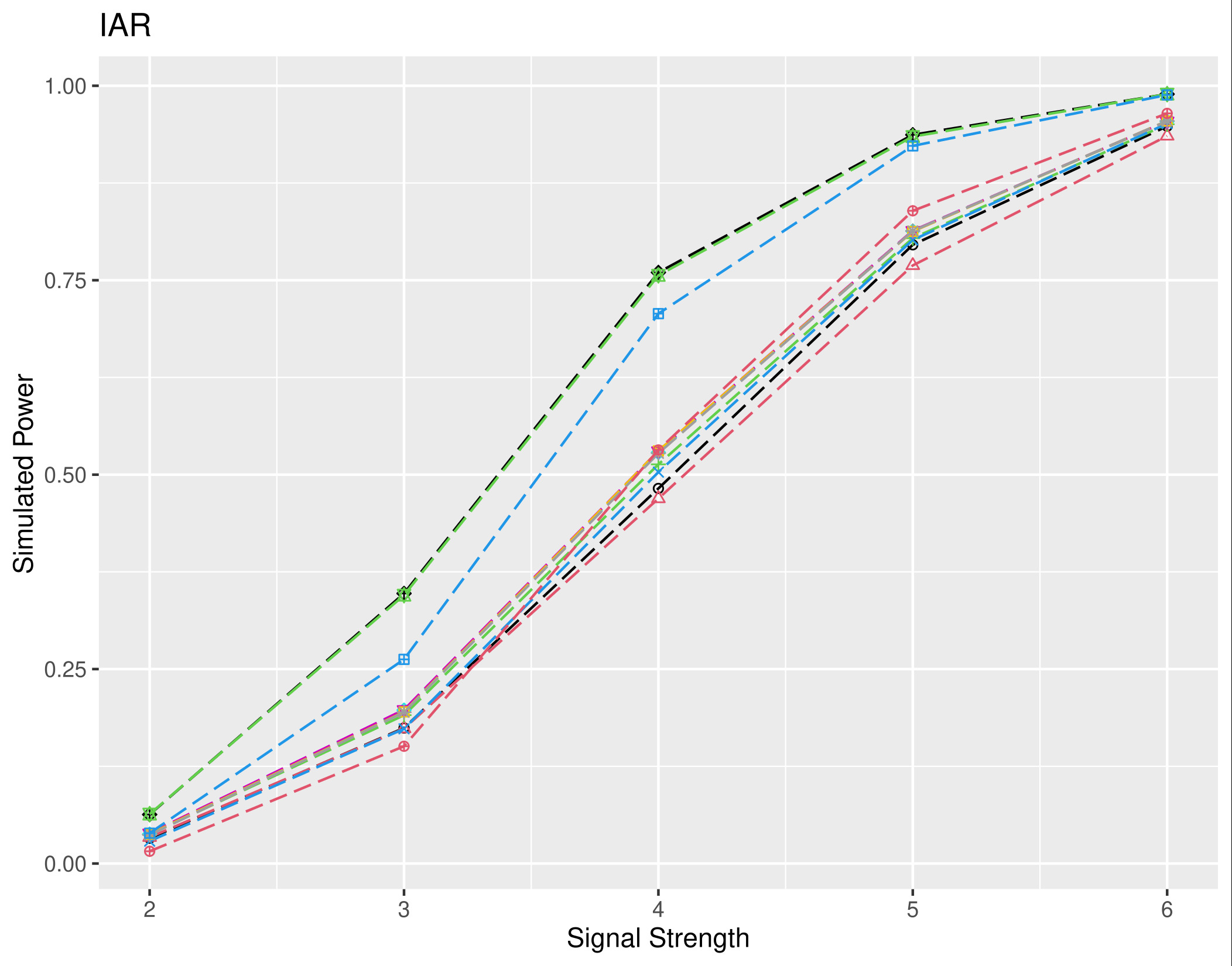} 
      \end{subfigure}  
  & \begin{subfigure}[b]{\linewidth}
      \centering
      \includegraphics[width=\linewidth]{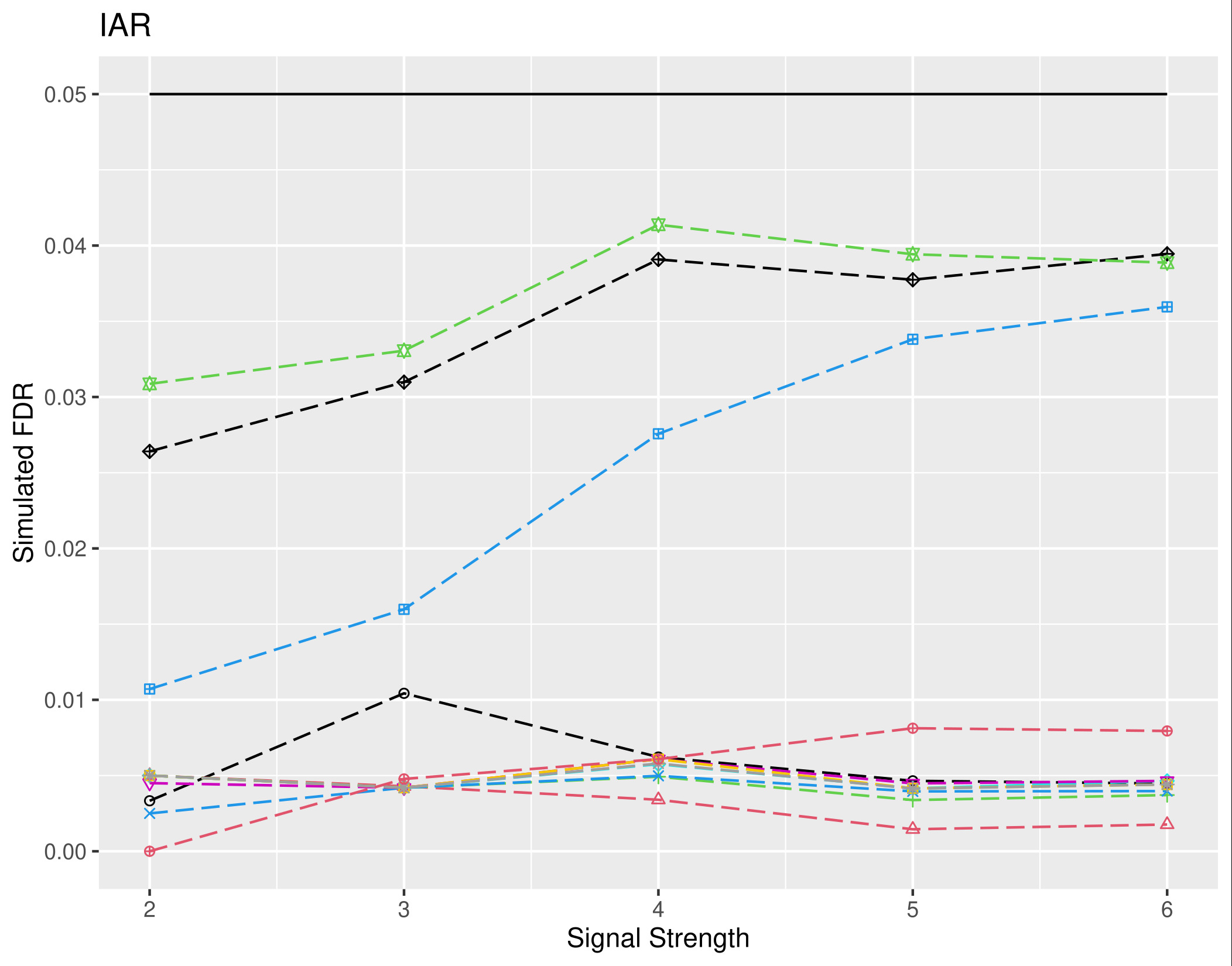} 
      \end{subfigure} 
  & \begin{subfigure}[b]{\linewidth}
      \centering
      \includegraphics[width=\linewidth]{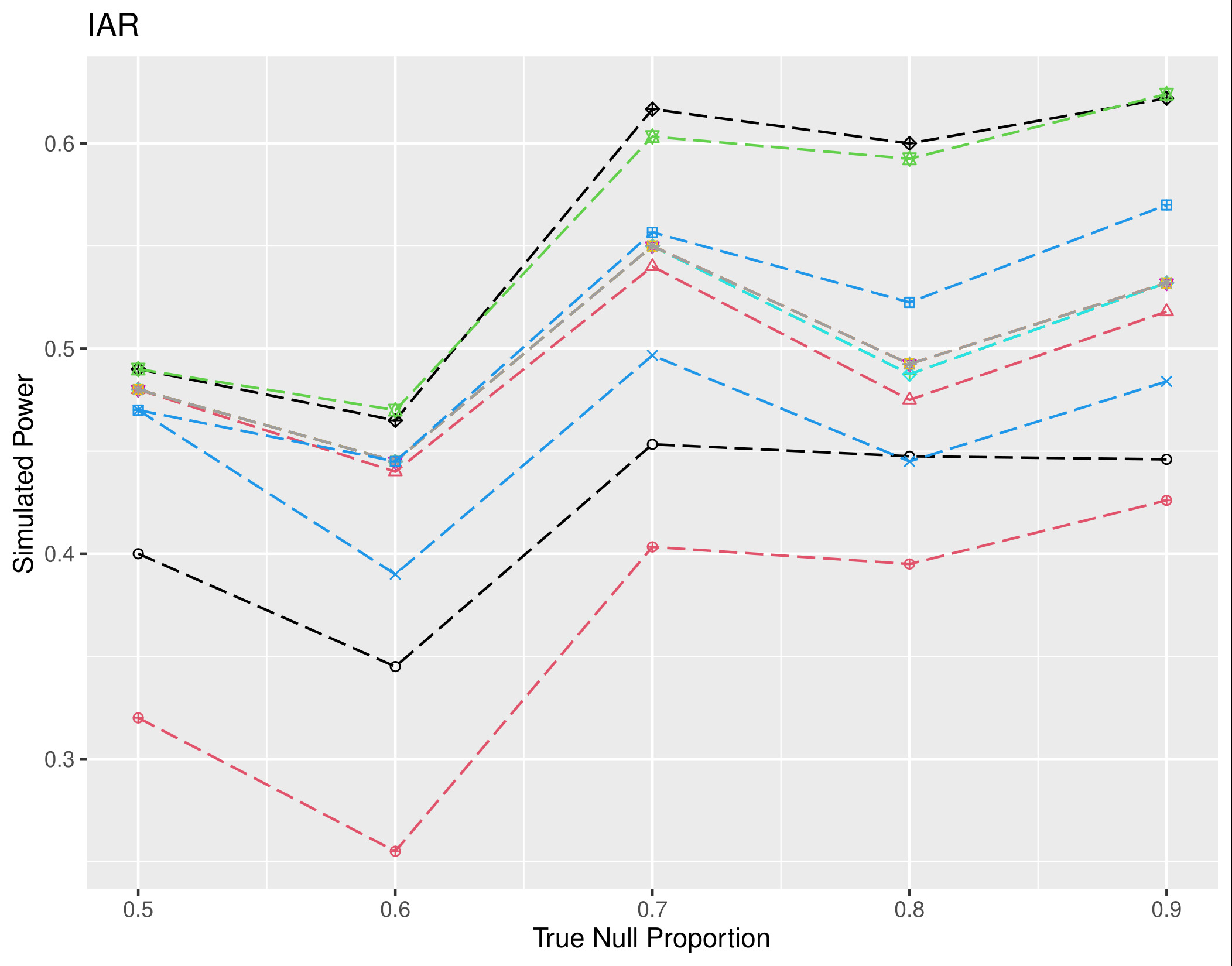} 
      \end{subfigure}    
  \end{tabular*} 
  \caption{Simulated Power (left column), simulated FDR (middle column) for fixed null proportion and simulated power (right column) for fixed signal strength, displayed for variable selection from $d=100$ parameters. Methods compared are SBH1 method (Circle and black), SBH2 method (Triangle point up and red), GSBH1 method (Plus and green), GSBH2 (Cross and blue), GSBH3 (Diamond and light blue), GSBH4 (Triangle point down and purple), GSBH5 (Square cross and yellow), GSBH6 (Star and grey), BH (Diamond plus and black), BY (Circle plus and red), dBH (Triangles up and down and green) and dBY (Square plus and blue)}
  \label{var_sel:figure5} 
\end{figure}

\begin{figure}[ht]
  \begin{tabular*}{\textwidth}{
    @{}m{0.5cm}
    @{}m{\dimexpr0.33\textwidth-0.25cm\relax}
    @{}m{\dimexpr0.33\textwidth-0.25cm\relax}
    @{}m{\dimexpr0.33\textwidth-0.25cm\relax}}
  \rotatebox{90}{Block Diagonal}
  & \begin{subfigure}[b]{\linewidth}
      \centering
      \includegraphics[width=\linewidth]{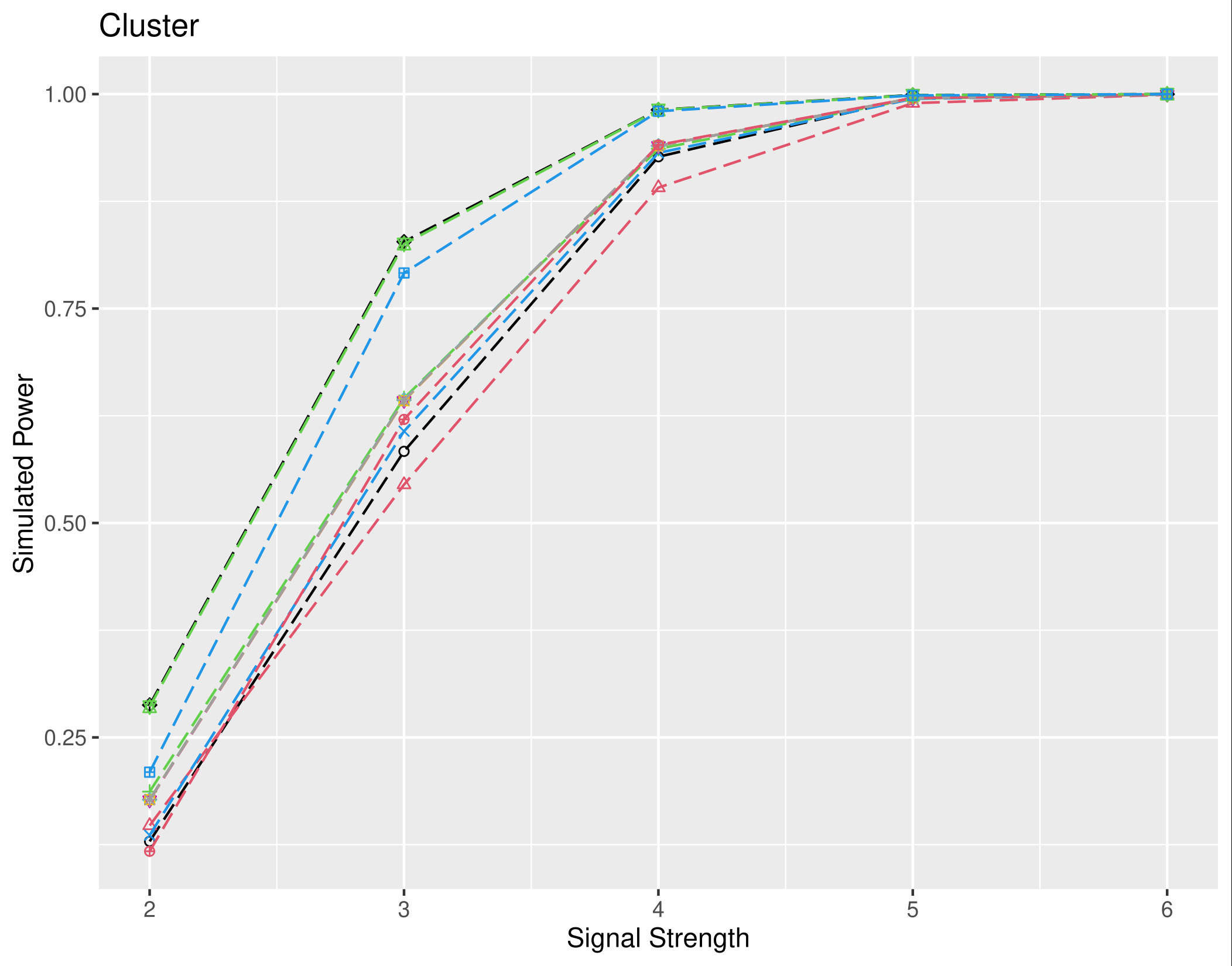} 
      \end{subfigure}  
  & \begin{subfigure}[b]{\linewidth}
      \centering
      \includegraphics[width=\linewidth]{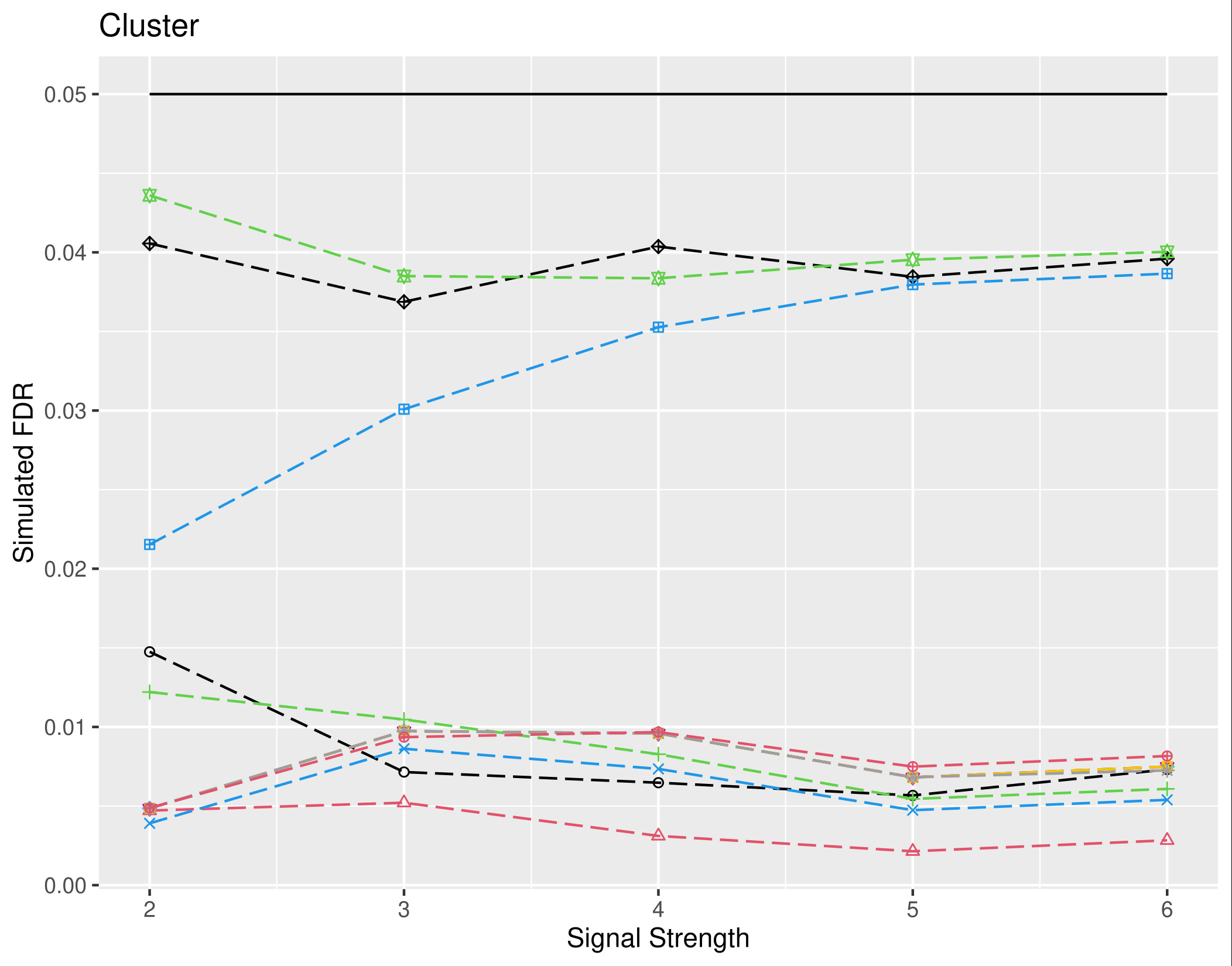} 
      \end{subfigure} 
  & \begin{subfigure}[b]{\linewidth}
      \centering
      \includegraphics[width=\linewidth]{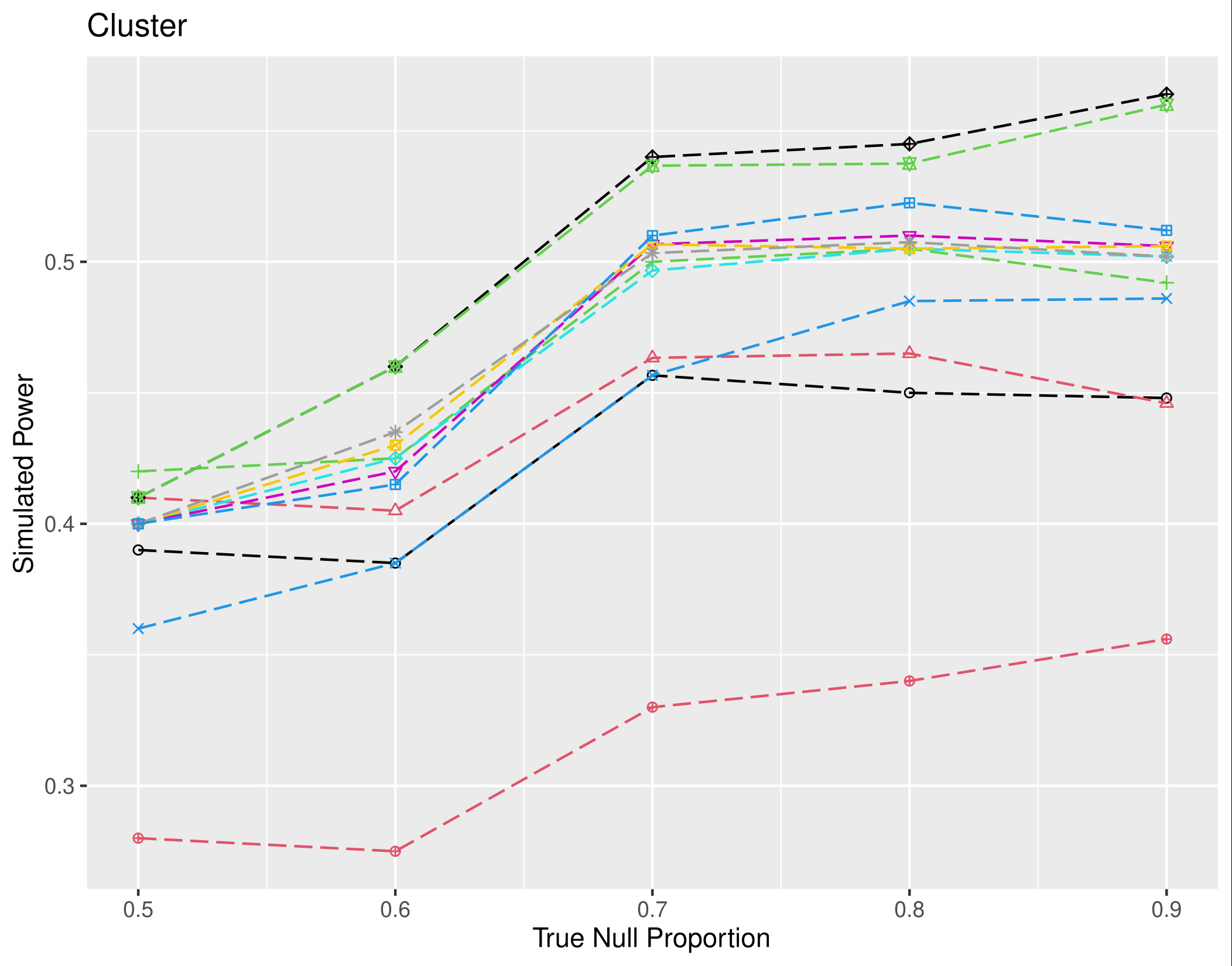} 
      \end{subfigure} \\
  \rotatebox{90}{Sparse} 
  & \begin{subfigure}[b]{\linewidth}
      \centering
      \includegraphics[width=\linewidth]{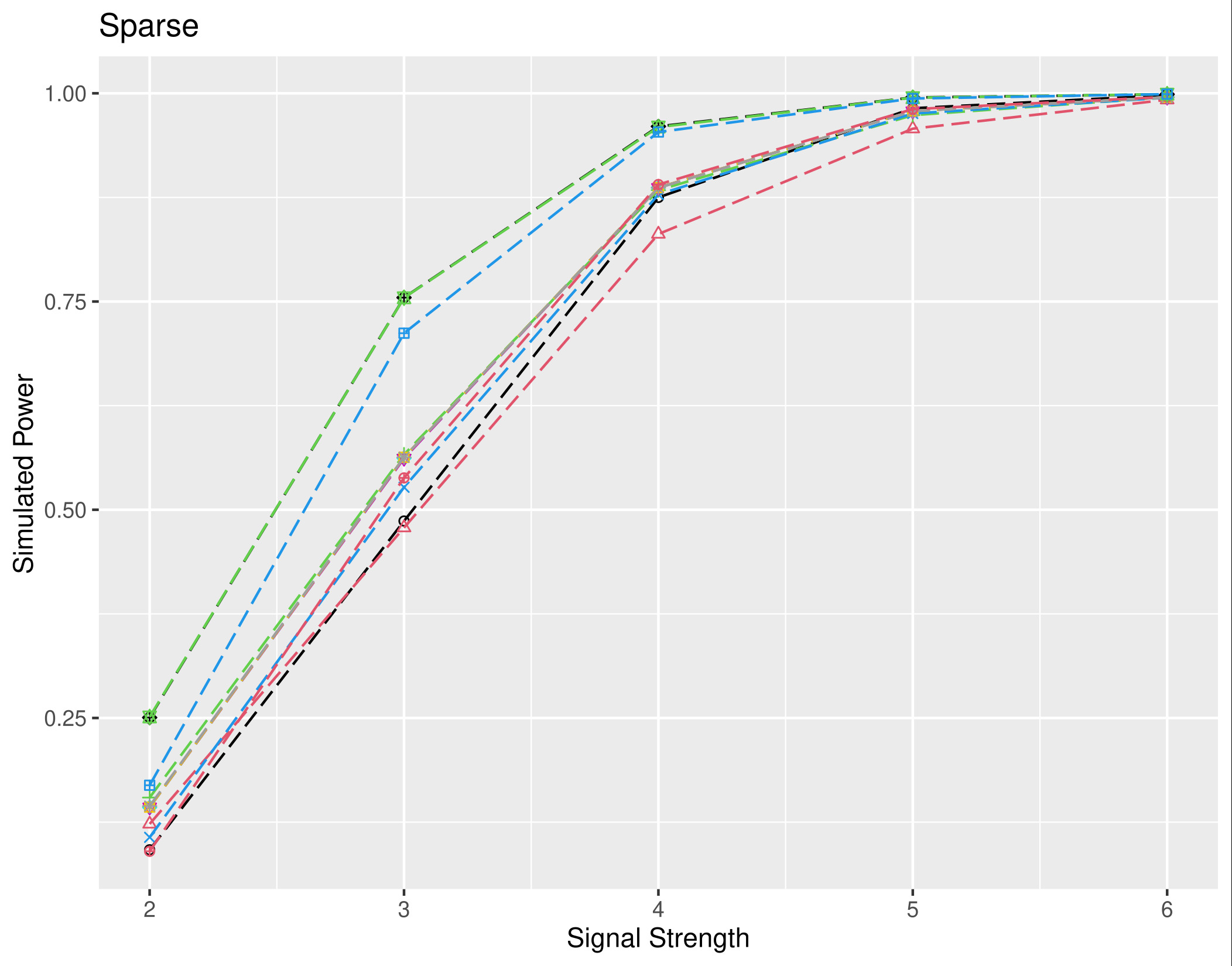} 
      \end{subfigure}  
  & \begin{subfigure}[b]{\linewidth}
      \centering
      \includegraphics[width=\linewidth]{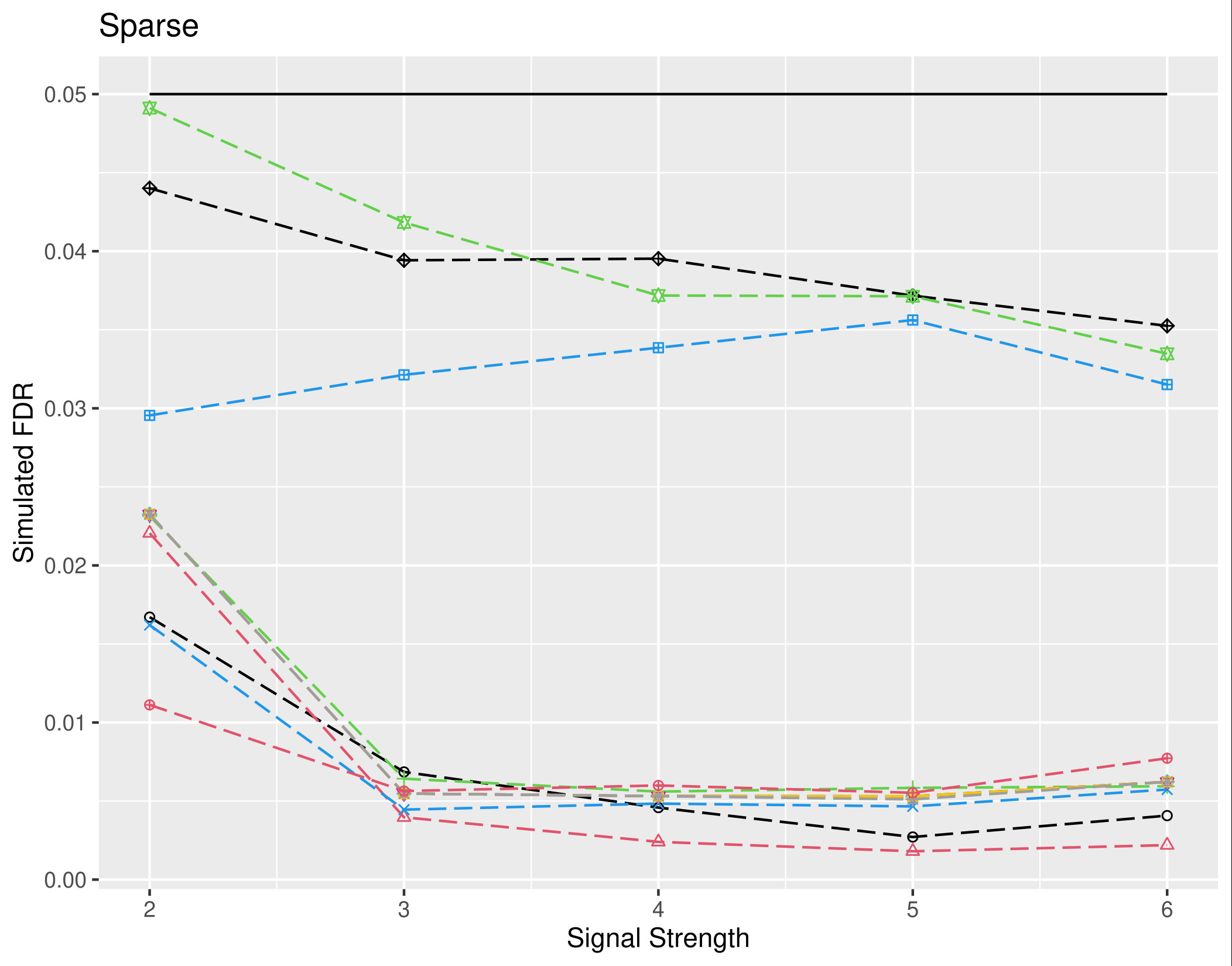} 
      \end{subfigure} 
  & \begin{subfigure}[b]{\linewidth}
      \centering
      \includegraphics[width=\linewidth]{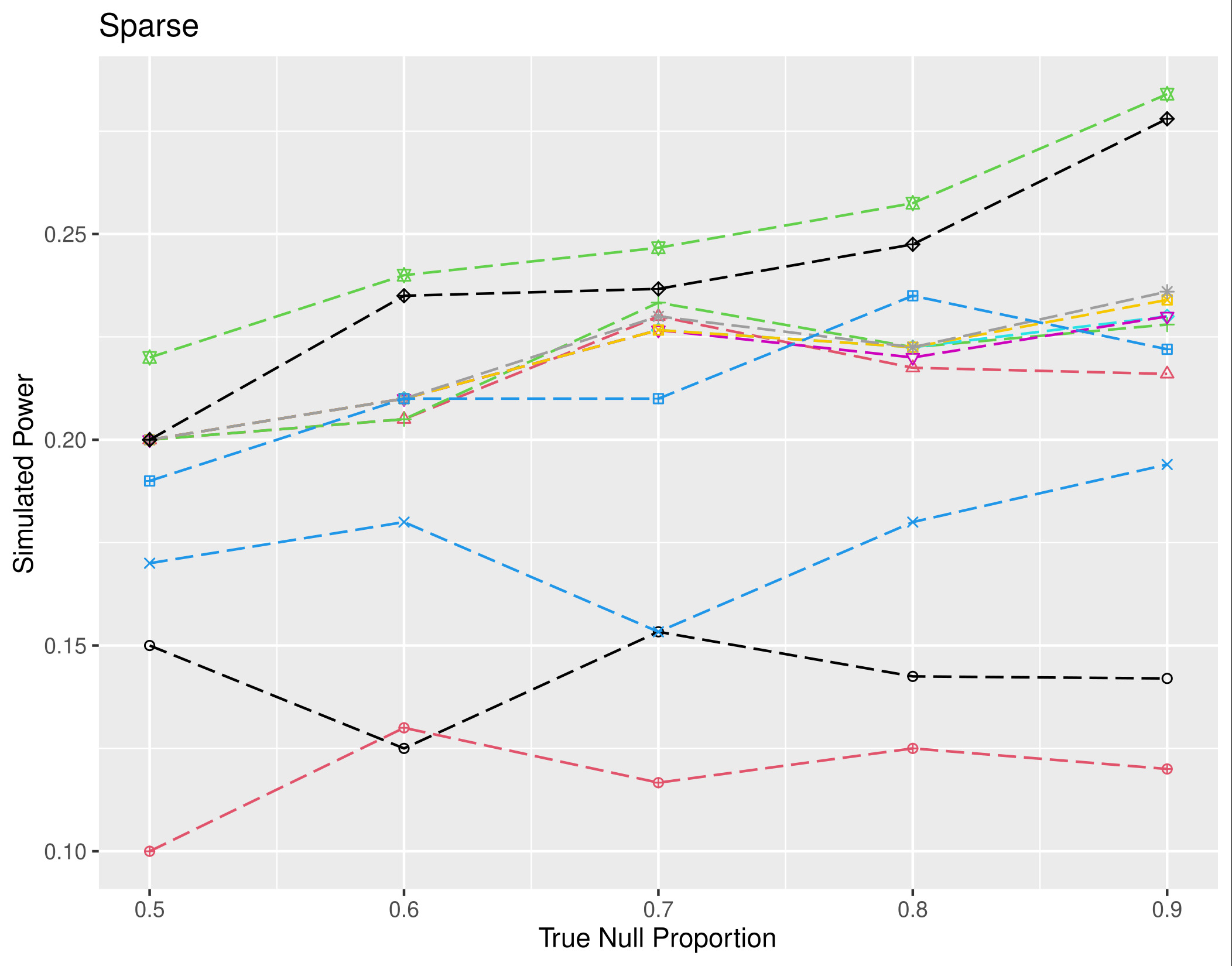} 
      \end{subfigure} \\
  \rotatebox{90}{Prefixed Corr 1} 
  & \begin{subfigure}[b]{\linewidth}
      \centering
      \includegraphics[width=\linewidth]{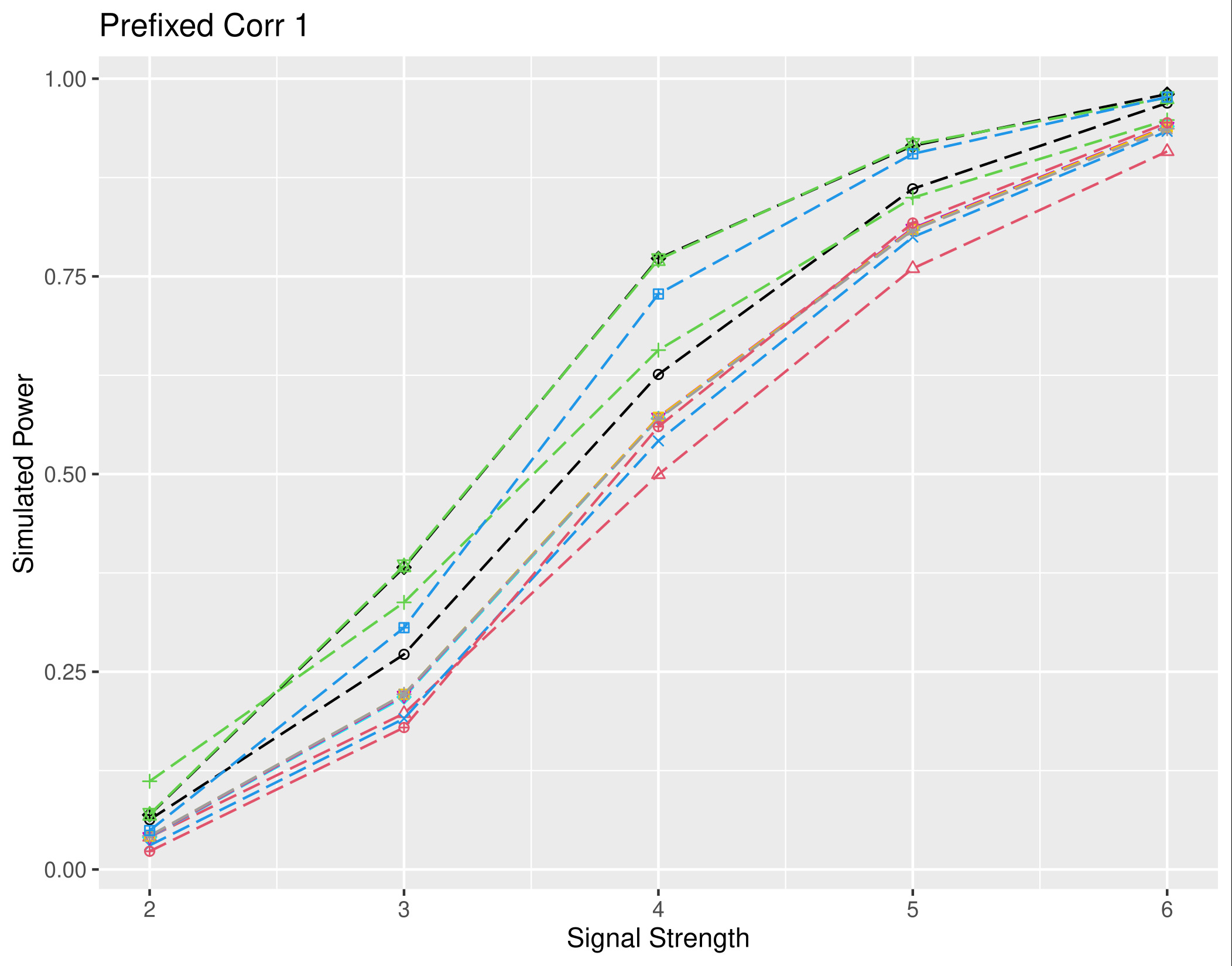} 
      \end{subfigure}  
  & \begin{subfigure}[b]{\linewidth}
      \centering
      \includegraphics[width=\linewidth]{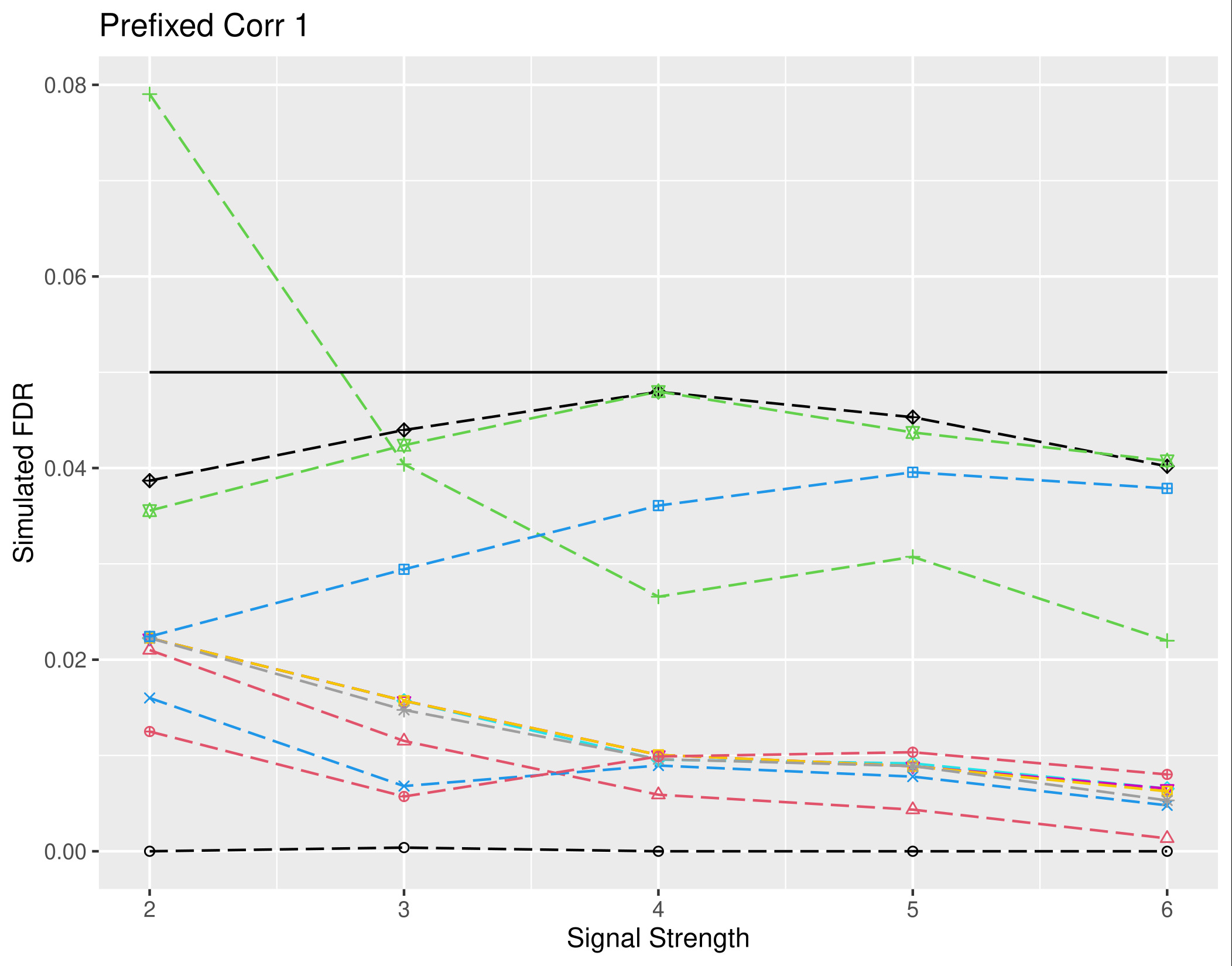} 
      \end{subfigure} 
  & \begin{subfigure}[b]{\linewidth}
      \centering
      \includegraphics[width=\linewidth]{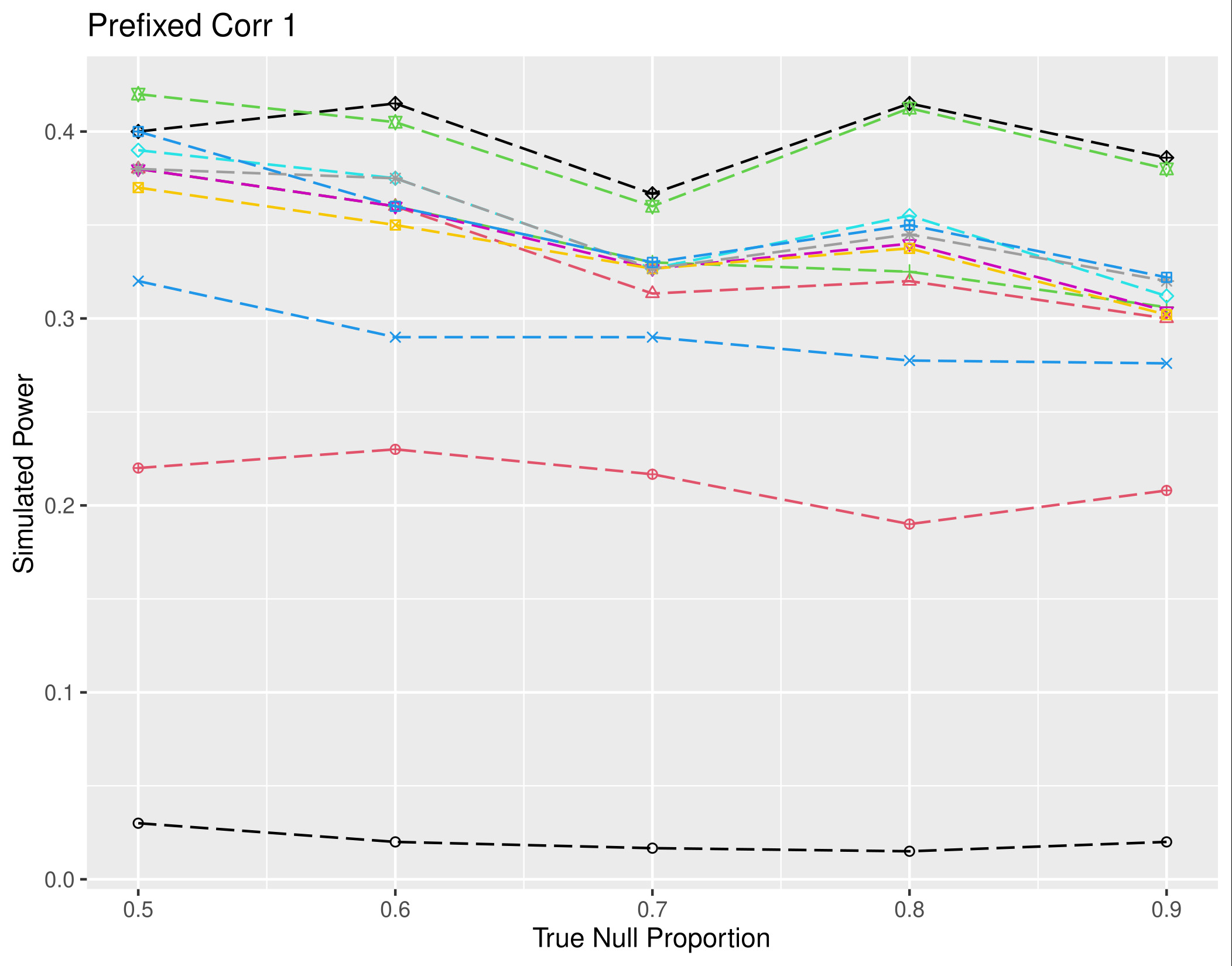} 
      \end{subfigure}    
  \end{tabular*} 
  \caption{Simulated Power (left column), simulated FDR (middle column) for fixed null proportion and simulated power (right column) for fixed signal strength, displayed for variable selection from $d=100$ parameters. Methods compared are SBH1 method (Circle and black), SBH2 method (Triangle point up and red), GSBH1 method (Plus and green), GSBH2 (Cross and blue), GSBH3 (Diamond and light blue), GSBH4 (Triangle point down and purple), GSBH5 (Square cross and yellow), GSBH6 (Star and grey), BH (Diamond plus and black), BY (Circle plus and red), dBH (Triangles up and down and green) and dBY (Square plus and blue)}
  \label{var_sel:figure6} 
\end{figure}

\begin{figure}[hbt!]
  \begin{tabular*}{\textwidth}{
    @{}m{0.5cm}
    @{}m{\dimexpr0.50\textwidth-0.25cm\relax}
    @{}m{\dimexpr0.50\textwidth-0.25cm\relax}}
  \rotatebox{90}{Equi(0.7)}
  & \begin{subfigure}[b]{\linewidth}
      \centering
      \includegraphics[width=0.8\linewidth]{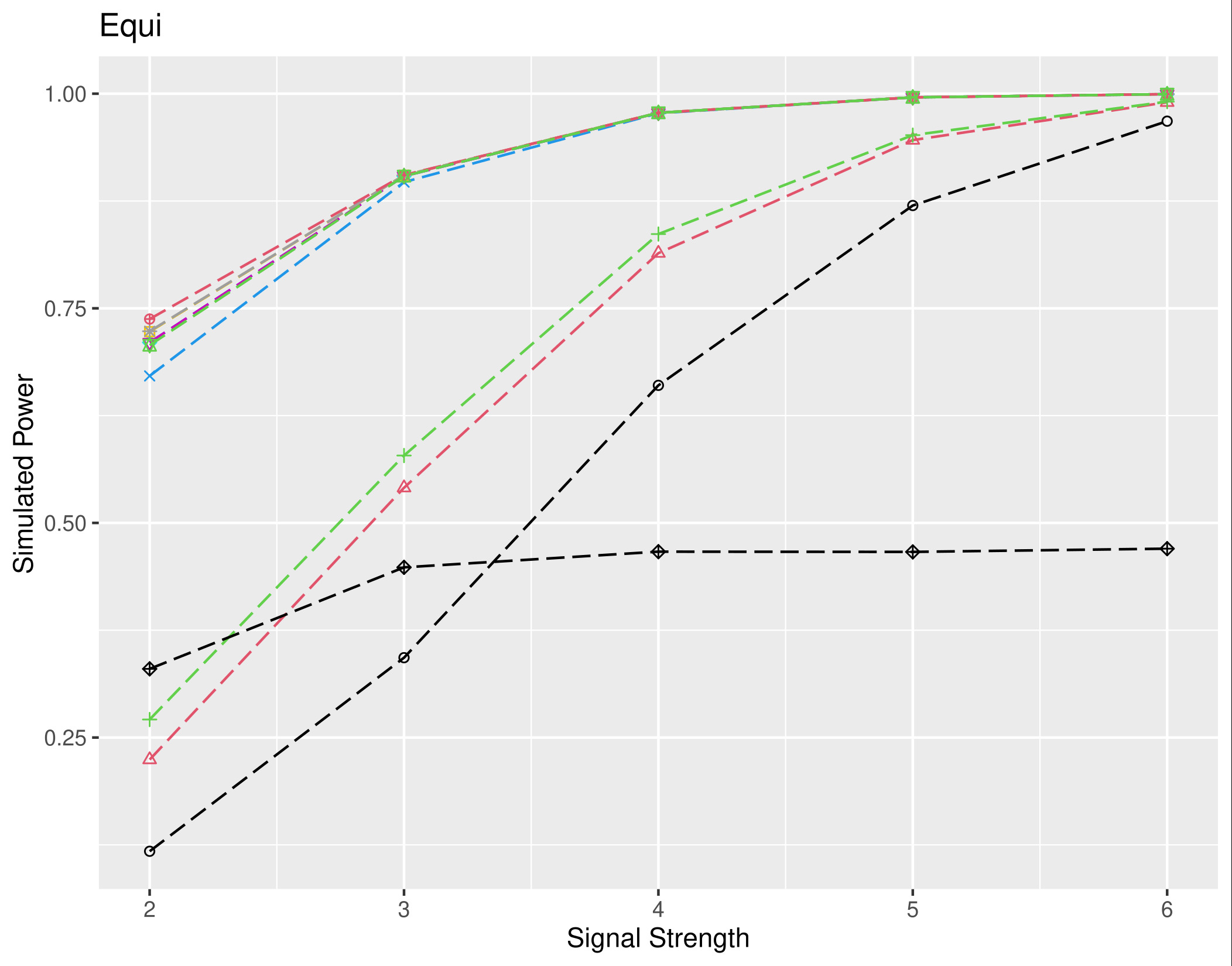} 
      \end{subfigure}  
  & \begin{subfigure}[b]{\linewidth}
      \centering
      \includegraphics[width=0.8\linewidth]{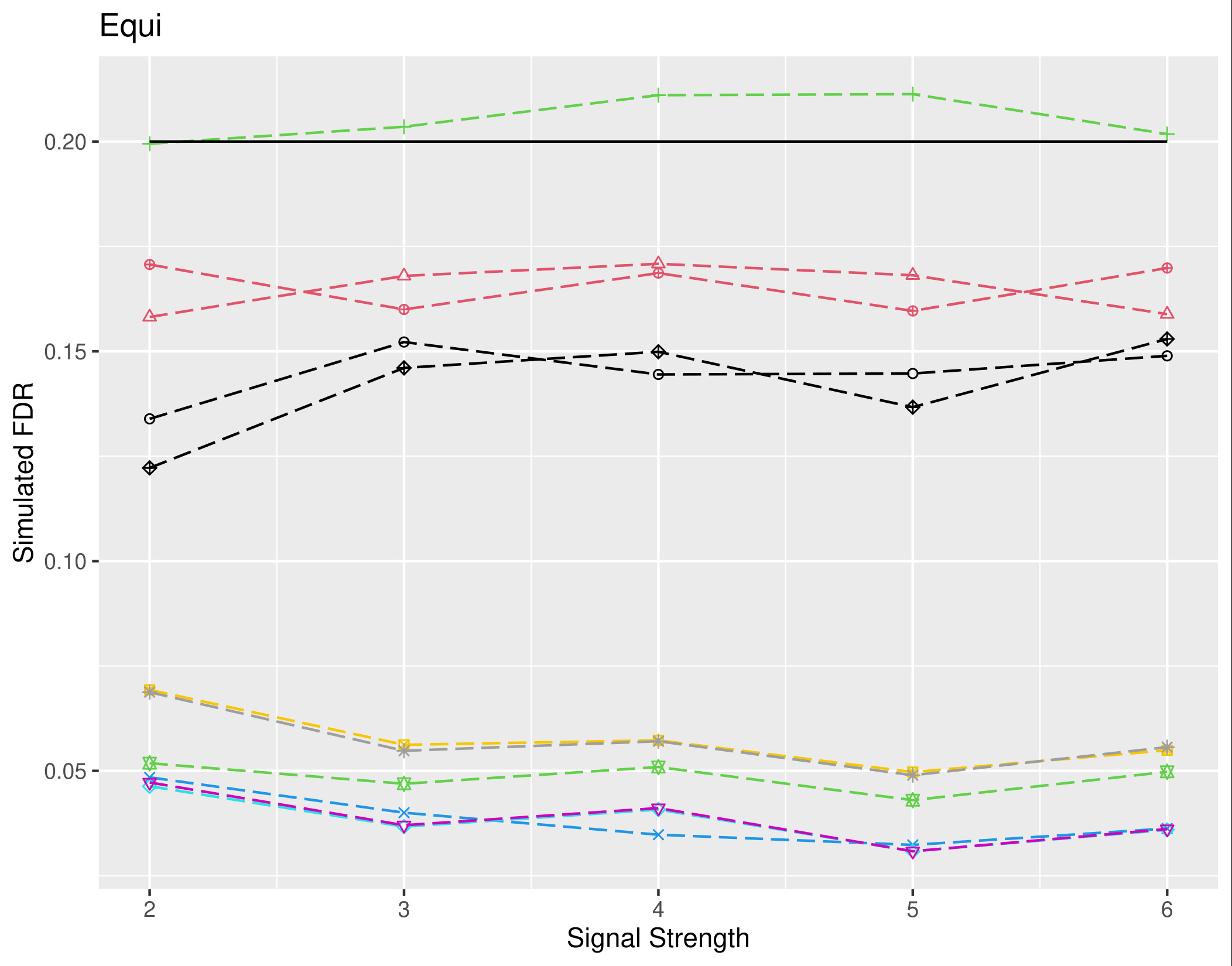} 
      \end{subfigure} \\
  \rotatebox{90}{AR(0.7)} 
  & \begin{subfigure}[b]{\linewidth}
      \centering
      \includegraphics[width=0.8\linewidth]{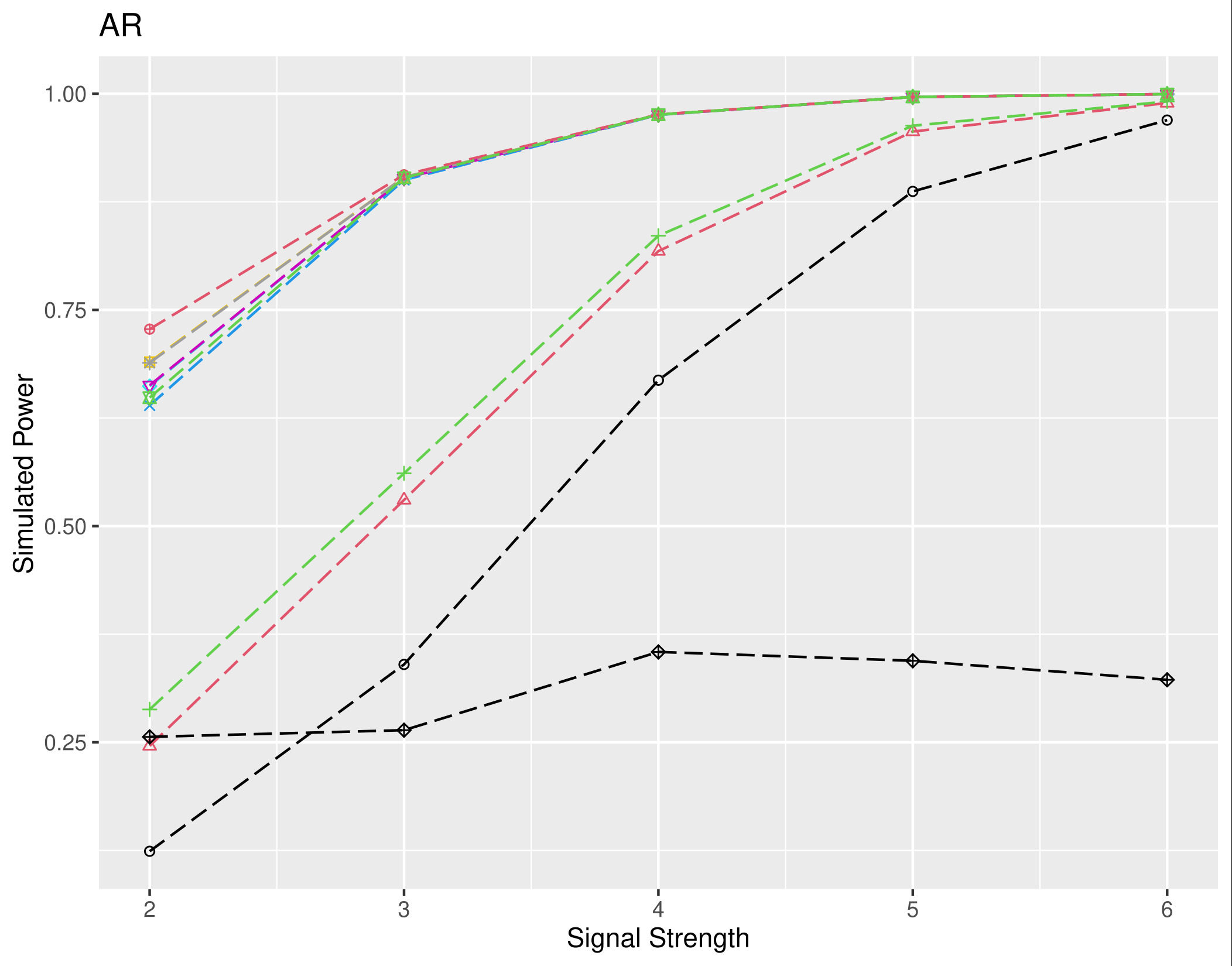} 
      \end{subfigure}  
  & \begin{subfigure}[b]{\linewidth}
      \centering
      \includegraphics[width=0.8\linewidth]{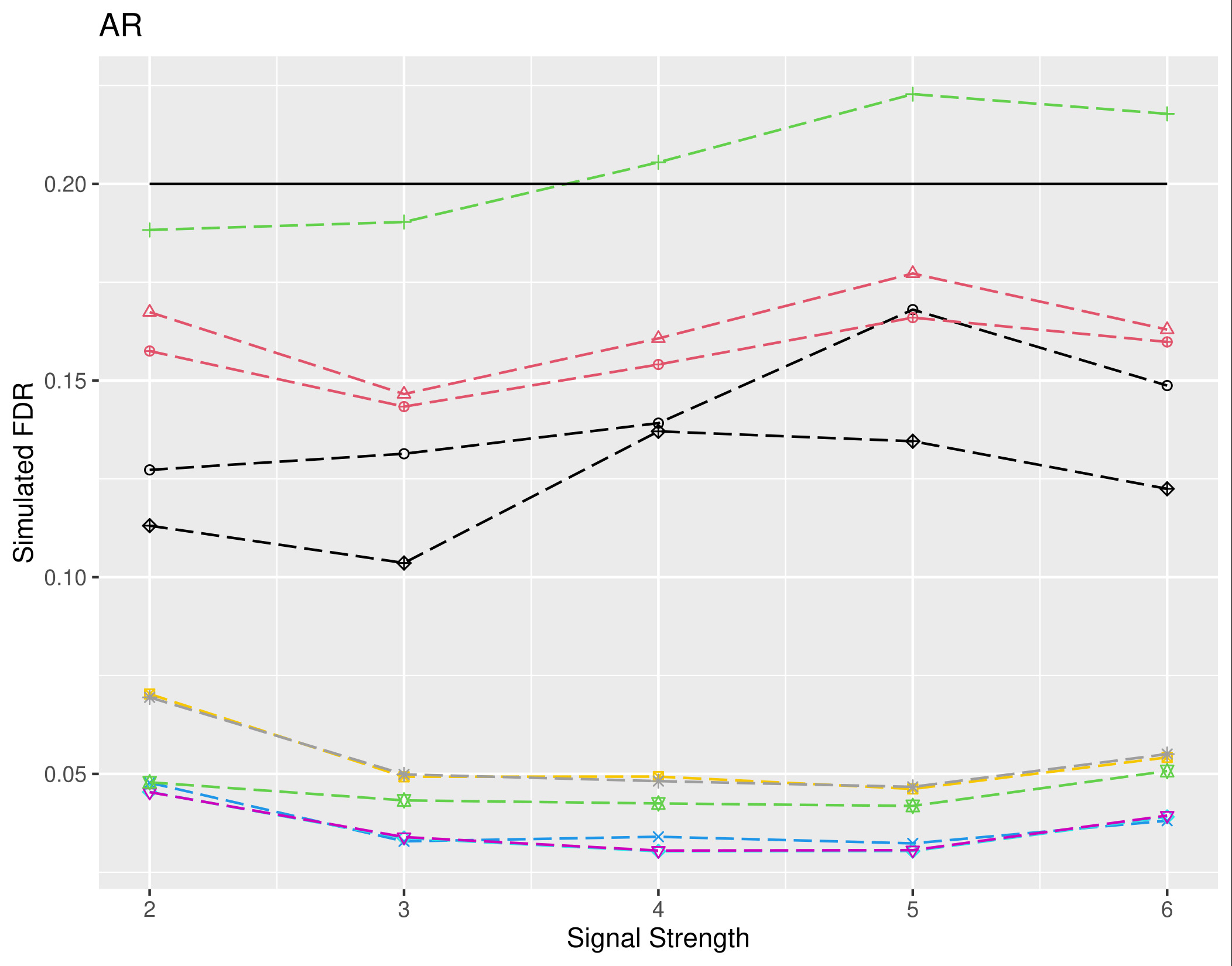} 
      \end{subfigure} \\
  \rotatebox{90}{IAR(0.7)} 
  & \begin{subfigure}[b]{\linewidth}
      \centering
      \includegraphics[width=0.8\linewidth]{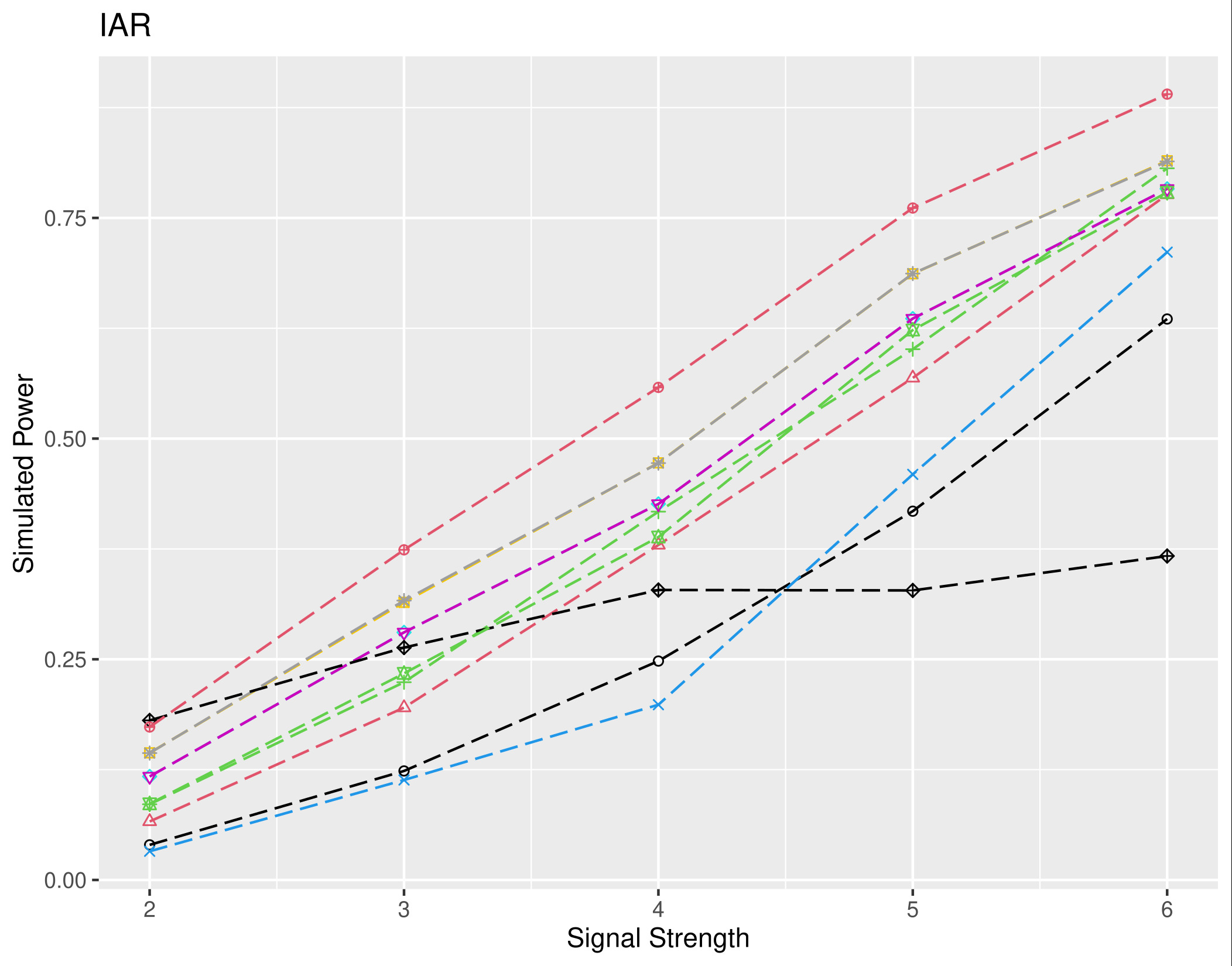} 
      \end{subfigure}  
  & \begin{subfigure}[b]{\linewidth}
      \centering
      \includegraphics[width=0.8\linewidth]{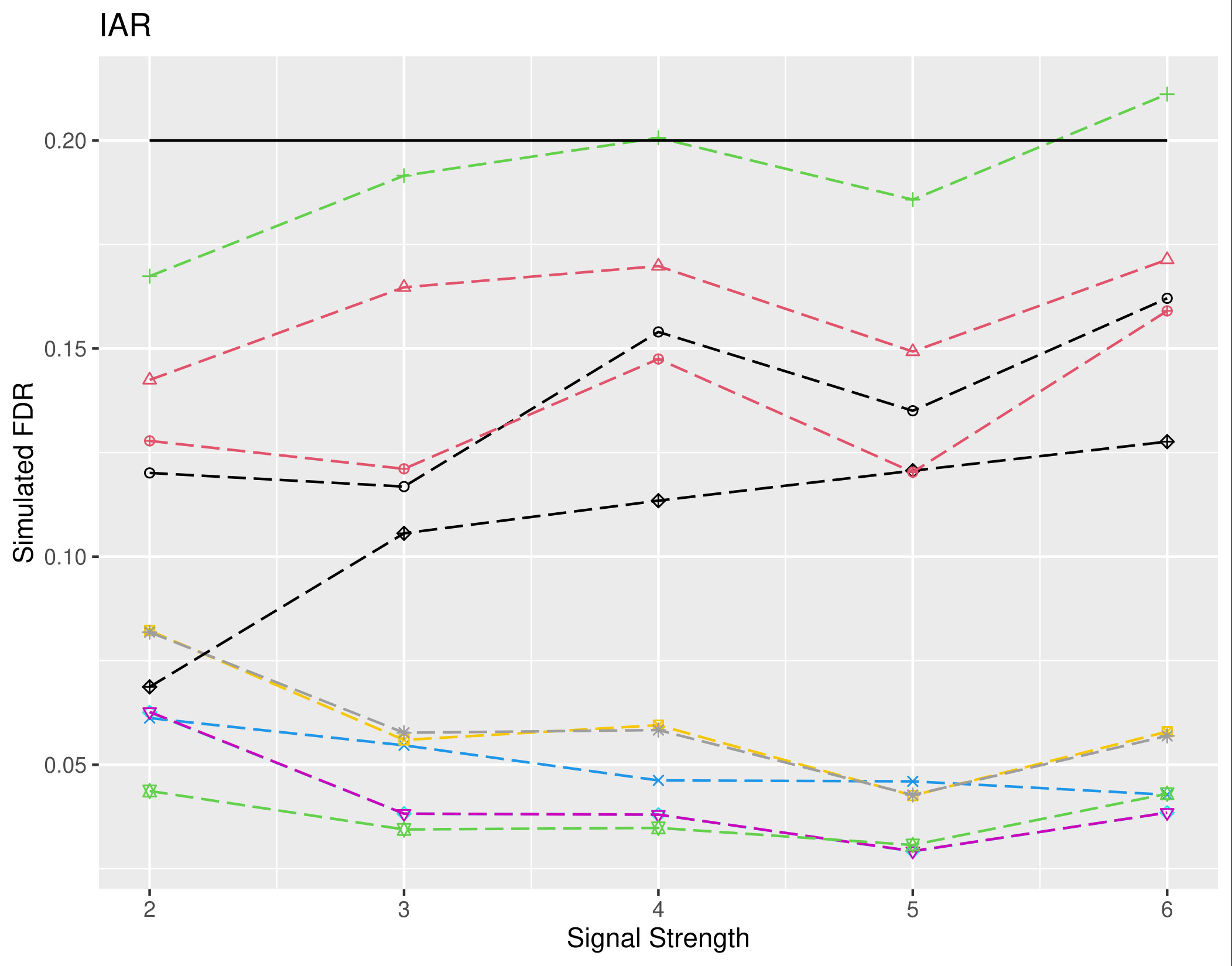} 
      \end{subfigure}    
  \end{tabular*} 
  \caption{Simulated Power (left column) and simulated FDR (right column) displayed for knockoff-assisted variable selection from $d=40$ parameters at a level of $\alpha=0.2$. Methods compared are BH method (Circle and black), BBH method (Triangle point up and red), ABBH method (Plus and green), SBBH1 (Cross and blue), SBBH2 (Diamond and light blue), SBBH3 (Triangle point down and purple), SBBH4 (Square cross and yellow), SBBH5 (Star and grey), Knockoff (Diamond plus and black), Rev-BBH (Circle plus and red) and BBY (Triangles up and down and green)}
  \label{knockoff:figure2} 
\end{figure}

\begin{figure}[hbt!]
  \begin{tabular*}{\textwidth}{
    @{}m{0.5cm}
    @{}m{\dimexpr0.50\textwidth-0.25cm\relax}
    @{}m{\dimexpr0.50\textwidth-0.25cm\relax}}
  \rotatebox{90}{Block Diagonal}
  & \begin{subfigure}[b]{\linewidth}
      \centering
      \includegraphics[width=0.8\linewidth]{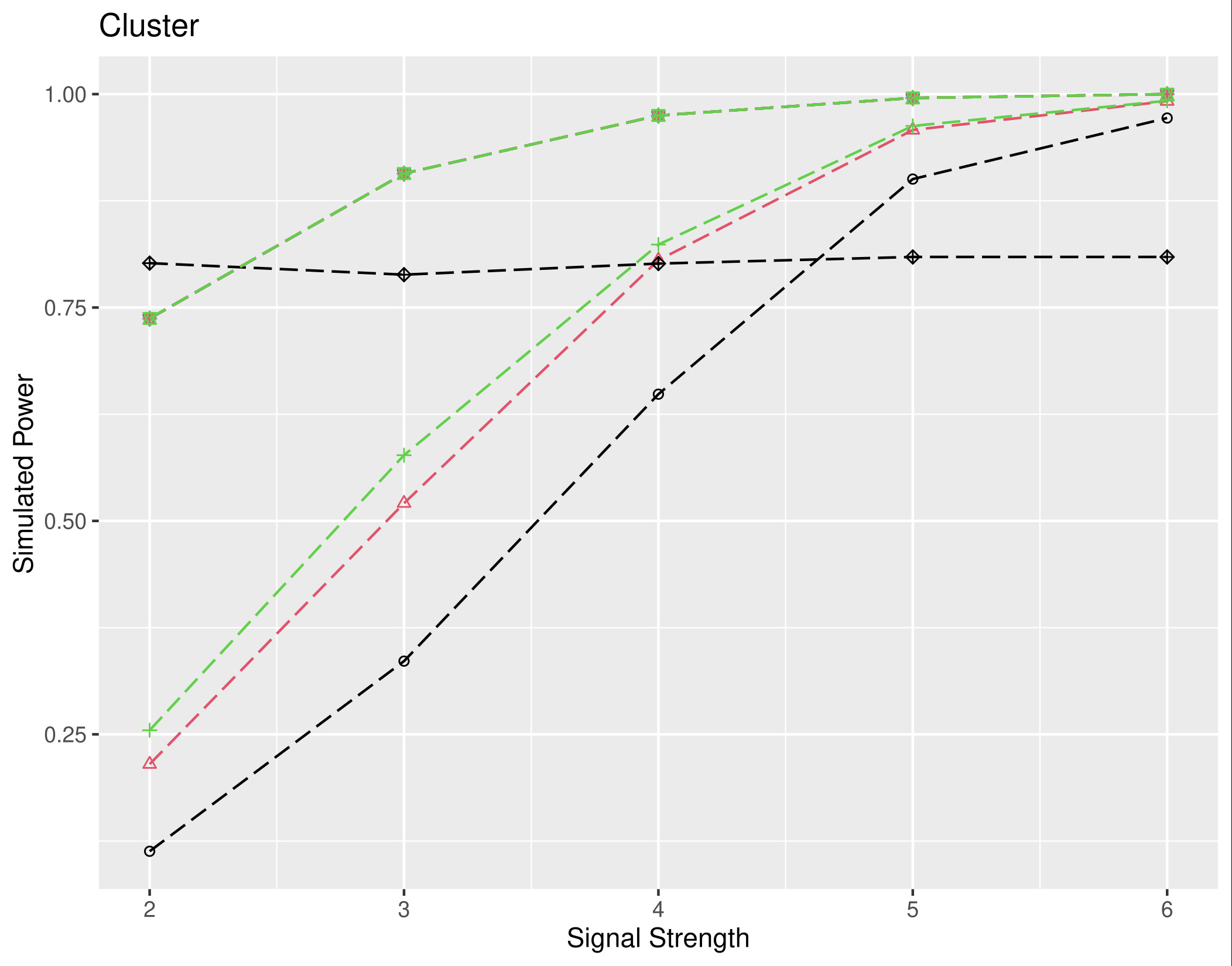} 
      \end{subfigure}  
  & \begin{subfigure}[b]{\linewidth}
      \centering
      \includegraphics[width=0.8\linewidth]{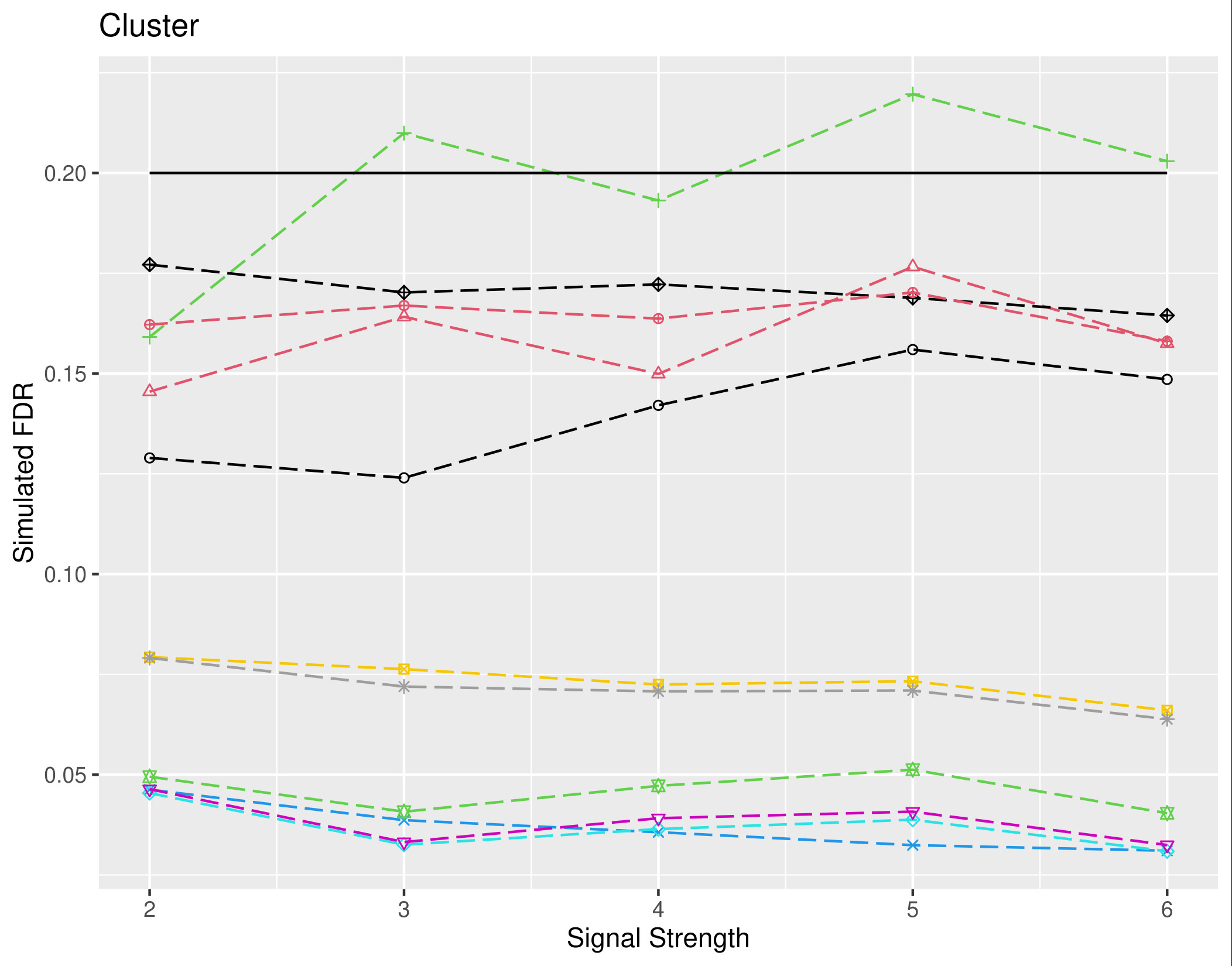} 
      \end{subfigure} \\
  \rotatebox{90}{Sparse} 
  & \begin{subfigure}[b]{\linewidth}
      \centering
      \includegraphics[width=0.8\linewidth]{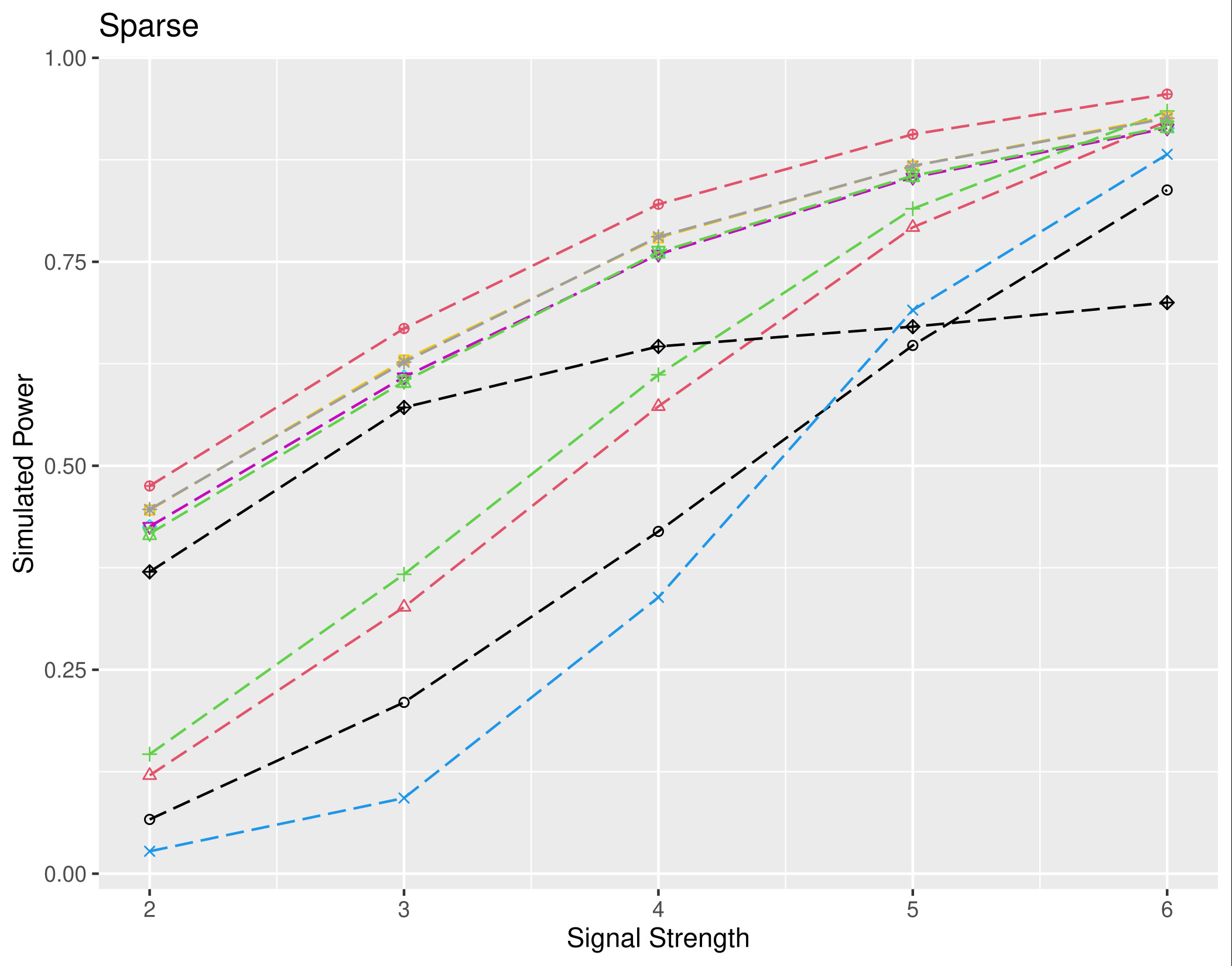} 
      \end{subfigure}  
  & \begin{subfigure}[b]{\linewidth}
      \centering
      \includegraphics[width=0.8\linewidth]{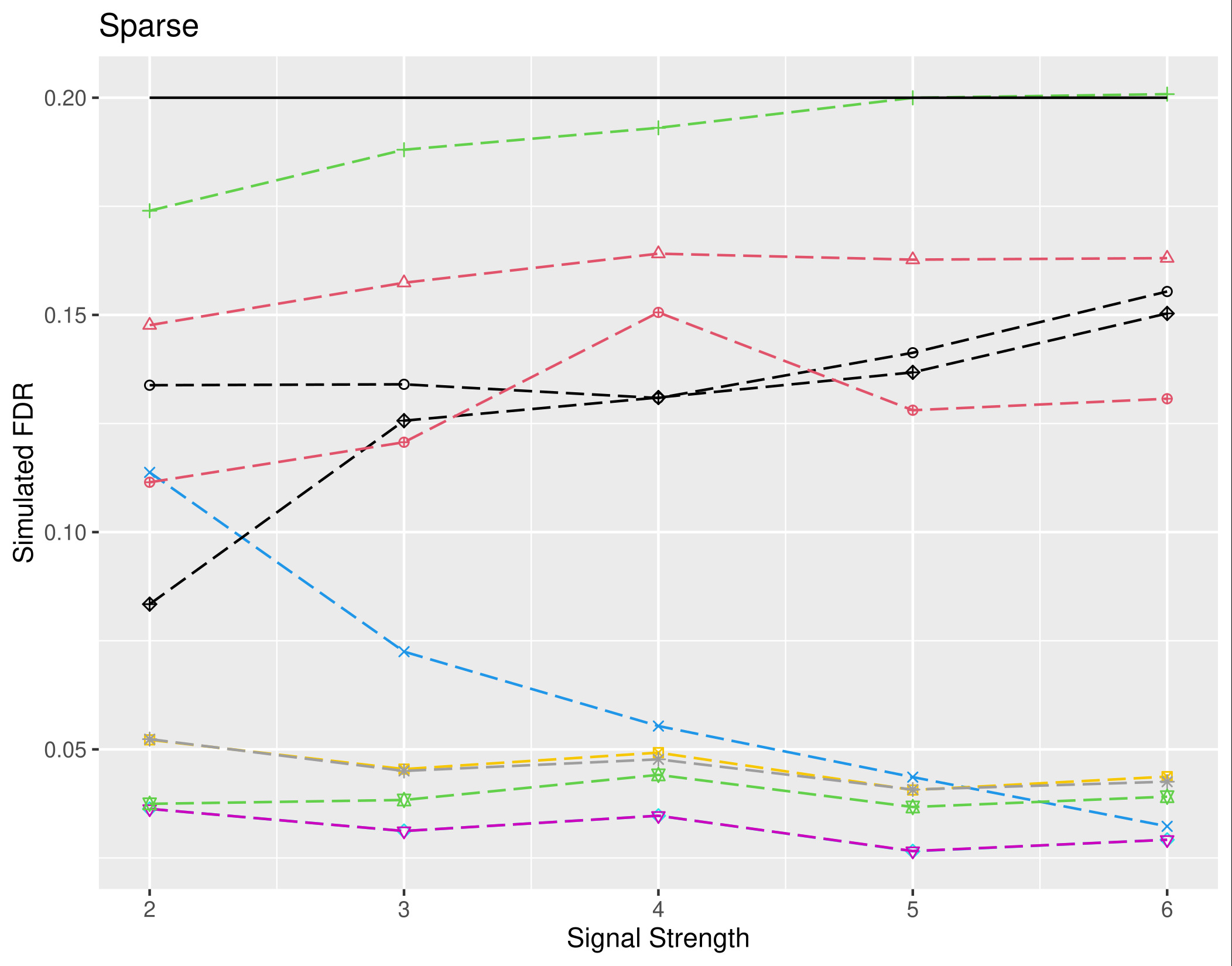} 
      \end{subfigure}   
  \end{tabular*} 
  \caption{Simulated Power (left column) and simulated FDR (right column) displayed for knockoff-assisted variable selection from $d=40$ parameters at a level of $\alpha=0.2$. Methods compared are BH method (Circle and black), BBH method (Triangle point up and red), ABBH method (Plus and green), SBBH1 (Cross and blue), SBBH2 (Diamond and light blue), SBBH3 (Triangle point down and purple), SBBH4 (Square cross and yellow), SBBH5 (Star and grey), Knockoff (Diamond plus and black), Rev-BBH (Circle plus and red) and BBY (Triangles up and down and green)}
  \label{knockoff:figure3} 
\end{figure}

\begin{figure}[hbt!]
  \begin{tabular*}{\textwidth}{
    @{}m{0.5cm}
    @{}m{\dimexpr0.50\textwidth-0.25cm\relax}
    @{}m{\dimexpr0.50\textwidth-0.25cm\relax}}
  \rotatebox{90}{Equi(0.3)}
  & \begin{subfigure}[b]{\linewidth}
      \centering
      \includegraphics[width=0.8\linewidth]{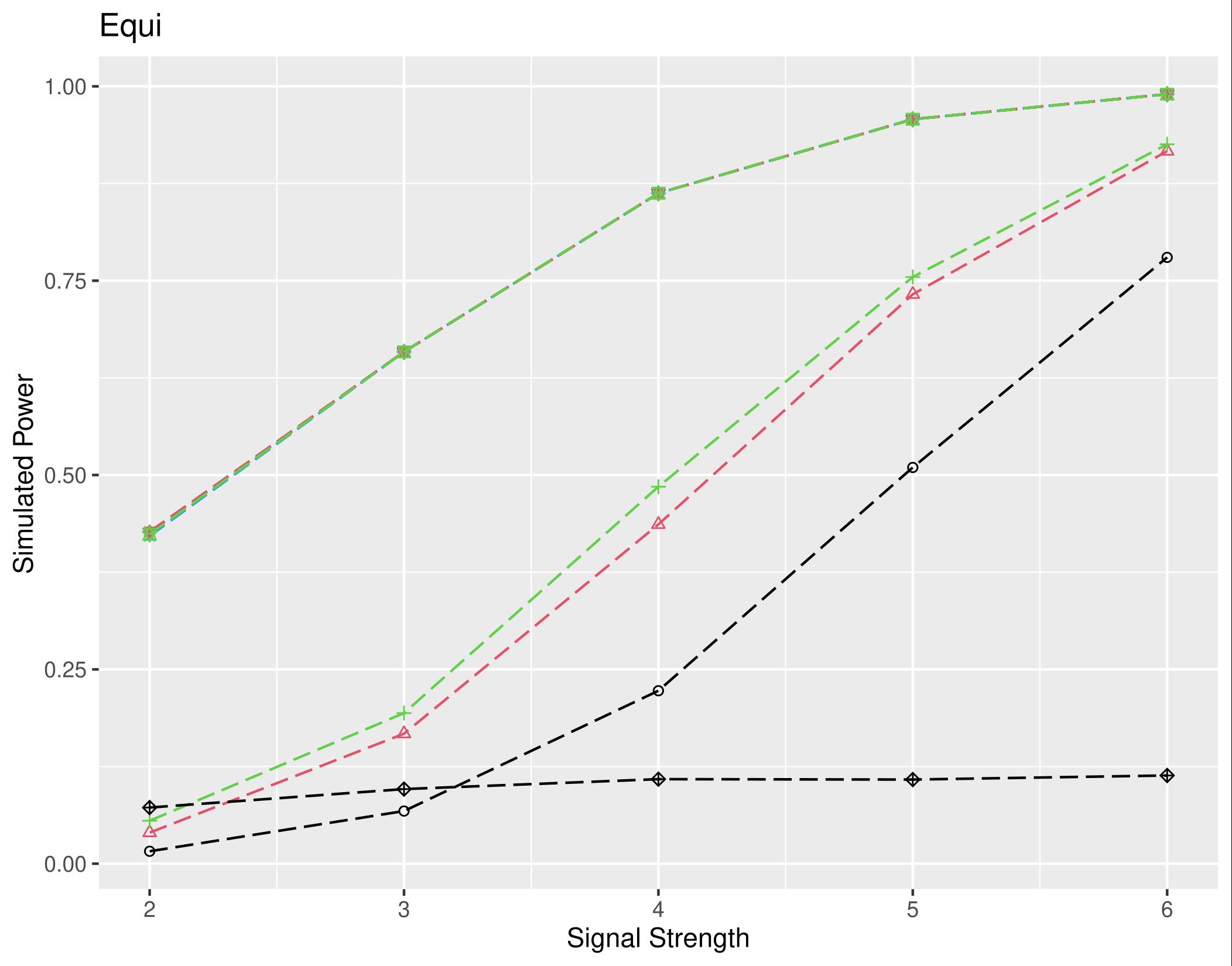} 
      \end{subfigure}  
  & \begin{subfigure}[b]{\linewidth}
      \centering
      \includegraphics[width=0.8\linewidth]{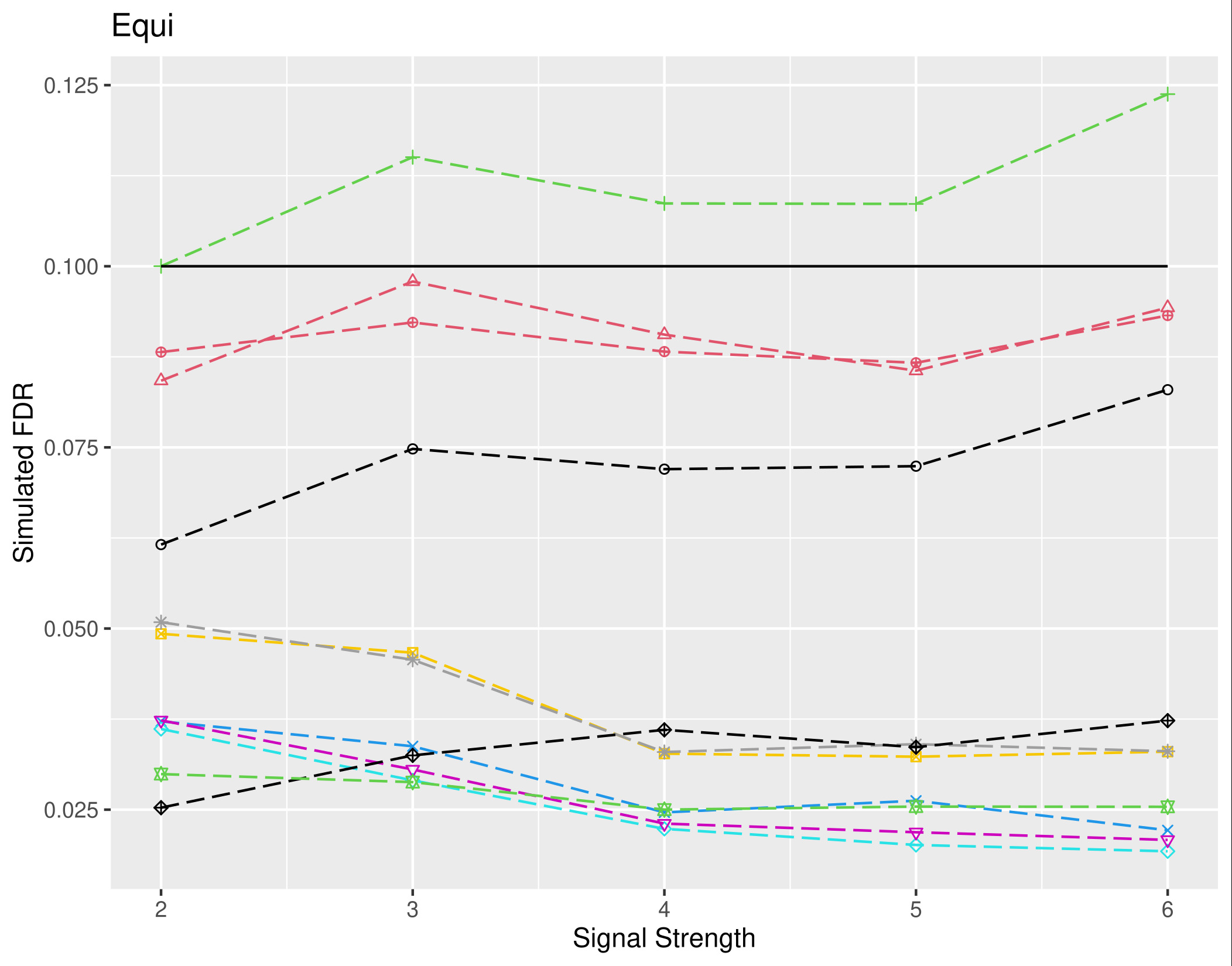} 
      \end{subfigure} \\
  \rotatebox{90}{AR(0.3)} 
  & \begin{subfigure}[b]{\linewidth}
      \centering
      \includegraphics[width=0.8\linewidth]{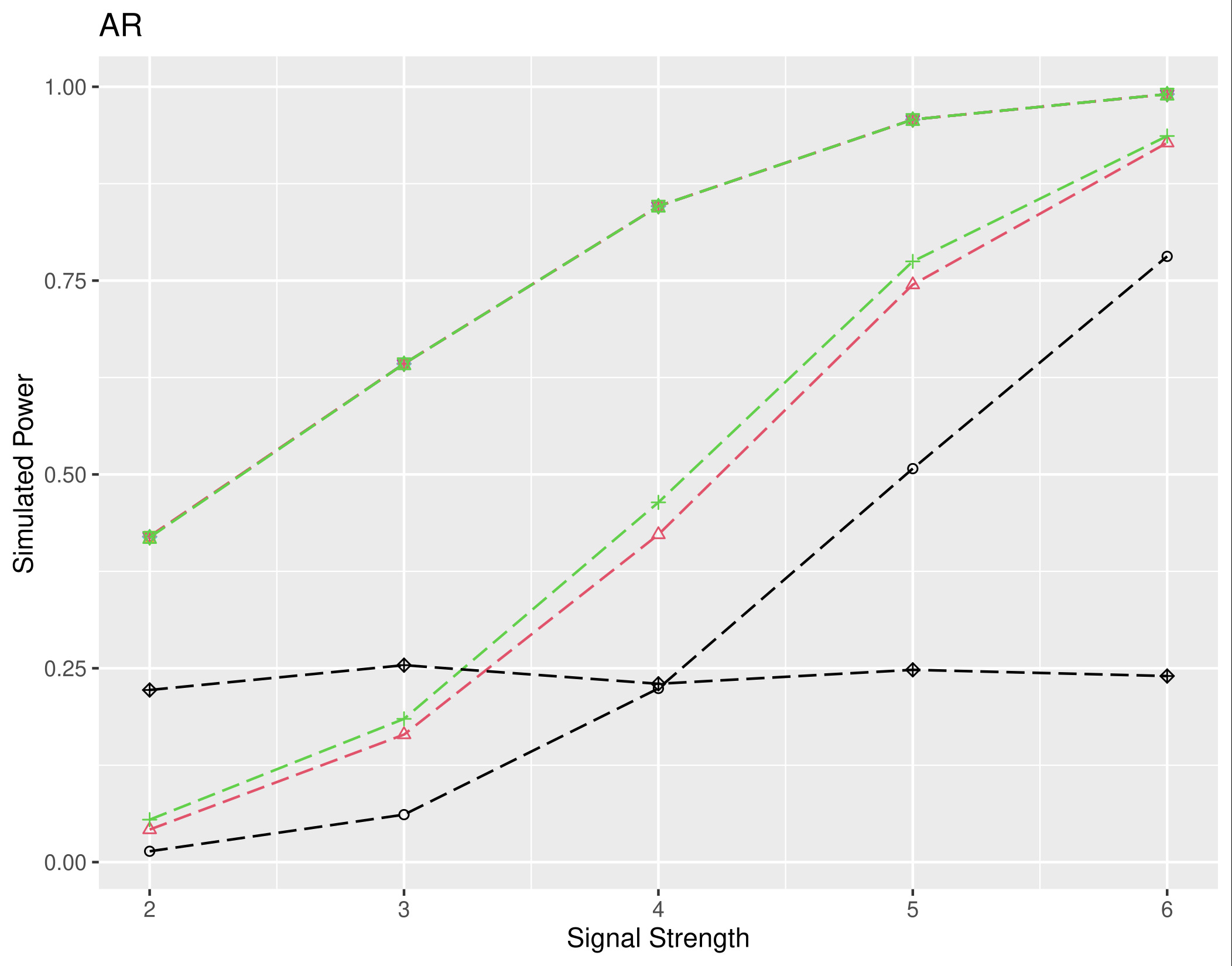} 
      \end{subfigure}  
  & \begin{subfigure}[b]{\linewidth}
      \centering
      \includegraphics[width=0.8\linewidth]{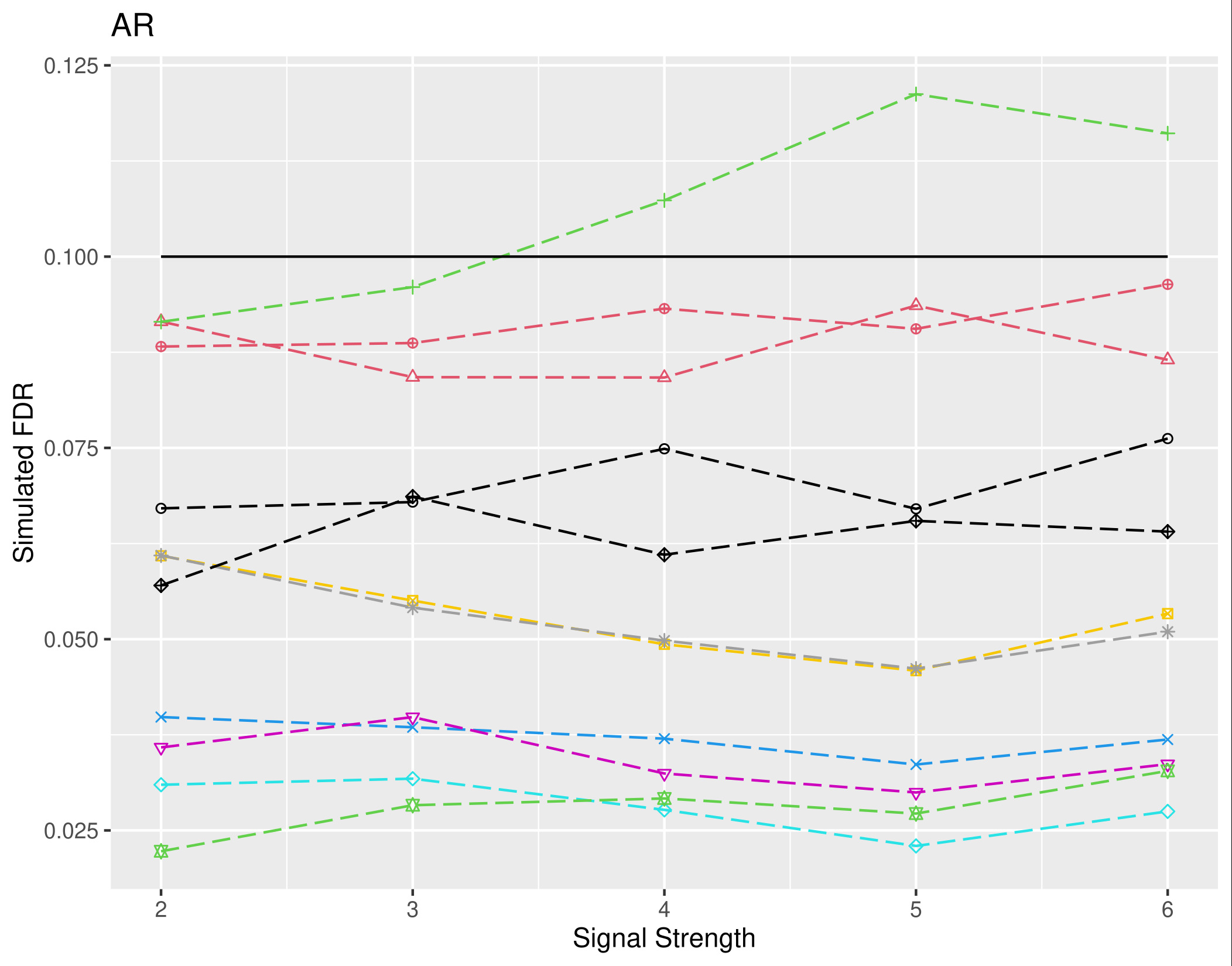} 
      \end{subfigure} \\
  \rotatebox{90}{IAR(0.3)} 
  & \begin{subfigure}[b]{\linewidth}
      \centering
      \includegraphics[width=0.8\linewidth]{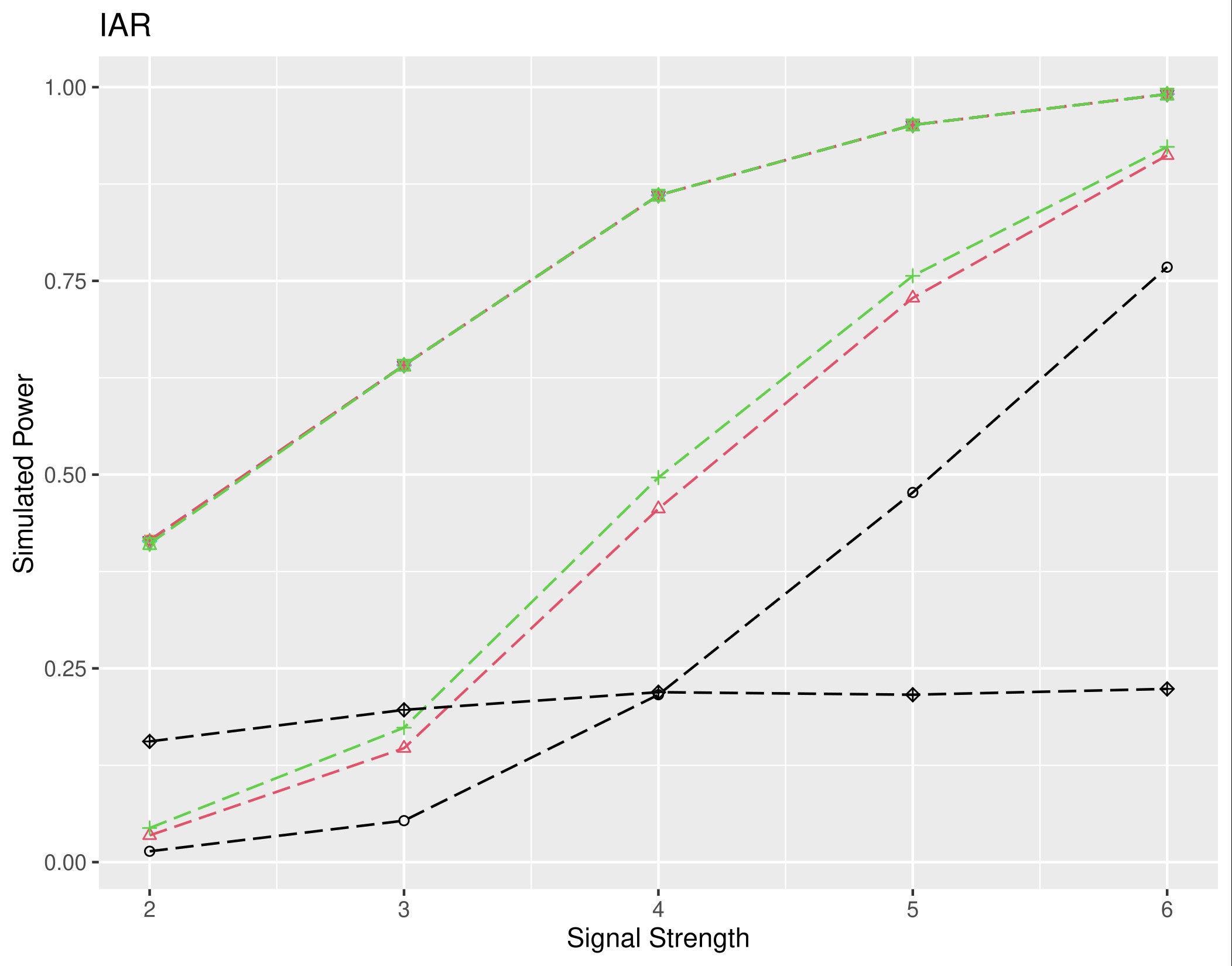} 
      \end{subfigure}  
  & \begin{subfigure}[b]{\linewidth}
      \centering
      \includegraphics[width=0.8\linewidth]{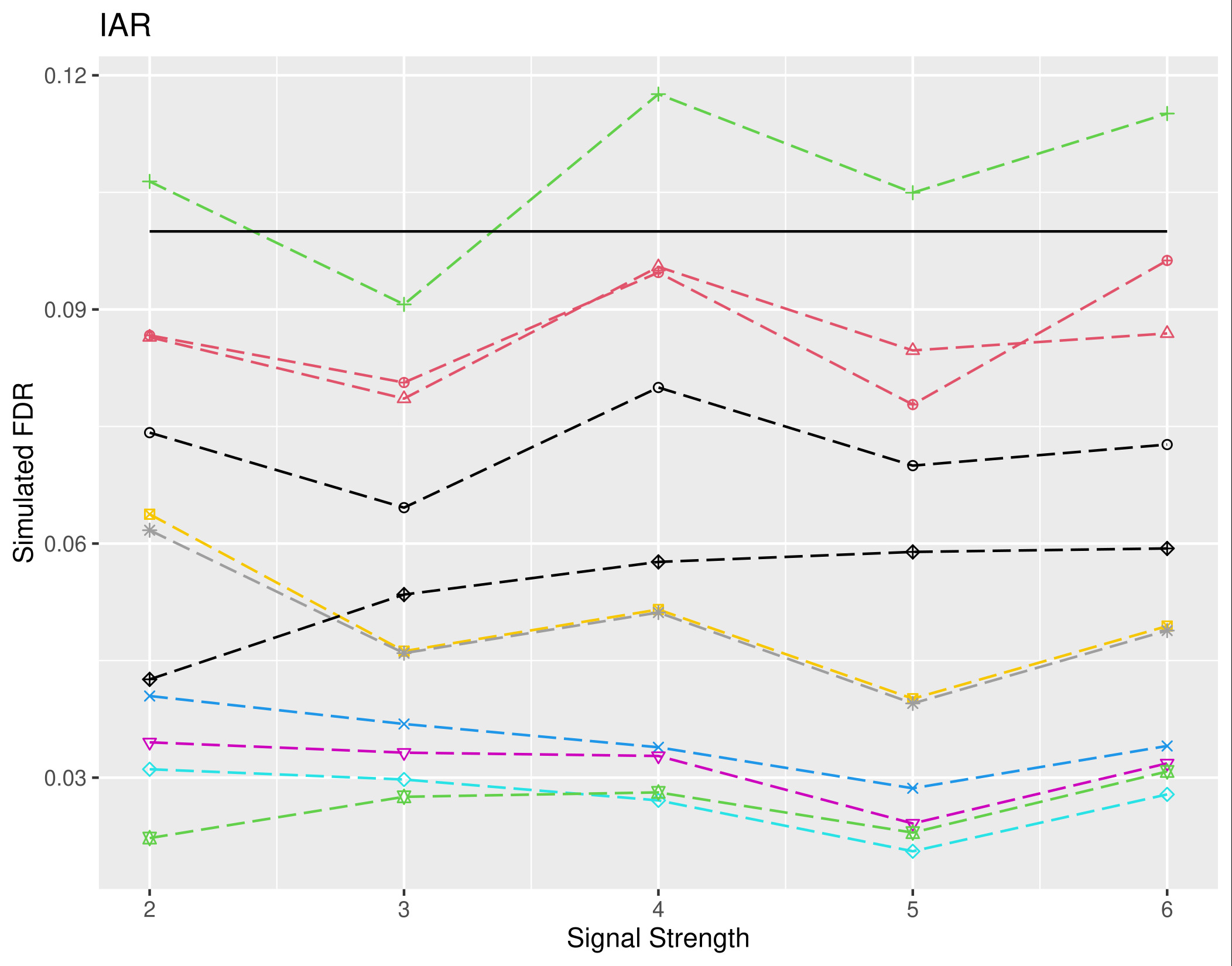} 
      \end{subfigure}    
  \end{tabular*} 
  \caption{Simulated Power (left column) and simulated FDR (right column) displayed for knockoff-assisted variable selection from $d=40$ parameters at a level of $\alpha=0.1$. Methods compared are BH method (Circle and black), BBH method (Triangle point up and red), ABBH method (Plus and green), SBBH1 (Cross and blue), SBBH2 (Diamond and light blue), SBBH3 (Triangle point down and purple), SBBH4 (Square cross and yellow), SBBH5 (Star and grey), Knockoff (Diamond plus and black), Rev-BBH (Circle plus and red) and BBY (Triangles up and down and green)}
  \label{knockoff:figure4} 
\end{figure}

\begin{figure}[hbt!]
  \begin{tabular*}{\textwidth}{
    @{}m{0.5cm}
    @{}m{\dimexpr0.50\textwidth-0.25cm\relax}
    @{}m{\dimexpr0.50\textwidth-0.25cm\relax}}
  \rotatebox{90}{Equi(0.7)}
  & \begin{subfigure}[b]{\linewidth}
      \centering
      \includegraphics[width=0.8\linewidth]{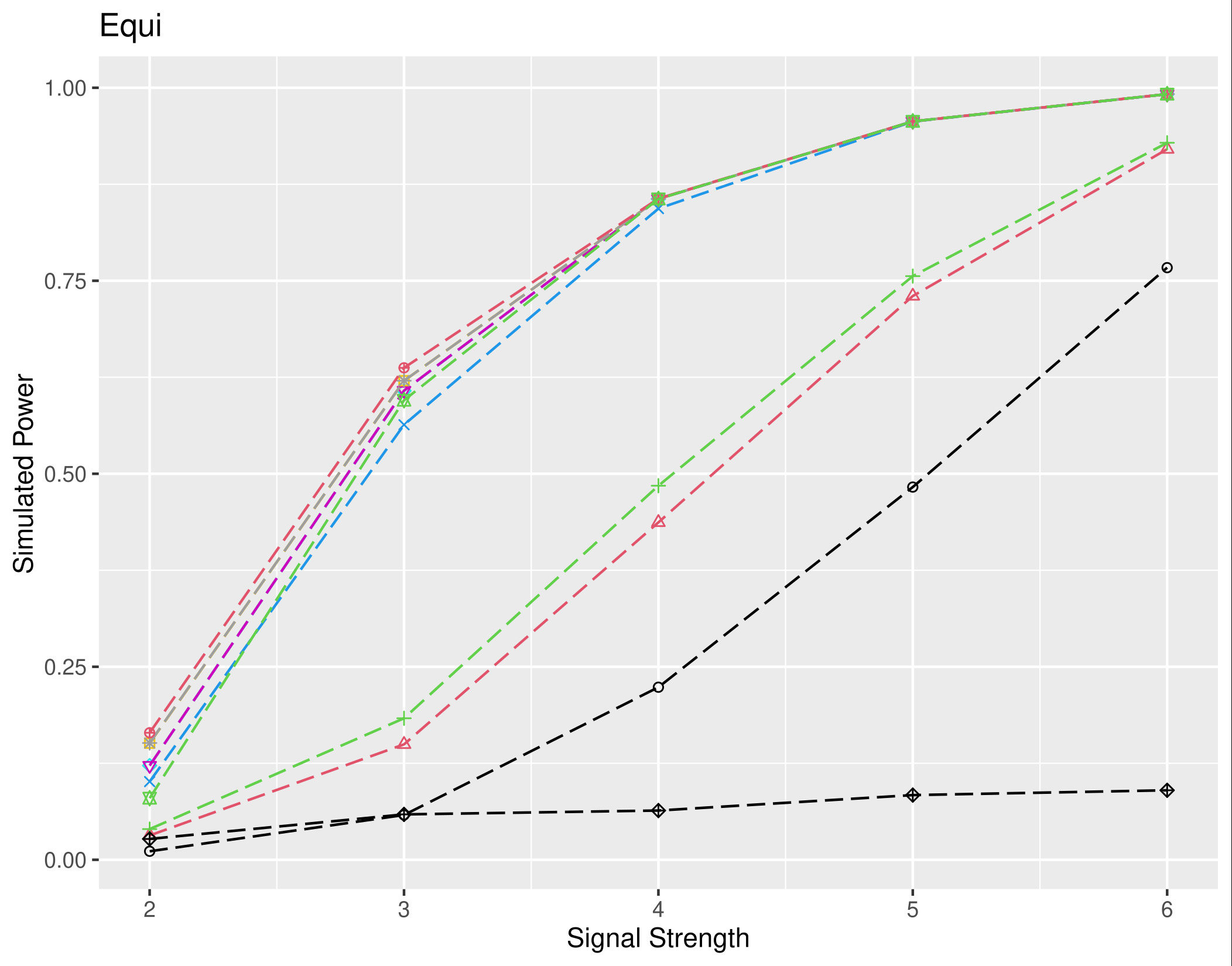} 
      \end{subfigure}  
  & \begin{subfigure}[b]{\linewidth}
      \centering
      \includegraphics[width=0.8\linewidth]{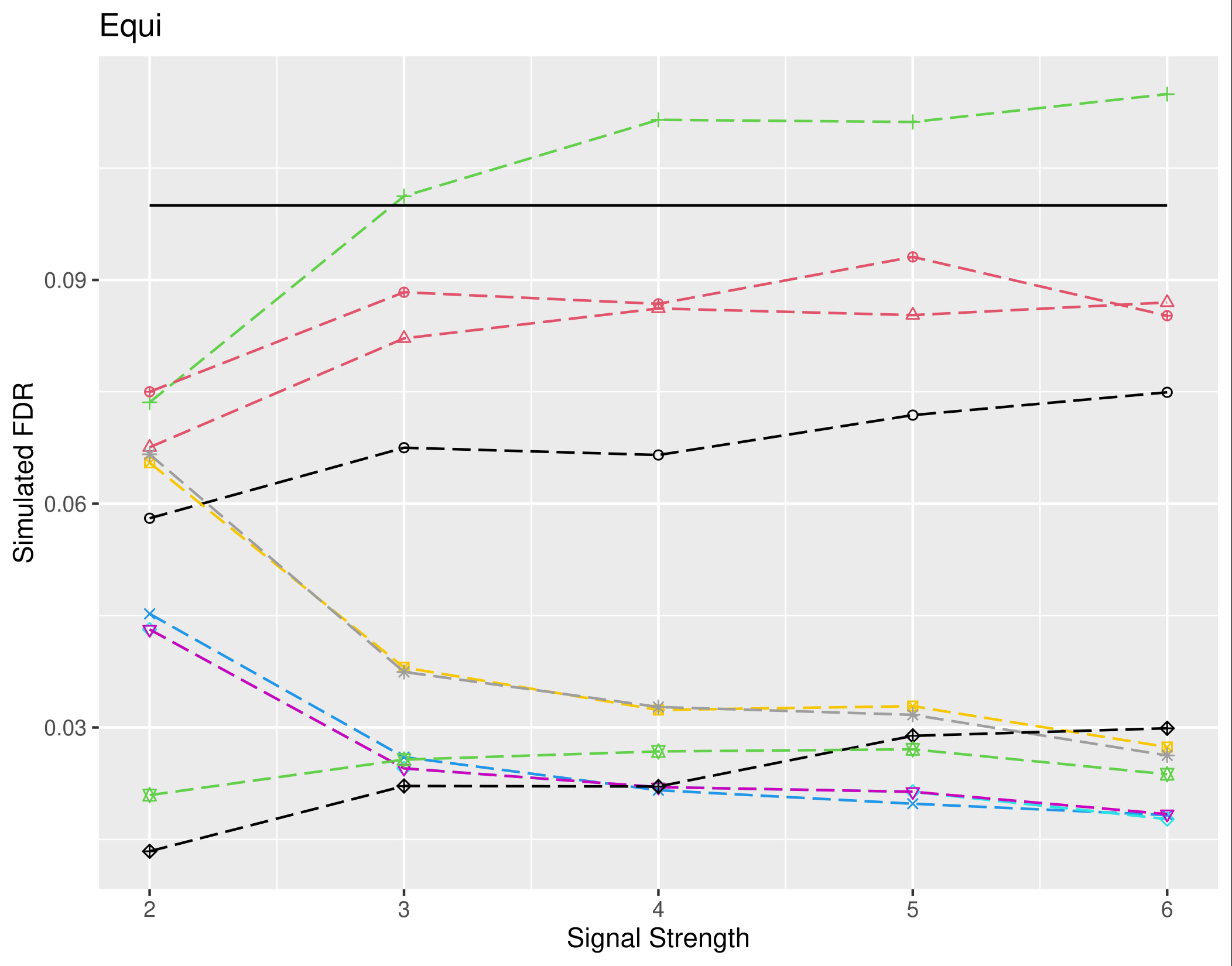} 
      \end{subfigure} \\
  \rotatebox{90}{AR(0.7)} 
  & \begin{subfigure}[b]{\linewidth}
      \centering
      \includegraphics[width=0.8\linewidth]{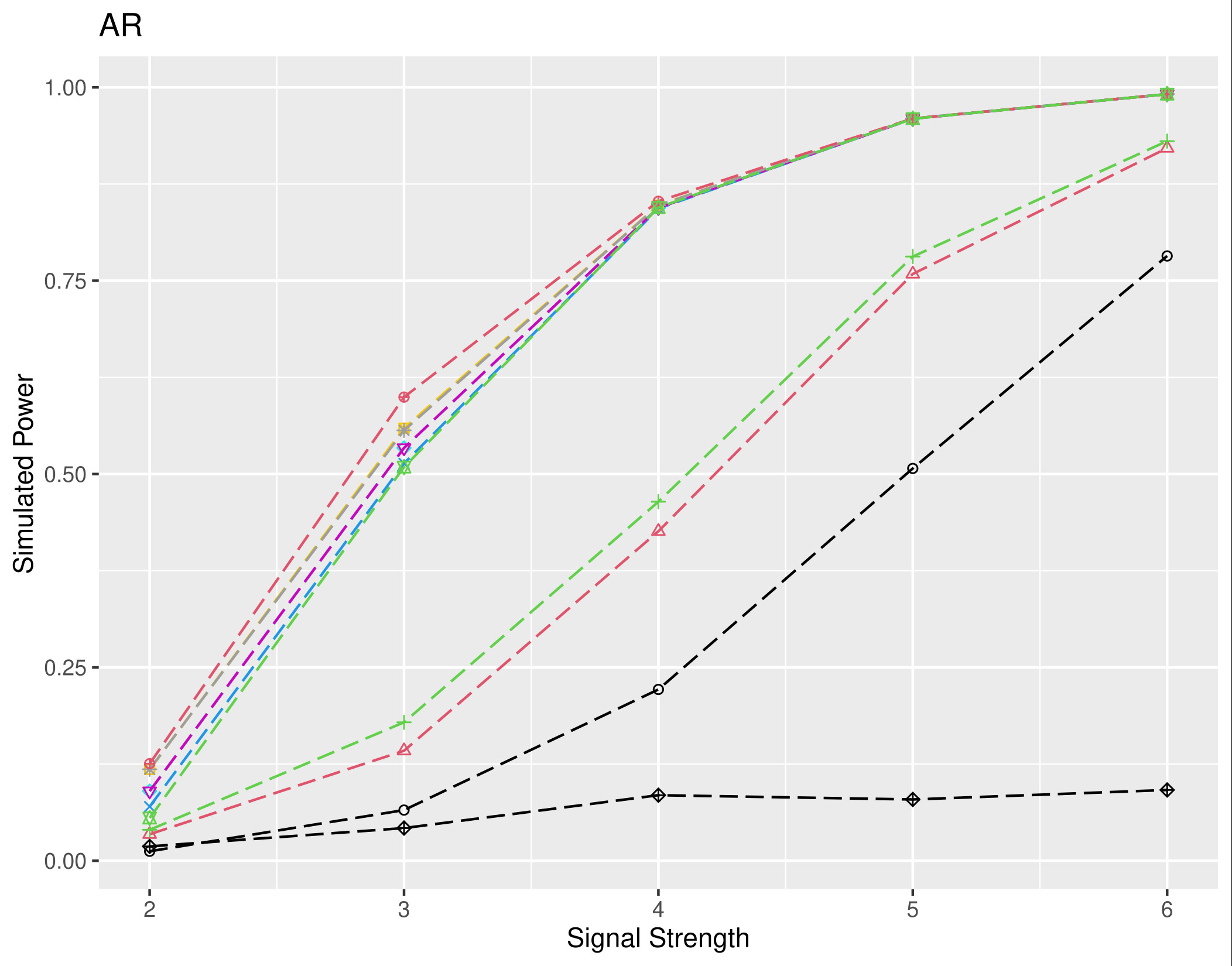} 
      \end{subfigure}  
  & \begin{subfigure}[b]{\linewidth}
      \centering
      \includegraphics[width=0.8\linewidth]{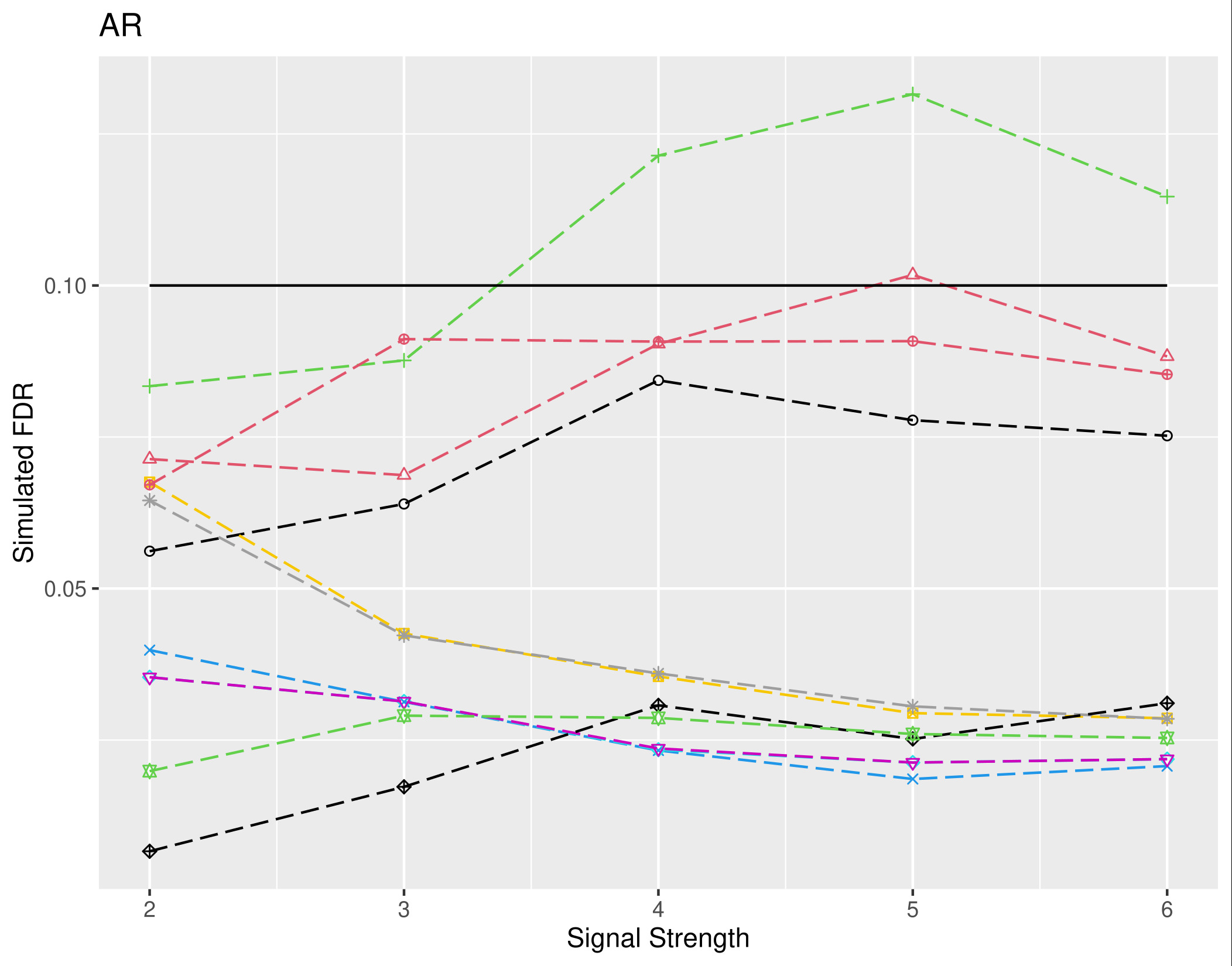} 
      \end{subfigure} \\
  \rotatebox{90}{IAR(0.7)} 
  & \begin{subfigure}[b]{\linewidth}
      \centering
      \includegraphics[width=0.8\linewidth]{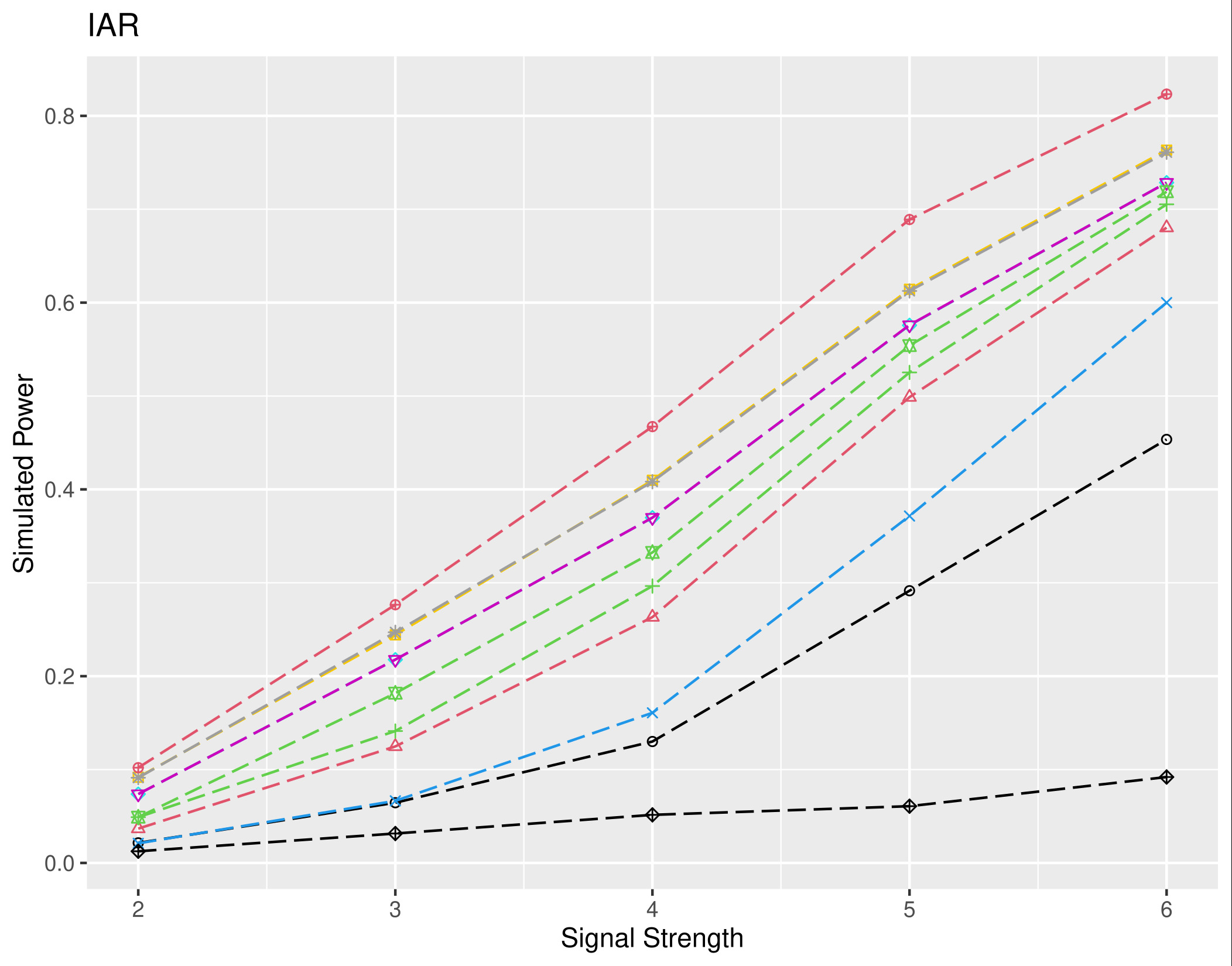} 
      \end{subfigure}  
  & \begin{subfigure}[b]{\linewidth}
      \centering
      \includegraphics[width=0.8\linewidth]{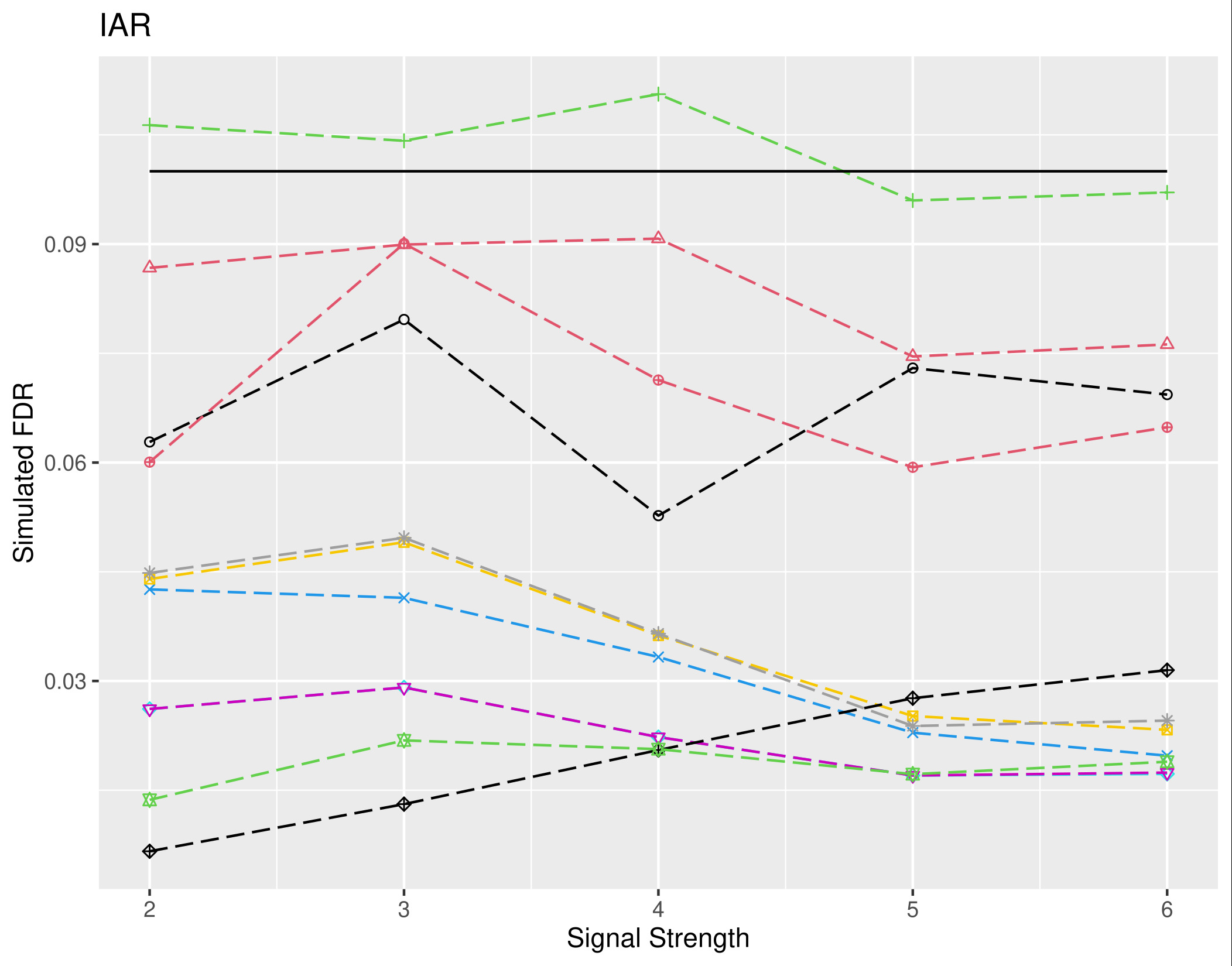} 
      \end{subfigure}    
  \end{tabular*} 
  \caption{Simulated Power (left column) and simulated FDR (right column) displayed for knockoff-assisted variable selection from $d=40$ parameters at a level of $\alpha=0.1$. Methods compared are BH method (Circle and black), BBH method (Triangle point up and red), ABBH method (Plus and green), SBBH1 (Cross and blue), SBBH2 (Diamond and light blue), SBBH3 (Triangle point down and purple), SBBH4 (Square cross and yellow), SBBH5 (Star and grey), Knockoff (Diamond plus and black), Rev-BBH (Circle plus and red) and BBY (Triangles up and down and green)}
  \label{knockoff:figure5} 
\end{figure}

\begin{figure}[hbt!]
  \begin{tabular*}{\textwidth}{
    @{}m{0.5cm}
    @{}m{\dimexpr0.50\textwidth-0.25cm\relax}
    @{}m{\dimexpr0.50\textwidth-0.25cm\relax}}
  \rotatebox{90}{Block Diagonal}
  & \begin{subfigure}[b]{\linewidth}
      \centering
      \includegraphics[width=0.8\linewidth]{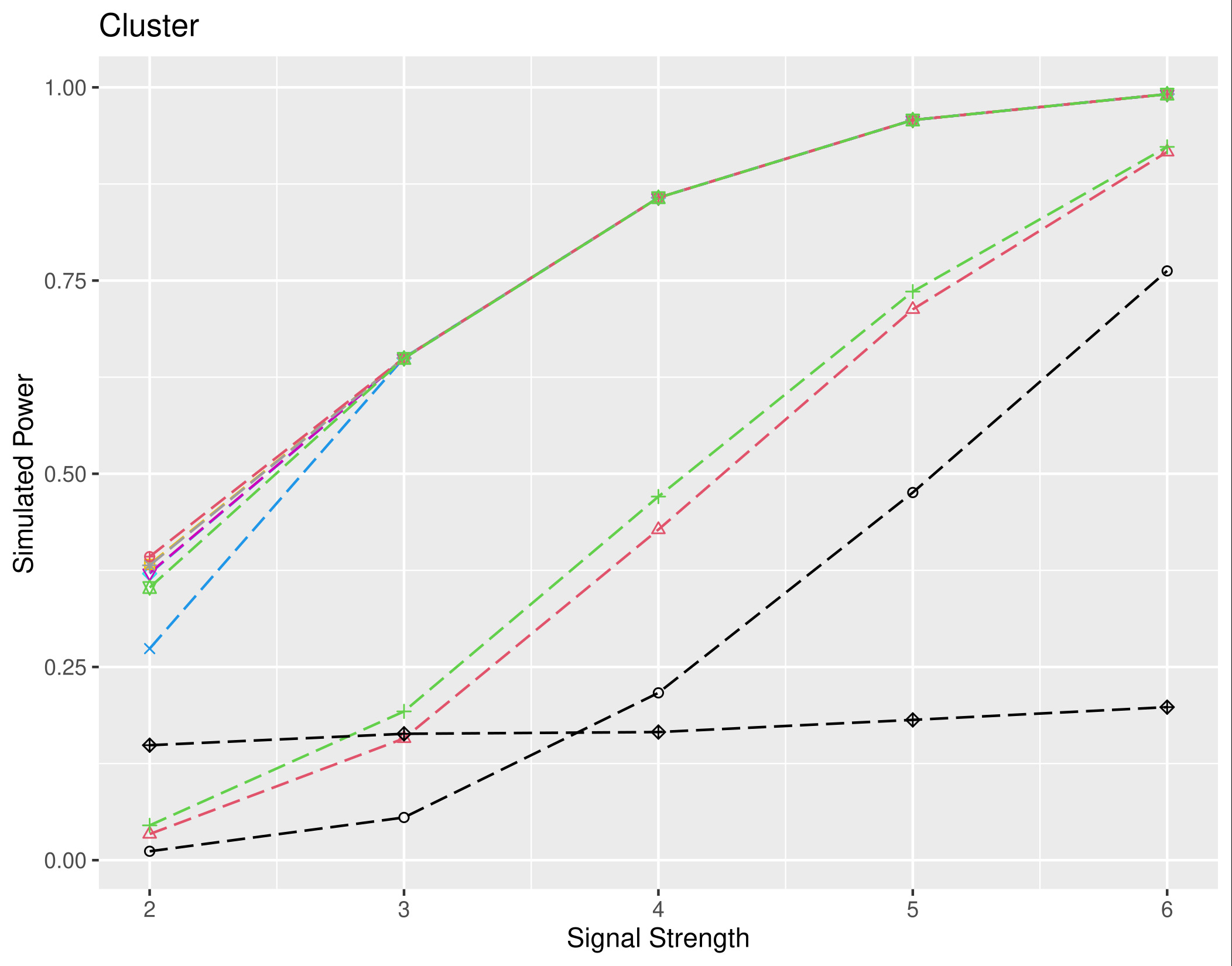} 
      \end{subfigure}  
  & \begin{subfigure}[b]{\linewidth}
      \centering
      \includegraphics[width=0.8\linewidth]{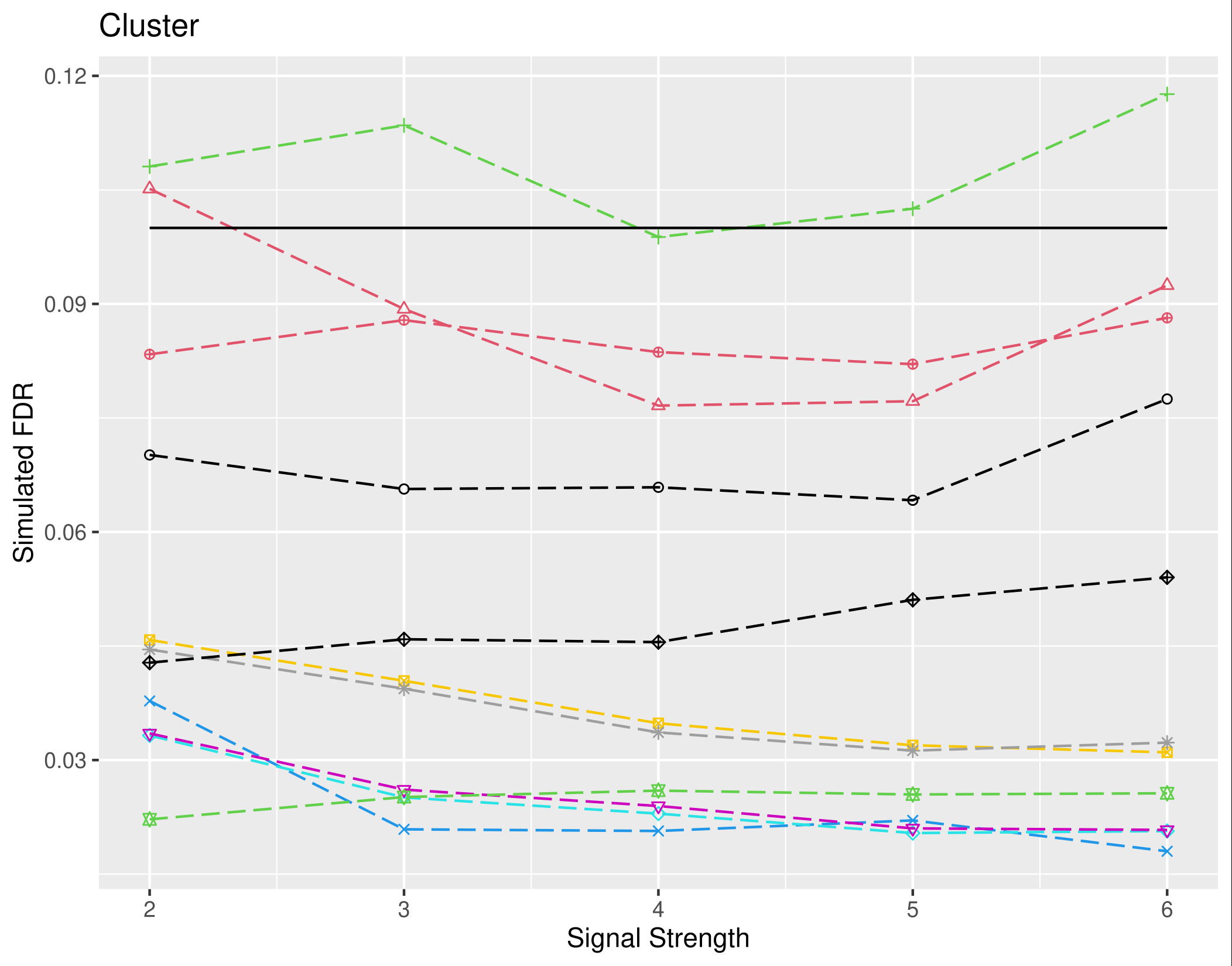} 
      \end{subfigure} \\
  \rotatebox{90}{Sparse} 
  & \begin{subfigure}[b]{\linewidth}
      \centering
      \includegraphics[width=0.8\linewidth]{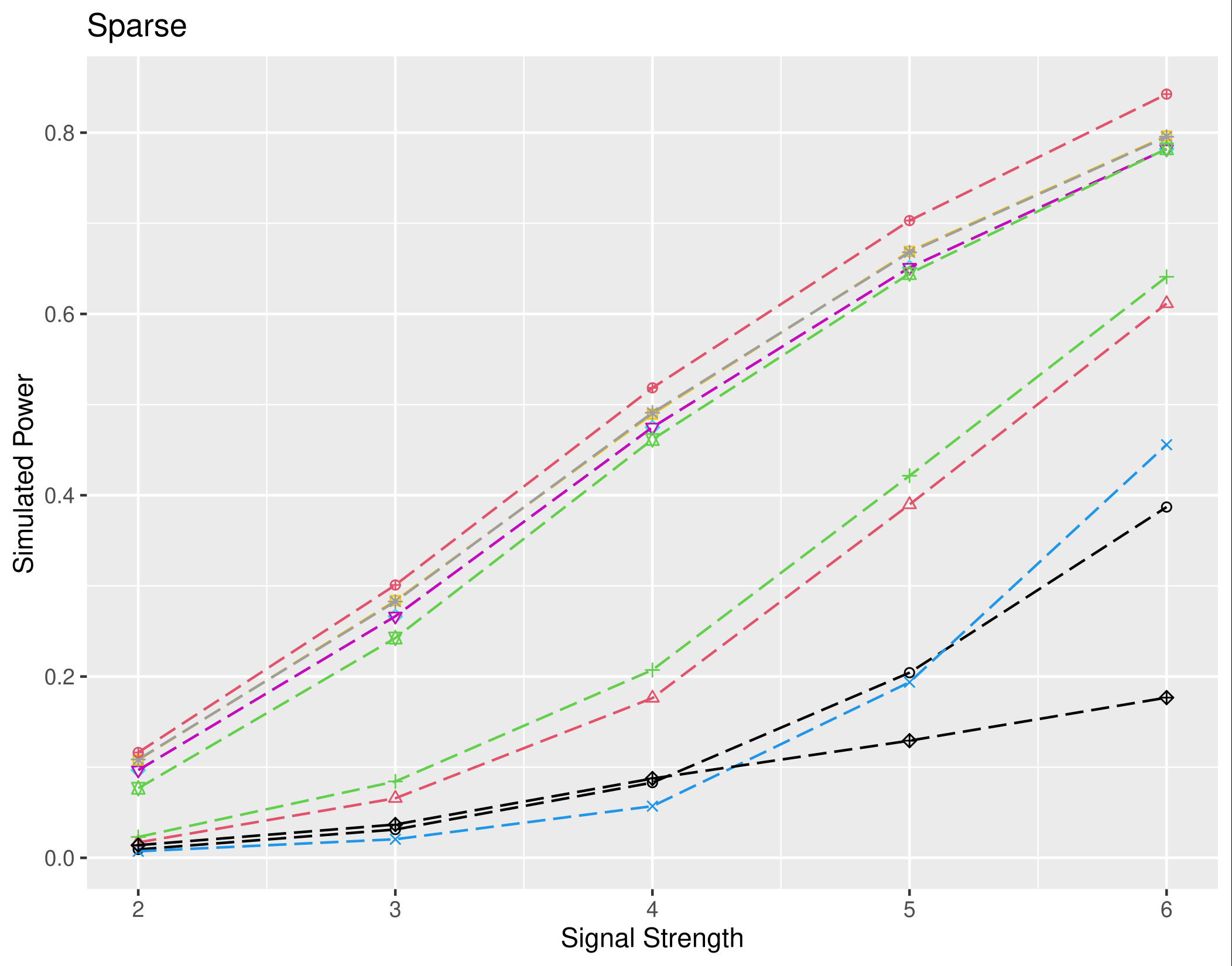} 
      \end{subfigure}  
  & \begin{subfigure}[b]{\linewidth}
      \centering
      \includegraphics[width=0.8\linewidth]{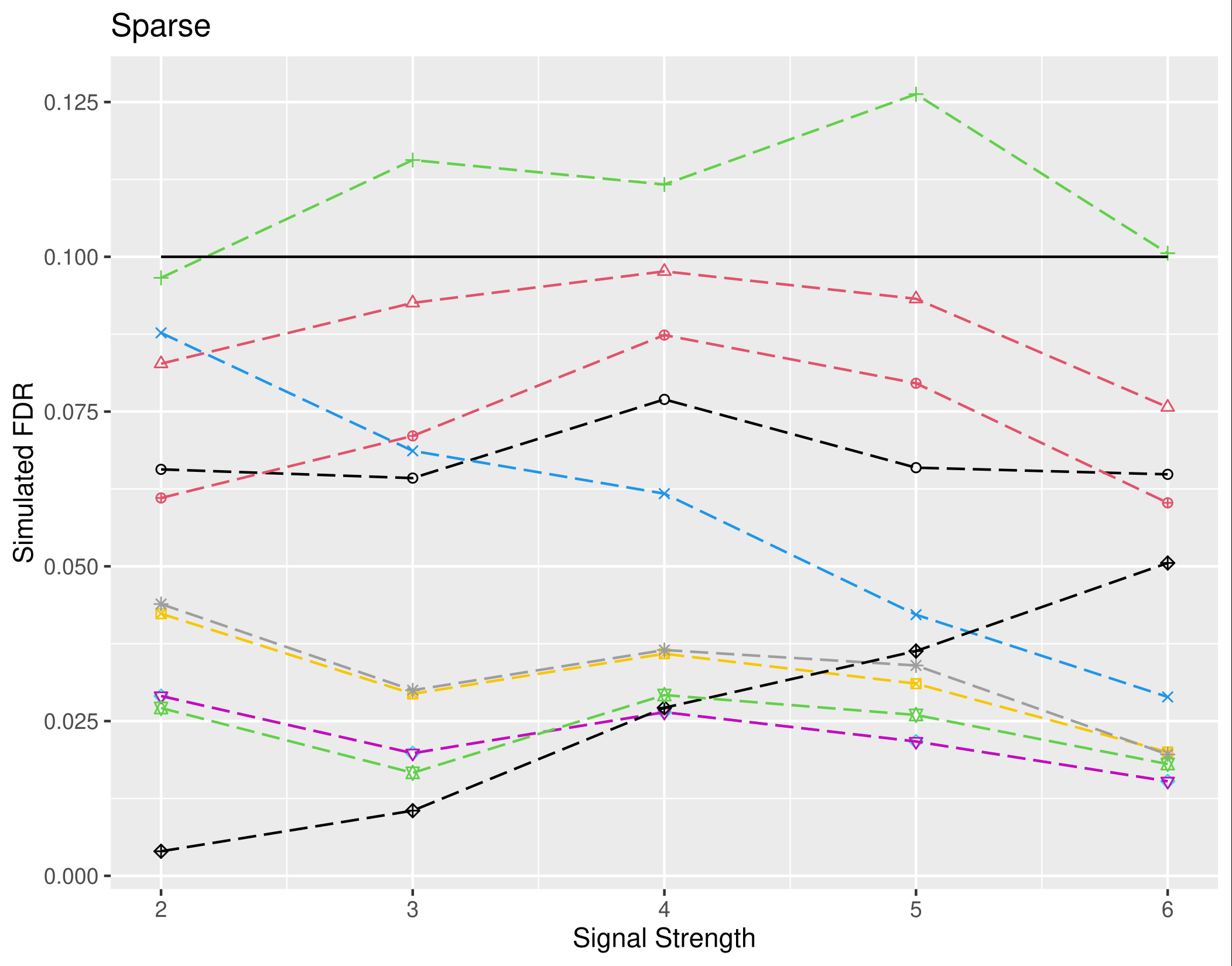} 
      \end{subfigure}   
  \end{tabular*} 
  \caption{Simulated Power (left column) and simulated FDR (right column) displayed for knockoff-assisted variable selection from $d=40$ parameters at a level of $\alpha=0.1$. Methods compared are BH method (Circle and black), BBH method (Triangle point up and red), ABBH method (Plus and green), SBBH1 (Cross and blue), SBBH2 (Diamond and light blue), SBBH3 (Triangle point down and purple), SBBH4 (Square cross and yellow), SBBH5 (Star and grey), Knockoff (Diamond plus and black), Rev-BBH (Circle plus and red) and BBY (Triangles up and down and green)}
  \label{knockoff:figure6} 
\end{figure}

\begin{figure}[hbt!]
  \begin{tabular*}{\textwidth}{
    @{}m{0.5cm}
    @{}m{\dimexpr0.50\textwidth-0.25cm\relax}
    @{}m{\dimexpr0.50\textwidth-0.25cm\relax}}
  \rotatebox{90}{Equi(0.3)}
  & \begin{subfigure}[b]{\linewidth}
      \centering
      \includegraphics[width=0.8\linewidth]{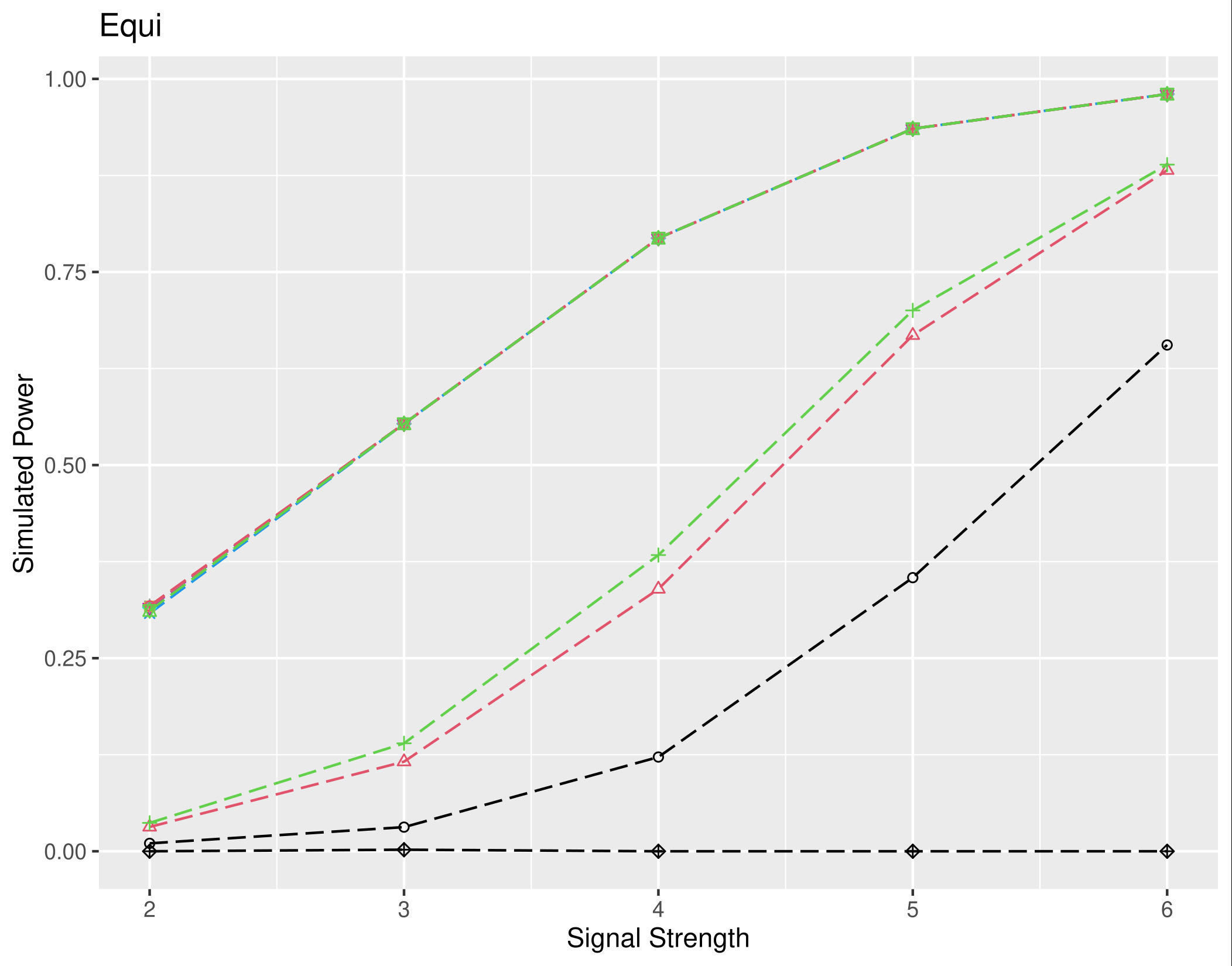} 
      \end{subfigure}  
  & \begin{subfigure}[b]{\linewidth}
      \centering
      \includegraphics[width=0.8\linewidth]{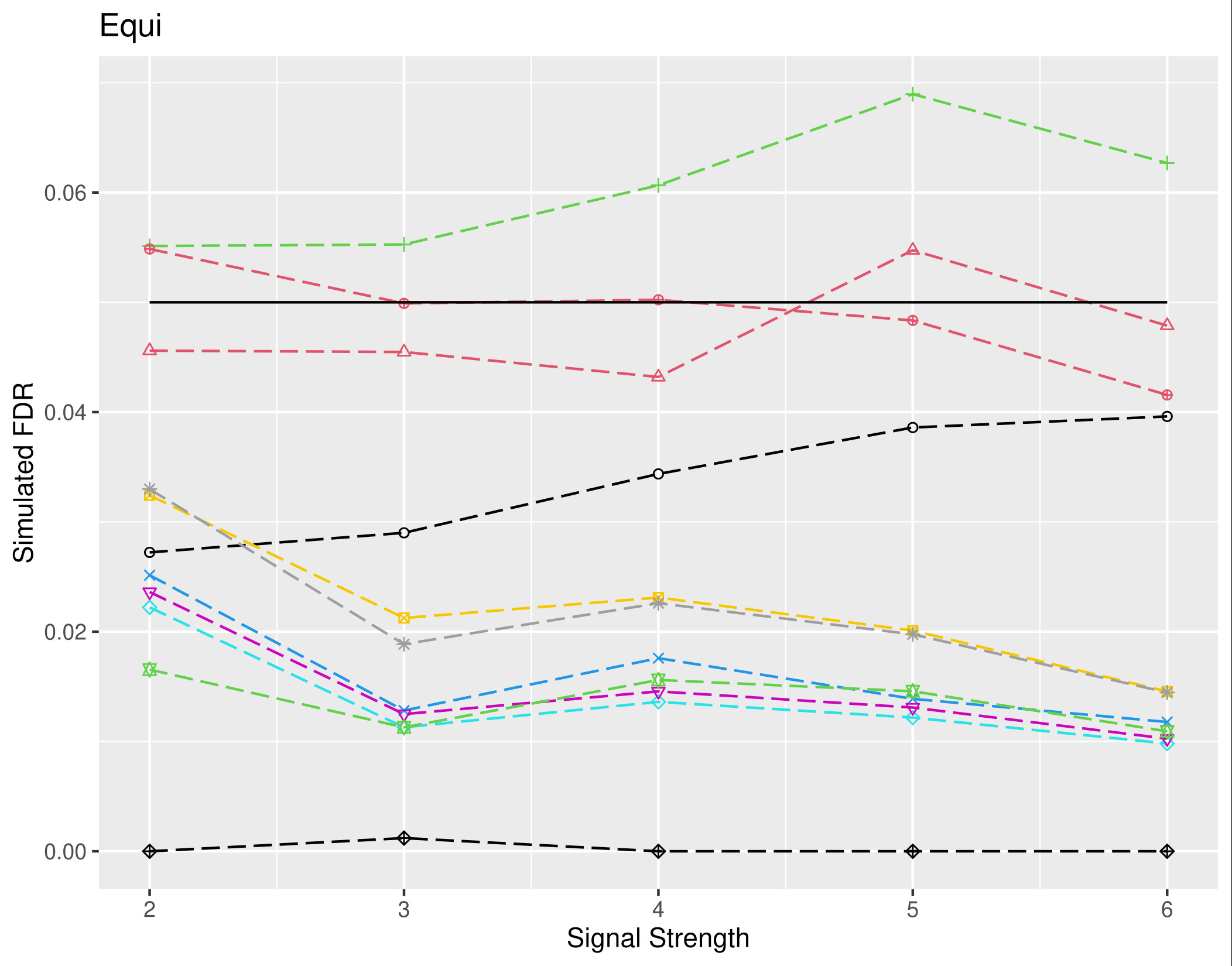} 
      \end{subfigure} \\
  \rotatebox{90}{AR(0.3)} 
  & \begin{subfigure}[b]{\linewidth}
      \centering
      \includegraphics[width=0.8\linewidth]{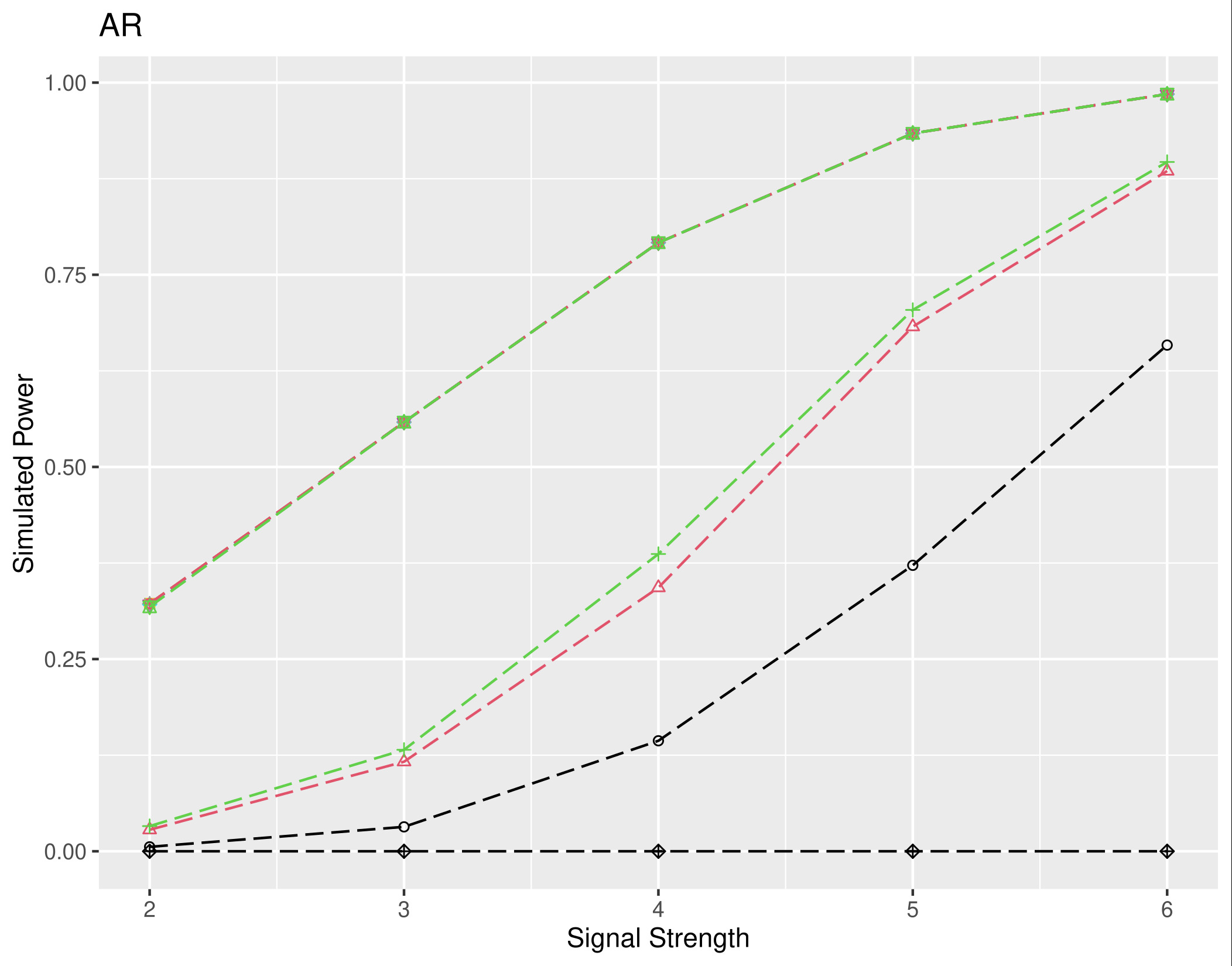} 
      \end{subfigure}  
  & \begin{subfigure}[b]{\linewidth}
      \centering
      \includegraphics[width=0.8\linewidth]{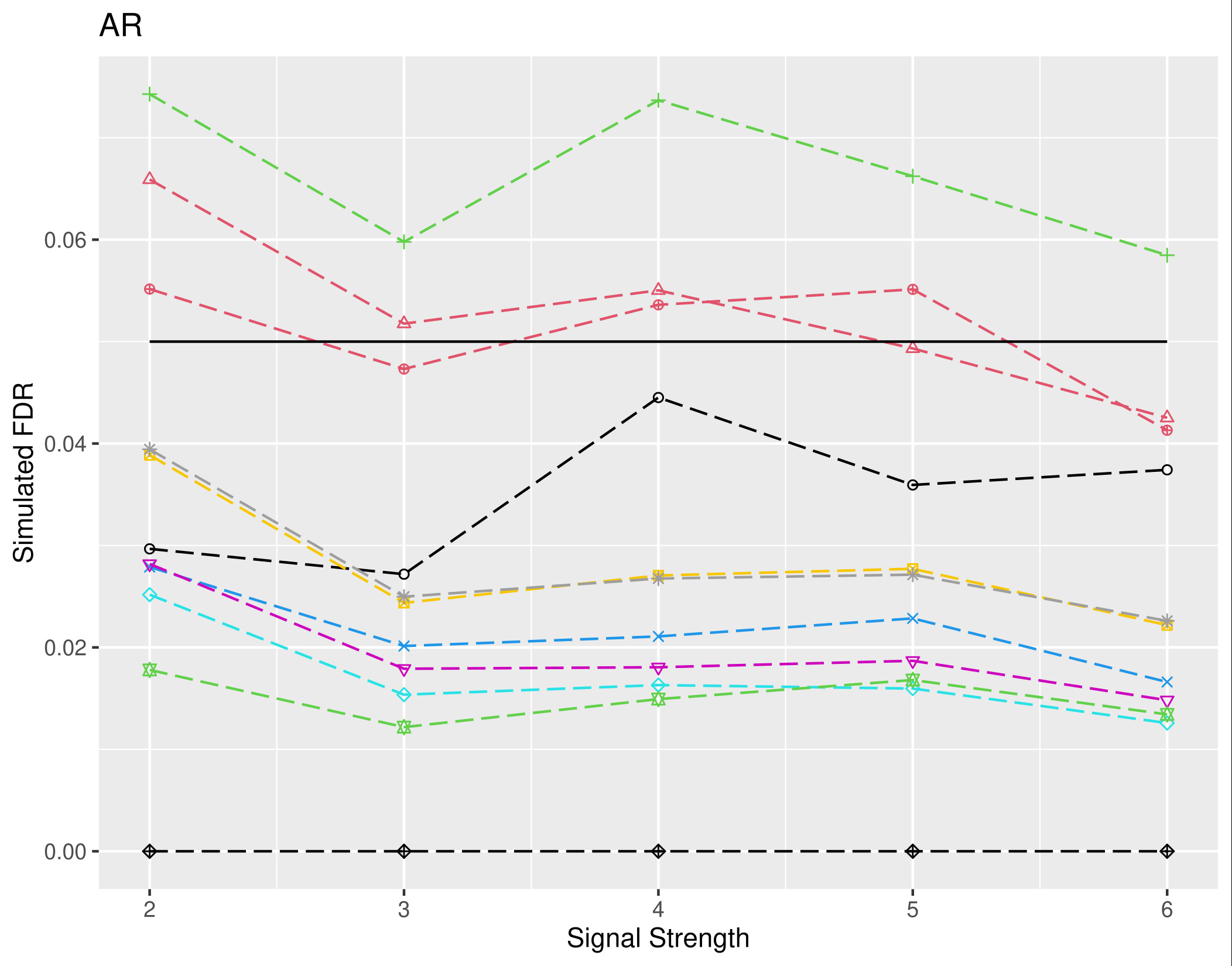} 
      \end{subfigure} \\
  \rotatebox{90}{IAR(0.3)} 
  & \begin{subfigure}[b]{\linewidth}
      \centering
      \includegraphics[width=0.8\linewidth]{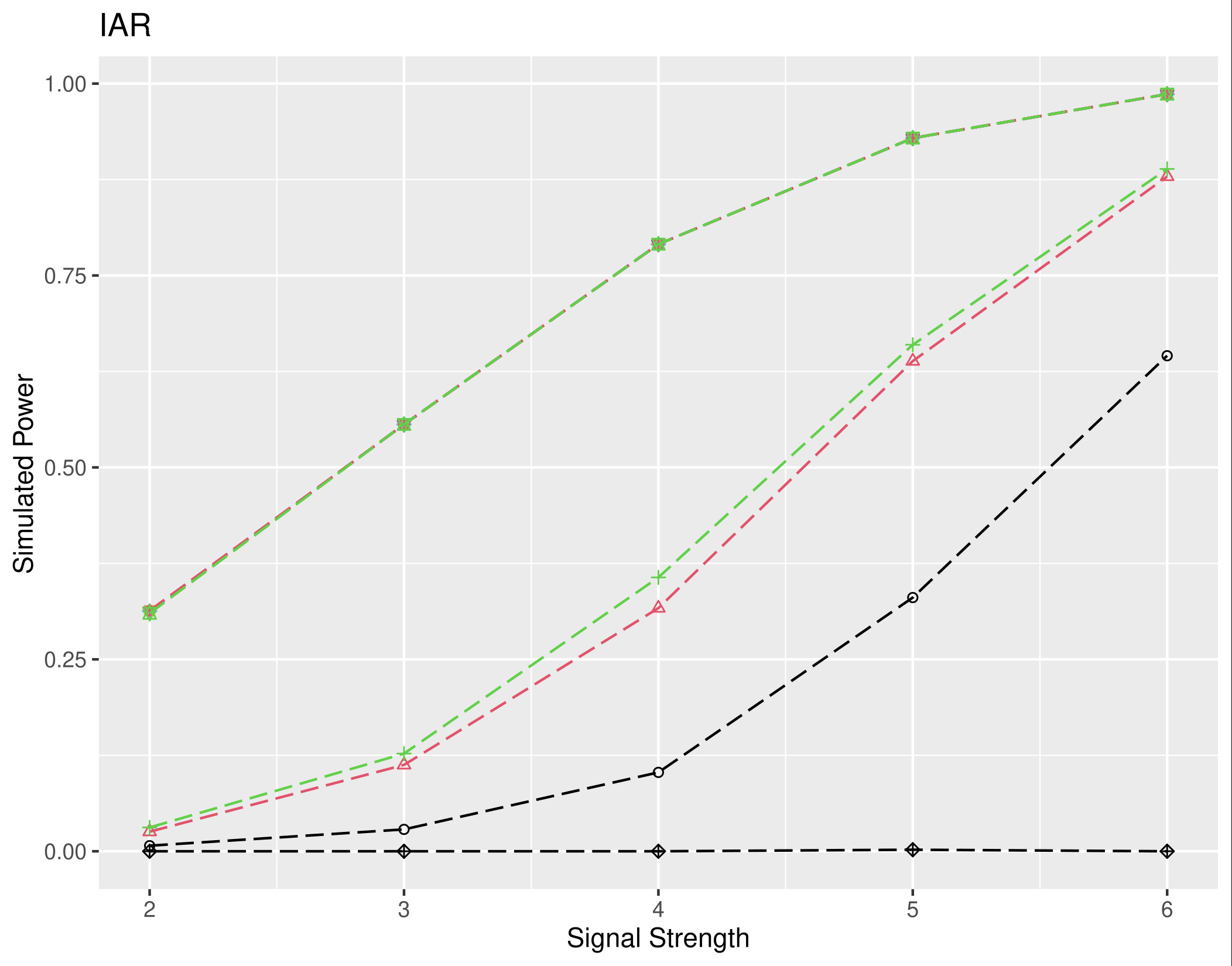} 
      \end{subfigure}  
  & \begin{subfigure}[b]{\linewidth}
      \centering
      \includegraphics[width=0.8\linewidth]{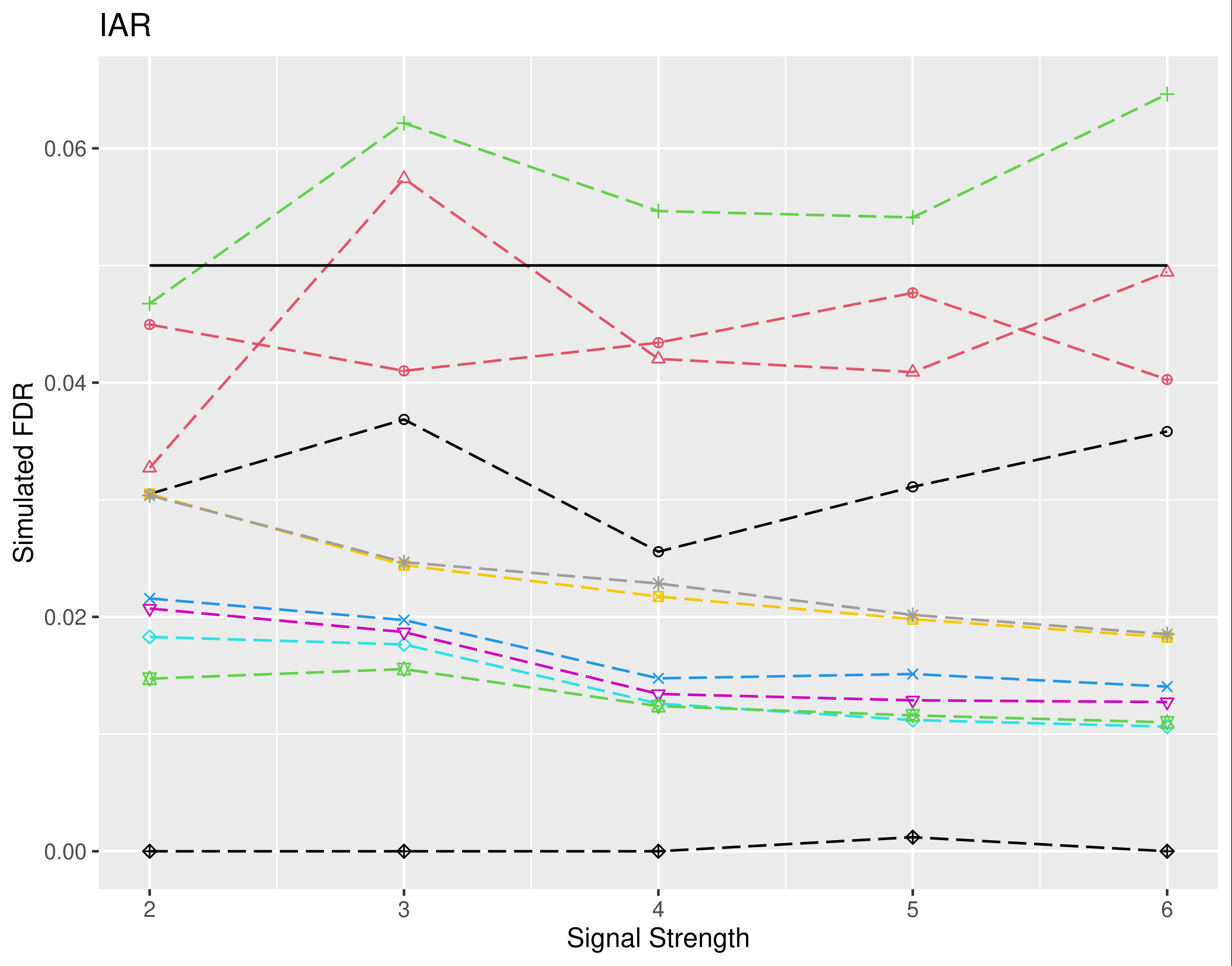} 
      \end{subfigure}    
  \end{tabular*} 
  \caption{Simulated Power (left column) and simulated FDR (right column) displayed for knockoff-assisted variable selection from $d=40$ parameters at a level of $\alpha=0.05$. Methods compared are BH method (Circle and black), BBH method (Triangle point up and red), ABBH method (Plus and green), SBBH1 (Cross and blue), SBBH2 (Diamond and light blue), SBBH3 (Triangle point down and purple), SBBH4 (Square cross and yellow), SBBH5 (Star and grey), Knockoff (Diamond plus and black), Rev-BBH (Circle plus and red) and BBY (Triangles up and down and green)}
  \label{knockoff:figure7} 
\end{figure}

\begin{figure}[hbt!]
  \begin{tabular*}{\textwidth}{
    @{}m{0.5cm}
    @{}m{\dimexpr0.50\textwidth-0.25cm\relax}
    @{}m{\dimexpr0.50\textwidth-0.25cm\relax}}
  \rotatebox{90}{Equi(0.7)}
  & \begin{subfigure}[b]{\linewidth}
      \centering
      \includegraphics[width=0.8\linewidth]{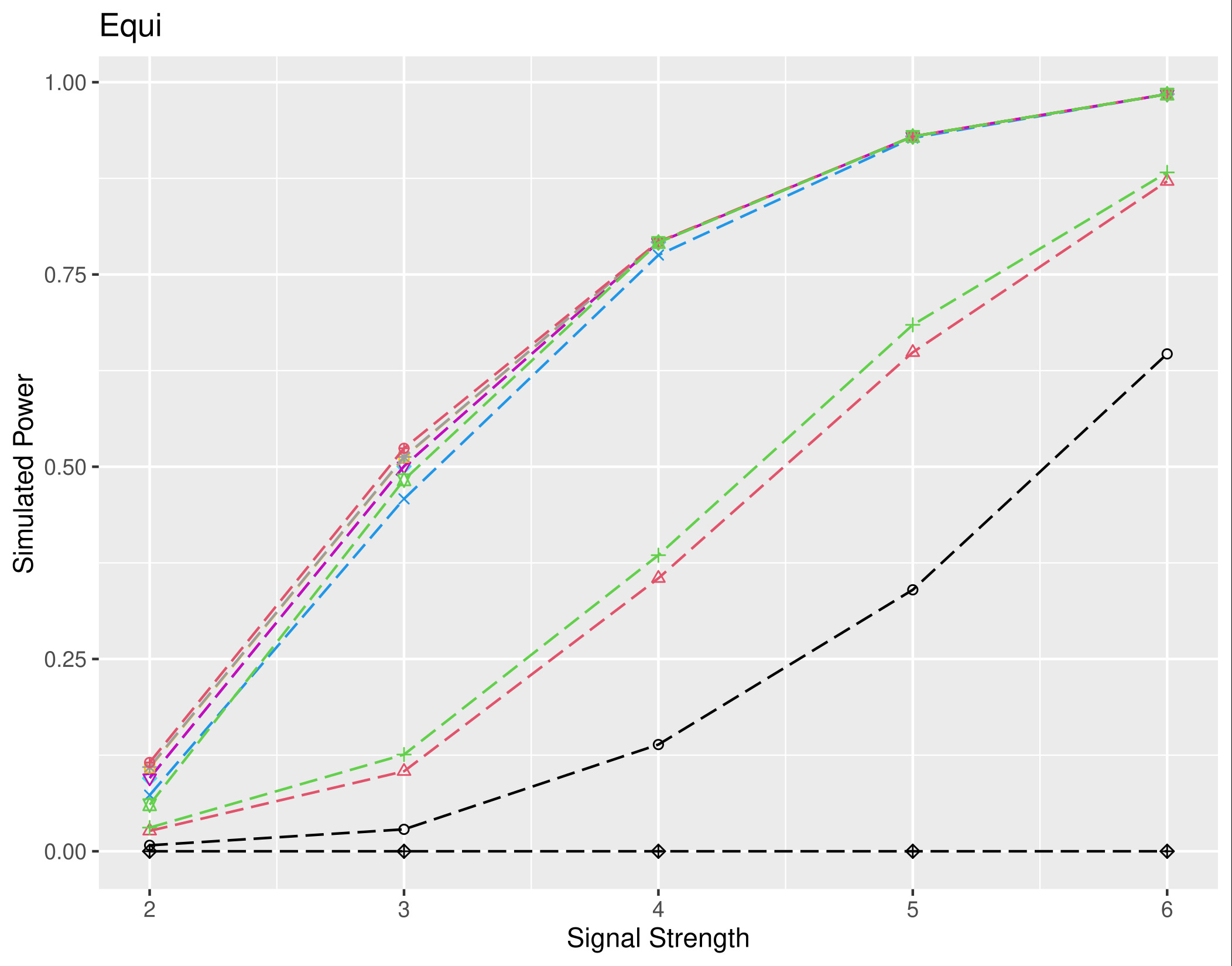} 
      \end{subfigure}  
  & \begin{subfigure}[b]{\linewidth}
      \centering
      \includegraphics[width=0.8\linewidth]{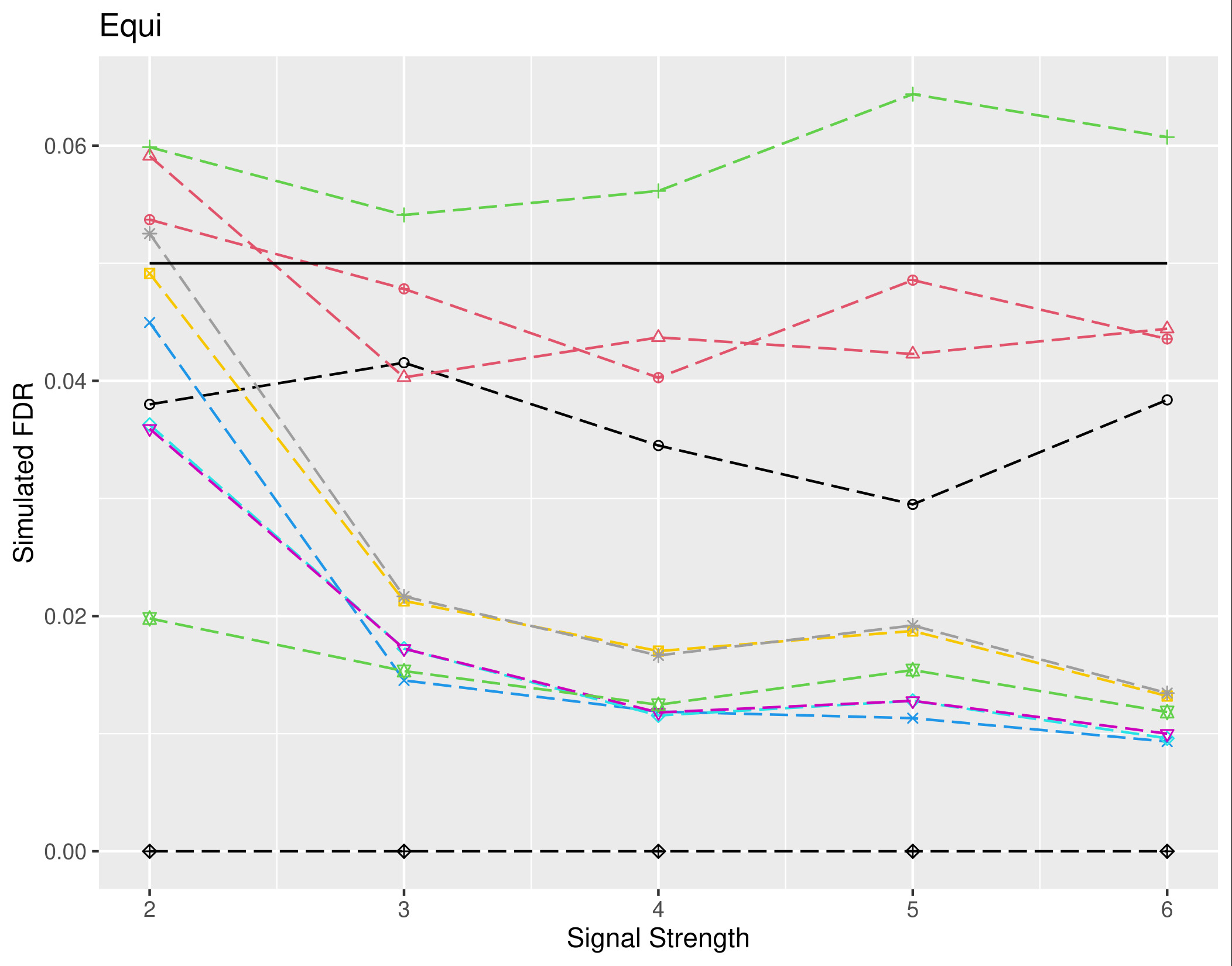} 
      \end{subfigure} \\
  \rotatebox{90}{AR(0.7)} 
  & \begin{subfigure}[b]{\linewidth}
      \centering
      \includegraphics[width=0.8\linewidth]{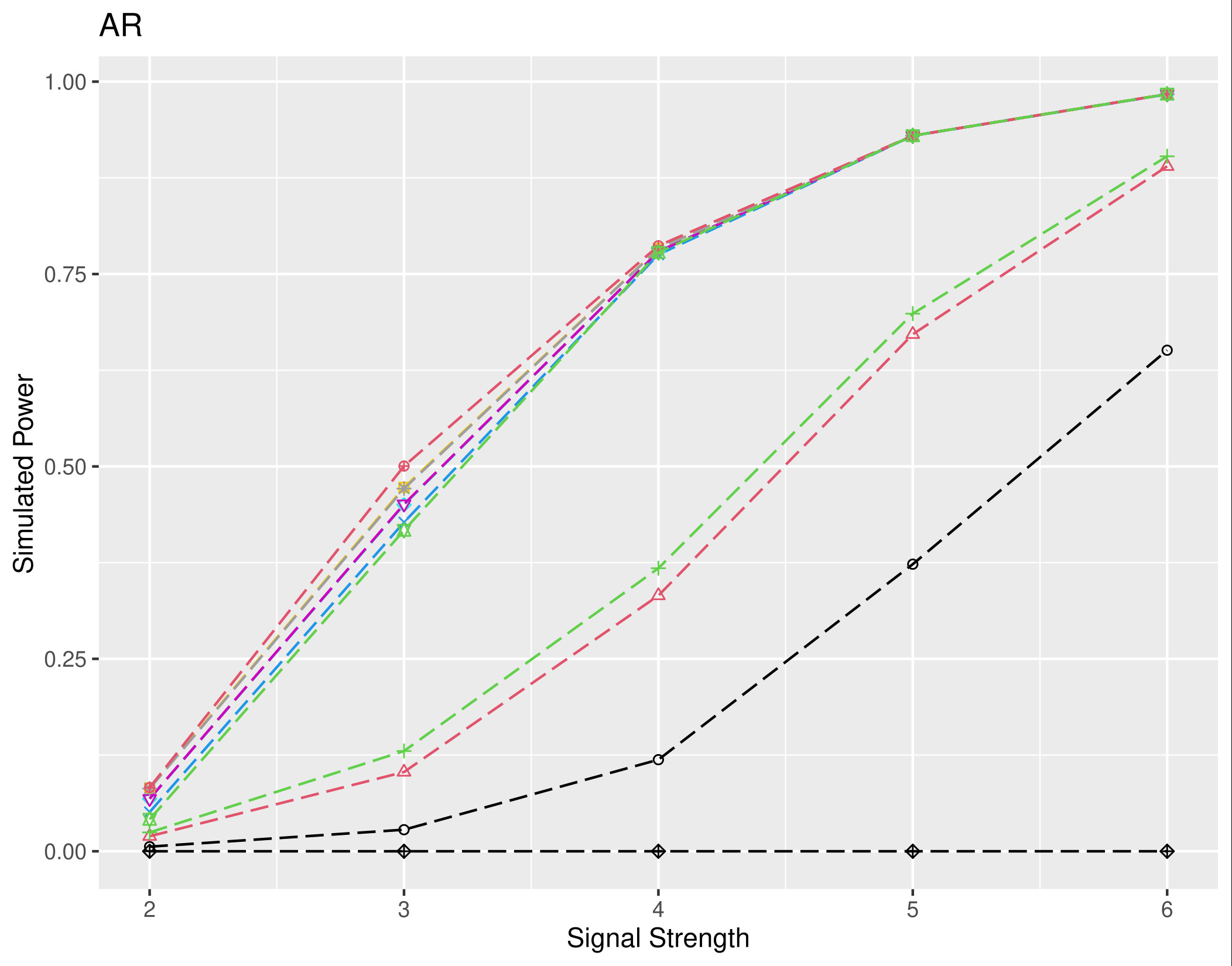} 
      \end{subfigure}  
  & \begin{subfigure}[b]{\linewidth}
      \centering
      \includegraphics[width=0.8\linewidth]{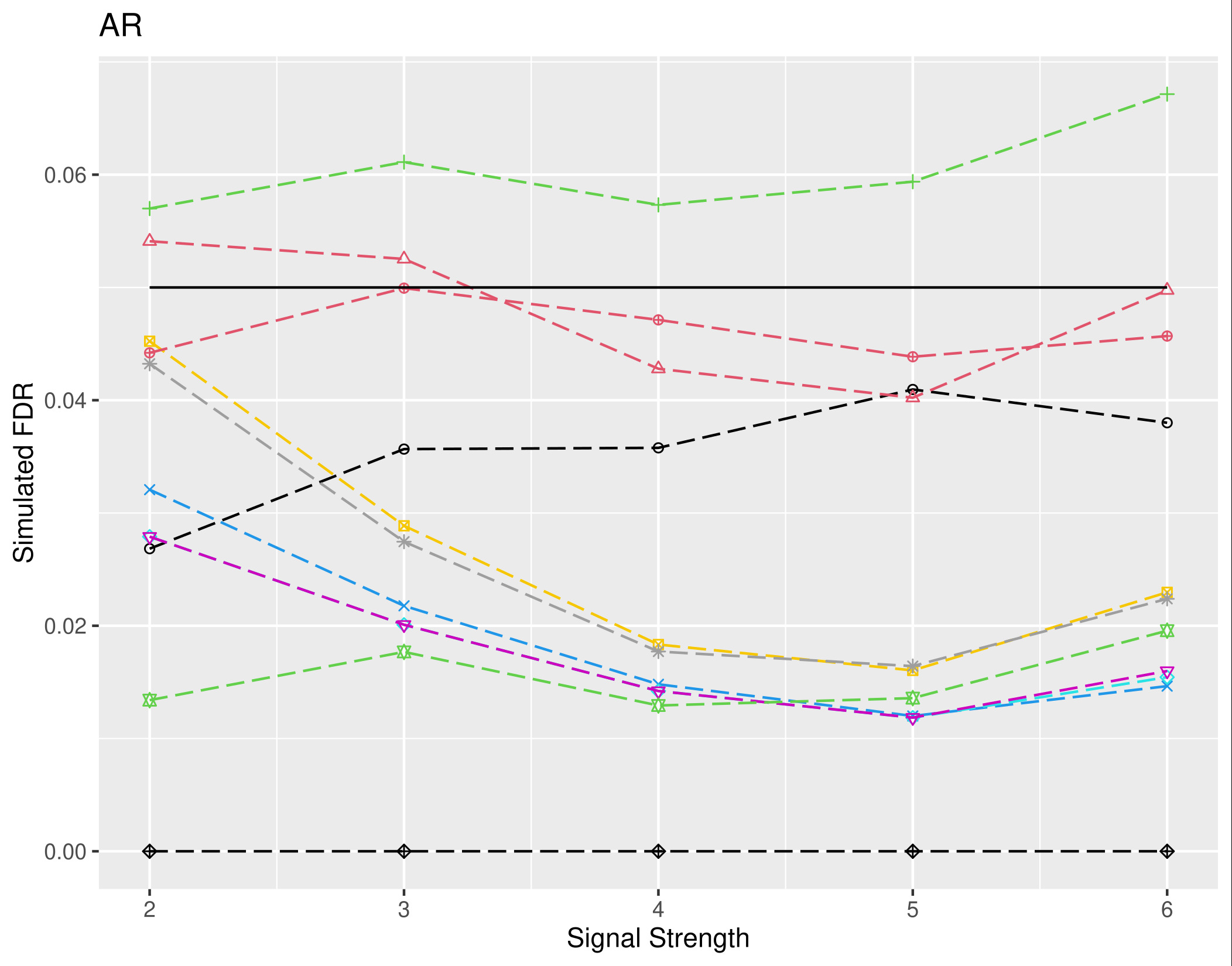} 
      \end{subfigure} \\
  \rotatebox{90}{IAR(0.7)} 
  & \begin{subfigure}[b]{\linewidth}
      \centering
      \includegraphics[width=0.8\linewidth]{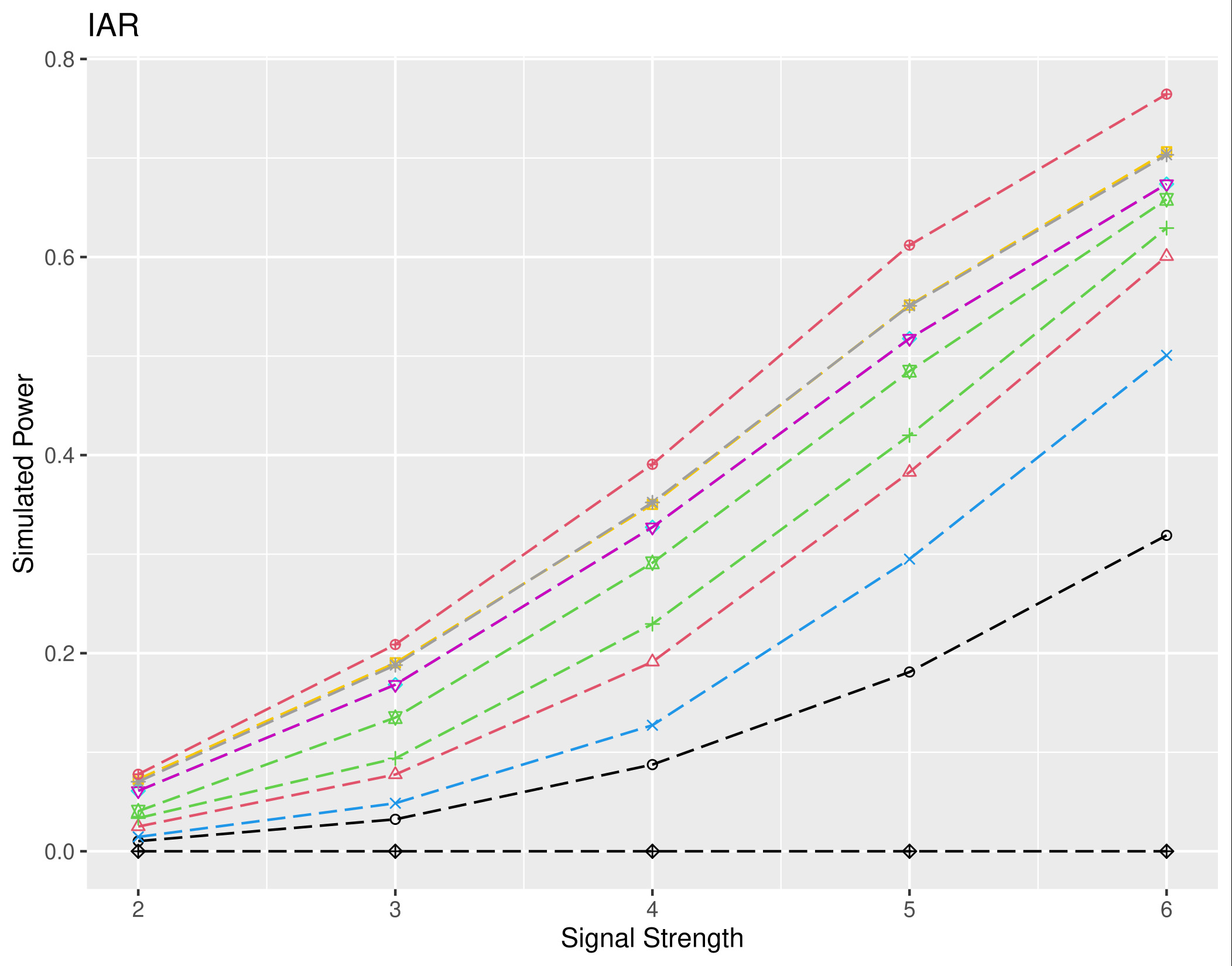} 
      \end{subfigure}  
  & \begin{subfigure}[b]{\linewidth}
      \centering
      \includegraphics[width=0.8\linewidth]{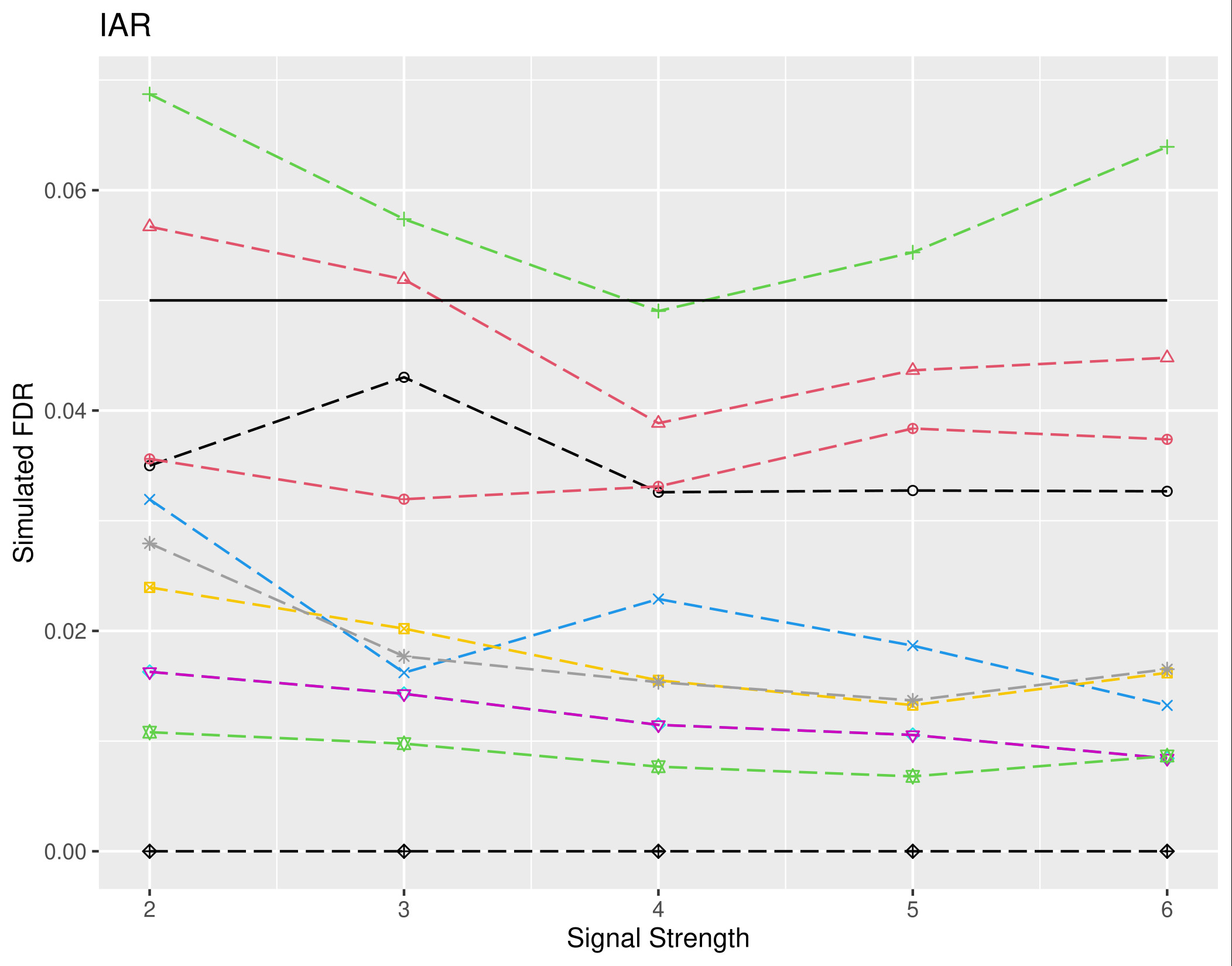} 
      \end{subfigure}    
  \end{tabular*} 
  \caption{Simulated Power (left column) and simulated FDR (right column) displayed for knockoff-assisted variable selection from $d=40$ parameters at a level of $\alpha=0.05$. Methods compared are BH method (Circle and black), BBH method (Triangle point up and red), ABBH method (Plus and green), SBBH1 (Cross and blue), SBBH2 (Diamond and light blue), SBBH3 (Triangle point down and purple), SBBH4 (Square cross and yellow), SBBH5 (Star and grey), Knockoff (Diamond plus and black), Rev-BBH (Circle plus and red) and BBY (Triangles up and down and green)}
  \label{knockoff:figure8} 
\end{figure}

\begin{figure}[hbt!]
  \begin{tabular*}{\textwidth}{
    @{}m{0.5cm}
    @{}m{\dimexpr0.50\textwidth-0.25cm\relax}
    @{}m{\dimexpr0.50\textwidth-0.25cm\relax}}
  \rotatebox{90}{Block Diagonal}
  & \begin{subfigure}[b]{\linewidth}
      \centering
      \includegraphics[width=0.8\linewidth]{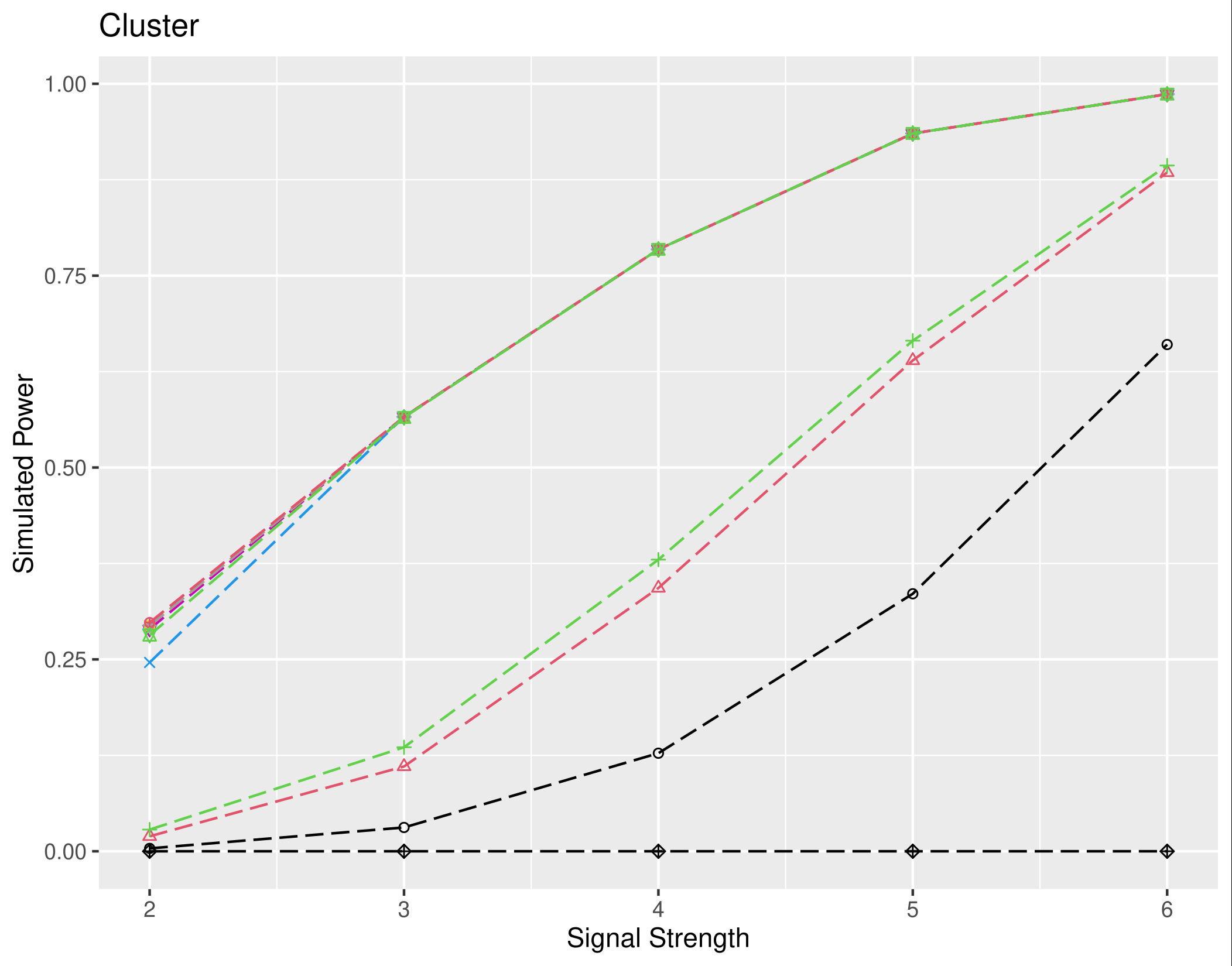} 
      \end{subfigure}  
  & \begin{subfigure}[b]{\linewidth}
      \centering
      \includegraphics[width=0.8\linewidth]{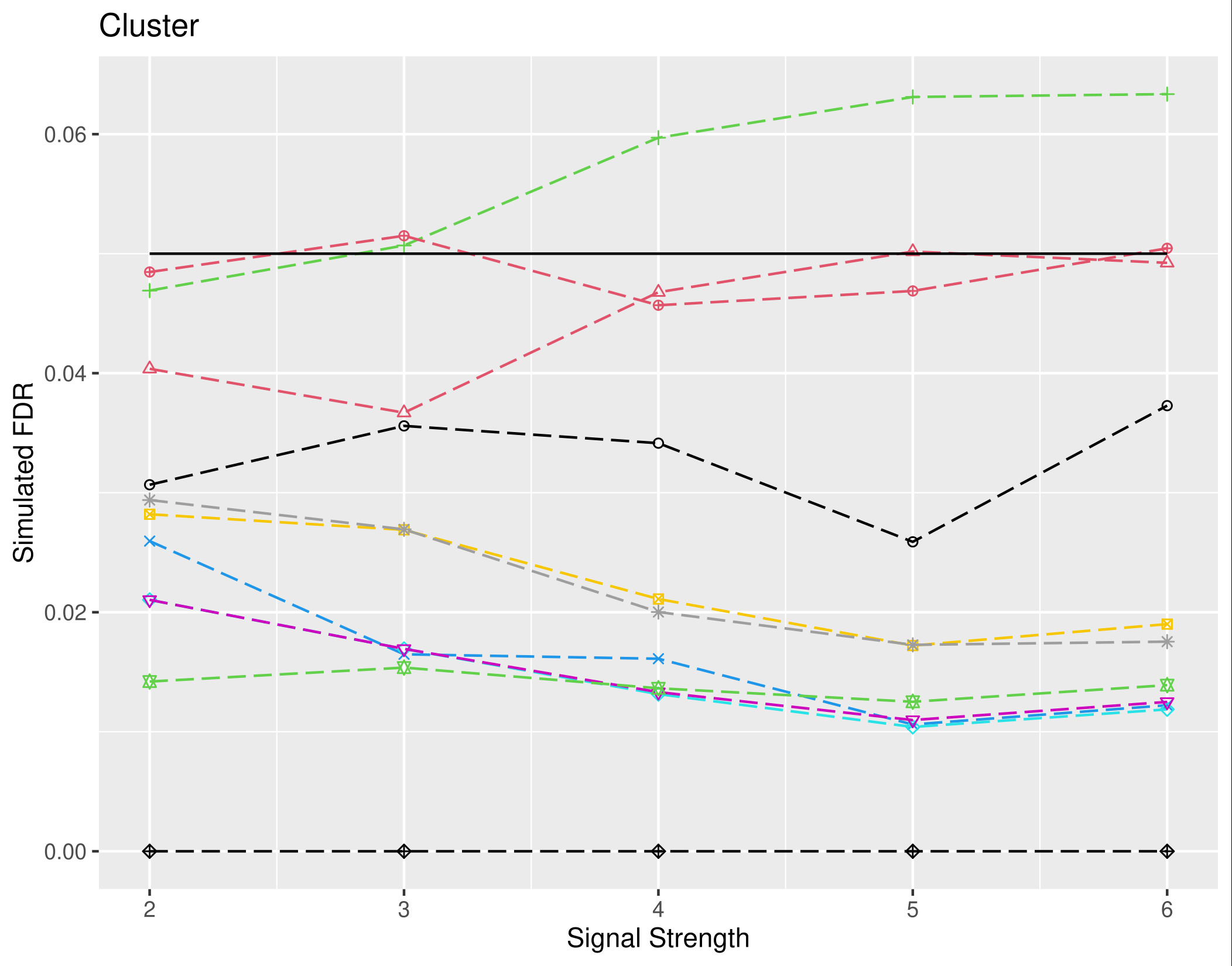} 
      \end{subfigure} \\
  \rotatebox{90}{Sparse} 
  & \begin{subfigure}[b]{\linewidth}
      \centering
      \includegraphics[width=0.8\linewidth]{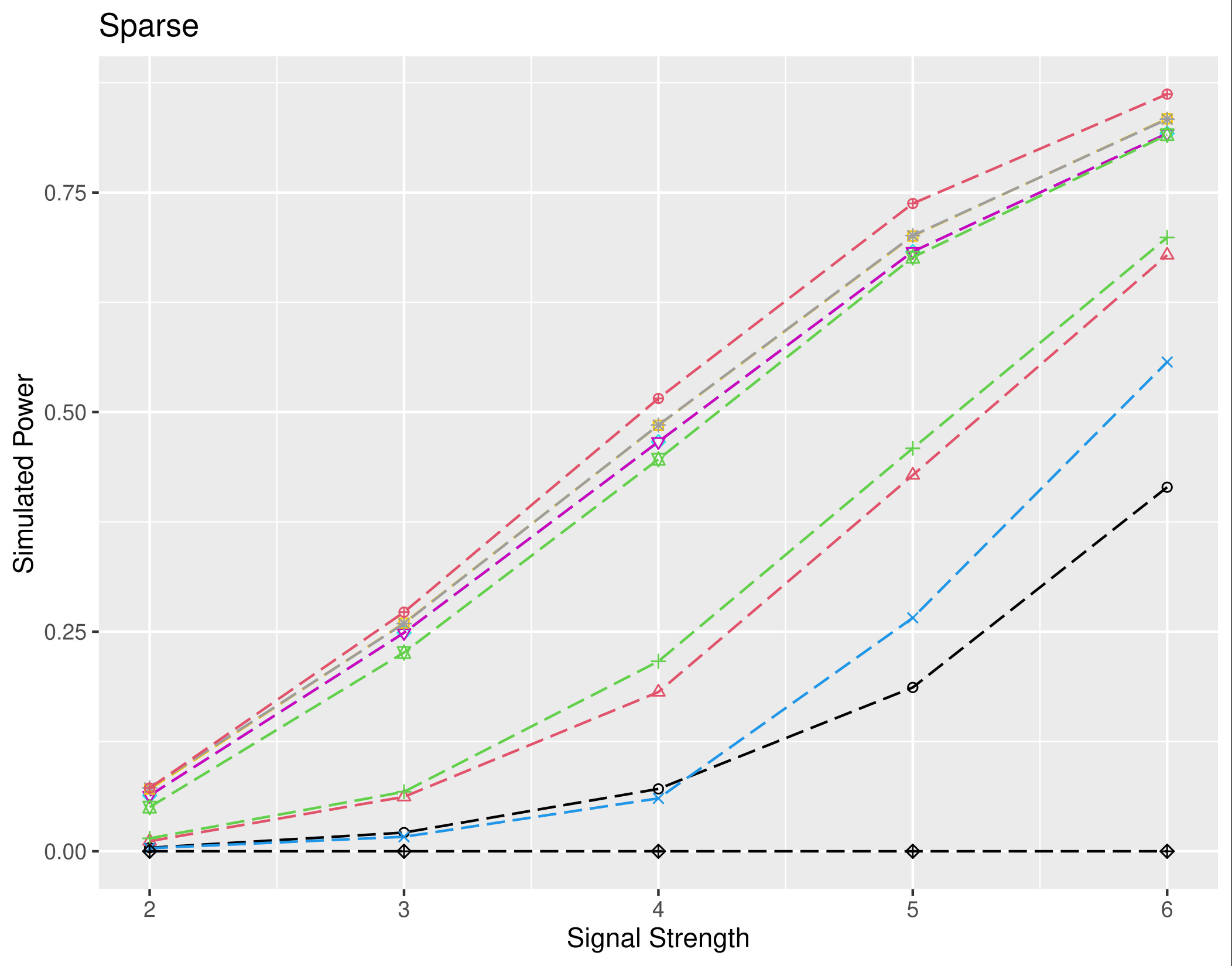} 
      \end{subfigure}  
  & \begin{subfigure}[b]{\linewidth}
      \centering
      \includegraphics[width=0.8\linewidth]{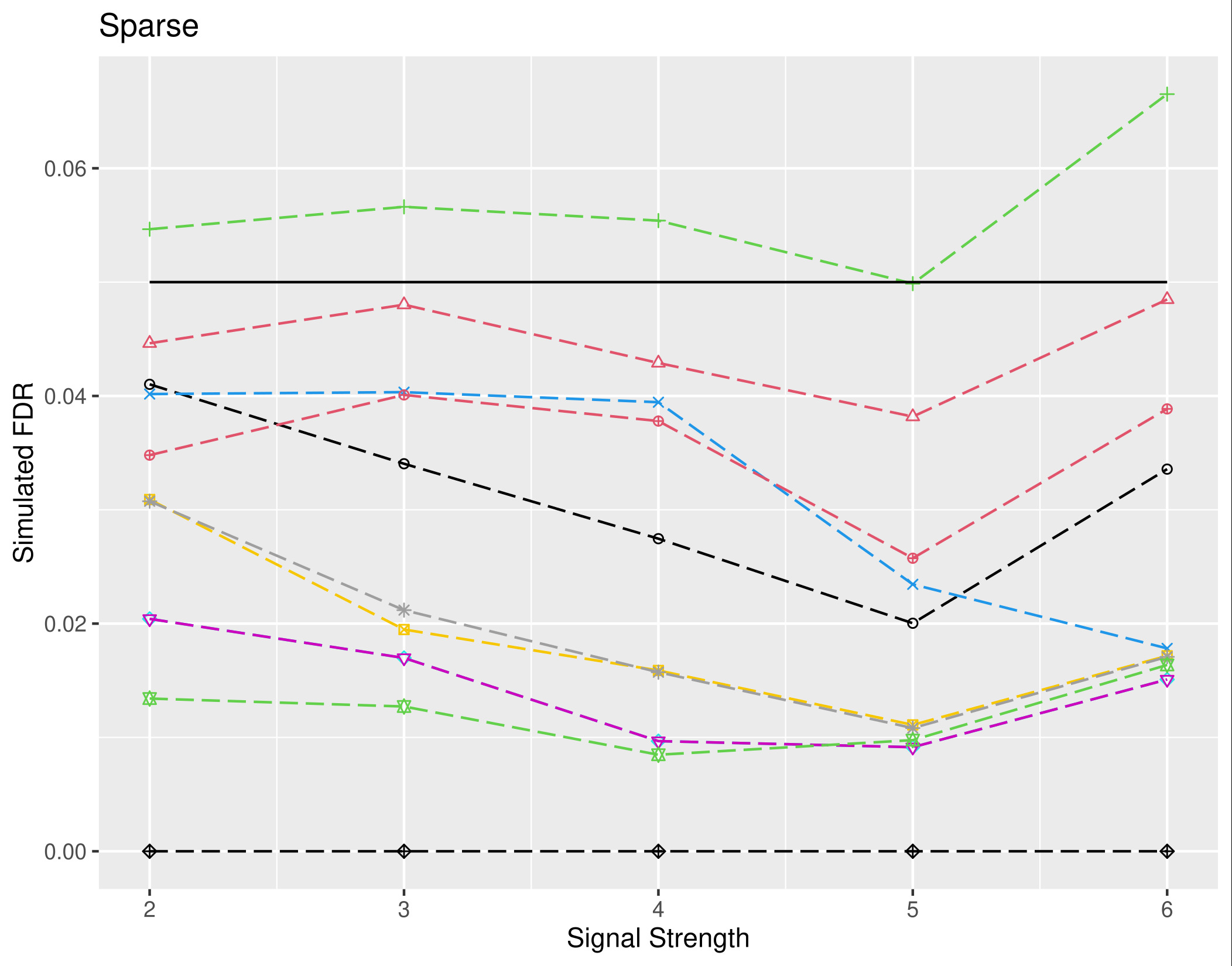} 
      \end{subfigure}   
  \end{tabular*} 
  \caption{Simulated Power (left column) and simulated FDR (right column) displayed for knockoff-assisted variable selection from $d=40$ parameters at a level of $\alpha=0.05$. Methods compared are BH method (Circle and black), BBH method (Triangle point up and red), ABBH method (Plus and green), SBBH1 (Cross and blue), SBBH2 (Diamond and light blue), SBBH3 (Triangle point down and purple), SBBH4 (Square cross and yellow), SBBH5 (Star and grey), Knockoff (Diamond plus and black), Rev-BBH (Circle plus and red) and BBY (Triangles up and down and green)}
  \label{knockoff:figure9} 
\end{figure}

\begin{figure}
    \begin{tabular*}{\textwidth}{
    @{}m{0.5cm}
    @{}m{\dimexpr0.33\textwidth-0.25cm\relax}
    @{}m{\dimexpr0.33\textwidth-0.25cm\relax}
    @{}m{\dimexpr0.33\textwidth-0.25cm\relax}}
  & \begin{subfigure}[b]{\linewidth}
      \centering
      \includegraphics[width=\linewidth]{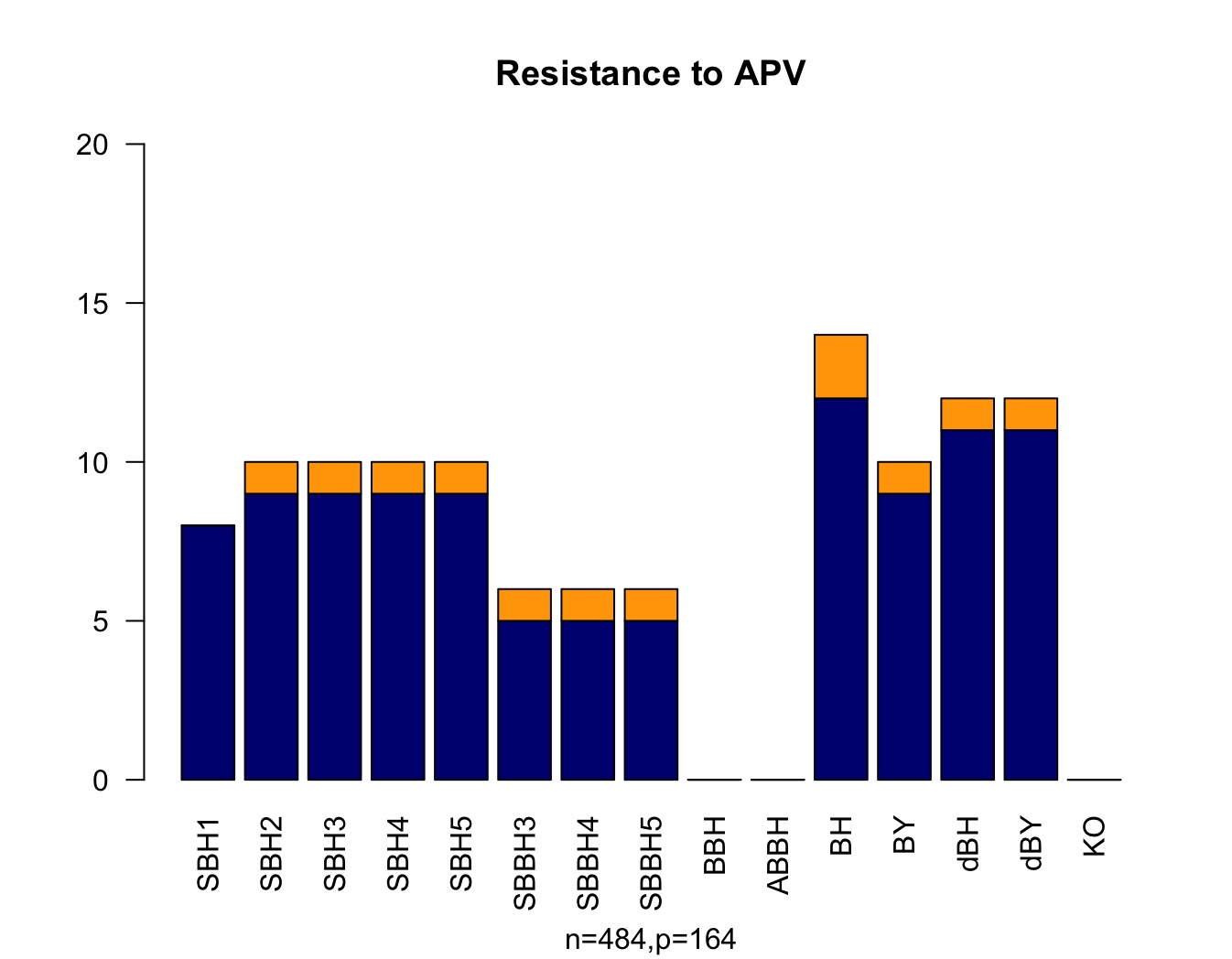} 
      \end{subfigure}  
  & \begin{subfigure}[b]{\linewidth}
      \centering
      \includegraphics[width=\linewidth]{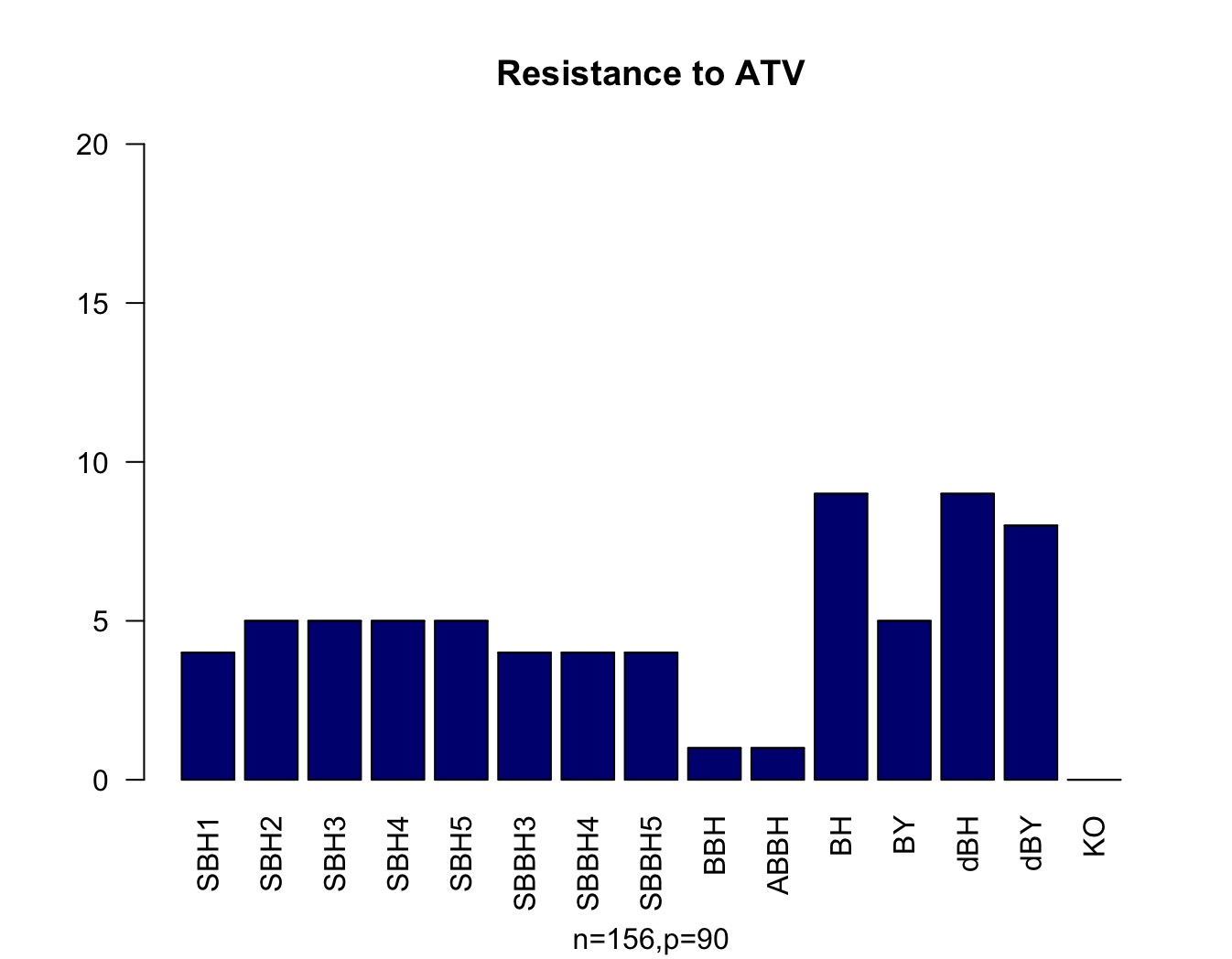} 
      \end{subfigure} 
  & \begin{subfigure}[b]{\linewidth}
      \centering
      \includegraphics[width=\linewidth]{draft_1_figs/real/0.05,idv.jpeg} 
      \end{subfigure} \\
  & \begin{subfigure}[b]{\linewidth}
      \centering
      \includegraphics[width=\linewidth]{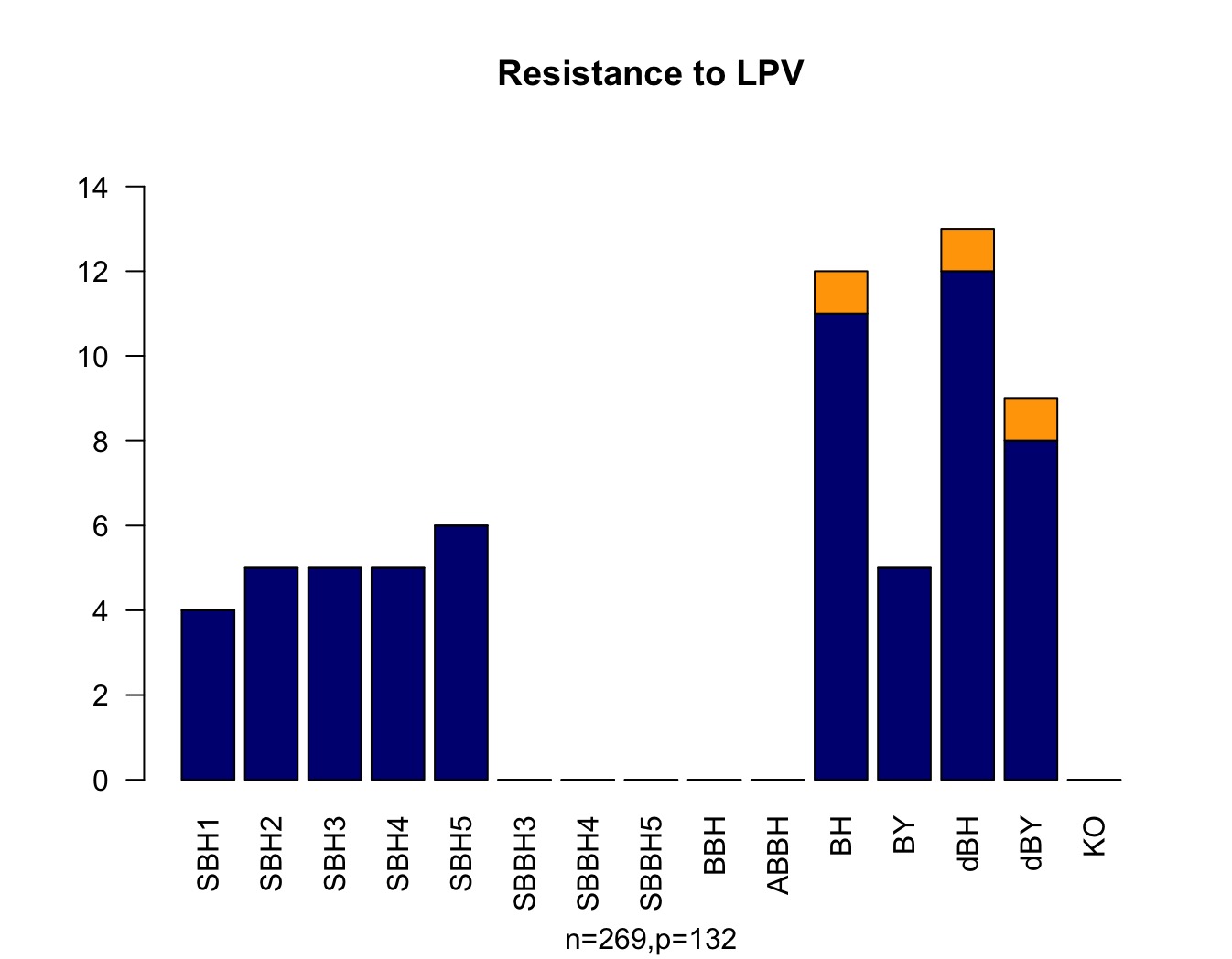} 
      \end{subfigure}  
  & \begin{subfigure}[b]{\linewidth}
      \centering
      \includegraphics[width=\linewidth]{draft_1_figs/real/0.05,nfv.jpeg} 
      \end{subfigure} 
  & \begin{subfigure}[b]{\linewidth}
      \centering
      \includegraphics[width=\linewidth]{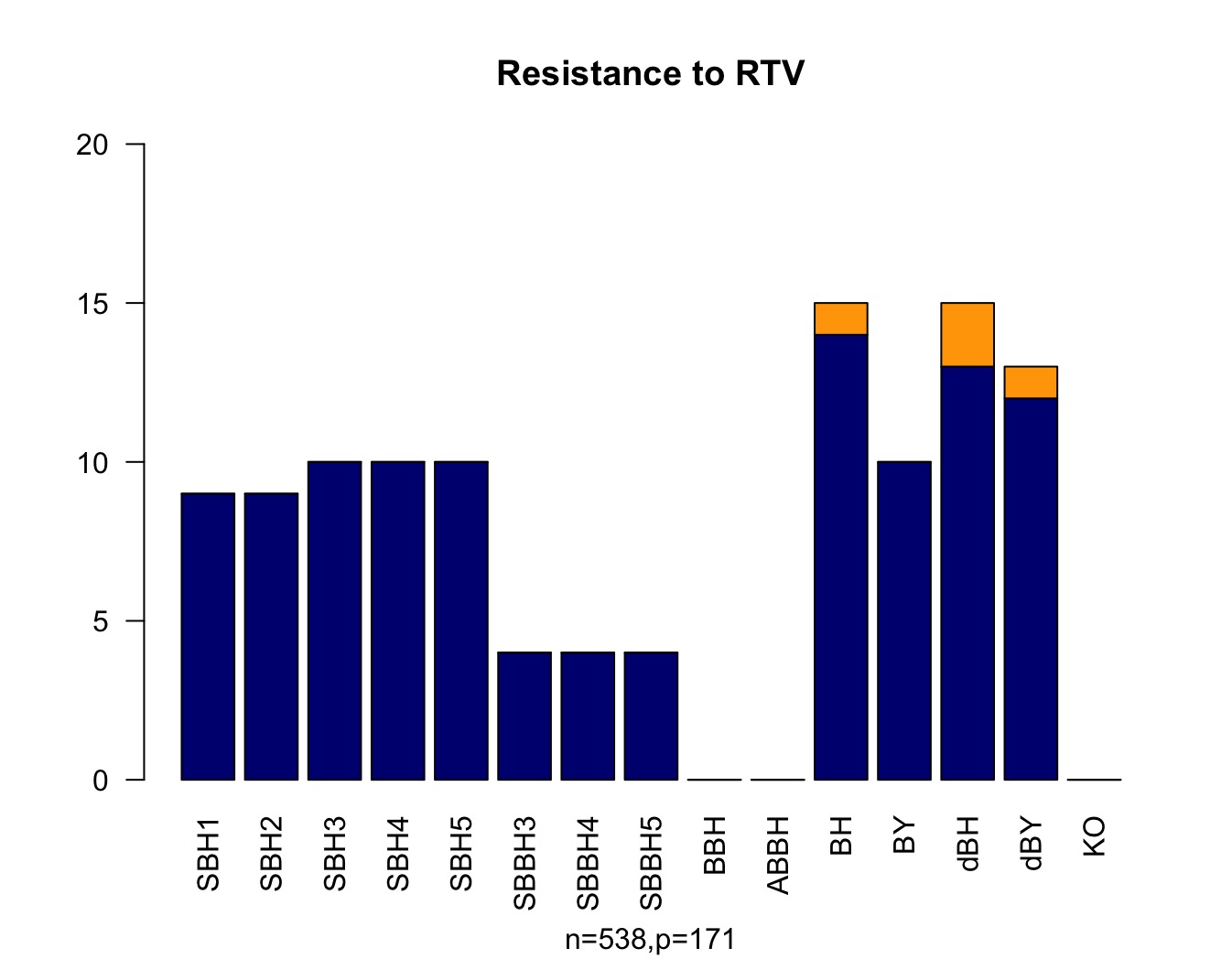} 
      \end{subfigure} \\
  & 
  & \begin{subfigure}[b]{\linewidth}
      \centering
      \includegraphics[width=\linewidth]{draft_1_figs/real/0.05,sqv.jpeg} 
      \end{subfigure} 
  &   
  \end{tabular*} 
    \caption{Dark blue indicates protease positions that appear in the treatment-selected mutation (TSM) panel for the PI class of treatments, while orange indicates positions selected by the method that do not appear in the TSM list. FDR was controlled at $\alpha=0.05$}
    \label{real:figure2}
\end{figure}

\begin{figure}
    \begin{tabular*}{\textwidth}{
    @{}m{0.5cm}
    @{}m{\dimexpr0.33\textwidth-0.25cm\relax}
    @{}m{\dimexpr0.33\textwidth-0.25cm\relax}
    @{}m{\dimexpr0.33\textwidth-0.25cm\relax}}
  & \begin{subfigure}[b]{\linewidth}
      \centering
      \includegraphics[width=\linewidth]{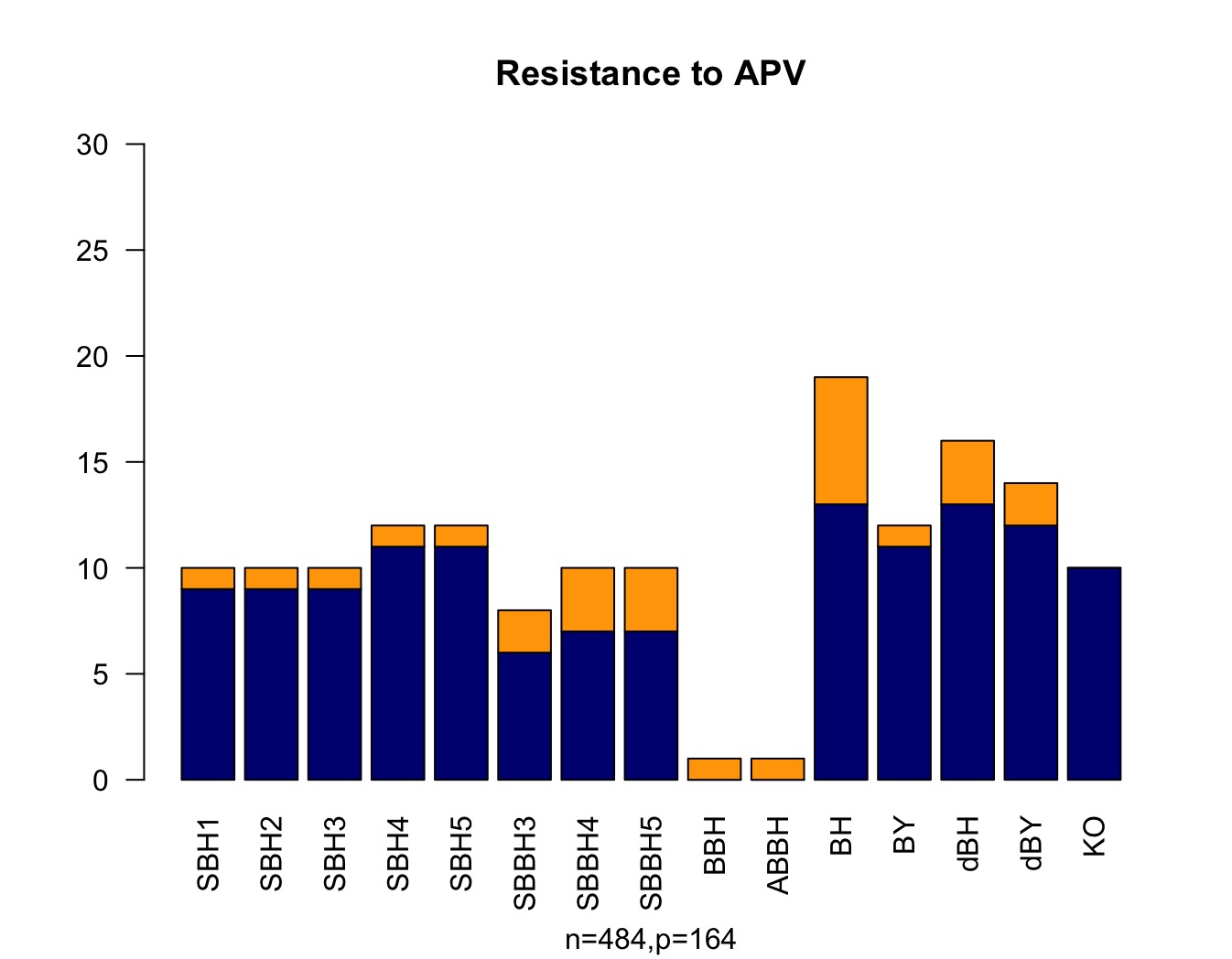} 
      \end{subfigure}  
  & \begin{subfigure}[b]{\linewidth}
      \centering
      \includegraphics[width=\linewidth]{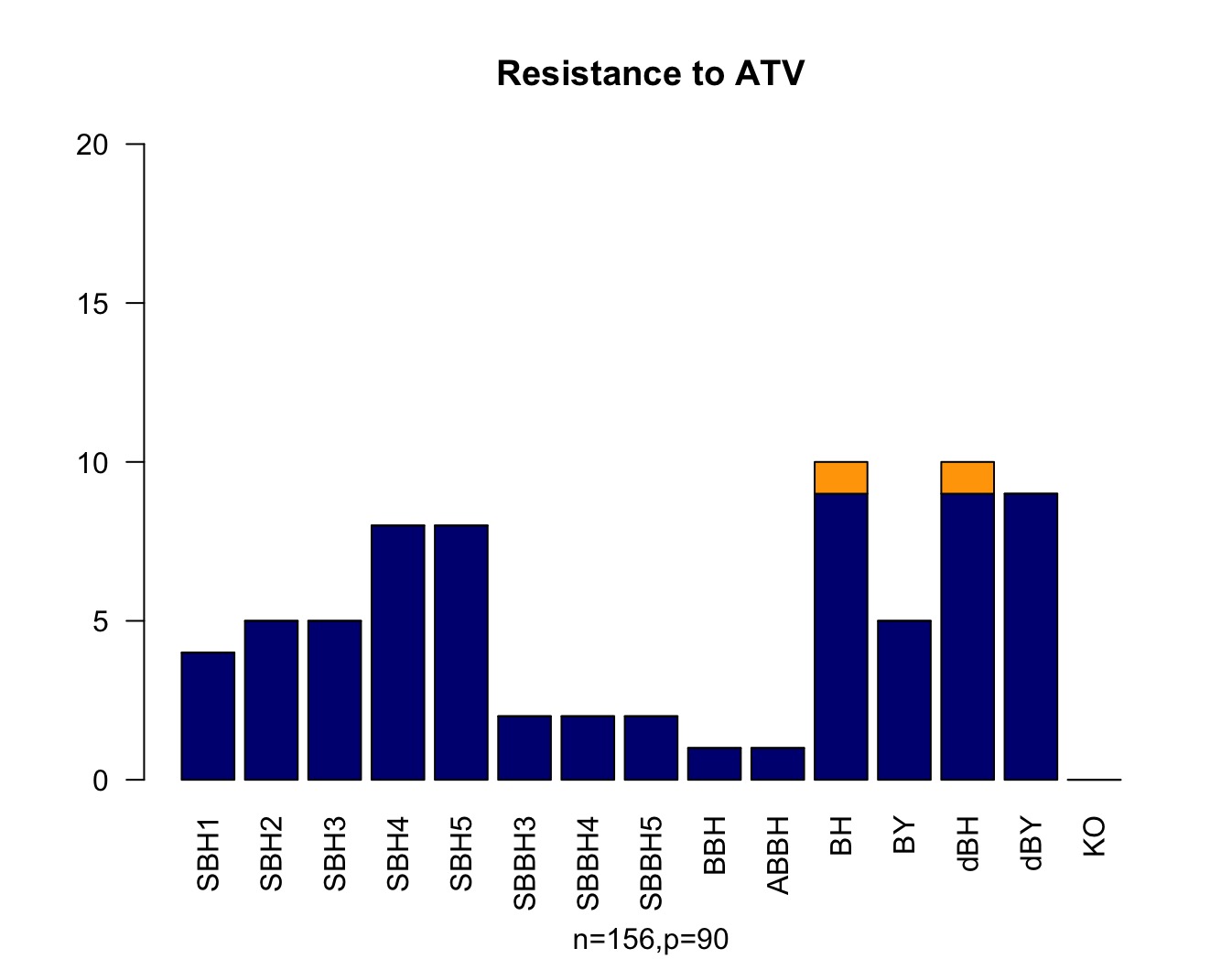} 
      \end{subfigure} 
  & \begin{subfigure}[b]{\linewidth}
      \centering
      \includegraphics[width=\linewidth]{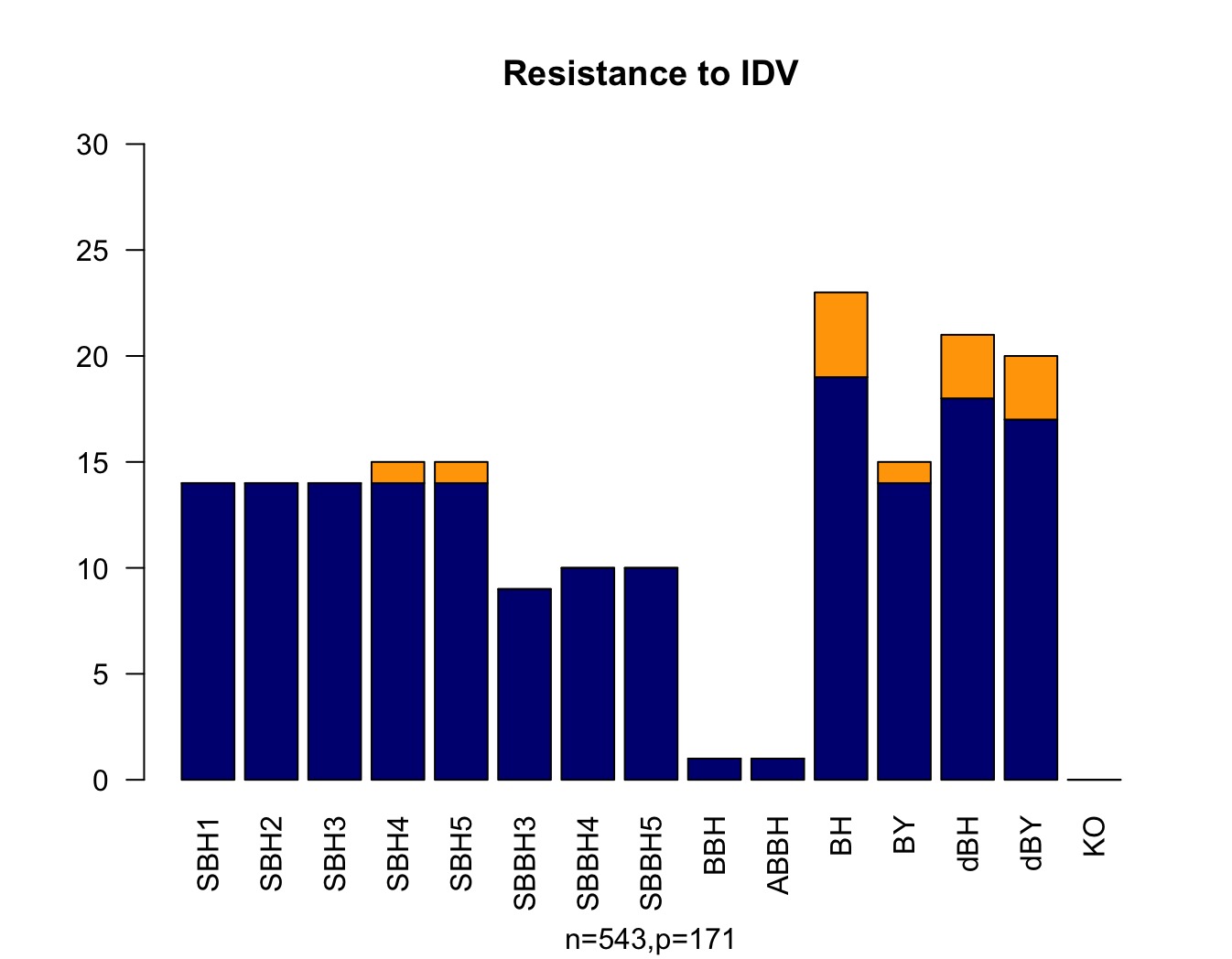} 
      \end{subfigure} \\
  & \begin{subfigure}[b]{\linewidth}
      \centering
      \includegraphics[width=\linewidth]{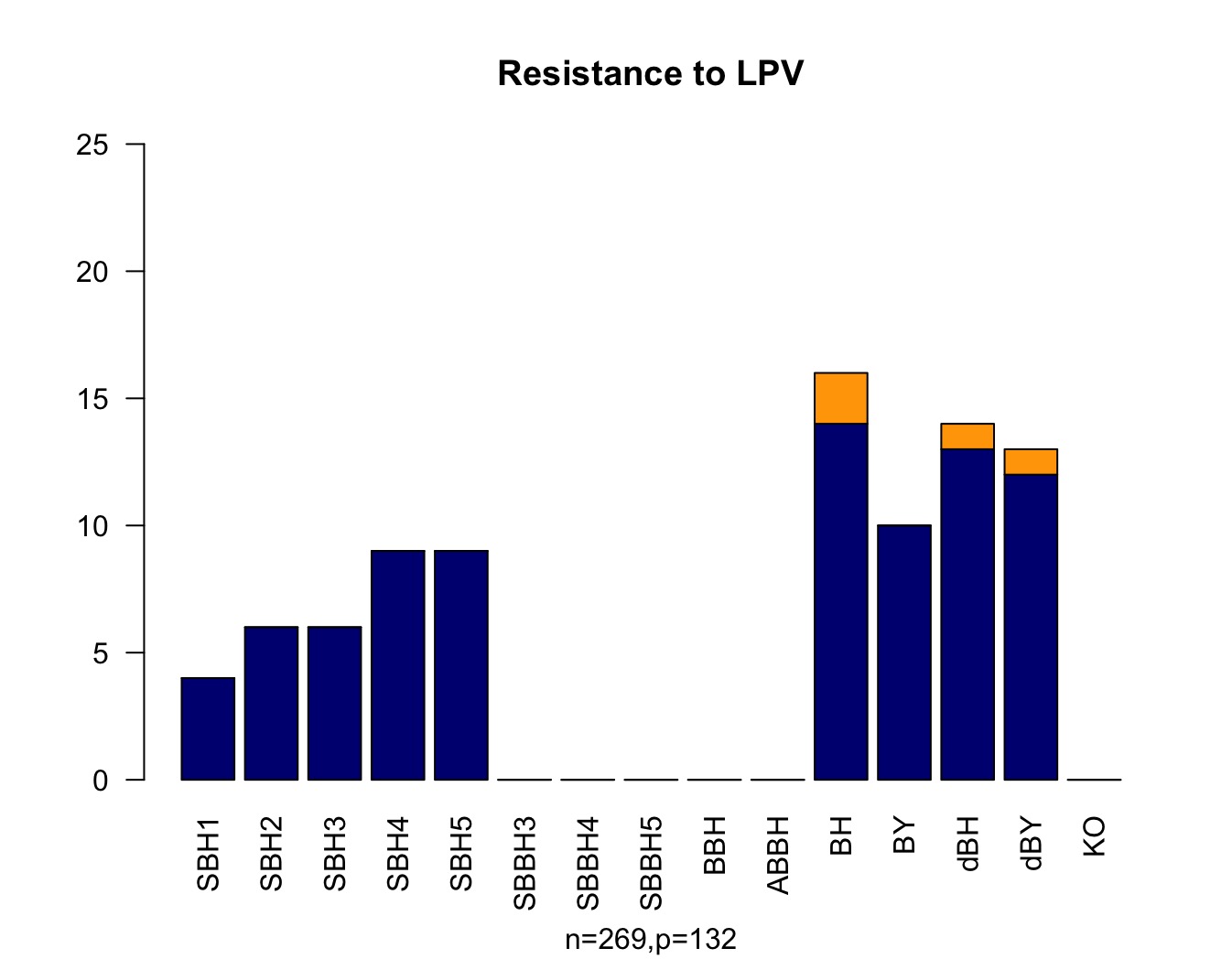} 
      \end{subfigure}  
  & \begin{subfigure}[b]{\linewidth}
      \centering
      \includegraphics[width=\linewidth]{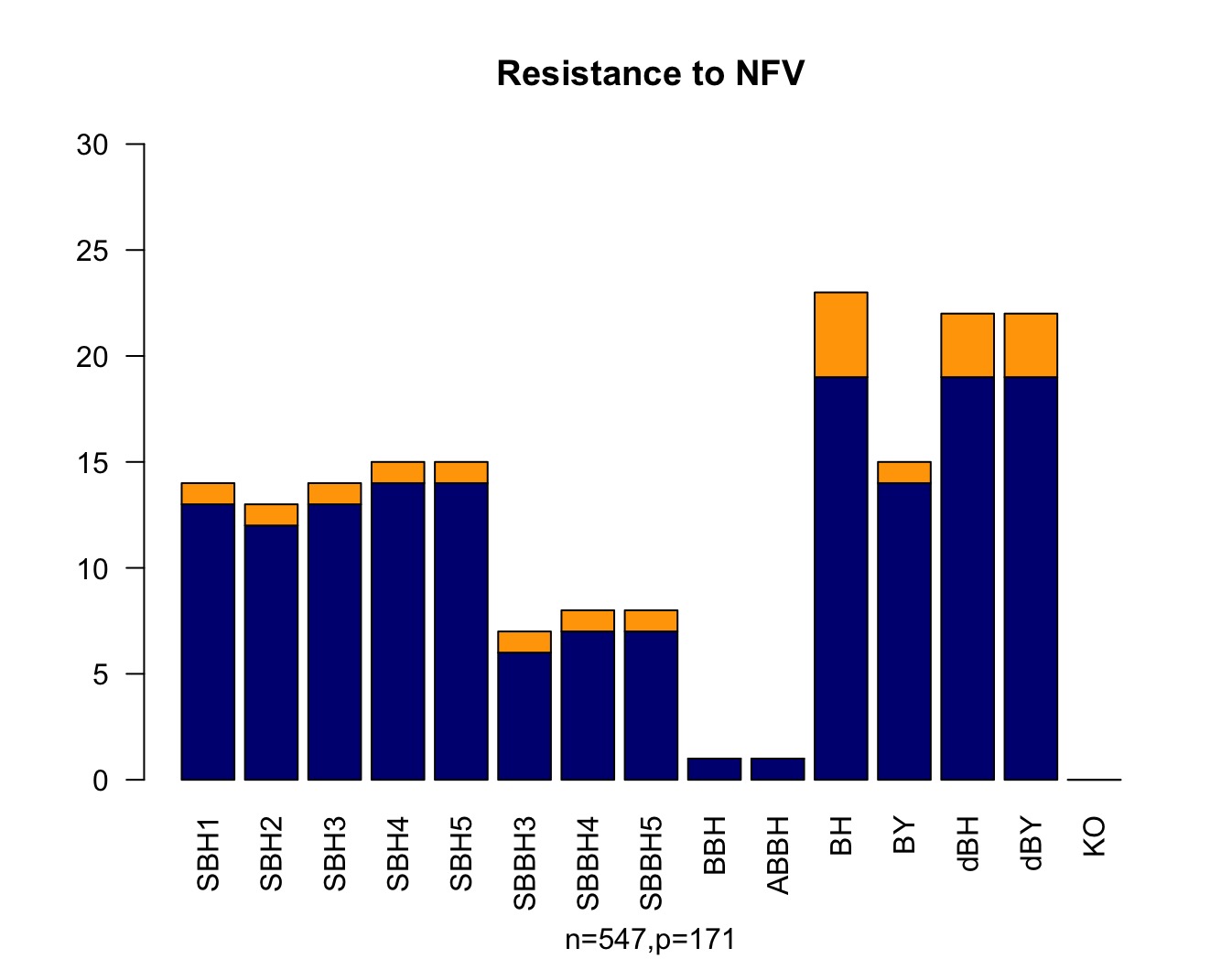} 
      \end{subfigure} 
  & \begin{subfigure}[b]{\linewidth}
      \centering
      \includegraphics[width=\linewidth]{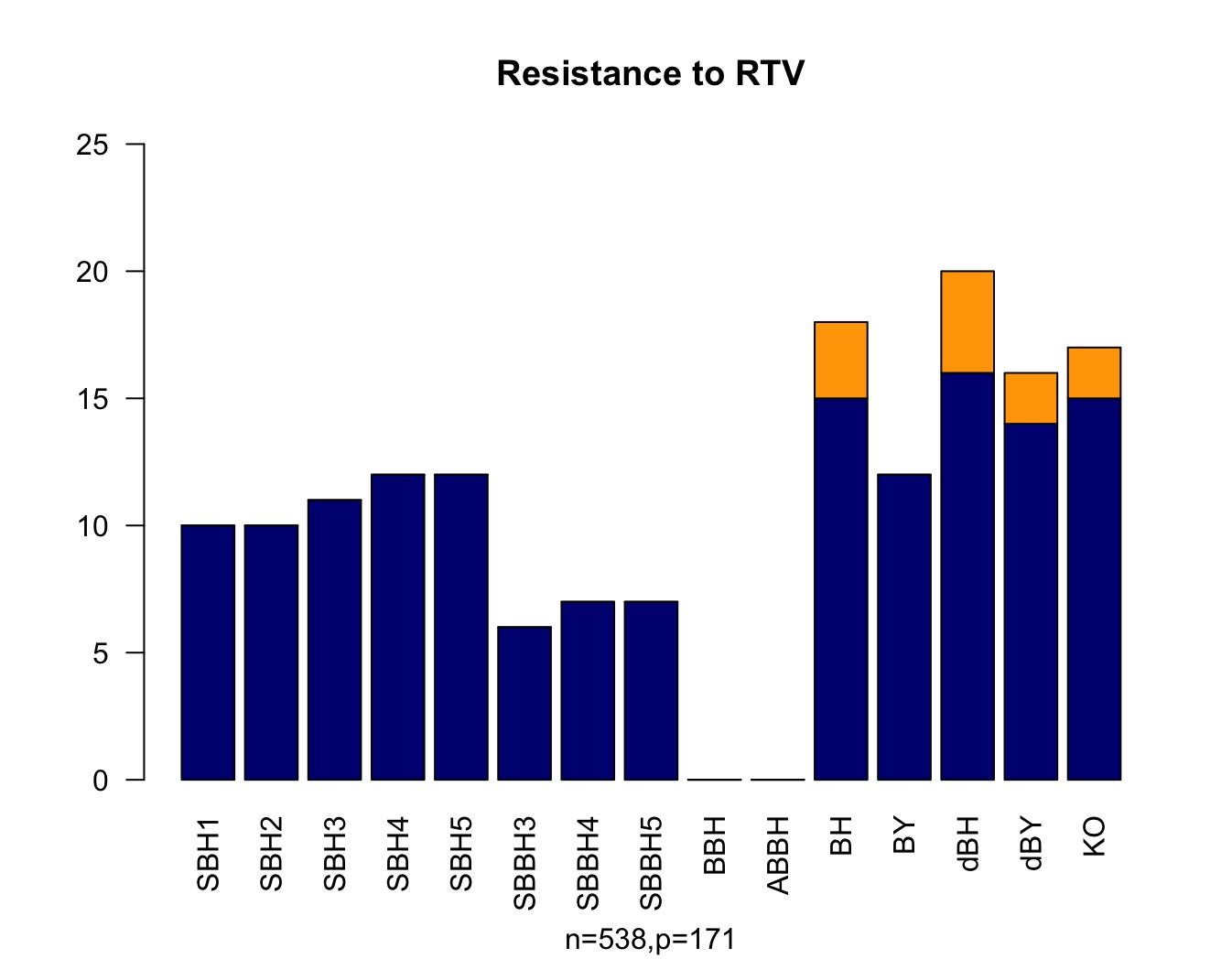} 
      \end{subfigure} \\
  & 
  & \begin{subfigure}[b]{\linewidth}
      \centering
      \includegraphics[width=\linewidth]{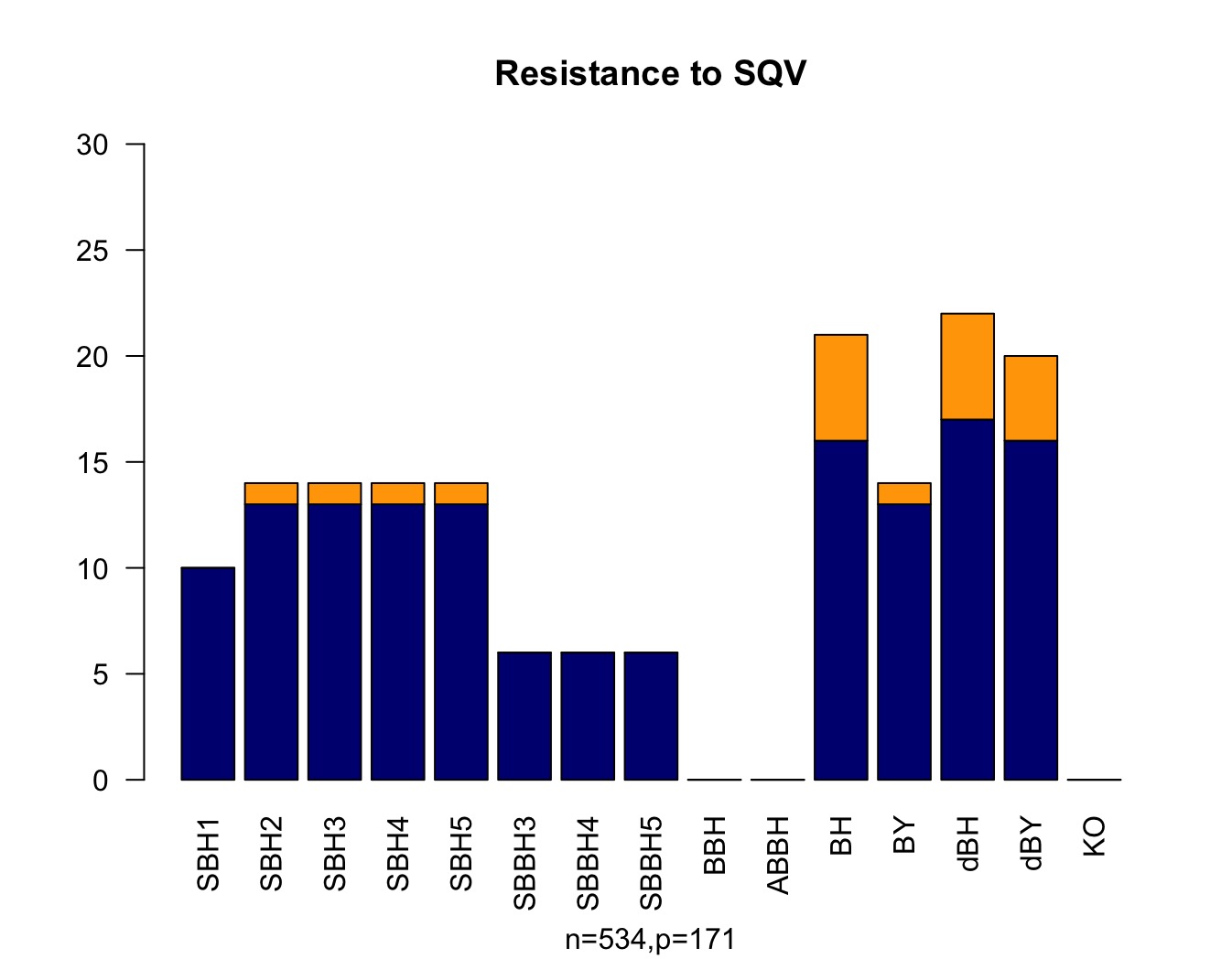} 
      \end{subfigure} 
  &   
  \end{tabular*} 
    \caption{Dark blue indicates protease positions that appear in the treatment-selected mutation (TSM) panel for the PI class of treatments, while orange indicates positions selected by the method that do not appear in the TSM list. FDR was controlled at $\alpha=0.1$}
    \label{real:figure3}
\end{figure}

\begin{figure}
    \begin{tabular*}{\textwidth}{
    @{}m{0.5cm}
    @{}m{\dimexpr0.33\textwidth-0.25cm\relax}
    @{}m{\dimexpr0.33\textwidth-0.25cm\relax}
    @{}m{\dimexpr0.33\textwidth-0.25cm\relax}}
  & \begin{subfigure}[b]{\linewidth}
      \centering
      \includegraphics[width=\linewidth]{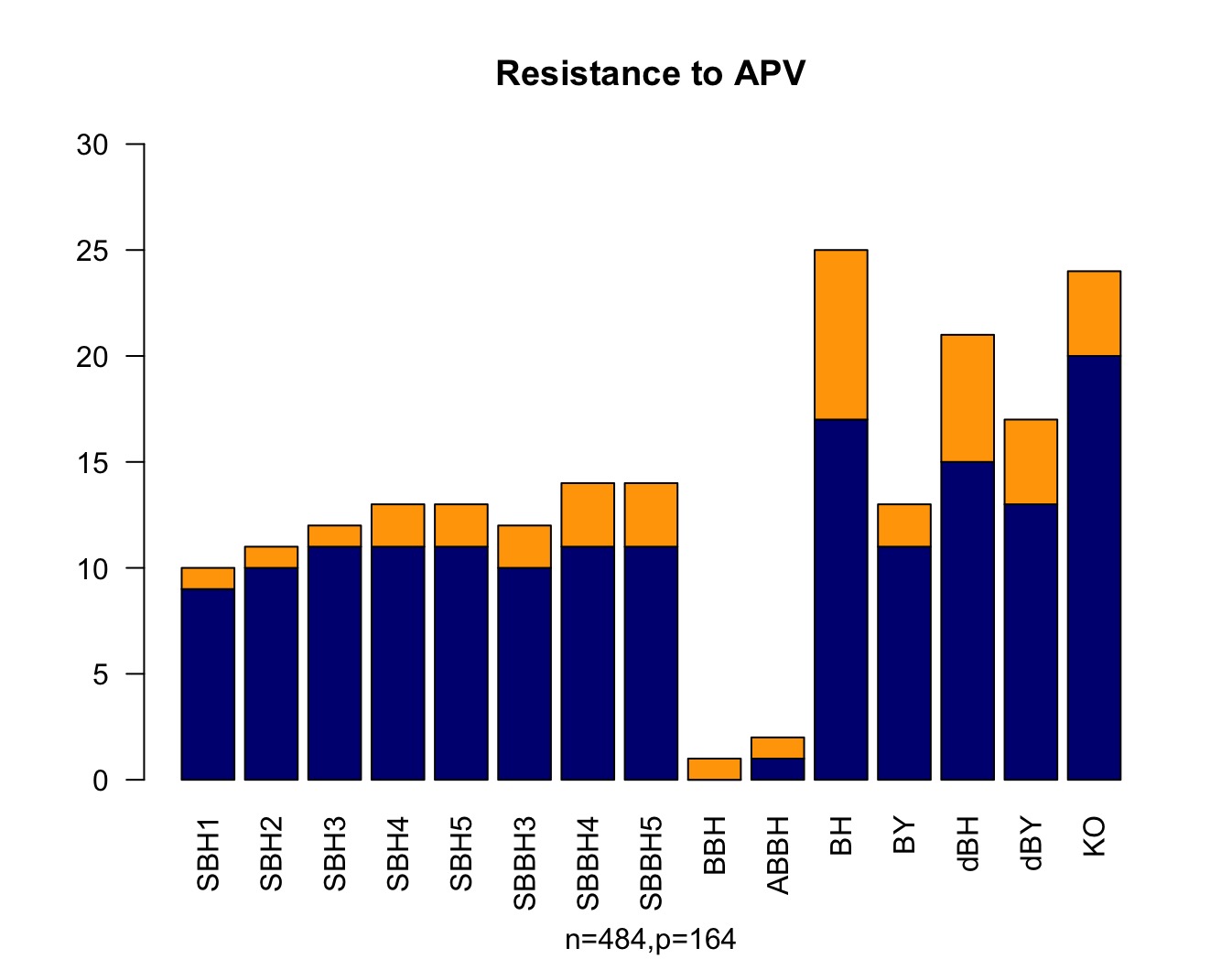} 
      \end{subfigure}  
  & \begin{subfigure}[b]{\linewidth}
      \centering
      \includegraphics[width=\linewidth]{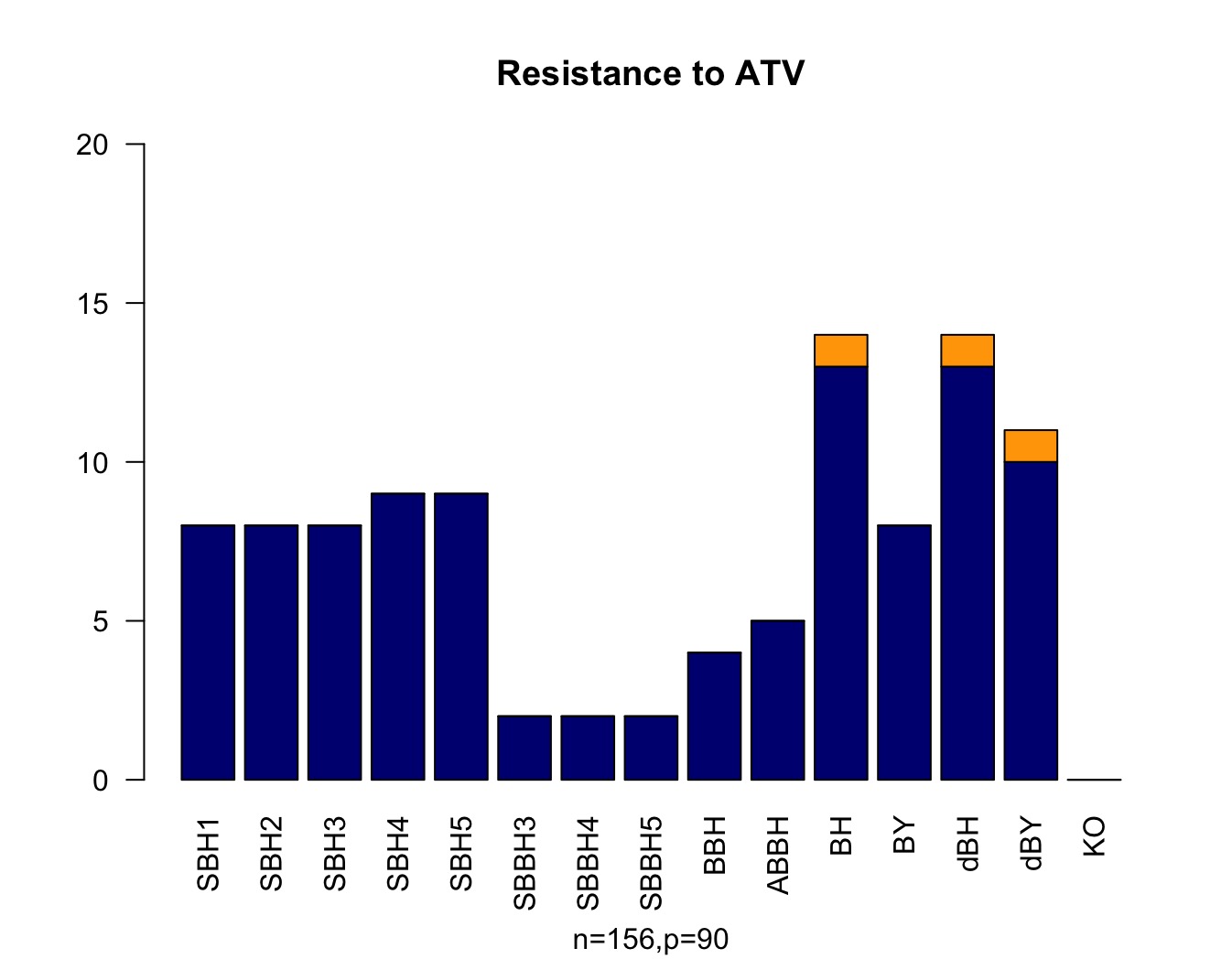} 
      \end{subfigure} 
  & \begin{subfigure}[b]{\linewidth}
      \centering
      \includegraphics[width=\linewidth]{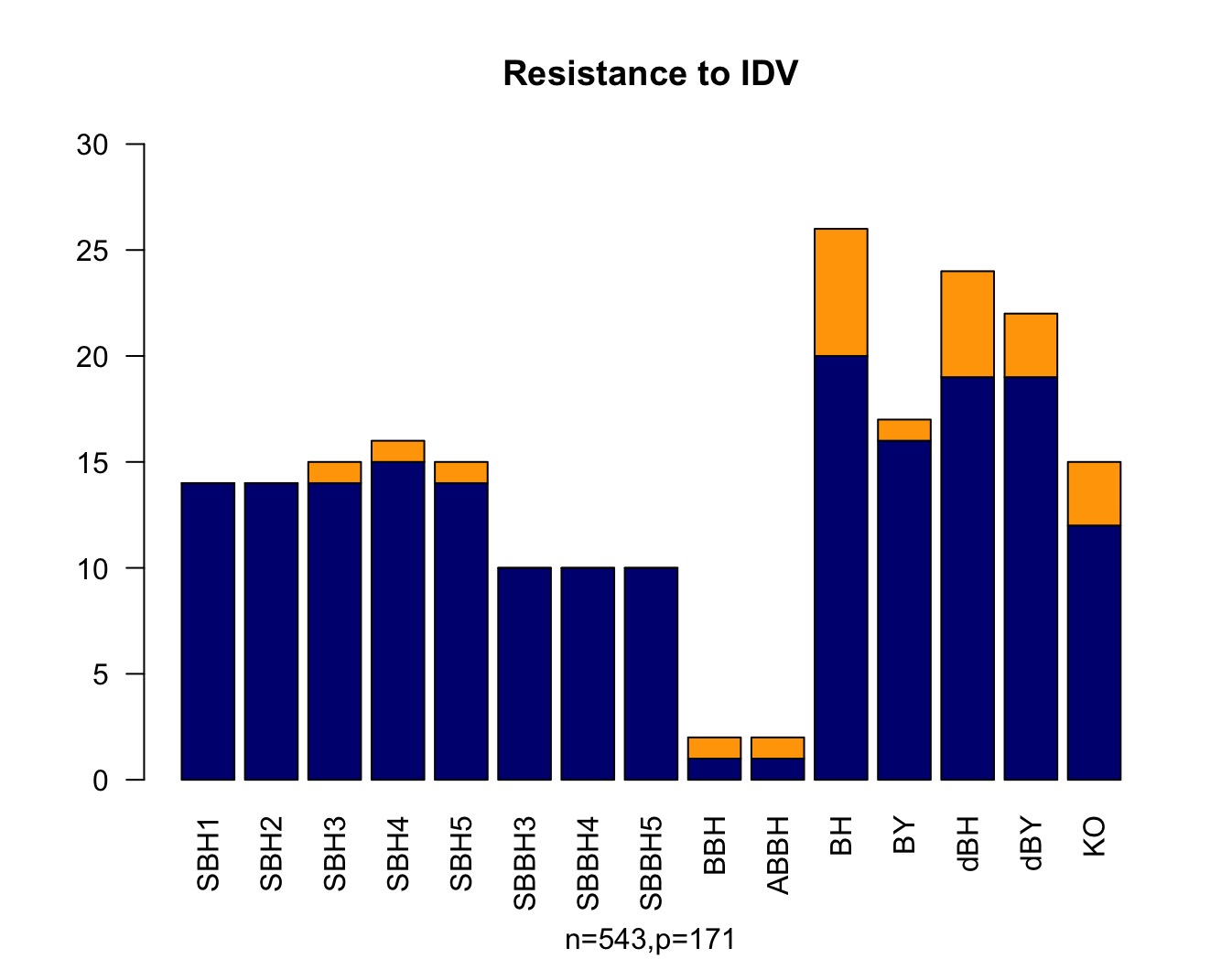} 
      \end{subfigure} \\
  & \begin{subfigure}[b]{\linewidth}
      \centering
      \includegraphics[width=\linewidth]{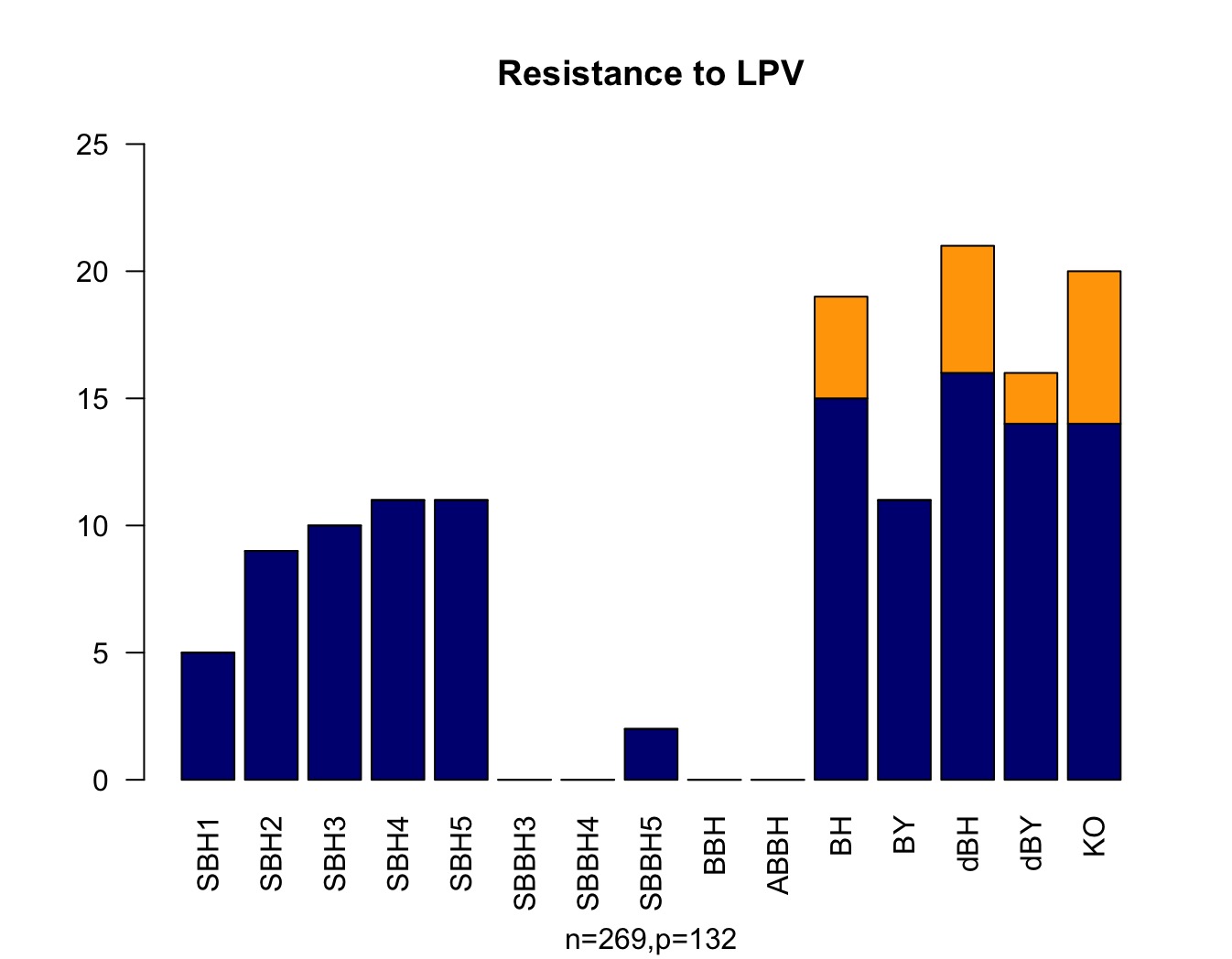} 
      \end{subfigure}  
  & \begin{subfigure}[b]{\linewidth}
      \centering
      \includegraphics[width=\linewidth]{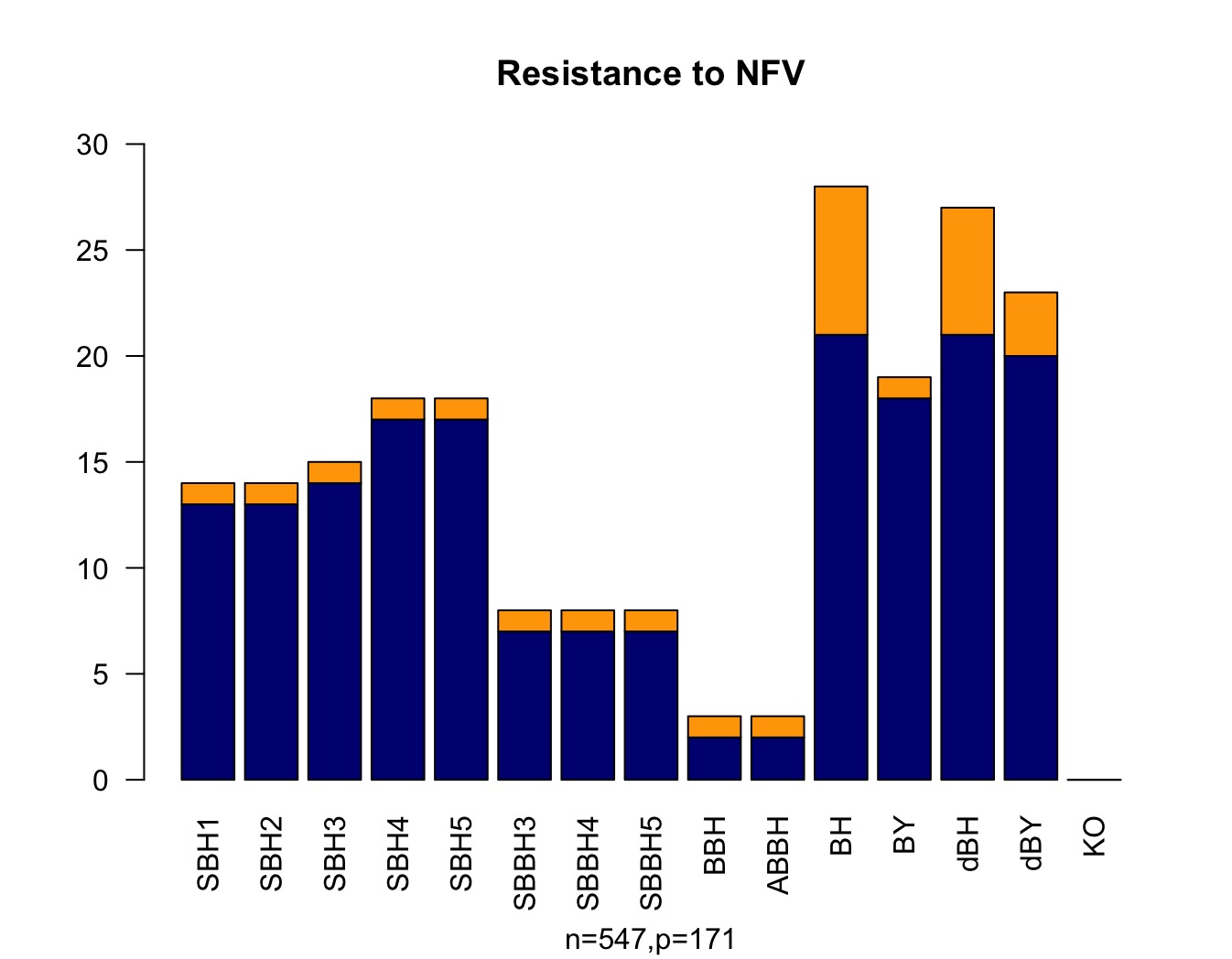} 
      \end{subfigure} 
  & \begin{subfigure}[b]{\linewidth}
      \centering
      \includegraphics[width=\linewidth]{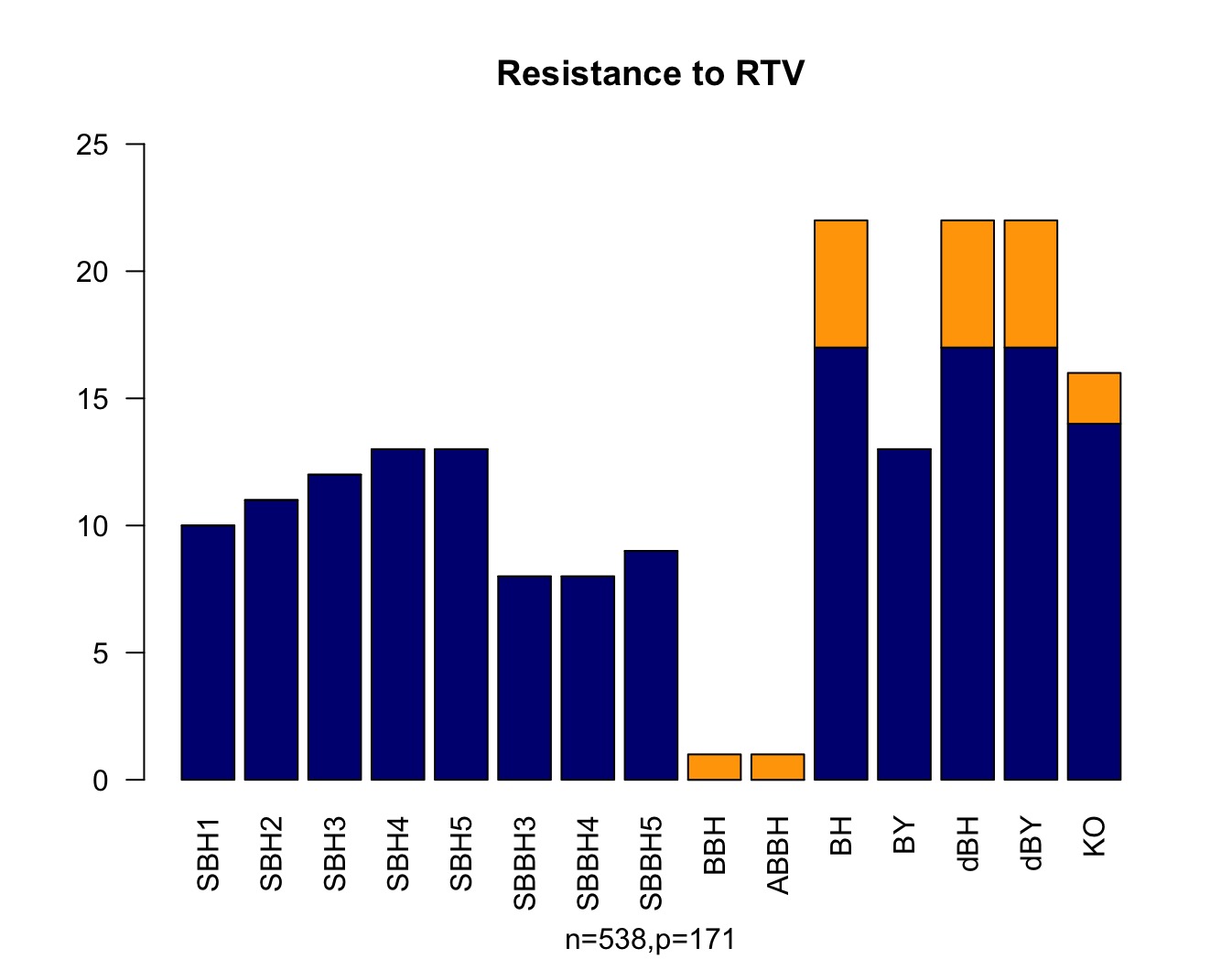} 
      \end{subfigure} \\
  & 
  & \begin{subfigure}[b]{\linewidth}
      \centering
      \includegraphics[width=\linewidth]{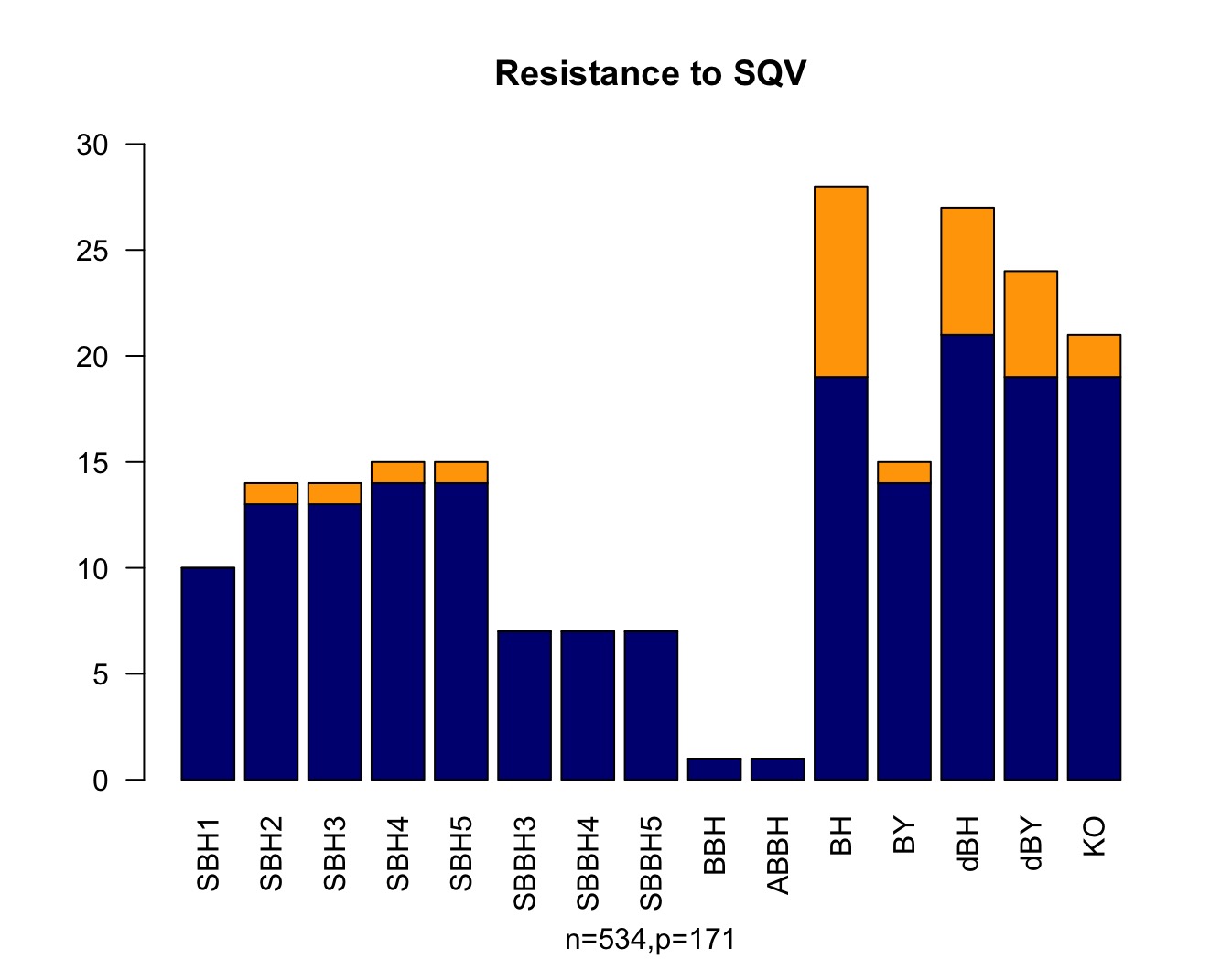} 
      \end{subfigure} 
  &   
  \end{tabular*} 
    \caption{Dark blue indicates protease positions that appear in the treatment-selected mutation (TSM) panel for the PI class of treatments, while orange indicates positions selected by the method that do not appear in the TSM list. FDR was controlled at $\alpha=0.2$}
    \label{real:figure4}
\end{figure}

\end{document}